\documentclass[hyper]{JHEP3}

\input{epsf}
\usepackage{epsfig}
\usepackage{amssymb}
\usepackage{amsfonts}
\usepackage{amsbsy}
\usepackage[all]{xy}
\usepackage{amsmath}
\usepackage{enumerate}

\usepackage{amssymb,amscd}
\usepackage{mathrsfs}
\usepackage{amsmath,amsthm}
\usepackage{xspace}
\usepackage{url}

\def\re{\mbox{Re }}
\def\im{\mbox{Im }}
\def\mod{{\rm mod}}
\def\pbar{{\bar \partial}}

\def\IA{\mathbb{A}}
\def\IC{\mathbb{C}}
\def\IF{\mathbb{F}}
\def\IH{\mathbb{H}}

\def\IP{\mathbb{P}}

\def\IR{{\mathbb{R}}}
\def\R{{\mathbb{R}}}
\def\IS{{\mathbb{S}}}
\def\IV{{\mathbb{V}}}
\def\IZ{{\mathbb{Z}}}

\def\CC {{\cal C}}

\def\CM {{\cal M}}

\def\CN{{\cal N}}
\def\CK {{\cal K}}
\def\CN {{\cal N}}
\def\CR {{\cal R}}
\def\CD {{\cal D}}
\def\CF {{\cal F}}
\def\CJ {{\cal J}}
\def\CP {{\cal P }}
\def\CL {{\cal L}}
\def\CV {{\cal V}}
\def\CW {{\cal W}}
\def\CX {{\cal X}}
\def\cX {{\cal X}}
\def\CY {{\cal Y}}

\def\cY {{\cal Y}}
\def\CO {{\cal O}}
\def\CZ {{\cal Z}}
\def\CE {{\cal E}}
\def\CG {{\cal G}}
\def\CH {{\cal H}}
\def\CB {{\cal B}}
\def\CS {{\cal S}}
\def\cS {{\cal S}}
\def\CA{{\cal A}}
\def\CK{{\cal K}}
\def\cK{{\cal K}}

\def\CU{{\cal U}}
\def\CZ{{\cal Z}}
\def\CT{{\cal T}}

\def\half{\frac{1}{2}}

\renewcommand{\Im}{{\rm Im }}
\renewcommand{\Re}{{\rm Re }}

\def\one{{\hbox{ 1\kern-.8mm l}}}

\def\vol{{\rm vol\,}}
\def\p{\partial}

\def\be{\bar{e}}

\def\half{\frac{1}{2}}

\def\hk{hyperk\"ahler\xspace}

\def\kahler{K\"ahler\xspace}
\newcommand{\abs}[1]{\lvert#1\rvert}

\newcommand{\ti}[1]{\textit{#1}}

\def\lieg{\mathfrak{g}}

\def\p{\partial}

\def\re{\mbox{Re }}
\def\im{\mbox{Im }}
\def\mod{{\rm mod}}

\def\IC{\mathbb{C}}

\def\IZ{{\mathbb{Z}}}
\def\IR{{\mathbb{R}}}
\def\IP{\mathbb{P}}

\def\CB {{\cal B}}
\def\CC {{\cal C}}

\def\CM {{\cal M}}

\def\CN{{\cal N}}
\def\CK {{\cal K}}
\def\CN {{\cal N}}
\def\CR {{\cal R}}
\def\CD {{\cal D}}
\def\CF {{\cal F}}
\def\CJ {{\cal J}}
\def\CP {{\cal P }}
\def\CL {{\cal L}}
\def\CV {{\cal V}}
\def\CW {{\cal W}}
\def\CX {{\cal X}}
\def\CY {{\cal Y}}
\def\CO {{\cal O}}
\def\CZ {{\cal Z}}
\def\CE {{\cal E}}
\def\CG {{\cal G}}
\def\CH {{\cal H}}
\def\CB {{\cal B}}
\def\CS {{\cal S}}
\def\CA{{\cal A}}
\def\CK{{\cal K}}

\def\CU{{\cal U}}
\def\CZ{{\cal Z}}
\def\CT{{\cal T}}

\def\half{\frac{1}{2}}

\def\be{\begin{equation}
}

\def\ee{\end{equation}}

\newcommand\I{{\mathrm i}}
\newcommand{\inprod}[1]{\langle#1\rangle}
\newcommand\N{{\cal N}}
\renewcommand\sf{{\mathrm{sf}}}

\newcommand\elec{{e}}

\newcommand\fro{{\overline{\underline{\Omega}}}}

\newcommand\WKB{{\mathrm{WKB}}}

\newcommand\de{{\mathrm{d}}}

\def\an{{\rm analytic} }
\newcommand\gh{{\mathrm{gh}}}
\newcommand\eff{{\mathrm{eff}}}

\newcommand\BPS{{\mathrm{BPS}}}

\newcommand\there{{\mathrm{there}}}

\newcommand\twod{\mathrm{2d}}

\newcommand{\hkq}{/\!/\!/}

\newcommand{\bS}{\mathbb S}

\DeclareMathOperator{\Aut}{{Aut}}

\DeclareMathOperator{\Pexp}{{Pexp}}

\DeclareMathOperator{\Tr}{{Tr}}
\DeclareMathOperator{\Hom}{{Hom}}

\newcommand{\insfig}[2]{\begin{figure}[htbp] \centering \includegraphics[scale=0.3]{figures/#1-crop.pdf} \caption{#2} \label{fig:#1} \end{figure}}
\newcommand{\insfigpng}[2]{\begin{figure}[htbp] \centering \includegraphics[scale=0.3]{figures/#1-crop.png} \caption{#2} \label{fig:#1} \end{figure}}

\title{Wall-Crossing in Coupled 2d-4d Systems}

\author{Davide Gaiotto$^1$, Gregory W. Moore$^2$, Andrew Neitzke$^3$\\
$^1$ School of Natural Sciences, Institute for Advanced Study,\\
Princeton, NJ 08540, USA\\
$^2$ NHETC and Department of Physics and Astronomy, Rutgers University,\\
Piscataway, NJ 08855--0849, USA\\
$^3$ Department of Mathematics, University of Texas at Austin,\\
Austin, TX 78712, USA\\
\\
{\tt dgaiotto@ias.edu, gmoore@physics.rutgers.edu, neitzke@math.utexas.edu} }

\abstract{We introduce a new wall-crossing formula which combines and generalizes
the Cecotti-Vafa and Kontsevich-Soibelman formulas for supersymmetric 2d and 4d systems respectively.
This 2d-4d wall-crossing formula governs
the wall-crossing of BPS states in an $\CN=2$ supersymmetric 4d
gauge theory coupled to a supersymmetric surface defect.  When the theory and defect are compactified on
a circle, we get a 3d theory with a supersymmetric line operator, corresponding to a hyperholomorphic connection
on a vector bundle over a \hk space.
The 2d-4d wall-crossing formula can be interpreted as a smoothness condition
for this hyperholomorphic connection.
We explain how the 2d-4d BPS spectrum can be
determined for 4d theories of class $\CS$, that is, for those theories
 obtained by compactifying the six-dimensional $(0,2)$ theory with a partial topological
 twist on a punctured Riemann surface $C$. For such theories there are canonical
 surface defects.    We illustrate
with several examples in the case of $A_1$ theories of class $\CS$.
Finally, we indicate how our results can be used to
produce solutions to the $A_1$ Hitchin equations on the Riemann surface $C$.}

\begin{document}

\bibliographystyle{utphys}

\section{Introduction: A guided tour of the paper}\label{sec:Intro}

\def\tf{2d-4d systems}
\def\PI{\cite{Gaiotto:2008cd}}
\def\PII{\cite{Gaiotto:2009hg}}
\def\PIII{\cite{Gaiotto:2010be}}

The present paper is a continuation of a project revolving around the
intertwined themes of wall-crossing, \hk geometry, BPS states,  and their
application to the rich mathematical physics associated with four-dimensional
field theories with $\CN=2$ supersymmetry \cite{Gaiotto:2008cd,Gaiotto:2009hg,Gaiotto:2010be}.
In this paper we introduce  several new ingredients into the emerging structure. All of these
new ingredients are associated in one way or another with \emph{surface defects}.

Several developments over the past decade  have amply demonstrated that
the inclusion of defects of decreasing codimension leads to physical
systems which possess increasing richness and beauty,   albeit at the
price of increasing complexity. For examples, in
2-dimensional conformal field theory the inclusion of defect lines
enhances the structure wonderfully. In four dimensional $\CN=4$ supersymmetric
field theory the line defects shed much light on S-duality \cite{Kapustin:2005py}
and play a central role in the geometric Langlands program \cite{Kapustin:2006pk}.
Deeper developments along this direction have come with the inclusion of
surface defects \cite{Gukov:2006jk,Witten:2007td,Gukov:2008sn}. Finally, in topological field theory,
a whole hierarchy of defects leads to a beautiful mathematical structure
involving higher category theory \cite{Freed:1993wb,Baez:1995xq,MR2555928,Freed:2009qp,Kapustin:2010ta,Kapustin:2010if}.

The present paper together with \cite{Kapustin:2007wm,Kapustin:2006hi,Gaiotto:2009fs} may be viewed as
part of an analogous line of development for theories with $\CN=2$ supersymmetry.\footnote{This
of course suggests that we should go on to investigate domain
walls in $\CN=2$ theories, but we will not address that interesting topic in the present work.}
For example, in \PIII\  we considered  defects of codimension three --- that is,
\emph{line defects} in $\CN=2$ theories.
This led naturally to a new class of BPS states, the framed BPS states.
The consistency of framed BPS indices gave a very simple and natural
derivation of the ``motivic'' Kontsevich-Soibelman wall-crossing formula
for the ``vanilla'' BPS states. (As in \cite{Gaiotto:2010be},
we refer to the BPS states of the original four-dimensional theory without
defects as ``vanilla.'')  In addition
it led to a natural quantization of the algebra of holomorphic functions on
Seiberg-Witten moduli spaces as well as to new connections with the mathematics
of laminations of punctured Riemann surfaces, quantized Teichm\"uller theory, cluster algebras,
and cluster varieties. In the present paper, as we have said, we move on
to investigate some aspects of codimension two defects, also known as
surface defects. Not surprisingly, this introduces a new layer of complexity.
The remainder of this introduction is an expository account of the
paper.

The surface defects we will consider preserve four of the eight supersymmetries
of the vanilla 4-dimensional theory (\S  \ref{subsec:kinematics} and Appendix
\ref{app:d2d4-multiplets}). Various UV constructions of such defects are
reviewed in \S \ref{subsec:examples-surface-defects}. One representative
construction proceeds by coupling a 1+1 dimensional theory $\CT_{2d}$ with (2,2) supersymmetry
on a surface $\bS$ to an ambient four-dimensional theory $\CT_{4d}$.  We assume that
$\CT_{2d}$ has a compact global symmetry group $G$ and moreover
we assume that in isolation $\CT_{2d}$ has a finite number of massive vacua, labeled $i \in \CV[\bS]$.\footnote{The symbol $\bS$ denotes the location of the surface defect, but
also sometimes the ``theory'' on the defect. It has no meaning without specifying the
ambient four-dimensional theory.}
The $G$-symmetry can then be gauged by
coupling to a $G$-gauge theory in $\CT_{4d}$. A second UV construction, due to Gukov and
Witten, involves just a 4d theory with $G$ gauge symmetry but with reduced structure group along the
surface $\bS$ \cite{Gukov:2006jk}.
In addition we can consider domain
walls welding together different theories $\CT_{2d}$ and $\CT_{2d}'$, or more generally
different defects on $\bS$ and $\bS'$. The domain wall can be viewed as
a line defect in four dimensions embedded in a surface. We will refer to
it as a supersymmetric interface.
We refer to the general systems we have just described as \emph{coupled 2d-4d systems} or
\tf\ for short.

\tf\ can support novel and nontrivial dynamics, qualitatively different from the
dynamics exhibited by, say,   $\CT_{2d}$ and $\CT_{4d}$ in isolation. They have a rich
set of BPS states. Besides the 4d vanilla BPS states,
there are BPS states of the type associated to the 2d theory $\CT_{2d}$.
Such BPS states were studied extensively by Cecotti and Vafa and collaborators in the
two-dimensional context \cite{Cecotti:1992qh,Cecotti:1992vy,Cecotti:1992rm}. The coupling to a 4d theory
$\CT_{4d}$ has an effect akin to the introduction of 2d twisted masses
for the $G$-symmetry \cite{Dorey:1998yh,Cecottiump,Tong:2008qd}. (``Twisted masses'' refers
to the parameters introduced in 2d  $(2,2)$ models in  \cite{Hanany:1997vm}.)
For example, there can be
infinitely many solitons interpolating between distinct vacua of the surface defect and carrying
four-dimensional gauge and ``flavor charges.''\footnote{In this
paper flavor charges play an important role. We define a
``flavor symmetry'' of a 4d $\CN=2$ theory to be a global symmetry commuting with the
supersymmetries which is not spontaneously broken on the Coulomb branch of vacua. States and
operators of the theory will transform in representations of the flavor symmetry group and
``flavor charges'' are characters of this group. We will in general limit our considerations
to flavor symmetry groups which are abelian.}
There can also be BPS states localized
on $\bS$ which do not interpolate between different vacua but rather reside in a single vacuum.
These states too can carry four-dimensional gauge and flavor charge.
Finally, in the presence of interfaces there are again framed BPS states.

In order to exhibit the curious antics of these new BPS states we must turn to
a low energy effective field theory description of the dynamics.
There is a description of the low energy dynamics of \tf \ generalizing the renowned
solution of pure four-dimensional theories initiated
by Seiberg and Witten \cite{Seiberg:1994rs,Seiberg:1994aj} (For reviews
see, for example, \cite{Lerche:1996xu,AlvarezGaume:1996mv}.)
This is described in some detail in \S \ref{sec:Physical-Interpret}, following and
developing further the story in \cite{Gaiotto:2009fs}.
The four-dimensional theory has a Coulomb branch of vacua $\CB$.\footnote{We do not venture onto
Higgs branches in this paper.} In \tf\ this is generalized to a ramified cover $\CB_{\bS}$ whose
sheets are in correspondence with vacua $\CV[\bS]$.
The dynamics at a vacuum $(u,i) \in \CB_{\bS}$ is
described by an effective four-dimensional abelian $\CN=2$ gauge theory coupled to a twisted
chiral effective 2d $(2,2)$ superpotential $\CW$, localized on $\bS$. (See equation \eqref{eq:4d-action}.)
Thus, the description of the low-energy dynamics of four-dimensional theories
by a prepotential $\CF$ is replaced in 2d-4d systems
by a description in terms of a pair $(\CF,\CW)$.  There is a simple IR picture of the
ground states of the surface defect in a definite charge sector:
the defect looks like a solenoid, outside of which there is a flat abelian gauge field.  If we describe the
4d IR abelian gauge theory by a self-dual gauge field
$\IF = \de \IA$, then the holonomy of $\IA$ around the solenoid is specified by
a symplectic vector $\nu$ (equation \eqref{eq:solenoid-gf}).
Upon choosing a duality frame, $\nu$  has electric and magnetic components
$\eta$ and $\alpha$, related to $\CW$ by equation \eqref{eq:IR-GW}:
\begin{equation}\label{eq:IR-GW-intro}
t = \eta + \tau \alpha = \frac{\p \CW}{\p a}.
\end{equation}
We refer to $(\eta,\alpha)$ as
(infrared) Gukov-Witten parameters.

When two defects are welded together by an interface
the analogous physical picture of the ground state in a definite charge sector
 is that of a pair of half-solenoids glued together on a BPS dyon,
rather like a nunchuk, or a boa which has swallowed an elephant
\cite{petitprince}. See
Figure \ref{fig:threaded-flux}.
The dyon allows flux to escape from the solenoids, and hence the field outside is no longer
flat. Moreover, the dyon charge need not satisfy Dirac quantization. Indeed we can
regard the worldsheets of the solenoids as those of Dirac strings which have materialized
into observable objects.

If we wish to describe our ground states with more precision we must be careful
about the description of the 4d charges of our BPS states. It turns out that this is no
simple matter. We begin with
the local system over $\CB$  of electromagnetic and flavor charges  $\Gamma$, and
its quotient system $\Gamma_g$ of electromagnetic charges. The latter is a
local system of symplectic lattices.
(See equation \eqref{eq:lattice-ext-seq}.) Associated to this is the Seiberg-Witten
moduli space
\begin{equation}\label{eq:M-intro}
\CM:= \Gamma_g^*\otimes \IR/(2\pi\IZ),
\end{equation}
 one of the central actors in \cite{Gaiotto:2008cd,Gaiotto:2009hg,Gaiotto:2010be}.
  Geometrically, the   holonomy $\tilde\nu$ of $\IA$ should be
viewed as a ``section'' of the ``mirror-dual'' moduli space
$\widetilde{\CM} = \Gamma_g \otimes \IR/(2\pi\IZ)$, but it will be important
to consider a lift $\nu \in \Gamma_g \otimes \IR$, thus assigning a definite
flux to the interior of the solenoid.  Of course, since $\Gamma_g$
is a nontrivial local system with monodromy there will be no globally well-defined
section $\nu$, only a multisection.
Generically the possible values of $\nu$ above $(u,i)$
will be a ``torsor'' for $\Gamma_u$
\footnote{A \emph{torsor} for a group $G$ is a principal homogeneous space ${\mathbf S}$ for $G$.
This means that there is a transitive $G$-action on ${\mathbf S}$ and moreover there are
no nontrivial stabilizers. That is, for any two elements $s_1, s_2\in  {\mathbf S}$
there is a $g\in G$ so that $s_2= g\cdot s_1$, and moreover, if $s=g\cdot s$ for any $s$
then $g=1$.  Thus, ${\mathbf S}$ is ``a copy of $G$,'' but there is no
distinguished element of ${\mathbf S}$ corresponding to the identity element of $G$.
A typical example of a $G$-torsor is the fiber of a principal $G$-bundle.}
and monodromies will act in an affine-linear way on $\nu$. It follows then from
\eqref{eq:IR-GW-intro} that the superpotential must have monodromy, and that
the set of possible effective superpotentials is a torsor for $\Gamma$. (This assumes 
that the ``mass parameters'' dual to the flavor charges are generic.) Physically this monodromy
can be traced back to the fact that at singular loci $\CB^{\rm sing}$ not only will
four-dimensional BPS states become massless, but also chiral multiplets on $\bS$ will
become massless. In \S \ref{subsec:BPS-Deg-WC}, Witten's famous computation
of effective twisted superpotentials in gauged linear sigma models
\cite{Witten:1993yc,Morrison:1994fr} is reinterpreted as computing
an extension of the central charge function $Z: \Gamma \to \IC$ to the torsors
associated to $\bS$.  A surprising physical implication
is that the number of chiral multiplets on the surface defect is not an absolute invariant
but rather more like a gauge choice in the description of the 2d-4d system. This is intimately related
to the fact that a 2d chiral multiplet can mix with 4d hypermultiplets with the same gauge and flavor charges.
We will denote the torsor associated to a vacuum $i$ as $\Gamma_i$.
The $\Gamma_i$ form a local system of torsors over $\CB_{\bS}$.\footnote{Actually,
truth be told, we will face situations where the torsors $\Gamma_i$ of effective superpotentials suffer
global twisting analogous to the notion of ``twisted vector bundles.'' We will return
to this subtlety later in this overview. Mercifully, the $\Gamma_{ij}$ introduced in the next paragraph,
corresponding to differences of
superpotentials, will always be honest torsors.}

If we consider 2d solitons between vacua $i$ and $j$, charge quantization is controlled by the difference in the
holonomies $\tilde \nu_i$ and $\tilde \nu_j$ associated to the two vacua, and hence the 4d charges of 2d solitons form a
$\Gamma$-torsor $\Gamma_{ij}$. With interfaces these are further generalized to $\Gamma_{ij'}$, where $i$, $j'$ are vacua of the two surface defects welded by the interface.
Charges in $\Gamma_{ij}$ will be denoted by symbols like $\gamma_{ij}$, and
satisfy addition rules like $\gamma_{ij} + \gamma_{jk}= \gamma_{ik}$. Mathematically this means we
define a \emph{groupoid of vacua} $\IV$ which will prove to be a useful concept when we
discuss wall-crossing. See \S\ref{subsec:Formal-Statement} for the formal construction.

The charge torsors described above grade the Hilbert spaces of 2d solitons as well as other
BPS states, framed and unframed. There are BPS bounds in all these charge sectors and hence
associated to our new BPS states are a host of new BPS indices. These include:

\begin{itemize}

\item $\mu(\gamma_{ij})$: These are the degeneracies of 2d solitons with 4d gauge charge $\gamma_{ij}$.
They generalize the degeneracies $\mu_{ij}$ studied by Cecotti and Vafa in the pure 2d context
\cite{Cecotti:1992rm}. See \S \ref{subsec:BPS-Deg-WC} and Appendix \ref{app:d2d4-multiplets}.

\item $\omega(\gamma,\gamma_i)$: These are the degeneracies of 2d BPS states in vacuum $i$ on
$\bS$ with 4d gauge charge $\gamma$. Their definition is extremely subtle, for reasons explained in
\S \ref{subsec:BPS-Deg-WC}, and these subtleties show that the degeneracy also depends on the
choice of $\gamma_i$, that is, the ``gauge choice'' of superpotential on $\bS$, or equivalently,
of the flux within the solenoid. The dependence on $\gamma_i$ is
affine-linear:
\begin{equation}
\omega(\gamma, \gamma_i + \gamma') = \omega(\gamma, \gamma_i  ) + \Omega(\gamma)\langle \gamma, \gamma'\rangle.
\end{equation}

\end{itemize}

In addition to $\mu$ and $\omega$ there are also generalizations of framed indices.
Moreover, all these have further generalizations which describe the response of BPS states to rotation around
the surface defect. This leads to a
``spin 2d-4d wall-crossing formula,'' analogous to the ``motivic''
wall-crossing formula for vanilla BPS states.  All of this is described in
\S \ref{sec:lineops}.

All of our new BPS states and their indices undergo wall-crossing phenomena analogous
to those already known for pure 2d and 4d theories.
In \S \ref{subsec:Formal-Statement} we state a precise 2d-4d wall-crossing
formula for \tf, whose existence was first suggested  in \cite{Gaiotto:2009fs}.
Our formula combines and generalizes the previous wall-crossing
formulae of Cecotti and Vafa \cite{Cecotti:1993rm} and Kontsevich and Soibelman \cite{ks1}.
The formula is a kind of ``matrix generalization'' of the Kontsevich-Soibelman formula.
The degeneracies $\mu$ are associated with certain finite-dimensional
non-diagonal matrices, called $\CS$-factors $\CS^\mu$ (equation \eqref{eq:cv-like}),
while the degeneracies
$\omega$ are associated with diagonal matrices of symplectomorphisms leading
to $\CK$-factors $\CK^\omega$ (equation \eqref{eq:ks-like}). Technically, these
are automorphisms of a noncommutative algebra associated to the vacuum groupoid.
They belong to a semidirect product of a group of matrix gauge transformations
and a group of Poisson morphisms, acting on matrix-valued functions on an algebraic Poisson torus $\Gamma^*\otimes \IC^\times$.
After stating our wall-crossing formula in \S \ref{subsec:Formal-Statement}, we
systematically analyze
its consequences in \S \ref{subsec:Examples-2d-4d-wcf}, and find that the
2d and 4d degeneracies influence each other's wall-crossing in intricate ways.
(When we come to apply these identities to some physical examples
we find that they lead to extremely rich spectra of BPS states
even in basic examples like those of \S \ref{subsec:WC-exple-su2}.)

The 2d-4d wall-crossing-formula is a mathematically natural synthesis of the known
formulae for 2d and 4d theories, but it remains to give it precise
physical meaning and justification. The physical meaning of the relevant
quantities that enter the formula has already been sketched above and is discussed in detail in
\S \ref{sec:Physical-Interpret}. The justification for the wall-crossing formula can, as with the
pure 4d case, be given in two ways:

\begin{itemize}

\item  First, one can study interfaces and
their framed BPS states. As in \PIII, invoking
the ``halo technique'' of \cite{Denef:2007vg,Gaiotto:2010be,Andriyash:2010qv},   the
consistency of the wall-crossing of the framed BPS indices implies the
wall-crossing formula for the unframed indices $\mu$ and $\omega$.  This story
can be found in \S \ref{sec:lineops}, which also draws on material in
Appendices \ref{app:LG-Radius} and \ref{app:Landau-Levels} for developing
the halo picture. Moreover, \S \ref{sec:lineops} begins the development of a
spin version of the 2d-4d WCF, although some details of that generalization
remain to be worked out.

\item Second, one can compactify the theory on a circle and study the
geometry implied by constraints of supersymmetry.

\end{itemize}

We now turn to a detailed discussion of the second justification of the 2d-4d WCF.

This approach was developed for purely 4-dimensional theories
in \PI . Let us briefly recall the reasoning here.
 The argument begins with the observation of
Seiberg and Witten \cite{Seiberg:1996nz}  that compactification of $\CT_{4d}$
on $\IR^3 \times S^1$ produces an effective low energy sigma model whose
target space is the space $\CM$ of \eqref{eq:M-intro}. Supersymmetry implies that $\CM$ carries a \hk\ metric.
If the circle has radius $R$, then when $R$ is
large the \hk\ metric on $\CM$ can be computed to exponentially good accuracy
to be a semiflat metric. (See \S 2.4 of \PI\  or \eqref{eq:sf-metric} below.)
The existence of BPS states in the 4-dimensional theory leads, at finite
but large values of $R$, to exponentially small corrections to this semiflat
metric. The continuity of these corrections across walls of marginal stability
in $\CB$ implies the Kontsevich-Soibelman WCF.  A key technical step in the argument is the observation that
there is a distinguished set of coordinates
$\CY_{\gamma}$ on $\CM$, obeying the (twisted) algebra of functions on the algebraic torus $\Gamma^*\otimes \IC^\times$,
in terms of which the holomorphic symplectic form on $\CM$ can be simply written:
$\{ \CY_\gamma, \CY_{\gamma'} \} \sim \langle \gamma, \gamma'\rangle \CY_{\gamma} \CY_{\gamma'}$.
In particular, upon choosing a basis for $\Gamma$ we obtain a system of Darboux coordinates on $\CM$.
It turns out that formulating the
quantum corrections of BPS states to the semiflat geometry can be formulated as a
Riemann-Hilbert problem whose solution is an integral equation for the $\CY_{\gamma}$.
(See equation \eqref{eq:int-old}.)  This equation is formally identical to the
Thermodynamic Bethe Ansatz equation of Zamolodchikov and, although in general no
known integrable system is associated to it, we will refer to this  equation and its
generalizations below as a TBA equation.

The line of argument we have just reviewed can also be generalized
to \tf, and this we proceed to do in \S\S \ref{sec:compactification} and
\ref{sec:LocalModel}, which follow in outline much of the development of
\PI . We compactify \tf\ by putting $\CT_{4d}$ on
$\IR^3\times S^1$ as before, but now we wrap one of the dimensions of $\bS$
on the circle to produce a 3d sigma model coupled to a 1d line defect.
The quantum mechanical system on the line defect has a vector space of
ground states which define a vector bundle $V_{\bS}$ over $\CM$.
Constraints of supersymmetry demand that this vector bundle carry a ``Berry connection''
which is hyperholomorphic. The latter epithet means that the curvature is of
type $(1,1)$ for all of the complex structures of the \hk\ manifold $\CM$.
This is a generalization of the notion of an instanton on a four-dimensional
\hk\ manifold, which has been studied in the
mathematics literature; see, for example,
\cite{MR1139657,Manton:1993aa,MR1486984,MR1995789,MR1815021,MR1919716}.
Once again, the geometry simplifies in the
semiflat limit $R\to \infty$. In this case each of the vacua $i\in \CV[\IS]$
defines a single line bundle $V_i \to \CM$ and the Berry connection on
$V = \oplus_i V_i$ becomes a diagonal semiflat connection (equation \ref{eq:a-sf}).
The semiflat connection has a natural geometrical interpretation in terms of
the relative Poincar\'e connection on the fiber product of $\CM \times_{\CB} \widetilde{\CM}$
over $\CB$. At finite $R$ the semiflat connection receives
quantum corrections from all our BPS states:
the vanilla BPS states of $\CT_{4d}$ as well as the new ones with degeneracies
$\omega$ and $\mu$. For example, the corrections associated with $\mu$ come from
the worldlines of solitons wrapping the circle in the cylindrical surface $\bS$.
Once again, analogs of ``Darboux coordinates'' $\CY_{\gamma_{i}}$ and
$\CY_{\gamma_{ij'}}$ can be introduced. As in \PIII\  they can
be interpreted physically in terms of expectation values of interfaces.
They can also be interpreted geometrically as  sections of $V$ and ${\rm Hom}(V,V')$, respectively,
which are holomorphic in all complex structures.  Finally, they satisfy
a Riemann-Hilbert problem analogous to that in \PI. As in that case we can
solve the RH problem through
a system of integral equations, \eqref{eq:int-old}-\eqref{eq:int-2},
allowing us to construct  $\CY_{\gamma_{i}}$ and $\CY_{\gamma_{ij'}}$ as an expansion
around the explicitly known semiflat sections \eqref{eq:xi-sf-vev}.  With these in
hand we can go on to construct the hyperholomorphic connection,
as explained in Appendix \ref{app:Twistor-HH}. As in the case of the \hk\ metric
on $\CM$, the 2d-4d WCF follows as a consistency condition on the system
of equations, \eqref{eq:int-old}-\eqref{eq:int-2},
  ensuring smoothness of the metric and connection.

In \PI\ an important role in the physical argument was played by a local analysis
of the quantum corrections coming from mutually local light BPS states. That analysis is
repeated and refined in the case of \tf\ in \S \ref{sec:LocalModel}.  A
classification of the possible singularities due to mutually
local massless 2d and 4d particles in a single vacuum $i$
is spelled out in \S \ref{subsec:1d-Coul-Branch},
which also contains a very explicit example of the construction of hyperholomorphic
connections on periodic Taub-NUT space. We must point out that the
argument of \PI\ also relied on a general analysis of anomalous Ward identities.
The analogous argument for \tf\ is omitted here, but we feel this gap can be
easily filled.

There is an interesting and important aspect of the geometry
which has thus far been suppressed in this exposition.
The bundles $V_{\bS}$ are, strictly speaking,
in general not bundles at all but rather ``twisted bundles.''
These objects have been encountered previously by  physicists, notably in the context of
D-branes in the presence of $B$-fields. There are analogous topological subtleties
(also suppressed in our exposition thus far)
related to the definitions of the charge torsors $\Gamma_i$. The essential point is that
the monodromy of the superpotentials, or equivalently of the $\nu_i$
 sometimes turns out to be fractional, leading to
ill-defined Aharonov-Bohm (AB) phases for test particles transported around the surface
defect. These ``anomalies'' are discussed abstractly in \S \ref{subsubsec:Potential-Anomaly}
and illustrated concretely in \S \ref{subsubsec:Example-Anomaly}. In the geometrical
construction of $V_{\bS}$ the twisting of $\Gamma_i$  leads to
a construction of twisted bundles over $\CM$, as explained in \S \ref{subsubsec:Twsted-Mirror}.
In order to have well-defined amplitudes we follow a strategy outlined in \S \ref{subsec:Anomaly-Cancelation}:
By gauging a suitable finite abelian group of flavor symmetries
we can twist the surface defects so that particles transported around them pick up compensating
phases, leading to well-defined AB phases. The cancellation mechanism is described in
detail in \S \ref{subsubsec:Anomaly Cancelation} and illustrated in a concrete class of
models in \S \ref{subsec:Lag-Desc}.
In addition, the relation of global symmetries of three-dimensional sigma
models to $B$-fields and  connections on twisted bundles is discussed in Appendix \ref{app:Flavor-Twisted},
which might prove to be of independent interest.

The somewhat abstract structure we have outlined is realized concretely
in the theories of class $\CS$ introduced in \cite{Witten:1997sc,Gaiotto:2009hg,Gaiotto:2009we}.
These are partially twisted $(2,0)$ superconformal theories compactified on $\IR^4 \times C$,
where $C$ is a punctured Riemann surface and certain codimension two defects are inserted at the
punctures.
As in our previous papers \PII\ and \PIII\, our later sections, \S\S \ref{sec:Hitchin-Expl}, \ref{sec:Detailed-Expl},
and \ref{sec:Solve-Hitchin}, are devoted to the $A_1$ theories of class $\CS$.  We leave,
yet once more, the full generalization to the higher rank $A_n$ cases for a future occasion.
The theories of class $\CS$ have the distinguishing property that the Seiberg-Witten moduli
spaces $\CM$ may be identified with moduli spaces of Hitchin systems on $C$. (See \S \ref{subsec:ClassS-3d}.) The
Seiberg-Witten curve $\Sigma$ is a ramified covering of $C$ and is identified with the
spectral curve of the Hitchin system, and thus sits in $T^*C$.  The Seiberg-Witten differential
$\lambda$ is then the restriction of the canonical trivialization of the symplectic form on $T^*C$.

As in \cite{Klemm:1996bj,Gaiotto:2009hg,Gaiotto:2010be} there are geometrical realizations of the new BPS
states in terms of ``WKB curves'' or, more generally, ``WKB networks.'' These
are certain networks of curves on $C$, satisfying a first-order differential equation determined by $\lambda$.
See \S \ref{subsec:WKB-networks}. The precise expression of the data of the
\tf\ discussed above in terms of the geometry of $\Sigma$,
$C$, and WKB networks is explained in \S \ref{sec:2d-4ddata}. As in \PII\ the basic
phenomenon of wall-crossing can be understood in terms of the discontinuous change of
WKB foliations of $C$, a topic discussed in \S \ref{subsec:WKB-2d-4d}.
In \tf\ there is a new wrinkle: Associated to a point $z\in C$
is a canonical surface defect $\bS_z$. (If we view theories of class $\CS$ as
decoupling limits of M5-branes on $\IR^4 \times C$, then the surface defects
arise from open membranes ending on $\bS \times \{ z \}$.)  In addition to
the dependence of our BPS states
on $u\in \CB$ we can also study their dependence on $z\in C$. There is wall-crossing, and
certain special WKB curves known as critical WKB curves play the role of walls of marginal
stability. Moreover, the ``vector bundles'' $V_{\bS_z}$ turn out to be the
canonically defined universal bundles over $\CM \times C$, restricted to $z$. Actually,
as discussed in \S \ref{subsec:ClassS-3d}, they are in fact twisted bundles. The
anomaly in AB phases from the twisting can be cancelled using the mechanism of gauging
discrete flavor symmetries, as explained in \S \ref{subsec:Lag-Desc}.

After explaining the general rules for understanding the BPS spectrum and its wall-crossing
for  $A_1$-type examples in
\S \ref{sec:Hitchin-Expl} we proceed in \S \ref{sec:Detailed-Expl} to
look at some detailed examples. We first consider Argyres-Douglas type superconformal
theories in \S \ref{subsec:AD-EXPL}.  Even the most trivial examples lead to interesting
wall-crossing of framed BPS states, as shown in \S \ref{subsubsec:AD-EXPL-N1}.
Many generic aspects of 2d-4d systems are nicely illustrated in the relatively
simple example of the $N=2$ AD theory in \S \ref{subsubsec:ADN=2}. In particular,
this carefully worked example illustrates the compatibility of the 2d-4d wall-crossing
formula with the monodromy of the local systems $\Gamma_{ij}$. Indeed, one could
use the 2d-4d wall-crossing formula to \emph{derive} the monodromy of the
local system, without going through the difficult task of analytic continuation
of periods! In \S \ref{subsubsec:AD-LargerN} we briefly look at 2d-4d wall-crossing
for AD theories with $N>2$. We argue that, while the spectrum of BPS states can
be very intricate, the essential qualitative wall-crossing phenomena have already been captured
in the examples with small $N$. In \S \ref{subsec:cp1-sigma} we turn to the
pure  two-dimensional $\IC \IP^1$ sigma model,
where we gain new insights though the four-dimensional theory is
empty. In particular, we reproduce known results on the BPS spectrum of this
model \cite{Dorey:1998yh} and clarify the relation of the marginal stability
phenomena in this model to those occurring in four-dimensional $SU(2)$ theory.
Indeed, this two-dimensional example is a useful warmup for the analysis
in \S \ref{subsec:WC-exple-su2}
of the $\IC \IP^1$ model coupled to the four-dimensional $SU(2)$ theory
without matter.  Even this relatively simple example turns out
to be rather nontrivial. There are two parameters which control the BPS
spectrum: $u$, which
controls the strength of the four-dimensional coupling, and $z$, which
controls the strength of the two-dimensional coupling. The BPS spectrum is a
very complicated function of $(u,z)$ and we require the
full power of the 2d-4d WCF to conquer it.  When   $u$
corresponds to strong 4d coupling the BPS spectrum is finite, and, as a function of $z$,
is similar to what one finds in the AD theories. However, when $u$ corresponds
to weak 4d coupling the spectrum depends sensitively on $z$. When $z$ corresponds
to strong 2d coupling there are infinitely many walls of marginal stability
and infinitely many chambers with different soliton spectra.
In each chamber the soliton spectrum is finite, but there is no upper
bound on the spectrum.
When $z$ corresponds to weak 2d coupling there is an explosion in the
complexity of the walls of marginal stability:  there are countably many such
walls, but uncountably many chambers. It would be very interesting to
check these predictions with a weak-coupling field theory analysis.

An application of our work is an algorithm for solving Hitchin's equations
\eqref{eq:HitchinEqs-redux},
\begin{equation}\label{eq:HitchinEqs-redux-intro}
\begin{split}
F + R^2 [\varphi , \bar\varphi ]  & = 0,\\
 \bar\partial_A \varphi & = 0,\\
 \partial_A \bar\varphi & = 0,
\end{split}
\end{equation}
with rank $1$ gauge group on a punctured Riemann surface $C$. (See \S \ref{subsec:ClassS-3d}
for notation.)  One begins with
the spectral double cover $\lambda^2 + \phi_2(z)=0$, where $\phi_2(z)$ is a
meromorphic quadratic differential on $C$. From the analysis of
WKB curves described in \S \ref{sec:Hitchin-Expl}, (or, in principle, using the spectrum generator
\eqref{spectrumgen}) one derives the BPS spectrum
$\mu$ and $\omega$. Then one writes down the integral equations of \S \ref{subsec:Integral-Equations}.
At large values of $R$ the integral equations can be solved by a rapidly convergent iteration.
Then, as explained in \S \ref{sec:Solve-Hitchin}, one expresses the solutions to the integral
equations as solutions to the linear problem $(\de + \CA) \Psi=0$, where
$\CA$ is given by \eqref{eq:hit-conn}:
\begin{equation} \label{eq:hit-conn-intro}
 \CA = R \frac{\varphi}{\zeta} + A + R \zeta \bar\varphi.
\end{equation}
From $\Psi$ one can determine $\CA$, and from $\CA$ one clearly obtains
a solution to the Hitchin equations.  It would be a very useful and interesting
check on our logic to work through
this algorithm in some explicit examples.  We have not yet done this.

We have indicated
just a few of the possible future lines of inquiry in \S \ref{sec:Future}.
We have said this before and might say it again: Despite the length of this paper, we feel we have
just scratched the surface of our topic.  Time and again,
the reader, who will require stamina and dedication to read what follows,
will find that we are forced to call a halt to an interesting line
of exploration, leaving unexplored fertile fields for future cultivation.

\section{Formal statement of wall-crossing formulae}\label{sec:Formal-Statements}

In this section we will formulate a new wall-crossing formula which combines features from both
the Cecotti-Vafa wall-crossing formula
for 2d $(2,2)$ theories and the Kontsevich-Soibelman wall-crossing formula for 4d $\CN=2$ theories.

\subsection{The 2d Cecotti-Vafa wall-crossing formula}\label{subsec:2d-CVWCF}

The data which enters the CV wall-crossing formula is a finite set $\CV$, a ``central charge function''
$Z: \CV \to \IC$,
and a ``degeneracy'' $\mu_{ij} \in \IZ$ associated to any pair $i,j \in \CV$ with $i\not=j$.
A physical interpretation of these data will be discussed in
\S \ref{sec:Physical-Interpret}, where elements of $\CV$ will be identified with vacua in an
$\CN=2$ supersymmetric $1+1$ dimensional QFT.

We consider the function $Z$ as allowed to vary, and the $\mu_{ij}$ as functions of $Z$, i.e. of
the values $Z_i$ for $i \in \CV$.
The dependence of the $\mu_{ij}$ on $Z$ is piecewise constant:
they are only allowed to jump when $Z$ crosses a ``wall
of marginal stability,'' where the values $Z_i$, $Z_j$, $Z_k$ for $i$, $j$, $k$ distinct become
collinear as points of $\IC$.
The precise way in which the $\mu_{ij}$ jump as $Z$ crosses such a wall is dictated by the
wall-crossing formula, which we now describe.

We begin by introducing formal variables $X_i$ and
$X_{ij}$, with the rule that $X_{ij}$ acts by left-multiplication on the $X_i$ as
\begin{equation}
X_{ij} X_k = \delta_{jk} X_i.
\end{equation}
(One could easily realize these product laws in terms of explicit matrices, but our abstract notation here will be convenient for the rest of the paper.)
Then we define the ``$\CS$-factor'' $\CS^\mu_{ij}$ to be a transformation which acts on the $X_k$ as
left-multiplication by $1-\mu_{ij}X_{ij}$, i.e.
\begin{equation} \label{eq:s-2d}
\CS^\mu_{ij}: X_k \to X_k - \delta_{jk} \mu_{ij} X_i.
\end{equation}
Next, to each pair $(i,j)$ with $\mu_{ij} \neq 0$ we associate a ``BPS ray'' $\ell_{ij}$
in the complex plane, with slope given by the phase of $Z_{ij} := Z_i - Z_j$:
\begin{equation}
 \ell_{ij} = Z_{ij} \IR_- \subset \IC.
\end{equation}
Now choose some convex angular sector $\sphericalangle$ in the complex plane, with apex at the origin.
The basic actor in the wall-crossing formula is the
composition $A(\sphericalangle)$ of all the $\CS$-factors corresponding to
BPS rays lying in $\sphericalangle$:
\begin{equation}
 A(\sphericalangle) = : \prod_{i,j: \ell_{ij} \subset \sphericalangle} \CS^\mu_{ij} :\ .
\end{equation}
Here the normal-ordering symbols mean that the
product is ordered by the phase of $Z_{ij}$, i.e.,
reading from left to right in the product we encounter the
factors associated with   the BPS rays $\ell_{ij}$ successively in the
counterclockwise direction.
See Figure \ref{fig:av-2d}.
\insfig{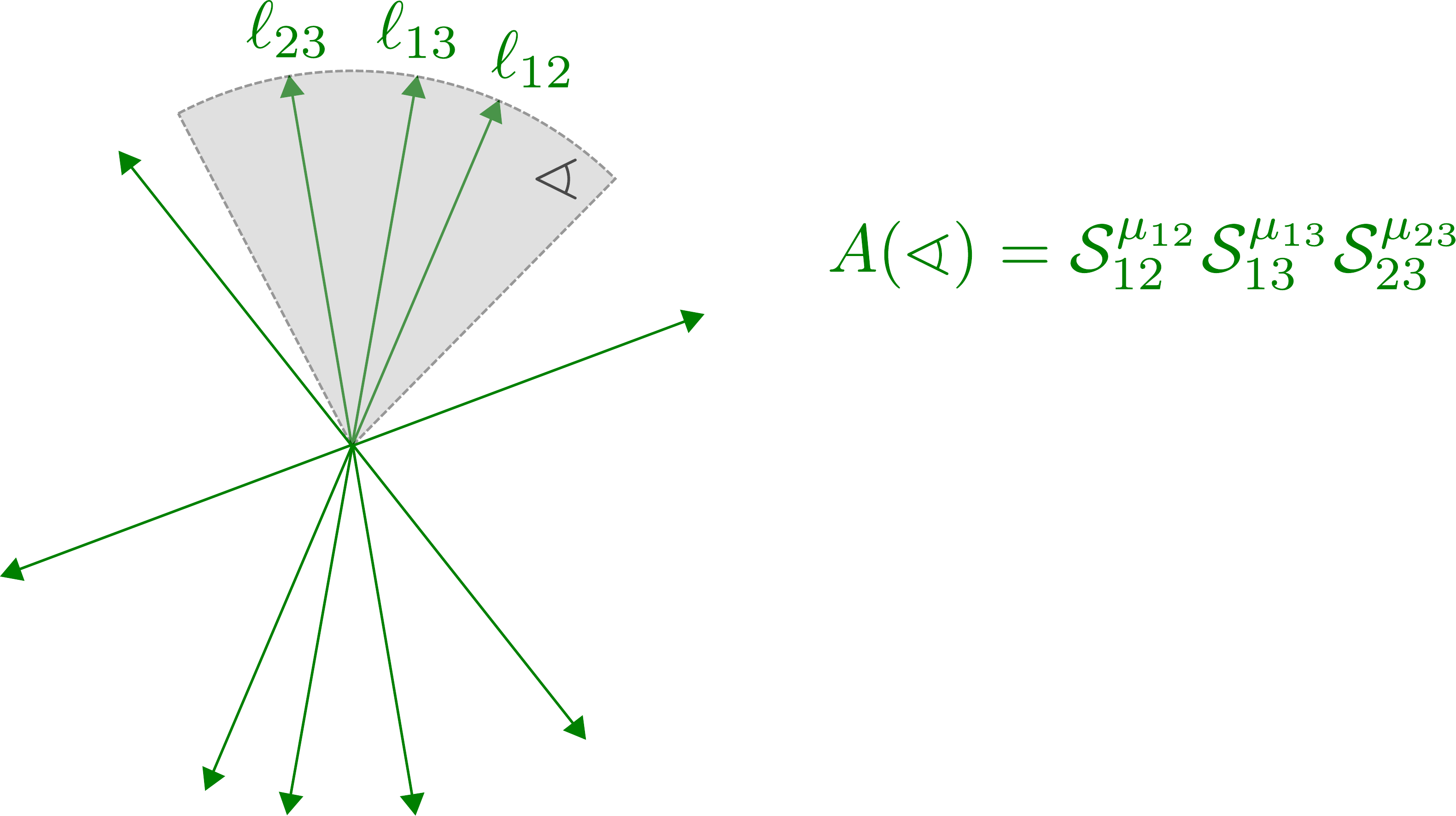}{The definition of $A(\sphericalangle)$ as a product of $\CS$-factors
associated to the BPS rays $\ell_{ij}$.}
The statement of the wall-crossing formula is:
\ti{$A(\sphericalangle)$ is constant under variation of $Z$, as long as no BPS rays
cross the boundary of $\sphericalangle$.}

Let us see how this WCF works in practice.
Suppose we have three rays $\ell_{ij}$, $\ell_{ik}$, $\ell_{jk}$ which are close together in the complex plane,
ordered say counterclockwise
(with $i$, $j$, $k$ pairwise distinct).
Choose the sector $\sphericalangle$ to include these three and no others.  Now suppose we vary $Z$ until we reach
a wall of marginal stability, where
$Z_{ij}$ and $Z_{jk}$ (and hence $Z_{ik}$) have the same phase.  Upon crossing this wall
the rays $\ell_{ij}$, $\ell_{ik}$, $\ell_{jk}$ switch from counterclockwise to
clockwise ordering.  Then the WCF says we should
compare the composition $\CS^\mu_{ij} \CS^\mu_{ik} \CS^\mu_{jk}$ on one side of the wall and $\CS^{\mu'}_{jk} \CS^{\mu'}_{ik} \CS^{\mu'}_{ij}$
on the other side:
\begin{align}
\CS^\mu_{ij} \CS^\mu_{ik} \CS^\mu_{jk} &=( 1-\mu_{ij}X_{ij})(1-\mu_{ik}X_{ik})(1-\mu_{jk}X_{jk})\notag \\ &= 1-\mu_{ij} X_{ij} - (\mu_{ik} - \mu_{ij} \mu_{jk}) X_{ik} - \mu_{jk}X_{jk}, \\
\CS^{\mu'}_{jk} \CS^{\mu'}_{ik} \CS^{\mu'}_{ij} &= ( 1-\mu'_{jk}X_{jk})(1-\mu'_{ik}X_{ik})(1-\mu'_{ij}X_{ij}) \notag \\
& = 1-\mu'_{ij} X_{ij} - \mu'_{ik}X_{ik} - \mu'_{jk}X_{jk}.
\end{align}
Hence we find
\begin{align}
\mu'_{ij} & = \mu_{ij}, \\
\mu'_{ik} & = \mu_{ik} - \mu_{ij} \mu_{jk},\\
\mu'_{jk} & = \mu_{jk}.
\end{align}
This is precisely the wall-crossing formula of Cecotti and Vafa \cite{Cecotti:1992rm}.
A simple example is $\mu_{ij} = \mu_{jk} = \mu_{ik} =1$,
which gives $\mu'_{ij}  = \mu'_{jk} =1$ but $\mu'_{ik}=0$.

Since $\mu_{ij}$ enters the wall-crossing formula only through $\mu_{ij} X_{ij}$ it is sometimes useful to give an
equivalent presentation of the wall-crossing formula, where some signs are shifted from the $\mu_{ij}$
to the definition of $X_{ij}$.

\subsection{The 4d Kontsevich-Soibelman wall-crossing formula}\label{subsec:4d-KSWCF}

The data which enters the KS wall-crossing formula is a lattice $\Gamma$ of ``charges''
 with an integer-valued bilinear
antisymmetric pairing $\langle \cdot, \cdot \rangle$ (the ``intersection product''),
a central charge function $Z: \Gamma \to \IC$, linear on $\Gamma$,
and a set of ``degeneracies'' $\Omega(\gamma) \in \IZ$ associated to charges $\gamma \in \Gamma$.

As in the 2d case, the $\Omega(\gamma)$ depend on the function $Z$ in a piecewise constant fashion.  They may jump
only when $Z$ crosses a ``wall of marginal stability'' where the phases of $Z_\gamma$ and $Z_{\gamma'}$
for two linearly independent $\gamma$, $\gamma'$ become aligned.  The wall-crossing formula
dictates precisely how the $\Omega(\gamma)$ jump when $Z$ crosses such a wall.

In order to formulate the wall-crossing formula it is useful to introduce formal variables $X_\gamma$
for $\gamma \in \Gamma$, with a twisted multiplication rule
\begin{equation}
X_\gamma X_{\gamma'} = (-1)^{\langle \gamma, \gamma' \rangle} X_{\gamma + \gamma'}.
\end{equation}
Then we introduce the ``$\CK$-factor'' $\CK^\Omega_\gamma$, which acts on the $X_{\gamma'}$
by\footnote{In previous papers on the subject we
defined $\CK_\gamma$ directly rather than $\CK_\gamma^\Omega$.  We do not do this here because $\CK_\gamma$
itself does not have a natural extension to the 2d-4d setting; $\CK_\gamma^\Omega$ turns out to be the
more natural object.}
\begin{equation} \label{eq:k-4d}
\CK^\Omega_{\gamma}: X_{\gamma'} \to (1-X_{\gamma})^{\langle \gamma', \gamma \rangle \Omega(\gamma)} X_{\gamma'}.
\end{equation}

To each $\gamma$ such that $\Omega(\gamma) \neq 0$ we associate a ``BPS ray'' $\ell_{\gamma}$
in the complex plane, with slope given by the phase of $Z_\gamma$:
\begin{equation}
 \ell_\gamma = Z_\gamma \IR_- \subset \IC.
\end{equation}
The basic actor in the wall-crossing formula is the composition $A(\sphericalangle)$
of all the $\CK$-factors corresponding
to BPS rays lying the sector $\sphericalangle$ of the complex plane:
\begin{equation}
 A(\sphericalangle) = \,\, : \prod_{\gamma: \ell_{\gamma} \subset \sphericalangle} \CK^\Omega_\gamma :\ .
\end{equation}
As in the 2d case, the ordering is such that if $\ell_{\gamma_1}, \ell_{\gamma_2}$ are in
counterclockwise order then $\CK^\Omega_{\gamma_1}$ is to the left of $\CK^\Omega_{\gamma_2}$.
The statement of the wall-crossing formula \cite{ks1} is:  \ti{$A(\sphericalangle)$ is constant under variation of $Z$, as long
as no BPS rays cross the boundary of $\sphericalangle$.}  All this is completely parallel to the 2d case.

For convenience of our discussion in \S\S \ref{sec:Hitchin-Expl} and \ref{sec:Detailed-Expl}
we now give two standard examples of this wall-crossing formula in action.
Consider the simple situation where we begin with two
BPS rays $\ell_\gamma$ and $\ell_{\gamma'}$, with $\Omega(\gamma) = \Omega(\gamma') = 1$, and no BPS rays
lying between $\ell_\gamma$ and $\ell_{\gamma'}$.  Then we adjust $Z$ until the two BPS rays collide and cross
one another.  After the collision we will have a more complicated set of BPS rays and degeneracies.
What we get precisely depends heavily on the value of $\inprod{\gamma, \gamma'}$.
First, suppose $\langle \gamma, \gamma' \rangle=1$.  In this case we have
\begin{equation} \label{eq:pentagon}
\CK^\Omega_{\gamma'} \CK^\Omega_{\gamma} = \CK^{\Omega'}_{\gamma} \CK^{\Omega'}_{\gamma+\gamma'}\CK^{\Omega'}_{\gamma'},
\end{equation}
where $\Omega'(\gamma) = \Omega'(\gamma') =\Omega'(\gamma+\gamma')=1$.
This identity is simple enough to be checked directly; just consider the action of $\CK^\Omega_{\gamma'} \CK^\Omega_{\gamma}$,
\begin{align}
(X_{\gamma};X_{-\gamma'}) \stackrel{\CK_{\gamma}^\Omega}{\mapsto} (X_\gamma;X_{-\gamma'}+X_{\gamma-\gamma'})  \stackrel{\CK_{\gamma'}^\Omega}{\mapsto} (X_{\gamma}+X_{\gamma+\gamma'};X_{-\gamma'}+X_{\gamma-\gamma'}+X_\gamma),
\end{align}
and compare it with the action of $\CK^{\Omega'}_{\gamma} \CK^{\Omega'}_{\gamma+\gamma'}\CK^{\Omega'}_{\gamma'}$,
\begin{align}
(X_{\gamma};X_{-\gamma'}) &\stackrel{\CK_{\gamma'}^{\Omega'}}{\mapsto} (X_{\gamma}+X_{\gamma+\gamma'};X_{-\gamma'}) \stackrel{\CK_{\gamma+\gamma'}^{\Omega'}}{\mapsto} (X_{\gamma}+X_{2\gamma+\gamma'}+X_{\gamma+\gamma'};X_{-\gamma'} + X_{\gamma})
\notag \\ &\stackrel{\CK_{\gamma}^{\Omega'}}{\mapsto} (X_\gamma + X_{\gamma+\gamma'};X_{-\gamma'} + X_{\gamma-\gamma'} + X_\gamma).
\end{align}
The fact that the two agree is equivalent to \eqref{eq:pentagon}.
For a more interesting example, suppose instead $\langle \gamma, \gamma' \rangle=2$.  Then
\begin{equation}
\CK^\Omega_{\gamma'} \CK^\Omega_{\gamma} = \CK^{\Omega'}_{\gamma} \CK^{\Omega'}_{2\gamma+\gamma'} \CK^{\Omega'}_{3\gamma+2\gamma'}\cdots \CK^{\Omega'}_{\gamma+\gamma'} \cdots \CK^{\Omega'}_{2 \gamma+3 \gamma'} \CK^{\Omega'}_{\gamma+2\gamma'}\CK^{\Omega'}_{\gamma'}
\end{equation}
where $\Omega'((n+1)\gamma+ n\gamma') = \Omega'(n\gamma+ (n+1)\gamma') = 1$
for all $n \ge 0$, $\Omega'(\gamma +\gamma') = -2$.
This identity is less trivial to prove; see  Appendix A of \cite{Gaiotto:2008cd} for an elementary proof.

\subsection{The 2d-4d wall-crossing formula}\label{subsec:Formal-Statement}

Now we are ready to give the new wall-crossing formula which is the subject of this paper.
It combines the features of the wall-crossing formulae of \S\S \ref{subsec:2d-CVWCF}
and \ref{subsec:4d-KSWCF}, but it involves some new features. Heuristically, it is
a kind of ``fiber-product'' of the two previously known wall-crossing formulae.

Our data include both the finite set $\CV$ and the lattice $\Gamma$ with
integral antisymmetric pairing $\langle \cdot, \cdot \rangle$.  In addition,
we will need new charge ``lattices,'' an extension of the central
charge function to these ``lattices,'' a new set of BPS degeneracies,
and, finally, a new $\IZ_2$-valued twisting function.  The physical interpretation of these data is
explained in \S \ref{sec:Physical-Interpret}.  We now describe the
new data more precisely.

The first piece of new data is a set of
distinct $\Gamma$-\emph{torsors}
 $\Gamma_i$, one for each $i \in \CV$.
We will usually
use the notation $\gamma_i$ to denote a generic element of $\Gamma_i$ (but the
reader should be warned that on several occasions $\gamma_i$ will refer to some
special or specific element of $\Gamma_i$). Each $\Gamma_i$ is both a left- and right-
$\Gamma$-torsor, with $\gamma + \gamma_i = \gamma_i + \gamma$.

We also introduce additional sets $\Gamma_{ij}$, defined as follows:  an element $\gamma_{ij} \in \Gamma_{ij}$ is a
formal difference $\gamma_i - \gamma_j$ of elements of $\Gamma_i$ and $\Gamma_j$, modulo
the equivalence $(\gamma_i + \gamma) - (\gamma_j + \gamma) = \gamma_i - \gamma_j$.
$\Gamma_{ij}$ is again a $\Gamma$-torsor, with $\gamma + \gamma_{ij} = \gamma_{ij} + \gamma$.
A convenient notation for this definition is
\begin{equation} \label{eq:diff-torsor}
 \Gamma_{ij} = \Gamma_i - \Gamma_j.
\end{equation}
One can identify $\Gamma_{ii}$ canonically with $\Gamma$,
and it will sometimes be convenient to do so.
\footnote{This leads to the slightly awkward notational point that
$\gamma_{ii}$ is not zero. It is generically $\gamma_i - \gamma_i'$ where
$\gamma_i, \gamma_i'$ are both in $\Gamma_i$, i.e. it is a generic element of $\Gamma$.}
Then, to an element $\gamma_{ij}\in \Gamma_{ij}$
we can associate a unique element $(-\gamma_{ij}) \in \Gamma_{ji}$ so
that $\gamma_{ij} +(- \gamma_{ij}) = 0 \in \Gamma_{ii}$.

There are naturally defined
addition operations $\Gamma_{ij} \times \Gamma_j \to \Gamma_i$ and $\Gamma_{ij} \times \Gamma_{jk}\to \Gamma_{ik}$.
We will write these operations as $+$, thus $\gamma_{ij} + \gamma_j \in \Gamma_i$,
but the reader should be warned that $+$ is not symmetric; for example, $\gamma_j + \gamma_{ij}$
is not defined when $i \neq j$.

\insfig{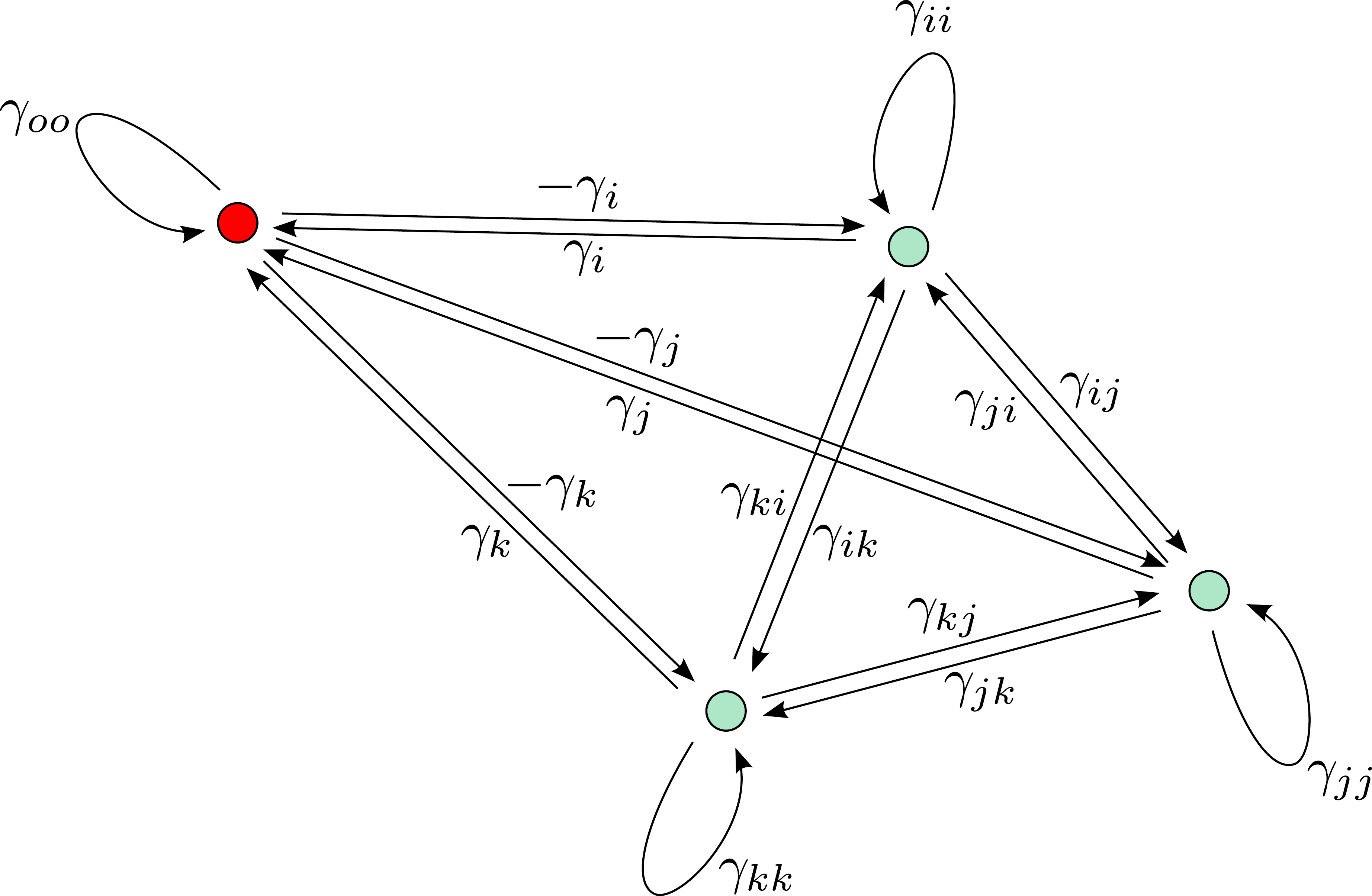}{A partial depiction of the groupoid $\IV$, in case $\CV$ has three elements, here labeled
$i$, $j$, $k$.  Each morphism space is a torsor for $\Gamma$, of which we have shown only a single element.}

The situation is nicely summarized by declaring that the first piece of
data is a   groupoid $\IV$ with the following properties. For each $i\in \CV$, $\IV$
has a corresponding  object.
The set of morphisms $\Hom(i,j)$ from $i$ to $j$ is  a $\Gamma$-torsor and identified
with $\Gamma_{ij}$.
The composition of morphisms is given by the addition laws we have written above:
In addition there is one more object denoted $o$,\footnote{$o$ will correspond physically to the
``empty surface defect.''} making $\IV$ into a pointed groupoid. We identify
 $\Hom(i,o) = \Gamma_i$. Note that   $\Hom(o,i)$ is another torsor which gives independent
meaning to elements
$-\gamma_i \in -\Gamma_i$ so that $\gamma_i + (-\gamma_j) \in \Gamma_{ij}$.
Thus, e.g. the composition of $\gamma_j \in \Gamma_j = \Hom(j,o)$ and $\gamma_{ij} \in \Gamma_{ij} = \Hom(i,j)$
is $\gamma_{ij} + \gamma_j \in \Gamma_i = \Hom(i,o)$ while $-\gamma_i + \gamma_{ij} \in - \Gamma_j$.
The automorphism
group $\Hom(i,i)$ of each object (including $i = o$) is canonically isomorphic to $\Gamma$.
See Figure \ref{fig:groupoid}.

Our second new piece of data is an extension of the central charge function $Z$ from $\Gamma$ to all
the $\Gamma_i$, which is compatible with the $\Gamma$-action. Thus we have
$Z: \amalg_i \Gamma_i \to \IC$ which is affine-linear in the sense that
$Z_{\gamma + \gamma_i} = Z_{\gamma} + Z_{\gamma_i}$.  Once we have this extension,
we automatically get a further extension of $Z$ to the $\Gamma_{ij}$, namely,
if $\gamma_{ij} = \gamma_i - \gamma_j$
then we define $Z_{\gamma_{ij}}:= Z_{\gamma_{i} } -  Z_{\gamma_j}$. Then we have
\begin{equation}
Z_{a+b} = Z_{a} + Z_{b}
\end{equation}
whenever $a+b$ is defined.
(Here and below, we use the notation $a, b, c, \dots$ in statements which are true
when $a, b, c, \dots$ are valued in any of $\Gamma$, $\Gamma_i$
or $\Gamma_{ij}$.)
In the categorical language mentioned above, giving $Z$ is equivalent to giving
a homomorphism from the groupoid $\IV$ to $\IC$.

Our third new piece of data is a new collection of ``BPS degeneracies''.  As in the 4d case
we have ``4d degeneracies'' $\Omega(\gamma)\in \IZ$ for $\gamma \in \Gamma$.
It is convenient to define a function
$\omega: \Gamma \times \Gamma \to \IZ$ by
\begin{equation}
\omega(\gamma, \gamma') := \Omega(\gamma)\langle \gamma, \gamma'\rangle.
\end{equation}
We then introduce new ``mixed degeneracies'' $\omega: \Gamma \times \amalg_i \Gamma_i \to \IZ$,
\footnote{In some situations we should relax the requirement that $\omega$ is $\IZ$-valued;
see the end of this subsection.}
subject to the condition
\begin{equation}\label{eq:omega-aff-lin}
\omega(\gamma, \gamma_i + \gamma') = \omega(\gamma, \gamma_i) + \Omega(\gamma)\langle \gamma, \gamma'\rangle.
\end{equation}
(Because of this condition, $\omega(\gamma, \cdot)$ is
determined once we give a single $\omega(\gamma, \gamma_i)$ for each $i$.)
We also define $\omega: \Gamma \times \amalg_{i,j} \Gamma_{ij} \to \IZ$ by
\begin{equation}
\omega(\gamma, \gamma_i - \gamma_j) := \omega(\gamma,\gamma_i) - \omega(\gamma,\gamma_j).
\end{equation}
Then for any two ``charges'' $a,b$, each belonging to $\Gamma$, $\Gamma_i$, or $\Gamma_{ij}$, we have
\begin{equation}
\omega(\gamma, a+b) = \omega(\gamma,a) + \omega(\gamma,b)
\end{equation}
so long as $a+b$ is defined.  In the categorical language,
for each $\gamma$, giving $\omega(\gamma, \cdot)$ is equivalent to giving
a homomorphism $\IV \to \IZ$.

Turning to 2d particles, we do not simply take over a set of ``2d degeneracies''
$\mu_{ij} \in \IZ$ as in the pure 2d story:  rather we promote these to a set of $\mu(\gamma_{ij}) \in \IZ$,
defined for each $\gamma_{ij} \in \Gamma_{ij}$ where $i\not=j$.

As in the pure 2d and 4d cases, the BPS degeneracies
$\mu$, $\omega$ depend on $Z$, but this dependence is completely determined by a
wall-crossing formula, to be stated below.

Our fourth, and final, piece of data is a ``twisting'' function $\sigma(a, b)$, defined whenever
$a+b$ is defined, and valued in $\{ \pm 1 \}$.
We require $\sigma$ to obey a cocycle condition,\footnote{$\sigma$
is very similar to the data one would associate to
a twisting of $K$ theory on the groupoid $\IV$. However, although
$\IV$ is equivalent to ${(pt)} /\!\!/\Gamma$ as a groupoid, the extra structure
of $Z$ does not respect this equivalence of categories. }
\begin{equation} \label{eq:cocycle}
 \sigma(a,b) \sigma(a+b, c) = \sigma(a,b+c) \sigma(b,c).
\end{equation}

In order to formulate the wall-crossing formula we introduce formal variables $X_a$, where,
as above, $a$ is an element of $\Gamma$, $\Gamma_i$ or $\Gamma_{ij}$.
We define a twisted multiplication rule:
\begin{equation} \label{eq:twisted-mult}
X_a X_b = \begin{cases} \sigma(a,b) X_{a+b} & \text{ if } a+b \text { is defined}, \\
                        0 & \text{ otherwise}.
          \end{cases}
\end{equation}
This product is associative thanks to \eqref{eq:cocycle}.
In general $a+b$ and $b+a$ are not
simultaneously defined, so the product is \emph{noncommutative}.
However, if $a+b$ and $b+a$ are both defined we assume $\sigma(a,b) = \sigma(b,a)$
so in this case $X_a X_b = X_b X_a$. We will also assume that $\sigma(\gamma, \gamma') = (-1)^{\langle \gamma, \gamma' \rangle}$ to insure compatibility with the pure 4d case.
In the categorical language mentioned above,
we are essentially considering the twisted groupoid algebra $\IC[\IV]$.\footnote{The $X_\gamma$ as defined
so far do not strictly belong to $\IC[\IV]$, but one can identify $X_\gamma$ with
$X_{\gamma_{oo}} + \sum_i X_{\gamma_{ii}} \in \IC[\IV]$, where each
$\gamma_{ii} \in \Gamma_{ii}$ (including $i = o$) is the one corresponding to $\gamma \in \Gamma$.  This
identification is compatible with everything we will do below.}

The wall-crossing formula controls the
dependence of the BPS degeneracies
$\mu$ and $\omega$ on $Z$.
The degeneracies are piecewise constant as functions of $Z$.  They jump only at the ``walls of marginal
stability'' where the phases of $Z_a$ and $Z_b$ become equal for some pair $(a,b)$ with $a\not=b$,
 where each of $a$, $b$
separately belongs either to $\Gamma$ or to some $\Gamma_{ij}$ (with the proviso that if both $a$ and $b$
belong to $\Gamma$ then they are linearly independent.)
We define rays  $\ell_{\gamma}$ and $\ell_{\gamma_{ij}}$ in the complex plane, with slopes
given respectively by the phases of $Z_\gamma$ or $Z_{\gamma_{ij}}$:
\begin{equation}
 \ell_\gamma = Z_\gamma \IR_-, \qquad \ell_{\gamma_{ij}} = Z_{\gamma_{ij}} \IR_-.
\end{equation}
We call $\ell_\gamma$ a \ti{BPS $\CK$-ray} if $\omega(\gamma, \cdot) \neq 0$, and likewise
$\ell_{\gamma_{ij}}$ a \ti{BPS $\CS$-ray} if $\mu_{\gamma_{ij}} \neq 0$.
BPS degeneracies can jump only when $Z$ is such that
two BPS rays become coincident.

As in the Cecotti-Vafa and Kontsevich-Soibelman WCF reviewed above, we will attach certain ``factors''
(automorphisms of $\IC[\IV]$) to the BPS rays.
The factors associated to the $\CS$-rays $\ell_{\gamma_{ij}}$, called ``$\CS$-factors'' and denoted
 $\CS^\mu_{\gamma_{ij}}$, are defined by
\begin{equation}\label{eq:cv-like}
\CS^\mu_{\gamma_{ij}}: X_a  \to  (1-\mu(\gamma_{ij}) X_{\gamma_{ij}}) X_{a} (1+\mu(\gamma_{ij}) X_{\gamma_{ij}}).
\end{equation}
In particular the $\CS$-factor acts trivially on $X_\gamma$ and
\begin{equation}
\CS^\mu_{\gamma_{ij}}: X_{\gamma_k} \mapsto
\begin{cases} X_{\gamma_k} & k \not=j  \\
X_{\gamma_j} - \mu(\gamma_{ij})\sigma(\gamma_{ij},\gamma_j) X_{\gamma_{ij} + \gamma_j} & k=j \\
\end{cases}
\end{equation}
\begin{equation}
\CS^\mu_{\gamma_{ij}}: X_{-\gamma_k} \mapsto
\begin{cases} X_{-\gamma_k} & k \not=i  \\
X_{-\gamma_i} + \mu(\gamma_{ij})\sigma(-\gamma_i,\gamma_{ij}) X_{-\gamma_i + \gamma_{ij}  } & k=i \\
\end{cases}
\end{equation}
which the reader should compare with \eqref{eq:s-2d}.
The factors associated to the $\CK$-rays $\ell_{\gamma}$, called ``$\CK$-factors''
and denoted $\CK^\omega_\gamma$, are defined by
\begin{equation} \label{eq:ks-like}
\CK^\omega_\gamma: X_a \to (1- X_\gamma)^{-\omega(\gamma, a) } X_a.
\end{equation}
Note that when $a = \gamma'$ this transformation
reduces to \eqref{eq:k-4d}:  this is the reason why our wall-crossing formula will
reduce to the 4d wall-crossing formula as far as the 4d degeneracies $\Omega$ are concerned.

We can now state our new wall-crossing formula in a way closely analogous to the 2d and 4d
cases.  For an angular sector $\sphericalangle$ we define an automorphism of the groupoid algebra,
\begin{equation}
A(\sphericalangle) =  \,\,  : \prod_{\gamma: \ell_{\gamma} \subset \sphericalangle} \CK_{\gamma}^\omega \prod_{\gamma_{ij}: \ell_{\gamma_{ij}}
\subset \sphericalangle} \CS_{\gamma_{ij}}^\mu :
\end{equation}
where the normal ordering symbol indicates that the factors --- be they of type $\CK$ or $\CS$ --- are
ordered so that reading from left to right we encounter factors associated with rays successively in
the counterclockwise direction.    Then the new wall-crossing formula says:  \ti{$A(\sphericalangle)$
is constant, as long as no BPS rays cross the boundary of $\sphericalangle$.}

In later sections of the paper we will describe physical realizations of this set of axioms.
We will also encounter a slightly ``twisted'' version of them, which originates from the observation
that one can consistently study the
wall-crossing of the quantities $\mu(\gamma_{ij})$, $\Omega(\gamma)$ and
$\omega(\gamma, \gamma_{ij})$ as $Z_{\gamma_{ij}}$ vary,
with no reference to the $Z_{\gamma_i}$, $\omega(\gamma, \gamma_i)$ or even $\Gamma_i$.
Indeed, in some physical examples, certain anomalies can prevent us from defining $\Gamma_i$, $Z_{\gamma_i}$,
$\omega(\gamma, \gamma_i)$ unambiguously,
even though $\Gamma_{ij}$, $Z_{\gamma_{ij}}$, $\omega(\gamma, \gamma_{ij})$, $\mu(\gamma_{ij})$, $\Omega(\gamma)$ are perfectly
well-defined and satisfy the 2d-4d wall-crossing axioms. See \S \ref{subsubsec:Potential-Anomaly} below.
In these situations it appears to be appropriate   simply to require integrality of
the $\omega(\gamma, \gamma_{ij})$ and in fact it might not be possible to define  $\omega(\gamma, \gamma_{i })$
as integers.

\subsubsection{Automorphisms of wall-crossing identities} \label{sec:automorphisms}

If $\alpha$ is an automorphism or anti-automorphism of the
groupoid algebra $\IC[\IV]$ then, given a wall-crossing
identity, we can conjugate the factors by $\alpha$ to produce
another wall-crossing identity.

The simplest example of this is to consider the transformation
$\alpha(X_a) = (-1)^{c_a} X_a$, which in general  changes $\sigma(a,b)$
by a coboundary but will be an automorphism if $c_{a+b} = c_a + c_b\,\mod\,2$.
In particular,  if degeneracies $\mu, \omega$ satisfy the wall-crossing
formula then so do $\tilde \mu, \omega$ where
\begin{equation}
\tilde \mu(\gamma_{ij}) = (-1)^{c_i - c_j} \mu(\gamma_{ij})
\end{equation}
for integer $c_i$.

As a second example, consider the  anti-automorphism of $\IC[\IV]$
defined by $\alpha: X_a \mapsto X_{-a}$.  (Note that if $a+b$ is
composable then $(-b)+(-a)$ is composable, so $\alpha$ must be an
anti-automorphism.) One can check that $\alpha \CK_\gamma^\omega \alpha^{-1}= \CK_{-\gamma}^{\tilde \omega}$
and
$\alpha \CS_{\gamma_{ij}}^\mu \alpha^{-1} = \CS_{-\gamma_{ij}}^{\tilde \mu}$
with
\begin{equation}
\tilde \mu(\gamma_{ij}) = - \mu(-\gamma_{ij}),
\qquad \tilde \omega(\gamma, a) = \omega(-\gamma, -a),
\end{equation}
giving a second automorphism of the wall-crossing formula.

In the physical realizations, the second automorphism above is closely related to CPT invariance.
CPT takes $a\to -a$, but also takes $Z \to \bar Z$, where $Z$ is the central charge. Therefore,
CPT reverses the clockwise ordering of BPS rays.
On the other hand, given a wall-crossing identity, written as an equality
 of a certain  word in $\CK$ and $\CS$ factors
with the identity operator,
we may always take the inverse of that word to produce another identity. Then the order of the BPS rays will
be reversed. Thus we may define the CPT conjugate of a wall-crossing identity as the one obtained by
taking the inverse and conjugating by $\alpha: X_a \mapsto X_{-a}$.

\subsection{Examples}\label{subsec:Examples-2d-4d-wcf}

In this section we give a few simple examples of wall-crossing formulae.
The general wall of marginal stability occurs when a ray of the form $\ell_{\gamma_{ij}}$
becomes collinear with a ray of the form $\ell_{\gamma_{kl}}$. We can consider
several examples according to how the indices $i,j,k,l$ are related.

First, if none of the indices $i,j,k,l$
coincide, then the charges $n \gamma_{ij}+m \gamma_{kl}$ are not defined, so there are only two
BPS rays in the game. We associate two commuting $\CS$-factors with these two rays,
and the wall-crossing is trivial:
\begin{equation} \label{eq:s-commuting}
 \CS^\mu_{\gamma_{ij}} \CS^\mu_{\gamma_{kl}} =  \CS^{\mu'}_{\gamma_{kl}} \CS^{\mu'}_{\gamma_{ij}}
\end{equation}
with $\mu'=\mu$.

Next, suppose there are three distinct indices and one pair of indices coincides. The cases
$j=k$ and $i=l$ are related by wall-crossing, and in this case the
wall-crossing formula is essentially identical to the
pure 2d case:  we must take into account a third BPS ray
associated to the charge  $\gamma_{il}:=\gamma_{ij} + \gamma_{jl}$,
and then we have
\begin{equation}\label{eq:example-1}
\CS^\mu_{\gamma_{ij}}  \CS^\mu_{\gamma_{il}} \CS^\mu_{\gamma_{jl}} =
\CS^{\mu'}_{\gamma_{jl}}  \CS^{\mu'}_{\gamma_{il}} \CS^{\mu'}_{\gamma_{ij}}
\end{equation}
with $\mu'(\gamma_{jl}) = \mu(\gamma_{jl})$, $\mu'(\gamma_{ij}) = \mu(\gamma_{ij}) $ and
$\mu'(\gamma_{il}) = \mu(\gamma_{il}) - \sigma(\gamma_{ij},\gamma_{jl}) \mu(\gamma_{ij}) \mu(\gamma_{jl})$.

If instead there are three distinct indices with $i=j$ or $j=l$, there cannot be a third
BPS ray, and the commuting $\CS$-factors pass through each other as in \eqref{eq:s-commuting}.

Next, suppose there are three distinct indices and, say, $i\not=j \not= k=l$,
so we can set $\gamma_{kk}=\gamma\in \Gamma$.
Now there can be intermediate rays associated to charges $n\gamma + \gamma_{ij}$ for $n>0$. The
wall-crossing formula then states that
\begin{equation}\label{eq:example-2}
\CK_\gamma^\omega \prod_{n\searrow } \CS^\mu_{\gamma_{ij}+n \gamma}=\prod_{n\nearrow} \CS^{\mu'}_{\gamma_{ij}+n\gamma} \CK_\gamma^{\omega'}
\end{equation}
where the range on $n$ is from $0$ to $+\infty$ and the product
$\prod_{n\nearrow}$ means that as we read from left to right $n$ is increasing.\footnote{In
fact, in this example the $\CS$-factors actually commute with each other.}  Applying this identity
on the generators $X_{\gamma_k}$ for $k\not=j$ shows that $\omega'(\gamma, \gamma_k) = \omega(\gamma,\gamma_k)$
for $k\not=j$ and $k\not=-i$.  Then applying the formula to $X_{\gamma_j}$ we find that in fact
$\omega'=\omega$ always and the $\mu'$ are determined in terms of $\omega$ and $\mu$.
To state the result it is useful to define a power series in $X_\gamma$:
\begin{equation}\label{eq:def-Sigma-ij}
\Sigma_{ij}:= \sum_{n=0}^\infty \frac{\mu(\gamma_{ij}+n \gamma)}{\sigma(\gamma_{ij},\gamma)^n}  X_\gamma^n,
\end{equation}
with a similar definition for $\Sigma'_{ij}$. Then we have
\begin{equation}\label{eq:tmn-law-Sig-prime}
\Sigma'_{ij} = \Sigma_{ij} (1-X_\gamma)^{-\omega(\gamma,\gamma_{ij})}.
\end{equation}

A few special cases are worth noting separately.  Suppose for example that we take
$\mu(\gamma_{ij}) = 1$ and $\omega(\gamma, \gamma_{ij}) = -1$ with all other $\mu$, $\omega$ vanishing.
Then $\Sigma_{ij} = 1$, and \eqref{eq:tmn-law-Sig-prime} says $\Sigma'_{ij} = 1 - X_\gamma$:
so \eqref{eq:example-2} specializes to
\begin{equation} \label{eq:example-2-spec}
\CK_\gamma^\omega \CS^\mu_{\gamma_{ij}} = \CS_{\gamma_{ij}}^{\mu'} \CS_{\gamma_{ij}+\gamma}^{\mu'} \CK_\gamma^{\omega'}
\end{equation}
where $\mu'(\gamma_{ij}) = 1$, $\mu'(\gamma_{ij} + \gamma) = -\sigma(\gamma_{ij},\gamma)$, and $\omega' = \omega$.
We will use this example, in the form \eqref{eq:ks-ssk} and \eqref{eq:sk-kss}, extensively in
\S \ref{sec:Detailed-Expl}.

Let us now suppose that there are only two distinct indices, i.e. two pairs of indices coincide. There
are three cases to consider. If
we have $\ell_{\gamma_{ij}}$ collinear with $\ell_{\gamma'_{ij}}$, then since $\gamma'_{ij}=\gamma_{ij}+\gamma$
for some $\gamma \in \Gamma$, it follows that $Z_{ \gamma}$ or $Z_{-\gamma}$ is collinear with $Z_{\gamma_{ij}}$,
and we return to the case of \eqref{eq:example-2}.

If $i=j$ and $k=l$, then we have BPS rays $\ell_{\gamma^1}$ and $\ell_{\gamma^2}$ crossing for $\gamma^1,\gamma^2\in \Gamma$.
In this case all the BPS rays are $\CK$-rays and the wall-crossing formula says
\begin{equation}\label{eq:example-3}
\prod_{\frac{n}{m}\nearrow} \CK^{\omega}_{n \gamma^1 + m \gamma^2}
= \prod_{\frac{n}{m}\searrow} \CK^{\omega'}_{n \gamma^1 + m \gamma^2}.
\end{equation}
Here $\sigma(a,b)$ will have to be generic, as $\langle \gamma^1, \gamma^2 \rangle$ can be odd. This will not affect the present calculation.
Acting on $X_{\gamma}$ for $\gamma \in \Gamma$ this reduces precisely to the 4d wall-crossing formula,
and hence implies the same relations between $\Omega$ and $\Omega'$ that we had in the pure 4d discussion.
As we have noted, though, $\omega$ and $\omega'$ contain a bit more information than $\Omega$ and $\Omega'$.
To determine the relation between $\omega$ and $\omega'$ we should act with both sides of \eqref{eq:example-3} on some
$X_{\gamma_i}$.
For example, consider again the case we discussed around \eqref{eq:pentagon},
where $\langle \gamma^1, \gamma^2 \rangle=1$ and $\Omega(\gamma^1) = \Omega(\gamma^2) = 1$, with
all other $\Omega(n \gamma_1 + m \gamma_2) = 0$.
Then the 4d wall-crossing formula says that $\Omega'(\gamma^1) = \Omega'(\gamma^2) = \Omega'(\gamma^1+\gamma^2)=1$, with
all other $\Omega'(n \gamma_1 + m \gamma_2) = 0$.
Suppose moreover that
the only nonvanishing $\omega$ are $\omega(\gamma^1, \cdot)$ and $\omega(\gamma^2, \cdot)$.  Then
\eqref{eq:example-3} is satisfied if and only if $\omega'$ are
\begin{align}
 \omega'(\gamma^1,\cdot) &= \omega(\gamma^1,\cdot), \\
 \omega'(\gamma^2,\cdot) &= \omega(\gamma^2,\cdot), \\
 \omega'(\gamma^1 + \gamma^2,\cdot) &= \omega(\gamma^1,\cdot)+ \omega(\gamma^2,\cdot)
\end{align}
with all other $\omega'(m \gamma^1 + n \gamma^2, \cdot)$ vanishing.
To prove this, first note that given \eqref{eq:omega-aff-lin} and
our knowledge of $\Omega$ and $\Omega'$, these relations
are equivalent to the ones where $\cdot$ is replaced by a single $\gamma_i^0 \in \Gamma_i$.
Then choose $\gamma_i^0$ with $\omega(\gamma^1, \gamma_i^0) = \omega(\gamma^2, \gamma_i^0)=0$.  (Concretely, such a $\gamma_i^0$ can be produced by starting with any generic $\gamma_i$
and then forming $\gamma_i^0 = \gamma_i - \omega(\gamma^1, \gamma_i) \gamma^2 + \omega(\gamma^2,\gamma_i) \gamma^1$.)  The check is then trivial, by comparing the action of both sides of \eqref{eq:example-3} on $X_{\gamma_i^0}$.
In many concrete examples a similar trick can be employed: find a convenient reference vector in each $\Gamma_i$ which simplifies $\omega$.

The third case with two distinct indices occurs when $\ell_{\gamma_{ij}}$ coincides with
$\ell_{\gamma_{ji}}$. Note that $\gamma_{ij} + \gamma_{ji} = \gamma \in \Gamma$ so that
all the rays  for $\gamma_{ij} + n \gamma$ and $\gamma_{ji} + n \gamma$ with $n \geq 0$
coincide, and generically there are no other collinear BPS rays.
\insfig{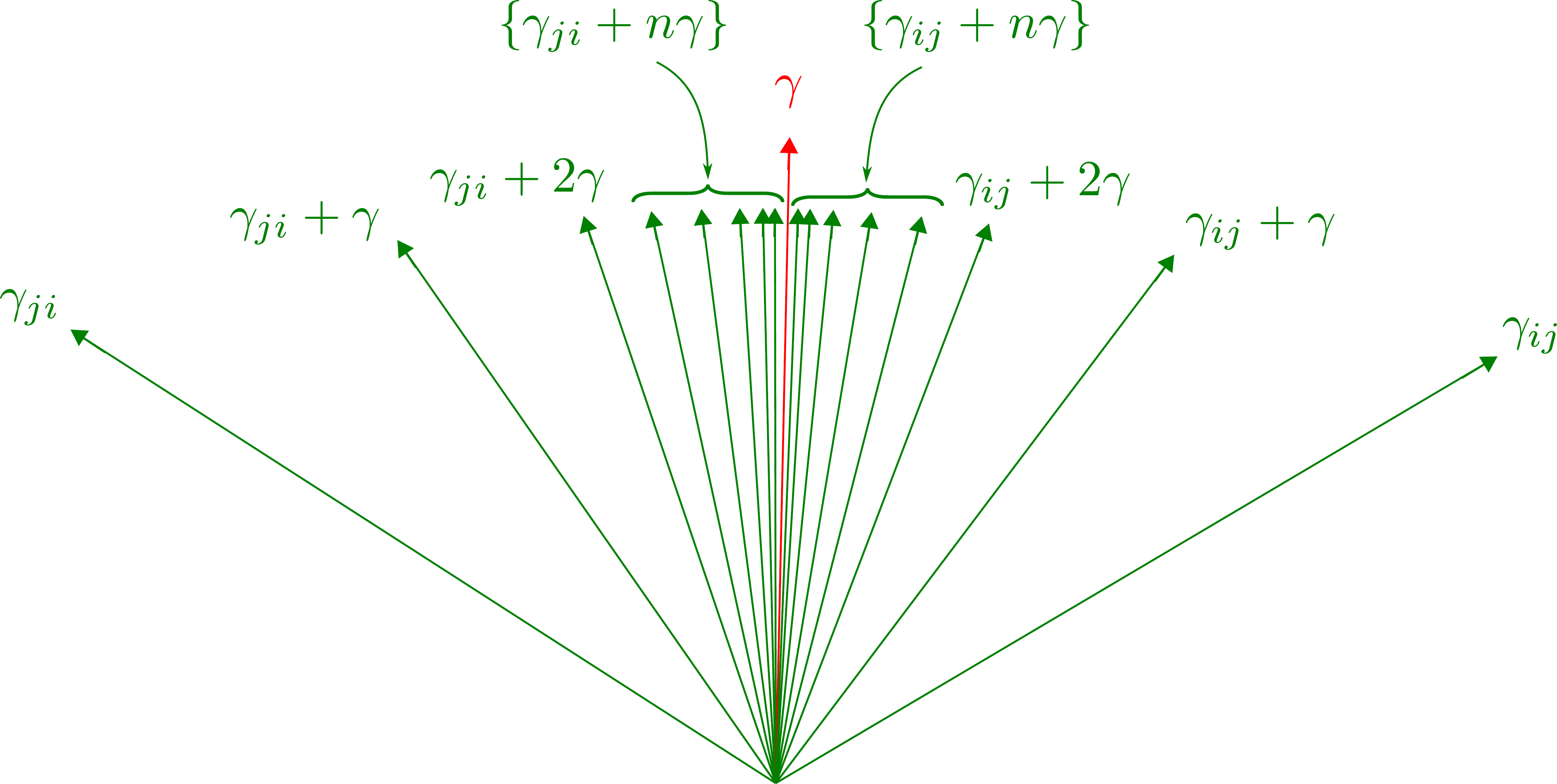}{The configuration of BPS rays which participate in the wall-crossing formula
\protect\eqref{eq:example-4}.  As we approach the wall, all of the rays shown here become aligned.
On the other side of the wall their ordering is reversed.}
The wall-crossing formula
therefore becomes
\begin{equation}\label{eq:example-4}
\prod_{n\nearrow} \CS^{\mu}_{\gamma_{ij}+n\gamma} \prod_{m=1}^\infty \CK_{m\gamma}^\omega \prod_{n\searrow} \CS^{\mu}_{\gamma_{ji}+n \gamma} = \prod_{n\nearrow} \CS^{\mu'}_{\gamma_{ji}+n \gamma}
\prod_{m=1}^\infty \CK_{m\gamma}^{\omega'} \prod_{n\searrow} \CS^{\mu'}_{\gamma_{ij}+n\gamma}
\end{equation}
Here we assumed $\gamma$ primitive, but the derivation goes through with minor modifications in the general case.
The general solution of \eqref{eq:example-4} can be given as follows.
Applying \eqref{eq:example-4} to $X_{\gamma_k}$ for $k\not=i,j$ reveals that
 $\omega'(m\gamma,\gamma_k) = \omega(m\gamma,\gamma_k)$ for $k\not=i,j$. Now
 choose some $\gamma_i\in \Gamma_i$ and define $\gamma_j := \gamma_{ji}+\gamma_i$.
 Then applying \eqref{eq:example-4} to $X_{\gamma_i}$ and $X_{\gamma_j}$
 shows that
\begin{equation}\label{eq:exple-4a}
\begin{split}
\Pi'_i & = \Pi_i + \sigma(\gamma_{ij},\gamma_{ji})\Sigma_{ij} \Sigma_{ji}X_\gamma \Pi_j ,\\
\Sigma'_{ij} \Pi'_i  & = \Sigma_{ij} \Pi_j,  \\
\Sigma'_{ji} \Pi'_i & = \Sigma_{ji} \Pi_j,  \\
\Pi'_j & = \Pi_j -  \sigma(\gamma_{ij},\gamma_{ji})\Sigma'_{ij}\Sigma'_{ji}X_\gamma \Pi'_i,
\end{split}
\end{equation}
where we recall that $\Sigma_{ij}$ was defined in \eqref{eq:def-Sigma-ij} (with similar
definitions for $\Sigma_{ij}', \Sigma_{ji}, \Sigma_{ji}'$) and we introduced the notation
\begin{equation}
\Pi_i := \prod_{m=1}^\infty (1-X_\gamma^m)^{-\omega(m\gamma,\gamma_i)}
\end{equation}
(with similar definitions for $\Pi_i', \Pi_j, \Pi_j'$.)  Given $\mu$ and $\omega$, these
equations determine $\mu'$ and $\omega'$ as follows. The first line can be viewed as
an upper triangular set of equations for $\omega'(m\gamma, \gamma_i)$, $m=1,2,\dots$.
Once we know that, the next two equations determine $\mu'(\gamma_{ij}+n \gamma)$ and
$\mu'(\gamma_{ji} + n \gamma)$. Finally, the fourth equation determines $\omega'(m\gamma, \gamma_j)$.
In fact, it is easy to see that it determines
\begin{equation}\label{eq:sple-sum-ex}
\omega'(m\gamma,\gamma_i) + \omega'(m\gamma, \gamma_j) =
\omega(m\gamma,\gamma_i) + \omega(m\gamma, \gamma_j).
\end{equation}
If, as is often the case, we are only interested in the differences $\omega(\cdot , \gamma_{ij})$, then
the transformation formulae can be written more directly as a transformation reminiscent of modular
forms:
\begin{equation}
\begin{split}
\Pi'_{ji} & = \Delta^{-2}  \Pi_{ij},  \\
\Sigma'_{ij} & = \Delta^{-1} \Sigma_{ij}, \\
\Sigma'_{ji} & =\Delta^{-1} \Sigma_{ji}, \\
\end{split}
\end{equation}
where $\Delta:=  \Pi_{ji} +   \sigma(\gamma_{ij},\gamma_{ji})\Sigma_{ij} \Sigma_{ji}X_\gamma$.

Two special cases of the above solution are worth noting:

\begin{enumerate}

\item  If $\sigma(\gamma_{ij},\gamma_{ji})\Sigma_{ij} \Sigma_{ji} = 1$
and $\omega(\gamma,\gamma_i) = \omega(\gamma, \gamma_j) - 1$
and $\omega(m\gamma,\gamma_i)= \omega(m\gamma,\gamma_j)=0$ for $m>1$, then we  have
\begin{equation}\label{eq:example-4-spec}
\CS^{\mu}_{\gamma_{ij}} \CK_\gamma^\omega \CS^\mu_{\gamma_{ji} }= \CS^{\mu'}_{\gamma_{ji}} \CK_\gamma^{\omega'} \CS^{\mu'}_{\gamma_{ij}}
\end{equation}
with $\Sigma'_{ij}= \Sigma_{ij}$, $\Sigma'_{ji}=\Sigma_{ji}$ and
 $\omega'(\gamma,\gamma_j) = \omega(\gamma,\gamma_i)$ and $\omega'(\gamma, \gamma_i) = \omega(\gamma,\gamma_j)$.

\item  If $\omega(\gamma, \gamma_i)=\omega(\gamma, \gamma_j)=0$ and
 $\omega(m\gamma,\gamma_i)= \omega(m\gamma,\gamma_j)=0$ for $m>1$ and also
$\sigma(\gamma_{ij},\gamma_{ji})\Sigma_{ij} \Sigma_{ji}=-1$,
 then we have $\omega'(\gamma,\gamma_i) = -1$,  $\omega'(\gamma, \gamma_j)=1$,
 $\omega'(m\gamma, \gamma_i) = \omega'(m\gamma, \gamma_j)=0$ for $m>1$ and
 \begin{equation}
 \begin{split}
 \Sigma'_{ij} & = \mu(\gamma_{ij}) (1-X_\gamma)^{-1} \\
 \Sigma'_{ji} & = \mu(\gamma_{ji})  (1-X_\gamma)^{-1}
 \end{split}
 \end{equation}
corresponding to
\begin{equation} \label{eq:wallcp1}
\CS^{\mu}_{\gamma_{ij}} \CS^\mu_{\gamma_{ji}} = \prod_{\nearrow} \CS^{\mu'}_{\gamma_{ji}+n \gamma} \CK^{\omega'}_\gamma \prod_{\searrow } \CS^{\mu'}_{\gamma_{ij} + n \gamma}.
\end{equation}
Some physical realizations of this formula are discussed in \S\ref{subsec:cp1-sigma} below.

\end{enumerate}

The cases where $3$ or $4$ indices $i,j,k,l$ coincide are subsumed by the above
examples.

\section{A physical interpretation}\label{sec:Physical-Interpret}

So far we have written out an abstract wall-crossing formula which we called the 2d-4d WCF.
In this section we describe a natural physical setting where the 2d-4d WCF arises, and thus
justify its name. The main idea is to study supersymmetric surface defects in
$d=4$, $\CN=2$ field theories.

In \S\S \ref{subsec:kinematics} and \ref{subsec:examples-surface-defects} we give basic
definitions of the relevant surface defects in the UV theory. Then \S \ref{eq:subsec:-IR-DESC}
discusses in detail the IR description of surface defects. The main upshot is to give a
physical realization of the vacuum groupoid $\IC[\IV]$ and in particular of the charge
torsors $\Gamma_i$. Then \S \ref{sec:gen-example} narrows the focus to certain classes
of surface defects, and illustrates some important subtleties that can arise in trying to
define the charge torsors $\Gamma_i$ in important physical situations. These subtleties
include certain ``anomalies'' which can prevent the surface defects from being physically
sensible.
In \S \ref{subsec:BPS-Deg-WC} we return to the general theory and give
definitions of the BPS data $\mu$ and $\omega$ in our physical context.
Finally, in \S \ref{subsec:Anomaly-Cancelation} we return to the potential
anomalies in defining surface defects and illustrate an anomaly cancellation
mechanism.

\subsection{Surface defects:  kinematics}\label{subsec:kinematics}

We consider an $\CN=2$ gauge theory in the presence of a surface defect $\bS$, satisfying the
same conditions as spelled out in \cite{Gaiotto:2009fs}.  The
surface defect is localized at $x^1=x^2=0$ in Minkowski space.  In the ultraviolet
it preserves a subalgebra of the superconformal $\CN=2$ algebra. Similarly to the
case of the line defects studied in \cite{Gaiotto:2010be}, the preserved subalgebra is the
fixed subalgebra of an involution. In this case we combine reflection in the $x^1$-$x^2$ plane,
i.e. a rotation by $\pi$ around the $x^3$ axis, with a rotation by $\pi$ around some
axis in ``R-symmetry space.'' The bosonic part of the superconformal algebra is
$so(2,2) \oplus so(2)_{12}  \oplus (u(1)_r \oplus u(1)_R)$, where
the first summand corresponds to
conformal transformations in the $x^0$-$x^3$ plane, the second summand corresponds to rotations
around the surface defect, and the third summand is a surviving $u(1) \oplus u(1)$ from the
$su(2)\oplus u(1)_R$ $R$-symmetry with $u(1)_r \subset su(2)$. The $u(1)_R$ subalgebra
is in general anomalous.  In the conventions of Appendix A of \cite{Gaiotto:2010be}, the
preserved Poincar\'e  supersymmetries are $Q_1^1$, $Q_2^2$, and their respective Hermitian conjugates
$\bar Q_{\dot 1}^2$ and $\bar Q_{\dot 2}^1$.

Thus the surface defect preserves $(2,2)$ SUSY on its ``worldsheet.''
The $d=4$ supermultiplets can be decomposed into
multiplets of this $d=2$ $(2,2)$ superalgebra.  Of central importance in what follows
is that part of the $d=4$ vectormultiplet can be used to define a twisted chiral multiplet
for this superalgebra.  See Appendix \ref{app:d2d4-multiplets} for details.

In the literature it is common to use the term ``surface operator,'' but we believe that  ``surface defect'' is a better name.
It is true that a surface defect wrapping a compact Euclidean surface acts as an operator on the Hilbert space of the bulk theory.
However, this operator carries less information than the surface defect itself.  For example, the operator does not detect
2d degrees of freedom living at the defect which do not interact with the bulk theory.

Finally, as in \cite{Gaiotto:2009fs}, we assume that in the IR the 4d Abelian
gauge fields exhaust all massless degrees
of freedom.  In particular there are no massless degrees of freedom confined to $\bS$.
Throughout this paper we let $n$ denote the number of vacua of $\bS$, and $r$ the rank of the 4d gauge group.

\subsection{Examples of surface defects}\label{subsec:examples-surface-defects}

How can surface defects of the desired sort be described?

The bulk vectormultiplet scalars
restricted to a two-dimensional subspace
transform as twisted chiral fields under the $(2,2)$ SUSY algebra (the bulk $u(1)_R$
becomes the axial $u(1)_A$ for the 2d SUSY algebra).
 So one way to write a Lagrangian
describing a surface defect is simply to build a twisted superpotential $\CW$ out of these
twisted chiral fields, and integrate it over two dimensions.
Another way of defining surface defects is to introduce some purely 2d degrees of freedom, with $(2,2)$ supersymmetry
and some continuous global ``flavor'' symmetries.  These degrees of freedom can then be coupled to the 4d theory
as follows:  the restriction of the 4d gauge fields to two dimensions gives 2d gauge fields,
and we can use these gauge fields to gauge the flavor symmetries.  In this case the 4d vectormultiplet
scalars behave as twisted masses from the 2d point of view.

Naively the $\CW$ we use would have to be gauge invariant.  However, there is an interesting alternative
\cite{Gukov:2006jk}.
One can reduce the gauge group $G$ at the defect down to a
subgroup $H$ whose center contains at least one $U(1)$ factor (more properly, a ``Levi subgroup'').
For simplicity here let us consider just the case of a single $U(1)$.
One can then write a simple FI-like twisted superpotential,
linear in the $U(1)$ part $\Phi^{U(1)}$ of the bulk vectormultiplet scalars:
\begin{equation} \label{eq:fi-sup}
\CW = t_{\rm uv} \, \Phi^{U(1)}.
\end{equation}

The effect of this superpotential can be conveniently expressed in terms of two real couplings ($\alpha$, $\eta$),
the ``Gukov-Witten parameters'' \cite{Gukov:2006jk}.  They are related to $t_{\rm uv}$ by
\begin{equation} \label{eq:gw-first}
t_{\rm uv} = \eta_{\rm uv} + \tau_{\rm uv} \alpha_{\rm uv}
\end{equation}
where $\tau_{\rm uv}$ is the complexified gauge coupling in the $U(1)$ we are considering.  (If $G$ is a simple group then
$\tau_{\rm uv}$ is just the coupling for $G$.)
$\eta_{\rm uv}$ is a 2d theta angle coupled to the magnetic field in the $U(1)$ factor.
$\alpha_{\rm uv}$ fixes a boundary condition
for the $U(1)$ part of the gauge connection near the defect, of the form
\begin{equation}
 A^{U(1)} \sim \alpha_{\rm uv}\,\de\varphi,
\end{equation}
where $\varphi$ is the angle in a system of polar
coordinates in the plane transverse to the defect.

So far we have implicitly worked with a cutoff field theory and
not considered the constraint of UV completeness.  Requiring the surface defect to have a good
continuum limit turns out to be a significant further
restriction.  The twisted superpotential $\CW$, for example, has to have dimension $1$,
and hence can only be linear in the 4d vectormultiplet scalars:
if the 4d theory is nonabelian, $\CW$ cannot include even the simplest gauge invariant
operators such as $\mathrm{Tr} \, \Phi^2$.  The Gukov-Witten defect just described gets around this problem
by breaking the gauge symmetry, and does make good sense in the UV.
Another possibility is to take a 2d $(2,2)$ theory which is defined in the UV (say, a gauged linear sigma model)
and couple it to the restrictions of the 4d gauge fields.

In the original context of  \cite{Gukov:2006jk}, ${\cal N}=4$ SYM, both $\eta_{\rm uv}$ and $\alpha_{\rm uv}$
are periodic, but in the current context of surface operators in ${\cal N}=2$ theories the parameter space is modified by 2d instanton effects
controlled by a 2d instanton factor $e^{2 \pi \I t_{\rm uv}}$.
Since the couplings run, and in particular   $\tau_{\rm uv}$ goes to $\I \infty$,
some renormalization is needed in defining $t_{\rm uv}$.

Finally, we should mention that the coupling of the surface defect to the bulk hypermultiplets
might  also be important. Unfortunately, it is also poorly understood.
When expanded in terms of 2d fields, the bulk hypermultiplets give rise only to chiral multiplets,
which cannot directly enter the twisted superpotential couplings which are usually the focus of our computations.
Still, the bulk hypermultiplets can affect the definition and properties of the surface defect in the UV
in more subtle ways.  Basic examples based on brane constructions suggest that some extra discrete structure
is hidden in the definition of a ``Gukov-Witten defect'' in the presence of matter hypermultiplets.
We will come back to some of these issues
when we specialize to theories and surface defects with a six-dimensional construction,
in \S \ref{sec:Hitchin-Expl} (see the introduction to  \S \ref{sec:Hitchin-Expl} and \S \ref{subsec:Lag-Desc}).
The possibility of gauging discrete flavor symmetries will play an important role.

\subsection{Low energy (IR) description of the 2d-4d system}\label{eq:subsec:-IR-DESC}

Now let us discuss what these combined 2d-4d systems look like in the IR.
We begin by recalling the 2d and 4d stories separately and then explain how they
can be combined.

\subsubsection{Massive 2d theories in the IR} \label{sec:2d-review}

We begin by briefly recalling some of the physics of
2d $\CN=(2,2)$ theories with no massless degrees of freedom in the IR, and
discrete vacua labeled by $i = 1, \dots, n$.

In such a theory one can consider solitons which interpolate between vacuum $i$ and vacuum
$j$ ($ij$-solitons for short).
Such a soliton obeys a BPS bound of the form
$M \ge \abs{Z_{ij}}$, where $Z_{ij}$ is the central charge, given by
\begin{equation} \label{eq:soliton-Z}
 Z_{ij} = \CW_i - \CW_j
\end{equation}
where $\CW_i$ and $\CW_j$ are the values of the IR superpotential in vacua $i$ and $j$ respectively.
In particular, BPS $ij$-solitons have mass exactly $\abs{Z_{ij}}$.

If the theory includes flavor symmetries, then there is a generalization of the above,
which will play an important role in this paper:  in addition to the topological charge
determined by the pair $(i,j)$ of vacua, particles can carry charge under the
flavor symmetries.  Letting $\gamma$ denote an element in the lattice $\Gamma$ of flavor charges,
the total charge of a particle is given by a triple $(i, j, \gamma)$.

We now introduce a bit of superfluous-looking notation which will be very convenient when we
treat coupled 2d-4d systems:
we let $\Gamma_{ij}$ denote the set of all charges
$\gamma_{ij} = (i, j, \gamma)$.  Of course,
$\Gamma_{ij}$ is trivially isomorphic to $\Gamma$, via the map
$(i,j,\gamma) \mapsto \gamma$.

Once we have flavor symmetries, the theory can be deformed by ``twisted masses''
\cite{Hanany:1997vm}, whose IR effect is to change the central charge to
\begin{equation} \label{eq:soliton-Z-2}
 Z_{\gamma_{ij}} = \CW_i - \CW_j + m \cdot \gamma,
\end{equation}
where $m$ is some linear functional on $\Gamma$.
Twisted masses can be rather important for the physics of a 2d system.
For example, in the presence of twisted masses
we might have BPS $ii$-solitons (or more simply ``BPS particles in vacuum $i$''),
carrying only the flavor charges.
These particles play an important role in the wall-crossing phenomena.
This contrasts with the behavior of 4d particles with pure flavor
charge, which are invisible to the 4d IR dynamics and wall-crossing.

When we consider 2d-4d coupling, we will
find an analogous but more intricate situation, where the lattice $\Gamma$ includes 4d charges
as well as 2d flavor charges.

\subsubsection{4d gauge theories in the IR} \label{sec:4d-review}

We now quickly review some standard facts about the IR behavior of 4-dimensional $\N=2$
gauge theories.

The vacua of the 4d theory are parameterized by the Coulomb branch $\CB$, an $r$-dimensional
complex manifold.  (Other branches
of the moduli space will play no role in our discussion.)  For a generic vacuum
$u\in \CB$, there is an unbroken abelian gauge group $\CG_{\rm ab} \simeq U(1)^r$.
There is a ``singular locus'' in $\CB$
where the description in terms of pure $\CG_{\rm ab}$ gauge theory breaks down; let $\CB^*$ be its complement.
There might also be a continuous group of flavor symmetries; for generic values of the mass parameters
of the theory, this flavor group will be of the form $U(1)^{n_f}$.\footnote{In general,
the full flavor group could have several disconnected components.
We elaborate on that in \S\ref{sec:flavor-phase}.}

There is a local system of lattices $\Gamma \to \CB^*$, of rank $2r + n_f$.
$\Gamma$ is the ``charge lattice'' of the theory.\footnote{As
explained in \cite{Gaiotto:2010be} this should be taken to be
the lattice of all IR charges $\Gamma_L$ where $L$ ranges over
a maximal set $\CL$ of mutually  local line defects in the theory. This
lattice was denoted $\Gamma_{\CL}$ in \cite{Gaiotto:2010be}.}
It arises as an extension
\begin{equation}\label{eq:lattice-ext-seq}
0 \rightarrow \Gamma_f \rightarrow \Gamma \rightarrow \Gamma_g \rightarrow 0,
\end{equation}
where $\Gamma_f$ and $\Gamma_g$ are respectively the lattices of charges under the
continuous flavor and gauge symmetries.
$\Gamma_f$ is fibered trivially over $\CB^*$, but $\Gamma_g$
generally is not:  the gauge charges can experience monodromy $\gamma_g \to M(\gamma_g)$,
for some $M \in \Aut(\Gamma_g)$, under parallel transport around a nontrivial loop in $\CB^*$.
Moreover, the sequence \eqref{eq:lattice-ext-seq} can be split locally but perhaps not globally.
What this means in more physical
terms is that the splitting of a charge $\gamma$ into flavor and
gauge charges, $\gamma = \gamma_f \oplus \gamma_g$,
is not globally well-defined:  monodromy around a loop in $\CB^*$ takes
$\gamma_f \to \gamma_f + N (\gamma_g)$ for some $N \in \Hom(\Gamma_g,\Gamma_f)$.
These phenomena were noted already at the birth of Seiberg-Witten theory
\cite{Seiberg:1994aj}.

The lattice $\Gamma_g$ is equipped with the Dirac-Schwinger-Zwanziger pairing of electric and magnetic charges:
this is an integral, anti-symmetric, monodromy invariant
pairing which we write as $\inprod{\cdot, \cdot}$.
We will assume that $\Gamma_g$ is self-dual under this pairing.\footnote{This
assumption can be justified in Lagrangian $\CN=2$ theories
at weak coupling. See in particular equation (2.13) of
\cite{Gaiotto:2010be}.  We believe it could be relaxed with some
minor adjustments to our analysis.}
We define an integral antisymmetric pairing $\inprod{\cdot,\cdot}$
on $\Gamma$, whose annihilator is the sublattice $\Gamma_f$, by projecting both arguments to
$\Gamma_g$ and using the nondegenerate pairing on $\Gamma_g$.

When writing local formulae it is often useful to choose a symplectic basis $e_I, e^I$
for $\Gamma_g$, with $\langle e^I, e_J \rangle = \delta^{I}_{~J}$, $I,J=1,\dots r$.
We then write IR gauge charges $\gamma_g \in \Gamma_g$ as
$\gamma_g = p^I e_I + q_I e^I$, with $p^I, q_I \in \IZ$.

An essential piece of IR data in the 4d theory is a globally defined section $Z \in \Hom(\Gamma, \IC)$, known as the central
charge function.  For each local section $\gamma$ of $\Gamma$, we thus have a function $Z_\gamma$ on a patch of $\CB^*$, obeying
\begin{equation} \label{eq:z-add}
 Z_{\gamma + \gamma'} = Z_\gamma + Z_{\gamma'}.
\end{equation}
BPS states of charge $\gamma$ have mass $\abs{Z_\gamma}$.

The $Z_{\gamma_f}$ for $\gamma_f \in \Gamma_f$ are actually \ti{constant} functions on
the whole $\CB^*$.  They can be organized
as $Z_{\gamma_f} = m \cdot \gamma_f$, where $m$ is a ($u$-independent) complex-valued
linear functional on $\Gamma_f$, i.e. $m \in \Gamma_f^* \otimes_\IZ \IC$.
$m$ is determined by the choice of mass parameters in the theory.
If we choose a duality basis and a splitting of \eqref{eq:lattice-ext-seq}, we
may decompose $\gamma = p^I e_I + q_I e^I + \gamma_f$, and correspondingly write
\begin{equation}
Z_{\gamma}(u) = p^I a_{D,I}(u) + q_I a^I(u) + m\cdot \gamma_f,
\end{equation}
for some local holomorphic functions $a^I$ and $a_{D,I}$.

As $u$ varies in $\CB$, $Z(u)$ varies in a Lagrangian subspace of $\Gamma^*\otimes_\IZ \IC$.
Once a duality basis has been chosen, we can write a local generating function $\CF(a^I)$ for this subspace
(also known as a prepotential), i.e. we have the standard relation
\begin{equation}\label{eq:a-dual}
 a_{D,I} = \frac{\partial \CF}{\partial a^I}.
\end{equation}

The $a^I$ have a simple physical interpretation, as follows.  Having chosen a duality basis over some patch of $\CB^*$,
we have also fixed a specific Lagrangian description of the effective
IR theory in that patch, as a $U(1)^r$ gauge theory.
The $a^I$ are the vacuum expectation values of the scalar
components of the $r$ $U(1)$ vector multiplets appearing in this Lagrangian.  The Lagrangian can
be written explicitly as
\begin{equation}\label{eq:4d-only}
\begin{split}
 S &= \int \de^4 \theta\,\de^4 x \, \CF \\
 &=  - \frac{1}{4 \pi} \int \left(Y_{IJ} F^I \star F^J    + X_{IJ} F^I F^J - Y_{IJ} D^I \star D^J \right) + \cdots
 \end{split}
 \end{equation}
where in the first line $\CF$ is evaluated on the superfields corresponding to $a^I$,
in the second line we have dropped terms which will not be used in explicit computations below,
and the matrix functions $X,Y$ of $a^I$ are the real and imaginary parts of the
coupling
\begin{equation}
  \tau_{IJ} = \frac{\partial a_{D,I}}{\partial a^J} = X_{IJ} + \I Y_{IJ}.
\end{equation}
The auxiliary fields $D^I_{AB}$ form an $su(2)_R$ triplet. Since we will eventually
add a surface defect which
breaks $su(2)_R \to u(1)_r$, one component, namely $D^I_{12}$, is distinguished, and is denoted
above simply by $D^I$.

\subsubsection{2d-4d theories in the IR: Local picture, Gukov-Witten parameters,
and the torsors $\Gamma_{ij}$}

Now we discuss the IR structure of the combined 2d-4d system.

A vacuum for the combined system is determined by a pair $(u, i)$ of
a point in the Coulomb branch $\CB$ of the bulk 4d theory
and a vacuum for the surface defect.
In a neighborhood of any specific $(u,i)$,
it is possible to write an IR Lagrangian for the
system in terms of a holomorphic prepotential $\CF$
and a holomorphic twisted superpotential $\CW_i$,
extending \eqref{eq:4d-only}:
\begin{equation} \label{eq:4d-action}
 S = \int \de^4 \theta\,\de^4 x \, \CF + \left( \frac{1}{2} \int \de^2 \tilde\vartheta\,\de x^3\,\de x^0\,\CW_i + \mathrm{c.c.} \right).
\end{equation}
The amount of supersymmetry we have here also allows unprotected D-terms integrated over the
defect.  We will restrict ourselves to considering objects which depend on the twisted F-terms only.\footnote{The
F-term integration measure can be written a bit more covariantly as the four-form in superspace
\begin{equation} n_{IJ} \de\theta^I_\alpha  \de\theta^J_\beta (C \gamma_{\mu \nu})^{\alpha \beta} \de x^\mu \de x^\nu,\end{equation}
where $I,J$ are $SU(2)_R$ doublet indices and $n_{IJ}$ is the direction of the $u(1)_V$ subalgebra in $SU(2)_R$.}

The twisted superpotential $\CW_i$ is a function of the associated twisted
chiral multiplet $\Upsilon$, defined in \eqref{eq:vm-to-tw-cm}.  It might also depend on
the mass parameters of the 4d theory, and possibly extra twisted mass parameters associated to
flavor symmetries of the UV surface defect.  Indeed, some flavor charges might well be carried
only by the 2d degrees of freedom, and be absent in the 4d theory.  The corresponding mass parameters
are twisted masses.

Notice that $\CW_i$ might become singular (and the simple Lagrangian description break down) at new singular loci in $\CB^*$, where the 2d physics becomes more intricate.  We will from now on redefine $\CB^*$ to exclude these loci.

One important place where the twisted superpotential plays a role is in the physics of BPS solitons.
So let us consider an $ij$-soliton.  We introduce the symbol $\gamma_{ij}$ for its total charge,
and let $\Gamma_{ij}$ denote the set of all possible charges $\gamma_{ij}$.
As we reviewed in \S \ref{sec:2d-review}, in a pure 2d theory one could decompose
any $\gamma_{ij}$ into the ``topological charge'' $(i,j)$
plus a pure flavor charge $\gamma$.  Here we have a similar story:
$\gamma_{ij}$ can be decomposed into the ``topological charge'' $(i,j)$
plus a charge $\gamma$, which now can
include 4d gauge charge as well as flavor charge.
BPS solitons obey a mass formula of the form $M = \abs{Z}$,
with $Z$ written as a sum of two pieces (cf. \eqref{eq:soliton-Z-2}):
\begin{equation} \label{eq:z-split}
Z_{\gamma_{ij}} = \CW_i - \CW_j + Z_\gamma.
\end{equation}
However, in contrast to the pure 2d case, we will discover momentarily that
the map from $\Gamma_{ij}$ to $\Gamma$
taking $\gamma_{ij} \mapsto \gamma$ is not canonically defined, and neither is the
decomposition \eqref{eq:z-split}.

Let us try to understand the precise relation between $\Gamma_{ij}$ and $\Gamma$.
Suppose we bring a bulk line defect, with some charge $\gamma'$, close to a soliton
with charge $\gamma_{ij}$.  When viewed from a long distance the resulting object
looks like a soliton carrying charge $\gamma' + \gamma_{ij}$; so
there is an addition operation $\Gamma \times \Gamma_{ij}  \to \Gamma_{ij}$ (and similarly
in the other order).  Moreover, bringing an $ij$
soliton and a $jk$ soliton together shows that there is an addition operation
$\Gamma_{ij} \times \Gamma_{jk} \to \Gamma_{ik}$.  Finally, consider
``$ii$ solitons,'' i.e. particles in the vacuum $i$.  If we have such a particle
we may put it onto a surface defect in vacuum $i$ with its spatial direction compactified.
At long distance we can neglect the circumference of this cylinder; it thus looks like a line defect,
so the charge it carries must lie in $\Gamma$ (which by definition is the lattice of
all charges of line defects in the 4d theory).  So we conclude that $\Gamma_{ii}$ is naturally
isomorphic to $\Gamma$, for every $i$.  It follows that each $\Gamma_{ij}$ is
a \ti{torsor} for $\Gamma$.\footnote{To see this, just choose any $\gamma_{ji} \in \Gamma_{ji}$:  then we get
a map $\Gamma_{ij} \to \Gamma_{ii} = \Gamma$ defined by adding $\gamma_{ji}$, which is injective and commutes
with the $\Gamma$ actions on both sides.}
The reader should note well that
the properties of $\Gamma_{ij}$ we have just verified are exactly the ones
which were used in formulating the 2d-4d WCF in \S \ref{subsec:Formal-Statement}.

We emphasize that in the above the charge $\gamma$ might include flavor contributions
as well as gauge charges.  In particular it might include charges for flavor symmetries
which act only on the surface defect $\bS$, not on the 4d theory.
We always extend the lattices $\Gamma_f$ and $\Gamma$ to include such
2d flavor charges as necessary; $Z_\gamma$
then includes the corresponding contribution from twisted mass couplings on $\bS$.

The charges measured by $\Gamma_{ij}$ are the same kind of charges measured by the bulk
charge lattice $\Gamma$ (they would be obtained by
integrating the appropriate currents).  From this point of view the difference between
$\Gamma_{ij}$ and $\Gamma$ is only that the charges in $\Gamma_{ij}$ need not be properly
quantized.  In other words, there is an embedding $\Gamma_{ij} \subset \Gamma_\IR$;
so $\Gamma_{ij}$ looks like a shifted copy of the charge lattice $\Gamma$.

Now we should explore some consequences of the Lagrangian couplings \eqref{eq:4d-action}.
Very similarly to \eqref{eq:a-dual}, differentiating $\CW_i$ once yields
\begin{equation}\label{eq:IR-GW}
\frac{\partial \CW_i}{\partial a^I} = t_I =: \eta_I + \tau_{IJ} \alpha^J,
\end{equation}
where $\eta_I, \alpha^I$ are real functions of the $a^I$, and
we suppress their $i$-dependence. We claim that these
parameters should be viewed as coordinates in a  symplectic vector space, so the object
\begin{equation}
 \nu := \alpha^I e_I + \eta_I e^I
\end{equation}
is duality invariant.  Moreover,
we claim that $\alpha^I, \eta_I$  should be interpreted as IR Gukov-Witten
parameters for the surface defect.\footnote{We emphasize that this is true
whether or not the surface defect was defined as a Gukov-Witten defect in the UV.  Moreover,
even if it were so defined, the UV Gukov-Witten parameters
are very different from
the ones we see in the IR unless everything is weakly coupled and instanton corrections are suppressed.}

In order to establish our claims we introduce the duality invariant gauge field $\IF = e_I F^I + e^I G_I$
(where $e_I, e^I$ are a symplectic frame as introduced in \S \ref{sec:4d-review}).
Then the relevant terms in the action \eqref{eq:4d-action} can be written as
\begin{equation}\label{eq:explct-action}
\begin{split}
& - \frac{1}{4 \pi} \int \left(Y_{IJ} F^I \star F^J    + X_{IJ} F^I F^J - Y_{IJ} D^I \star D^J \right) + \\
& +
 \int \de x^0 \de x^3 \left( (\eta_I + X_{IJ} \alpha^J) F_{03}^I   + \alpha^I Y_{IJ} (F_{12}^J - D^J) \right).
\end{split}
\end{equation}
The equations of motion
following from \eqref{eq:explct-action} admit a cylindrically symmetric solution
$F_{12}^I = 2\pi \delta \alpha^I$ and $F_{03}^I =0$, where  $\delta := \delta(x^1) \delta(x^2) $.
This implies that $\alpha^I$ determines the holonomy of the gauge field $A^I$ according
to the Gukov-Witten prescription.  Moreover, properly dualizing the gauge field in
the presence of electric sources gives
\begin{equation}\label{eq:G-with-source}
\begin{split}
G_{12I} & = Y_{IJ} F_{03}^J - X_{IJ} F_{12}^J + 2\pi\delta  (\eta_I + X_{IJ} \alpha^J), \\
G_{03I} & = - Y_{IJ} F_{12}^J - X_{IJ} F_{03}^J + 2\pi\delta  Y_{IJ}  \alpha^J,
\end{split}
\end{equation}
and hence the duality invariant gauge field is
\begin{equation}\label{eq:solenoid-gf}
\IF_{12} = 2 \pi \delta \nu,
\end{equation}
with
all other components   $\IF_{mn}=0$. In particular, $\eta_I$ determines the holonomy
of the dual gauge field $A_{D,I}$.
Integrating out the auxiliary term and substituting the solution \eqref{eq:solenoid-gf} one
finds that the on-shell action is \emph{precisely zero}, thus showing that the configuration
is in fact supersymmetric.
Thus, in the IR the surface defect appears to be an infinitely thin solenoid with
field configuration \eqref{eq:solenoid-gf}.

\subsubsection{Superpotentials and their shifts: the torsors $\Gamma_i$}\label{subsubsec:SuperPot-Shift}

We must stress that the Gukov-Witten parameters used in \eqref{eq:IR-GW} and
\eqref{eq:solenoid-gf} are real valued rather than periodic,
that is, $\nu$ is a local section
of $\Gamma_g \otimes \IR$ over $\CB^*$. This raises an apparent puzzle since large
gauge transformations outside of the solenoid induce shifts
\begin{equation}\label{eq:lgt-shft}
\nu \to \nu + \gamma_g
\end{equation}
with $\gamma_g \in \Gamma_g$, and hence our superpotentials are
not gauge invariant.

Usually, the shift \eqref{eq:lgt-shft} is not considered a puzzle.
It is usually said that the  physical effects of the solenoid
are only measurable through Aharonov-Bohm phases of test particles transported
around loops linking the solenoid. Recall that the
wavefunction of a test particle with charge
$\gamma^{\rm test}$ which is adiabatically transported around the surface defect picks up a phase
$e^{2 \pi \I \langle\nu, \gamma_g^{\rm test}\rangle}$ where
$\gamma_g^{\rm test}$ is the projection of $\gamma^{\rm test}$
in \eqref{eq:lattice-ext-seq}.\footnote{For the moment we are assuming that
test particles carrying pure flavor charge, $\gamma = \gamma_f$, do not pick up any phase; we will loosen this
restriction in \S\S \ref{subsubsec:Alg-Int-Mirr} and \ref{sec:flavor-phase} below.}
Of course, the Aharonov-Bohm
phase determines $(\alpha^I, \eta_I)$ only up to shifts $(\alpha^I, \eta_I) \to (\alpha^I + m^I, \eta_I + n_I)$ with $m^I, n_I \in \IZ$,
or equivalently up to \eqref{eq:lgt-shft}.

In our story, by contrast, we need $\nu \in \Gamma_g \otimes \IR$ and by
\eqref{eq:IR-GW} it follows that the shifts \eqref{eq:lgt-shft}
correspond to  shifting $\CW_i \to \CW_i + n_I a^I + m^I a_{D,I}$, i.e.
\begin{equation} \label{eq:w-shift}
\CW_i \to \CW_i + Z_{\gamma'}
\end{equation}
where we have now lifted $\gamma'_g \in \Gamma_g$ to some $\gamma' \in \Gamma$, i.e.
we have included a further possible constant shift coming from the flavor charges.
The resolution of the apparent puzzle is that a specific choice of superpotential
is like a specific gauge choice.

How does this ambiguity of $\CW_i$ affect the central charge functions as given in \eqref{eq:z-split}?  For a fixed
charge $\gamma_{ij}$, $Z_{\gamma_{ij}}$ is a physical quantity which should have no ambiguity.  The shift
\eqref{eq:w-shift} leaves $Z_{\gamma_{ij}}$ invariant
only if it is compensated by shifting the $\gamma$ in \eqref{eq:z-split} by
$\gamma \to \gamma - \gamma'$.  This makes sense:  by shifting $\nu$ we are changing our prescription
for the flux through the solenoid, but the flux sourced by the BPS
soliton outside the solenoid should be invariant, so we need to shift the (quantized) total flux.
This shift explains our earlier remark that the map from $\Gamma_{ij}$ to $\Gamma$ which
takes $\gamma_{ij} \mapsto \gamma$ is not canonical:  it depends on how we choose to fix
the ambiguities of $\CW_i$ and $\CW_j$.

If one is considering the physics of massive BPS particles charged under the abelian gauge fields,
the shifts $(\alpha^I, \eta_I) \to (\alpha^I+ m^I, \eta_I+ n_I)$ change the boundary conditions for the
wavefunctions of the charged particles at the solenoid.  We  discuss the consequences in more detail in \S
\ref{app:Landau-Levels}.

The precise form \eqref{eq:w-shift} of the ambiguity of $\CW_i$ suggests
a notation which will turn out to be very useful below.  We let $\Gamma_i$ be a set
parameterizing possible choices of $\CW_i$.
Then for any $\gamma_i \in \Gamma_i$,
we denote the corresponding $\CW_i$ as $Z_{\gamma_i}$.
The possibility of a shift \eqref{eq:w-shift}
then just means that for any $\gamma_i \in \Gamma_i$ and $\gamma \in \Gamma$
we can define a new object $\gamma_i + \gamma \in \Gamma_i$,
and we have an equation parallel to \eqref{eq:z-add},
\begin{equation}
 Z_{\gamma_i + \gamma} = Z_{\gamma_i} + Z_\gamma.
\end{equation}
These are not the only possible modifications of $\CW_i$:
after all, we can always shift all the $\CW_i$ by the same constant, without changing any of the physical
central charges $Z_{\gamma_{ij}}$.
For our purposes to follow, though, we will not need to include all possible choices of $\CW_i$ in
$\Gamma_i$.
We will only need that $\Gamma_i$ is closed under all
allowed shifts \eqref{eq:w-shift}.  We will therefore
take $\Gamma_i$ to be a $\Gamma$-torsor.

Earlier we considered another $\Gamma$-torsor $\Gamma_{ij}$ which keeps track of the charges
of $ij$-solitons.  It follows from \eqref{eq:z-split} that the
$\Gamma_i$ we just introduced is simply related to $\Gamma_{ij}$:
\begin{equation} \label{eq:diff-torsor-2}
 \Gamma_{ij} = \Gamma_i - \Gamma_j.
\end{equation}
(We defined this notation around \eqref{eq:diff-torsor}.)
This suggests a nice physical interpretation of $\Gamma_i$:  it is a set of possible charges
for \ti{boundaries} of the surface
defect $\bS$ in vacuum $i$.  At least as far as the IR charges are concerned, then, \eqref{eq:diff-torsor-2}
says that a general interface between $\bS$ and $\bS'$ could
be obtained by bringing a boundary of $\bS$ close to a boundary of $\bS'$.
We will explore the properties of boundaries
and interfaces further in \S \ref{sec:lineops}.

\subsubsection{2d-4d theories in the IR:  Global picture}\label{subsubsec:2d-4d-GlobalPicture}

It is well known that a single physical system can generally
be described by many possible prepotentials $\CF$, corresponding to different electromagnetic duality frames.  We have seen that a similar phenomenon occurs for the superpotentials $\CW_i$.
In general, the data $(\CF, \CW_i)$ can be chosen \ti{locally}, but there is no global description
of the theory valid everywhere on the Coulomb branch.  Rather, we need to glue together different
local descriptions.  In this section we describe the global structure obtained from
this gluing.

Let us first consider $\bS$ with a single vacuum ($n = 1$).  Call this vacuum $i$.
Recall that globally the charge lattice $\Gamma$ is a local system over $\CB^*$, with monodromies
\begin{equation}
\gamma \to M(\gamma)
\end{equation}
as we go around nontrivial loops in $\CB^*$.
Similarly, $\Gamma_i$ is a local system of $\Gamma$-torsors over $\CB^*$.
The monodromy around nontrivial loops in $\CB^*$ of $\Gamma_i$ is similar to that of $\Gamma$, but
can involve an additional shift by an element of $\Gamma$.  To write this concretely we could
choose an origin in $\Gamma_i$ and thus identify $\Gamma_i$ with $\Gamma$; then the
monodromies would be represented by \emph{affine-linear} maps of the form
\begin{equation}
\gamma_i \to M(\gamma_i) + \gamma.
\end{equation}
We could also ask about the torsor $\Gamma_{ii}$; as we have noted, this is canonically
isomorphic to $\Gamma$, so its monodromies are exactly those of $\Gamma$.

For surface defects with $n > 1$ vacua, there is a further complication.
Globally, in addition to the kinds of monodromy we discussed above, the vacua $i$ might
be permuted among themselves.
Locally we have $n$ distinct $\Gamma$-torsors $\Gamma_i$.
The $\Gamma_i$ do not really exist globally over $\CB^*$,
but we can continue to use them locally, or we can talk about their direct sum $\Gamma_\bS$, which does
exist globally over $\CB^*$.
Upon choosing an origin in each $\Gamma_i$, we could
represent the monodromy of $\Gamma_\bS$ in the form
\begin{equation} \label{eq:gamma-i-monodromy}
\gamma_i \to M(\gamma_{P(i)}) + \gamma^{(i)}
\end{equation}
where $P$ is a permutation of the vacua $i = 1, \dots, n$.  Similarly, locally we have $n^2$ distinct $\Gamma$-torsors $\Gamma_{ij}$, which also do not exist
globally over $\CB^*$.  Their monodromies are of the form
\begin{equation}
\gamma_{ij} \to M(\gamma_{P(i) P(j)}) + \gamma^{(ij)}
\end{equation}
where $\gamma^{(ij)} = \gamma^{(i)} - \gamma^{(j)}$, and in particular $\gamma^{(ii)} = 0$,
reflecting the fact that $\Gamma_{ii} \simeq \Gamma$ canonically.

We will refer to the above structure as a ``local system of torsors,'' although
this is a slight abuse of terminology.  More precisely, we can
 combine the $\Gamma_i$ into a single local system of torsors $\Gamma_{\bS}$ defined
 over the ramified cover $\CB_{\bS}\to \CB$ defined in \S \ref{sec:Intro},
 while $\Gamma_{ij}$ combine into a single local system of torsors $\Gamma_{\bS\bS}$  over the fiber product
$\CB_{\bS} \times_\CB \CB_{\bS} \to \CB$.

By now we have introduced almost all of the structure that appeared in the 2d-4d WCF we formulated in \S
\ref{subsec:Formal-Statement}.
It only remains to comment on one important --- but unfortunately rather obscure --- ingredient,
the sign $\sigma(a,b)$.  Already in the 4d WCF the analogous sign was rather subtle.  Based on the halo picture of wall-crossing,
we believe that the physical interpretation of the sign has to do with the fact that the electro-magnetic fields sourced by a pair of well-separated bulk
particles of charges $\gamma$ and $\gamma'$ carry an extra contribution $\langle \gamma, \gamma' \rangle$ to the fermion number.

Locally in moduli space, $\sigma(a,b)$ does not really have more information than the bulk $\sigma(\gamma, \gamma') = (-1)^{\langle \gamma, \gamma' \rangle}$.
Indeed, we can always pick an origin $\gamma_i^0$
in each $\Gamma_i$, thus giving ``trivializations'' $\Gamma_i \simeq \Gamma$ and $\Gamma_{ij} \simeq \Gamma$.
Having done so, we may define $\inprod{a,b}$ for any $a,b$ using these
trivializations and the $\inprod{\cdot,\cdot}$ on $\Gamma$.
Then, given any choice of $\sigma(a,b)$,
it is possible to make a sign redefinition of the $X_a$ to bring $\sigma(a,b)$ to the form
$(-1)^{\langle a, b \rangle}$.
To find the correct sign redefinition we just require that
$X_{\gamma_i^0 + \gamma} = X_\gamma X_{\gamma_i^0}$
and $X_{\gamma_i^0 - \gamma_j^0 + \gamma} X_{\gamma_j^0} = X_{\gamma_i^0 + \gamma}$.
To preserve the wall-crossing identities, such changes of sign should be accompanied by sign changes for
$\mu(\gamma_{ij})$, as discussed in \S \ref{sec:automorphisms}.
We cannot exclude the possibility that this local redefinition might be impossible globally.
We leave such global issues for later investigation in general, but we will give a sign prescription
for the $A_1$ examples in \S \ref{sec:2d-4ddata}.

\subsubsection{Algebraic integrable systems, mirror symmetry, and surface defects}\label{subsubsec:Alg-Int-Mirr}

There is another, rather elegant way of describing the IR data associated to a surface defect
in terms of a certain Lagrangian subvariety. This will give a concise summary of the
considerations of the previous section.

Recall from \S \ref{sec:Intro} that we defined
\begin{equation}\label{eq:old-fibration}
\CM = \Gamma_g^* \otimes_\IZ \IR/(2\pi \IZ) \rightarrow \CB.
\end{equation}
As discussed in \cite{Gaiotto:2008cd}, this is a \hk\ manifold. Over
$\CB^*$ it is a torus fibration, and with the special complex structure
denoted $\zeta=0$ in \cite{Gaiotto:2008cd}
the fibration \eqref{eq:old-fibration} is a holomorphic fibration
making it a completely integrable system.
Closely allied to this is the system
\begin{equation}
\widetilde{\CM} := \Gamma_g \otimes_\IZ \IR/(2\pi \IZ) \rightarrow \CB.
\end{equation}
In fact,
$\CM$ and $\widetilde{\CM}$ are mirror dual manifolds and define
mirror conformal field theories upon compactification of the 4d theory to
2d along a two-dimensional torus.  In particular  $\widetilde{\CM}$ is \hk.
Moreover, since $\Gamma_g$ is self-dual, the manifold is self-mirror,
i.e. $\CM$ and $\widetilde{\CM}$ are actually isomorphic.
This is very natural since the sigma model with target space
$\CM$ arises from compactification of the 4d theory on a torus,
and exchange of the circles induces mirror symmetry.

Now, the discussion of the previous subsection \S \ref{subsubsec:2d-4d-GlobalPicture} implies that,
given a surface defect $\bS$, the corresponding GW parameters
$\tilde\nu$ define local sections of $\widetilde{\CM}$.
Since the vacua form a ramified cover over $\CB$ we cannot in general
extend $\tilde\nu$ to a global section, although it will define
a multisection of $\widetilde{\CM}$ with a finite number of values in
any fiber.

Next, we claim that this multisection is a holomorphic Lagrangian submanifold
of $\widetilde{\CM}$ in complex structure $\zeta=0$.

To demonstrate this
note that any local section $\gamma_g$ of $\Gamma_g$ canonically defines a
local $U(1)$-valued function on $\CM$. We denote this function as   $e^{ \I \theta_{\gamma_g}}$.
Upon choosing a local framing $\{\gamma_g^j\}$ for $\Gamma_g$ we obtain a set of
coordinate functions   $\theta_{\gamma_g^j}$ on the torus fibers of $\CM$.
The total space of the fibration carries a natural symplectic form
\begin{equation}
\langle \de Z, \de \theta \rangle := C_{ij} \de Z_{\gamma_g^i} \wedge \de \theta_{\gamma_g^j}
\end{equation}
which at $\zeta=0$ is holomorphic symplectic. Here $C_{ij}$ is inverse to $C^{ij} = \langle \gamma_g^i,
\gamma_g^j \rangle$.

Next, given a local section $\tilde\nu$ of  $\widetilde{\CM}$ we can define a
corresponding local section of $\CM$ whose coordinates at $u\in \CB^*$ are given by
\begin{equation}\label{eq:mult-sec-def}
e^{\I \theta_{\gamma_g}(u)} = e^{2 \pi \I \langle\tilde \nu(u), \gamma_g \rangle}.
\end{equation}
This subvariety is clearly Lagrangian with respect to $\langle \de Z, \de \theta \rangle$.
Moreover it is  holomorphic at $\zeta=0$. We conclude that the same properties
hold for the global multisection $\tilde\nu$ associated to $\bS$.
Moreover, by \eqref{eq:IR-GW} its generating function is $\CW$ for an
appropriate choice of superpotential.

Thus, we have arrived at the most parsimonious description of the IR
data of the surface defect $\bS$: It is a holomorphic Lagrangian multisection
of $\CM$ with generating function $\CW$.

\bigskip
\textbf{Remarks}

\begin{enumerate}

\item In the presence of a continuous flavor symmetry group
 it is natural to
consider a slight twisting of this story, as follows.
Since $ \Gamma_f^* \otimes_\IZ \IR/(2\pi \IZ)$ is trivially fibered
over $\CB$ we can define the continuous flavor symmetry group $\CF$ to
be the fiber. We can then weakly gauge this global symmetry, and
consequently define surface defects with flat connection in the
plane transverse to the defect whose holonomy is given by $\theta_f\in \CF$.
The Aharonov-Bohm phases of test particles transported around
 such a surface defect involve
the flavor charges of the test particles, which can mix with the gauge charges under monodromy around
paths in $\CB$; in consequence,
such a surface defect does not quite correspond to a section of $\widetilde{\CM}$.
Instead we should consider a torus fibration modeled on the full charge lattice $\Gamma$, i.e.,
$\Gamma^* \otimes_\IZ \IR/(2\pi \IZ)$.
This space sits naturally in an exact sequence of torus fibrations, fiberwise Pontryagin dual to \eqref{eq:lattice-ext-seq},
\begin{equation}\label{eq:torus-ext-seq}
0 \rightarrow \Gamma_g^* \otimes_\IZ \IR/(2\pi \IZ) \rightarrow \Gamma^* \otimes_\IZ \IR/(2\pi \IZ)
 \rightarrow \Gamma_f^* \otimes_\IZ \IR/(2\pi \IZ)\rightarrow 0.
\end{equation}
The inverse image of $\theta_f$
is a twisted version $\CM_{\theta_f}$ of $\CM$.  On $\CM_{\theta_f}$ the coordinate transformations of
the angular coordinates
on the fiber are affine-linear rather than linear.  A surface defect with flavor monodromy $\theta_f$
gives a multi-section of this twisted integrable system.

\item In \S \ref{sec:compactification} we will associate a hyperholomorphic vector
bundle $V_{\bS}\to \CM$ to a surface defect $\bS$. Roughly speaking, this vector bundle is
the mirror dual to $\tilde\nu$.  We discuss this further in \S \ref{subsubsec:Twsted-Mirror}
and Appendix \ref{app:Flavor-Twisted}.

\end{enumerate}

\subsection{Surface defect (twisted) chiral rings: periods as superpotentials} \label{sec:gen-example}

In \cite{Gaiotto:2009fs} a close relation was conjectured between
surface defects with a non-trivial parameter space $C$ of marginal deformations and
Seiberg-Witten-like descriptions of the IR bulk physics. (Here $C$ is a complex
variety, possibly of dimension larger than one.)
The basic idea is that the marginal deformations of the parameters of the surface defect,
$z \to z + \delta z$, are associated to twisted chiral operators $\hat x$,
and the vevs $x$ of $\hat x$ define a 1-form $\lambda = x\,\de z$ on an $n$-fold cover $\Sigma$ of $C$.
Let us stress that the symbol $z$ used here is part of the UV definition of the surface defect!

The points $(x_i, z)$ lying over $z \in C$ correspond to the IR vacua of the surface defect
with parameter $z$.
Moreover, the charge lattice $\Gamma$ should be a
subquotient of the homology lattice $H_1(\Sigma,\IZ)$ of closed paths in $\Sigma$,
and the central charge function $Z_\gamma$
should be just
\begin{equation} \label{eq:per}
 Z_\gamma = \frac{1}{\pi} \oint_\gamma \lambda.
\end{equation}
So $\Sigma$ is a (possibly higher-dimensional) analog of
the Seiberg-Witten curve, and $\lambda$ of the Seiberg-Witten differential.
The Coulomb branch $\CB$ should be a subset of the space of possible
complex structure deformations of the covering $\Sigma$ of the fixed complex manifold $C$.
The lattice $\Gamma$
should be the quotient by the kernel of the map \eqref{eq:per} of
some subset of the homology lattice closed under all the monodromies
around the singular locus of $\CB$.

The language developed in this paper is well suited to understand the origin of these statements.
Indeed, in this formalism, we have a lot of information available about the effective twisted superpotential;
we have \cite{Gaiotto:2009fs}
\begin{equation} \label{eq:w-dt}
x = \pi \partial_z \CW.
\end{equation}
We can use this to compute the \ti{difference} $(\Delta \CW)_{ij}$ between
two IR vacua of the surface defect,
$i = (x_i, z)$ and $j = (x_j, z )$.  According to \eqref{eq:w-dt}, we should
just integrate the SW differential $\lambda = x\,\de z$ along an open path $\gamma_{i j}$
on the SW curve from $(x_i, z)$ to $(x_j, z)$:
\begin{equation}
(\Delta \CW)_{ij} = \frac{1}{\pi} \int_{\gamma_{ij}} \lambda.
\end{equation}
It is immediately clear that this formula has an ambiguity, namely, different
choices of the path $\gamma_{ij}$ give different $(\Delta \CW)_{ij}$.  The difference is a period
of $\lambda$ along some closed path $\gamma$ in $\Sigma$. Thus we should really
use the notation $(\Delta \CW)_{\gamma_{ij}}$.

We propose to identify the superpotential differences
 $(\Delta \CW)_{\gamma_{ij}}$ with the $Z_{\gamma_{ij}}$ controlling
BPS solitons between vacua $i$ and $j$.  In particular we
provisionally identify the set $\Gamma_{ij}$ of soliton charges
with the set of possible paths $\gamma_{ij}$ (considered modulo boundaries of 2-chains on $\Sigma$).
The fact that we can shift an open path by adding a closed
 path means that $\Gamma_{ij}$ has a natural action of $\Gamma$.
This action corresponds to adding a gauge charge to the soliton charge;
the annoying ambiguity of $\Delta \CW$ is thus identified with the important
fact that the solitons carry gauge charges!

More precisely, to be consistent with our general discussion we require that $\Gamma_{ij}$
should be a torsor for the 4d gauge charge lattice $\Gamma$, i.e. any two paths in $\Gamma_{ij}$
should be related by adding an element of $\Gamma$.  Since $\Gamma$ is not quite the same as the
full $H_1(\Sigma, \IZ)$, this implies that $\Gamma_{ij}$ should not consist of \ti{all} paths from $x_i$
to $x_j$, but only some subset of paths closed under all the monodromies around the singular locus of $\CB$.

Our discussion has a natural extension:  we can consider
two different surface defects $\bS$ and $\bS'$, labeled by parameters $z$ and $z'$.
If they are joined (in physical space) by an   \ti{interface} then it makes
sense to discuss the difference of their superpotentials.
This will be given by $(\Delta \CW)_{\gamma_{i j'} }= \frac{1}{\pi} \int_{\gamma_{ij'}} \lambda$
where $\gamma_{ij'}$ is a path connecting $(x_i, z)$ to $(x_{j'}, z')$.
The integral only depends on the   relative homology class (with fixed endpoints)
 defined by $\gamma_{ij'}$
and, as in the above discussion the allowed set of homology classes
will form a    $\Gamma$-torsor $\Gamma_{ij'}$.  For further discussion see  \S
\ref{sec:lineops}.

\textbf{Example \ref{sec:gen-example}}.
Consider the case where $C$ and $\Sigma$ are Riemann surfaces.
A typical example of a singularity in $\CB^*$ arises from the presence of a single light hypermultiplet,
whose charge is represented by a vanishing cycle $\tilde \gamma$ in $\Sigma$.
 The monodromy of the local system $\Gamma$ around this singularity
is of the Lefschetz type, $\gamma \to \gamma + \langle \tilde \gamma, \gamma \rangle \tilde \gamma$.
Let us now derive the monodromy of $\Gamma_{ij}$.  The torsor $\Gamma_{ij}$
is represented by elements of a relative homology group, paths on $\Sigma$ which end on the preimages of $z$.
These paths have a well defined intersection pairing with elements
of another homology group $H_1(\Sigma - \pi^{-1} \{z\},\IZ)$.
In the region of $\CB^*$ sufficiently close to the singularity, we can lift uniquely $\tilde \gamma$ to
 a well defined element $\tilde \gamma'$ of this enlarged homology group:
 $\tilde \gamma'$ is the representative of the vanishing cycle which
 can shrink at the singular locus without crossing the preimages of $z$. Then the monodromy will be given by the
same formula
\begin{equation}\label{eq:hyper-mono}
\gamma_{ij} \to \gamma_{ij} + \langle \tilde \gamma', \gamma_{ij} \rangle \tilde \gamma.
\end{equation}

Another generic type of singularity arises when $z$ hits a branch point
of the covering $  \Sigma\to C$.
Monodromy around a branch point which exchanges two sheets
$i$ and $j$ will result in a monodromy exchanging the torsors $\Gamma_{ik}$ and $\Gamma_{jk}$.
 More precisely, if we denote as $\gamma_{ij}^0$ the path in $\Gamma_{ij}$ which goes from
$(x_i,z)$ to $(x_j,z)$ through the branch point
(this path is unique in a neighborhood of the singular locus),
then we have monodromies $\gamma_{jk} \to \gamma^0_{ij}+ \gamma_{jk}$ and
$\gamma_{ik} \to - \gamma^0_{ij} + \gamma_{ik}$, and of course
$\gamma_{ji} \to \gamma^0_{ij} + \gamma_{ji}+\gamma^0_{ij} $
and $\gamma_{ij} \to - \gamma^0_{ij} + \gamma_{ij}-\gamma^0_{ij}$.\footnote{Note
that it would be wrong to ``simplify'' $\gamma^0_{ij} + \gamma_{ji}+\gamma^0_{ij}$
to  $ \gamma_{ji}+2\gamma^0_{ij}$. The latter makes no sense!}
This is the monodromy expected around a locus where a single 2d particle
of charge $\gamma_{ij}^0$ becomes massless:
 it is the unique reflection which
sends $\gamma_{ij}^0\to - \gamma_{ij}^0$.

\subsubsection{A potential anomaly}\label{subsubsec:Potential-Anomaly}

We would like to define the superpotential $\CW_i$ in a \ti{given} vacuum,
rather than the differences $(\Delta \CW)_{ij}$.  In general it might
not be possible to find suitable superpotentials $\CW_i$ for
both mathematical and physical reasons. In this section we
indicate some of the possible reasons for such obstructions.

The first potential problem is that it might be impossible to find
an appropriate local system of torsors $\Gamma_i$
over $\CB C^*$. (Here $\CB C^*$ is $\CB \times C$ with certain
divisors removed, namely $\CB^{\rm sing}\times C$ and
the branch locus of the covering $\Sigma \to C$.)  We are trying to solve a cohomology problem
by splitting $\Gamma_{ij} = \Gamma_i - \Gamma_j$ where $\Gamma_i$ are
$\Gamma$-torsors.  (We consider all three terms concretely as
subsets of $\Gamma \otimes \IR$ rather than just abstract torsors.)
In general, there can be an obstruction to this splitting.\footnote{The
precise obstruction lies in $H^2(\CB^*, \Gamma)$.
To see this we begin with  local splittings
$\Gamma_{ij} = \Gamma_i - \Gamma_j$.  Starting from splittings over
each patch of $\CB^*$, one can always glue together over the intersections of two
patches $\CU_{\alpha\beta}$.  However at the intersection of three patches $\CU_{\alpha\beta\gamma}$
the composition of the three gluing maps is in general an automorphism of a
local splitting over the triple overlap.  But the automorphisms of the
splitting are just given by shifts $\gamma_i \rightarrow \gamma_i + \gamma$.
This defines a \v{C}ech 2-cocycle
valued in $\Gamma$.}

Even when torsors $\Gamma_i$ can be found, we must then also define a suitable
extension of $Z$ from ${\rm Hom}(\Gamma_{ij},\IC)$ to ${\rm Hom}(\Gamma_i,\IC)$.
Concretely, we must find $Z_{\gamma_i}$ so that $Z_{\gamma_{ij}} = Z_{\gamma_i}- Z_{\gamma_j}$,
when $\gamma_{ij}= \gamma_i - \gamma_j$,
$Z_{\gamma_i}$ have monodromy shifts
only around the singularities in $\CB C^*$ (and in no other places),
and these shifts are by $Z_{\gamma}$, $\gamma\in \Gamma$.

Even if such candidate superpotentials $Z_{\gamma_i}$ can be found, we should then worry
about possible non-uniqueness. For example, given one solution
one could shift $Z_{\gamma_i}  \to  Z_{\gamma_i}  + \CW$ where
$\CW$ has suitable monodromies. (For example, this might be possible
if the theory has a surface defect with a single vacuum.)
Indeed, in a given physical situation we might wish to put
further conditions on the $Z_{\gamma_i}$, including properties under
the deck transformations permuting the vacua or on the asymptotic
behavior in weak coupling limits. We expect that in a physical
problem the $Z_{\gamma_i}$, if they exist, should be unique, up
to trivial redefinitions such as shifts by constants.

We will see explicit examples in
\S \ref{subsubsec:Example-Anomaly} and \S \ref{sec:Hitchin-Expl} below where
suitable collections of superpotentials $Z_{\gamma_i}$
in fact do \emph{not} exist.

In order to illustrate what can go wrong, still working in rather
general terms, let us consider fibrations $\Sigma$
with the special property that the sum of $\lambda$ over the sheets above
any point $z$ is zero.  (This makes sense for a finite covering
and in particular it turns out to be true of the examples in \S \ref{sec:Hitchin-Expl},
obtained from six-dimensional engineering.)  Now,  choose some set of local sections
$\gamma_{ij'} \in \Gamma_{ij'}$ and $\gamma_{ij} \in \Gamma_{ij}$ such that
$\gamma_{ij}+\gamma_{jk}=\gamma_{ik}$ and $\gamma_{ij} + \gamma_{jk'} = \gamma_{ik'}$.
Then one could consider the sum
\begin{equation}\label{eq:tilde-Zi-def}
\tilde Z_i = \frac{1}{n}\sum_{j'=1}^n Z_{\gamma_{ij'}}.
\end{equation}
Notice that $\partial_{z'} \tilde Z_i = 0$, so $\tilde Z_i$ really depends only on $z$. Furthermore,
\begin{equation}
\tilde Z_i -\tilde Z_k = \frac{1}{n}\sum_{j'=1}^n \left( Z_{\gamma_{ij'}} -Z_{\gamma_{kj'}} \right) = Z_{\gamma_{ik}}.
\end{equation}

It is tempting to use the $\tilde Z_i$ to define $Z_{\gamma_i}$, as follows.
First we embed $\Gamma_{ij}$ in $\Gamma \otimes \IR$.
Then, after choosing representatives $\gamma_{ij} \in \Gamma_{ij}$ such that
$\gamma_{ij} + \gamma_{jk} = \gamma_{ik}$, we define
\begin{equation} \label{eq:prop}
\Gamma_i = \left( \frac{1}{n}\sum_{j=1}^n \gamma_{ij} \right) + \Gamma.
\end{equation}
This satisfies $\Gamma_{ik} = \Gamma_i - \Gamma_k$ as desired. Then we define $Z_{\gamma_i}$ by linearity.

Unfortunately, there is a problem with this definition.
Suppose we change $\gamma_{ij} \to \gamma_{ij} + \tilde \gamma^{(ij)}$
where $\tilde \gamma^{(ij)} \in \Gamma$. Then $\Gamma_i$ shifts by
\begin{equation}\label{eq:tilde-gammaij-shft}
\frac{1}{n}\sum_j \tilde \gamma^{(ij)}
\end{equation}
which is only guaranteed to live in $\frac{1}{n}\Gamma$. Unless a set of
$\gamma_{ij}$ exists for which the monodromies around loops in $\CB^*$
only act by shifts in $\Gamma$, this naive proposal will run into trouble.
As we indicate in \S\S \ref{subsubsec:Example-Anomaly} and \ref{sec:Hitchin-Expl},
there are indeed examples where the proposal fails: Shift happens.

While the above attempt failed to produce a suitable torsor $\Gamma_i$, the
problem is ``only'' that there is a shift by a fractional charge in $\Gamma$.
Locally we could embed  $\Gamma_i$ in $\Gamma \otimes \IR$ and then
take the smallest subset closed under monodromy to produce a discrete
covering space of $\CB C^*$. Thus, in some sense, the $\Gamma_i$ exist, and we can call these $\Gamma_i$
\emph{twisted torsors}. We will indicate in \S \ref{subsec:Anomaly-Cancelation}
how surface defects with such twisted torsors of charges
can be made nonanomalous by gauging finite flavor symmetry
groups. This is reassuring since the choice \eqref{eq:tilde-Zi-def}
is very natural in the $A_n$ theories of class $\CS$, considered in \S \ref{sec:Hitchin-Expl}.

\subsubsection{Example: The GW surface defect in pure $SU(2)$ gauge theory}\label{subsubsec:ExampleSU2-CP1}

In this section we consider in detail an example of the general type discussed above.
Further discussion of this example can be found in \S\S \ref{sec:Hitchin-Expl} and \ref{sec:Detailed-Expl}.

The pure $SU(2)$ gauge theory coupled to a GW surface defect was studied in \cite{Gaiotto:2009fs}.
The infrared theory is controlled by two order parameters:
a bulk vev $u = \langle \half \mathrm{Tr} \, \Phi^2\rangle$ and the vev of a
twisted chiral operator $x = \langle \frac{1}{2 \I} \Phi^{U(1)} \rangle$.
Classically, in the vacuum we will have $x^2 = - 2 u$.
The quantum-corrected chiral ring relation is conjectured \cite{Gaiotto:2009fs} to become
\begin{equation} \label{eq:4dbundle}
x^2 = \Lambda^2 e^{ t} - 2 u + \Lambda^2 e^{- t}.
\end{equation}
We caution the reader about some potential notational confusions:
We stress that $t$ is a parameter defined in the UV, and should not be confused
with the IR GW parameter of \eqref{eq:IR-GW}, which will be denoted $t_{IR}$ in this subsection.
In certain regimes, explained below, $t$ can be identified with a UV GW parameter
normalized such that
$ t = 2\pi \I t_{\rm uv}$, when compared with \eqref{eq:gw-first}. Moreover, $t$ in this
subsection is denoted $z$ above, but because of its dual interpretation as a 2d instanton
amplitude we prefer to use $t$.

We take $C$ to be a cylinder, identified
as the quotient of the  complex $t$-plane by translation by $2\pi \I \IZ$. Physically, the cylinder
$C$ is the one used in Witten's M-theory  construction of
the SW model of \cite{Witten:1997sc}, where $t=x^6 + \I x^{10}$.   We will use the fundamental
domain $-\pi < \Im\,t \leq \pi$ to illustrate $C$, as in Figure \ref{fig:ccuts}.

Equation \eqref{eq:4dbundle} should   be identified with the defining
equation of the  Seiberg-Witten curve of the 4d theory,
thus identifying $\Sigma$ in the above discussion with that ramified covering of $C$.
Likewise, $\lambda$ as described above becomes identified with the   Seiberg-Witten differential
$\lambda =  x\,\de t$. At fixed $u$ there are four  branch points
of the covering $\lambda \to t$. Two are at $\Re\,t \to \pm \infty$.
There are also branch points where $x=0$, i.e. for $t$ such that
\begin{equation}\label{eq:branch-locus}
\Lambda^2 \cosh t = u.
\end{equation}
There will only be two distinct branch
points in $C$ which we denote $t_\pm(u)$. Representing $C$ by its
fundamental domain we take $t_+(u)$ to be the solution
with positive real part, when that real part  is nonzero. Note that $t_-(u)=-t_+(u)$.
For generic $u$ the Seiberg-Witten curve $\Sigma$ is a twice-punctured torus.
The two punctures are the preimages of the branch points at $\Re(t)=\pm \infty$.
As a function of $u$, the   two branch points coincide when $u=\pm \Lambda^2$. These are the
monopole and dyon points in the $u$-plane, $\CB$, and
we define $\CB^*$ to be the complex $u$-plane with $\pm \Lambda^2$ removed.

\insfig{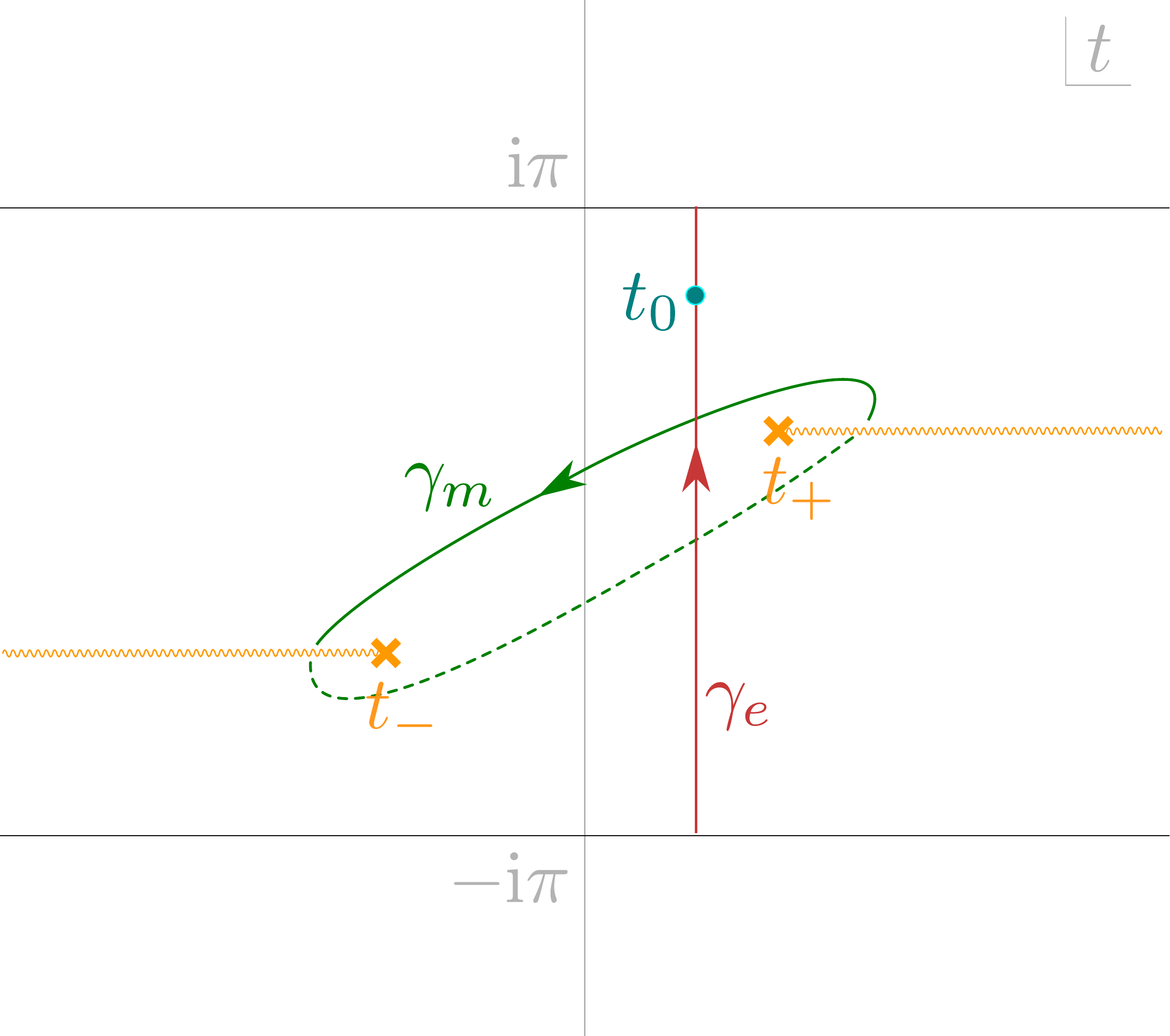}{The Riemann surface $C$, together with a basis of
cycles generating $\Gamma_u$, for $u$ in the weak coupling domain.
We choose a square root $x \sim \sqrt{-2u}$ on the
complement of the cuts. Paths on this sheet are drawn with
solid lines while the paths on the sheet with  $x \sim -\sqrt{-2u}$ are
drawn with dashed lines.  Wiggly orange lines denote branch cuts emanating
from the branch points $t_\pm$.}

\insfig{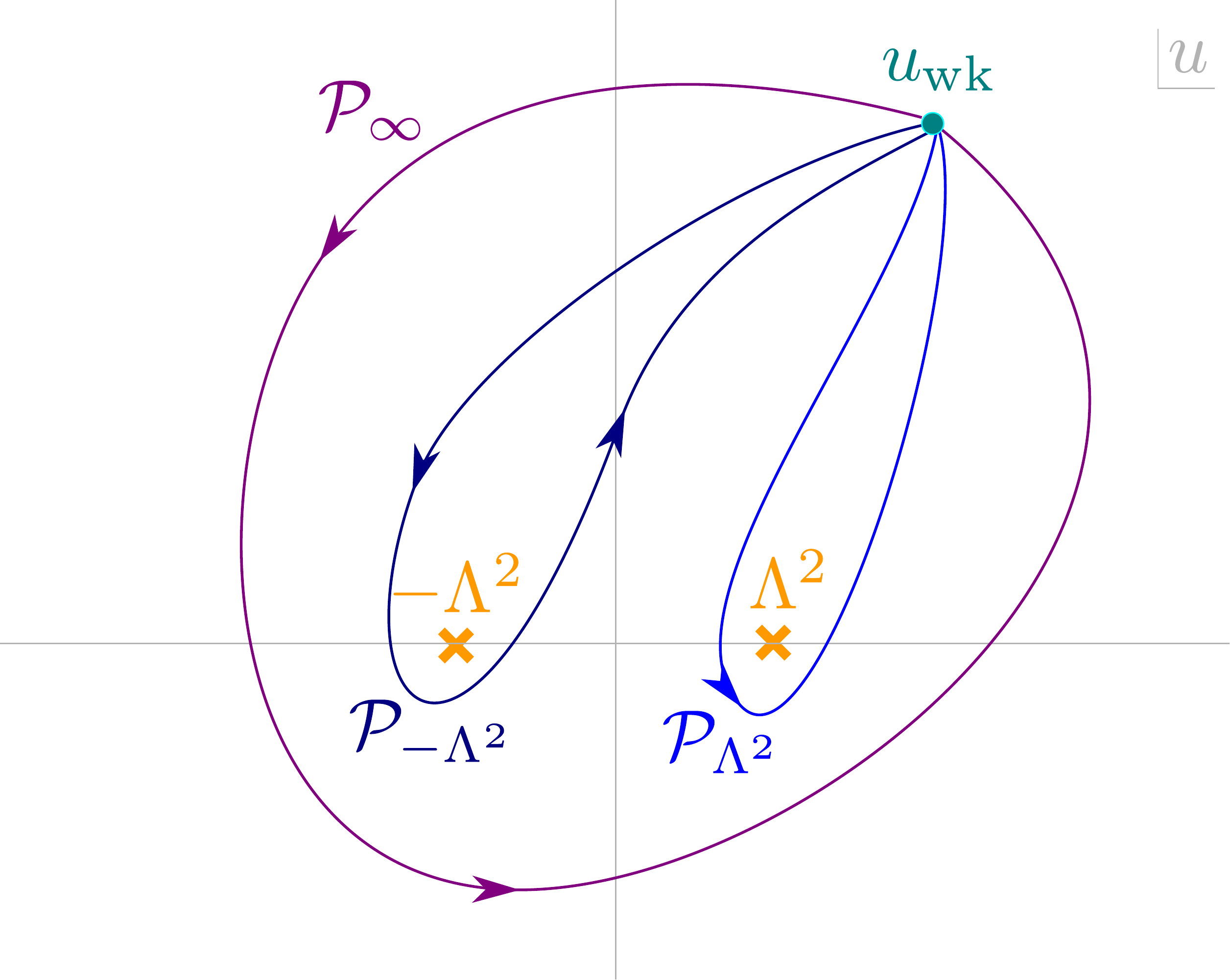}{The $u$-plane for $SU(2)$ Seiberg-Witten theory. The paths
$\CP_{\pm \Lambda^2}$ are based at $u$ in the weak-coupling regime.
Define $\CP_{\infty}$ as the path $u\to e^{2\pi \I} u$. It is homotopic
to the path given by traversing first $\CP_{-\Lambda^2}$ and then  $\CP_{\Lambda^2}$.}

Let us first review the (very well known) structure of the local system $\Gamma \to \CB^*$
in this presentation.  Recall that, physically,   $\Gamma$
is the set of IR charges $\Gamma_{\CL}$ associated to a maximally local
set of line defects in the   theory. As described in \cite{Gaiotto:2010be} there are three
choices for $\Gamma$ depending on which set of mutually local line defects we
choose to incorporate.
We take the theory with gauge group $SU(2)$. In this case
$\Gamma = H_1(\bar \Sigma;\IZ)$, where $\bar\Sigma$ is the torus with no punctures.
As we have noted there are two singular points $u_\pm = \pm \Lambda^2$ in $\CB$
and we can describe $\Gamma$
by giving the monodromy around the paths $\CP_{\pm \Lambda^2}$ shown
in Figure \ref{fig:uplane}. It will be convenient to choose a basis of the fiber of $\Gamma$
over a point $u$ in the weak coupling region   $|u| \gg |\Lambda^2|$.
Therefore, we choose  a system of cuts for the covering $\Sigma \to C$ as in
Figure   \ref{fig:ccuts}.
We then choose a cycle $\gamma_e$ to be given by taking a path in $\Sigma$ whose
image on $C$ is  $t(s) = \left( t_0 + 2\pi \I s \right) \mod ~ 2\pi \I\IZ$,
 $0 \leq s \leq 1$, where $t_0$ is a point
with $|\Lambda^2 \cosh t_0 \vert \ll \vert u\vert $. Note that $\gamma_e$ remains on
one of the two sheets of the cover and is not (anti)invariant under the deck transformation
of the cover. See Figure \ref{fig:ccuts}. However, the image of $\gamma_e$ in  $H_1(\bar \Sigma;\IZ)$
\emph{is} anti-invariant.
We next choose a cycle $\gamma_m$ on $\Sigma$ whose image on $C$ is a path   circling around the two branch points,
as in Figure \ref{fig:ccuts}.  Note that $\langle \gamma_e, \gamma_m \rangle=1$. These cycles
generate the lattice $\Gamma$ at $u$. Physically the cycle $\gamma_m$ corresponds to a
magnetic monopole and $2\gamma_e = \gamma_W$ is the charge of the  $W$-boson in the weak-coupling
region.

The periods of the cycles $\gamma_e, \gamma_m$ are:
\begin{equation}
\begin{split}
a :=  &  \frac{1}{\pi} \oint_{\gamma_e}  \lambda,  \\
a_D := &  \frac{1}{\pi} \oint_{\gamma_m}  \lambda. \\
\end{split}
\end{equation}
Extending by linearity we obtain the central charge function $Z: H_1(\Sigma;\IZ)\to\IZ$.
The kernel of $Z$ coincides with the kernel of the map $H_1(\Sigma;\IZ)\to H_1(\bar\Sigma;\IZ)$
(this is generated by the small circles around the two punctures of the torus)
so $Z$ descends to an element of ${\rm Hom}(\Gamma, \IC)$.

It is well known that the monodromy of the periods $a,a_D$ around the paths
$\CP_{\pm \Lambda^2}$ in Figure \ref{fig:uplane} is:
\begin{equation}
\CP_{-\Lambda^2}: \qquad \begin{cases}
a    \rightarrow   & a- a_D,  \\
a_D   \rightarrow  &  a_D. \end{cases}
\end{equation}
\begin{equation}
\CP_{\Lambda^2}: \qquad \begin{cases}
a    \rightarrow   & 3a- a_D,  \\
a_D    \rightarrow  & 4a- a_D.  \\\end{cases}
\end{equation}

It will be useful for what follows to have weak coupling expressions
for these periods. Since $|\Lambda^2 \cosh t_0 \vert \ll \vert u\vert$,
we easily estimate
\begin{equation}\label{eq:e-period}
Z_{\gamma_e} = \frac{1}{\pi} \oint_{\gamma_e}  \lambda  \approx  2 \I \sqrt{- 2 u}.
\end{equation}
For the other period note that  at weak coupling the two branch points are approximately
\begin{equation}
t_{\pm}(u) \approx \pm   \log \frac{2 u}{\Lambda^2}
\end{equation}
and hence
\begin{equation}\label{eq:m-period}
Z_{\gamma_m} = - \frac{2}{\pi} \int_{t_-}^{t_+} \sqrt{2 \Lambda^2 \cosh t' -2 u}~ \de t'
  \approx - \frac{4}{\pi} \sqrt{-2u} \log \frac{2 u}{\Lambda^2}
\end{equation}
up to instanton corrections. From \eqref{eq:e-period} and \eqref{eq:m-period}
we easily check that the monodromy under $\CP_{\infty}$, given by $u \to e^{2\pi \I} u$
for $\vert u \vert \gg \vert \Lambda^2 \vert$, is $a \to -a$ and $a_D \to 4 a - a_D$.
One can check that this is indeed the composition of the transformations around $\CP_{-\Lambda^2}$
and then $\CP_{\Lambda^2}$.

Now let us add a surface defect at a point $t_0 \in C$. For definiteness,
we will choose $\Re(t_0)$ large and positive.
We will focus on  two physically
distinct regimes for the physics in the $u$-plane.
The first regime is defined
by   $|\Lambda^2 \cosh t_0\vert \ll \vert u\vert $   and  $|\Lambda^2 \vert \ll \vert u\vert $.
We call this the  \emph{semiclassical GW regime}, because in this regime the 4d theory
is weakly coupled and the surface defect is well-described as a Gukov-Witten
defect with UV GW parameter $t_0$.
The second regime is defined by  $|\Lambda^2 \vert \ll \vert u\vert \ll |\Lambda^2 \cosh t_0\vert   $.
In this regime
the first term in the twisted chiral ring will become important, while the third term will remain subdominant.
In this intermediate region, the twisted chiral ring is well approximated by
\begin{equation} \label{eq:cpbundle}
x^2 = \Lambda^2 e^{t} - 2 u
\end{equation}
which can be identified with the twisted chiral ring of a
$(2,2)$ $\IC \IP^1$ sigma model, with twisted mass parameter for the $SU(2)$ flavor symmetry $m^2 = 8 u$.
Indeed the surface defect is more usefully described in this second regime
 as the $(2,2)$ $\IC \IP^1$ sigma model weakly coupled to the
bulk fields. The 2d dynamics become strongly coupled when the
first term in the twisted chiral ring relation dominates over the second.
Therefore, we will call this second regime the \emph{weakly coupled sigma model regime}.  Of course there is also
a corresponding weakly coupled sigma model regime when $t_0$ has large negative real part,
since the chiral ring has $t\to -t$ symmetry.

In the semiclassical GW regime it follows from \eqref{eq:4dbundle} that we can   write $x$     as a double
 expansion
\begin{equation}
x = \pm \sqrt{-2u} \cdot \sum_{n\geq 0} \sum_{p=-n}^\infty c_{n,p}\left( \frac{\Lambda^4}{u^2}\right)^n
 \left( \frac{\Lambda^2}{u} e^{ t_0}\right)^p,
 \end{equation}
Physically, the two expansion factors are interpreted as   4d and 2d instanton effects.  The expansion breaks
down when $t_0$ is near a branch point, i.e. near a zero of $x$. As we will explain in
\S \ref{sec:Hitchin-Expl} this is due to the existence of light BPS states, modeled on
paths running from $t_0$ to a branch point. Therefore, there will be a singularity in
the physics along the divisor \eqref{eq:branch-locus}. In particular, if we fix $t_0$
there is a singularity in $\CB$ at
$u=u_0$ defined by
\begin{equation}\label{eq:u0-def}
u_0 = \Lambda^2 \cosh t_0
\end{equation}
where  the branch point  $t_+$ collides with $t_0$. As in our general
discussion above, we will henceforth redefine
$\CB^*$ to be the complex $u$-plane with $\pm \Lambda^2$ \emph{and} $u_0$ removed.

\insfig{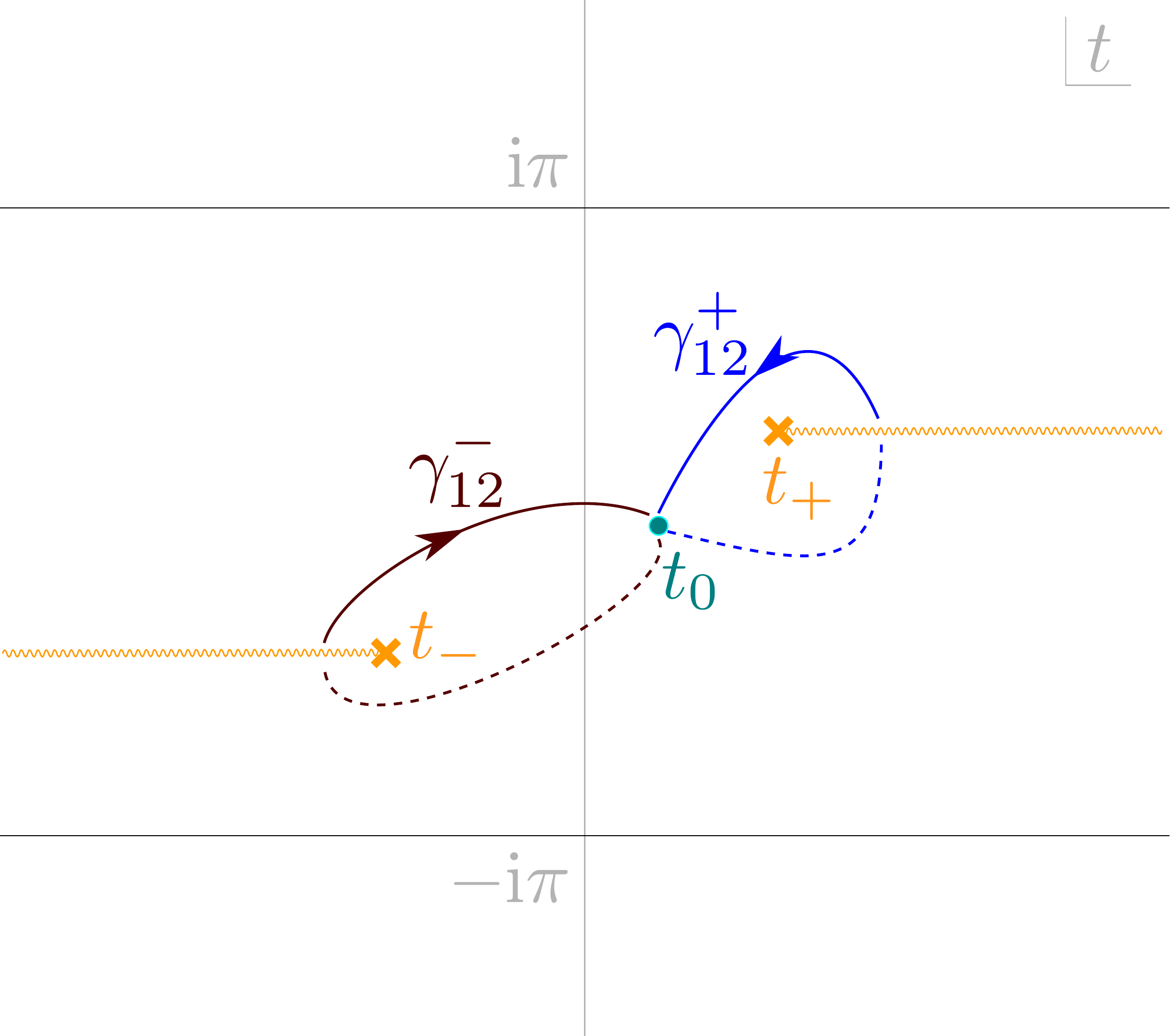}{When $u=u_{\rm gw}$ is in the semiclassical GW regime we can
define open paths  $\gamma_{12}^+$ and $\gamma_{12}^-$ on $\Sigma$, which belong to
the fiber of $\Gamma_{12}$ above $u_{\rm gw}$.}

\insfig{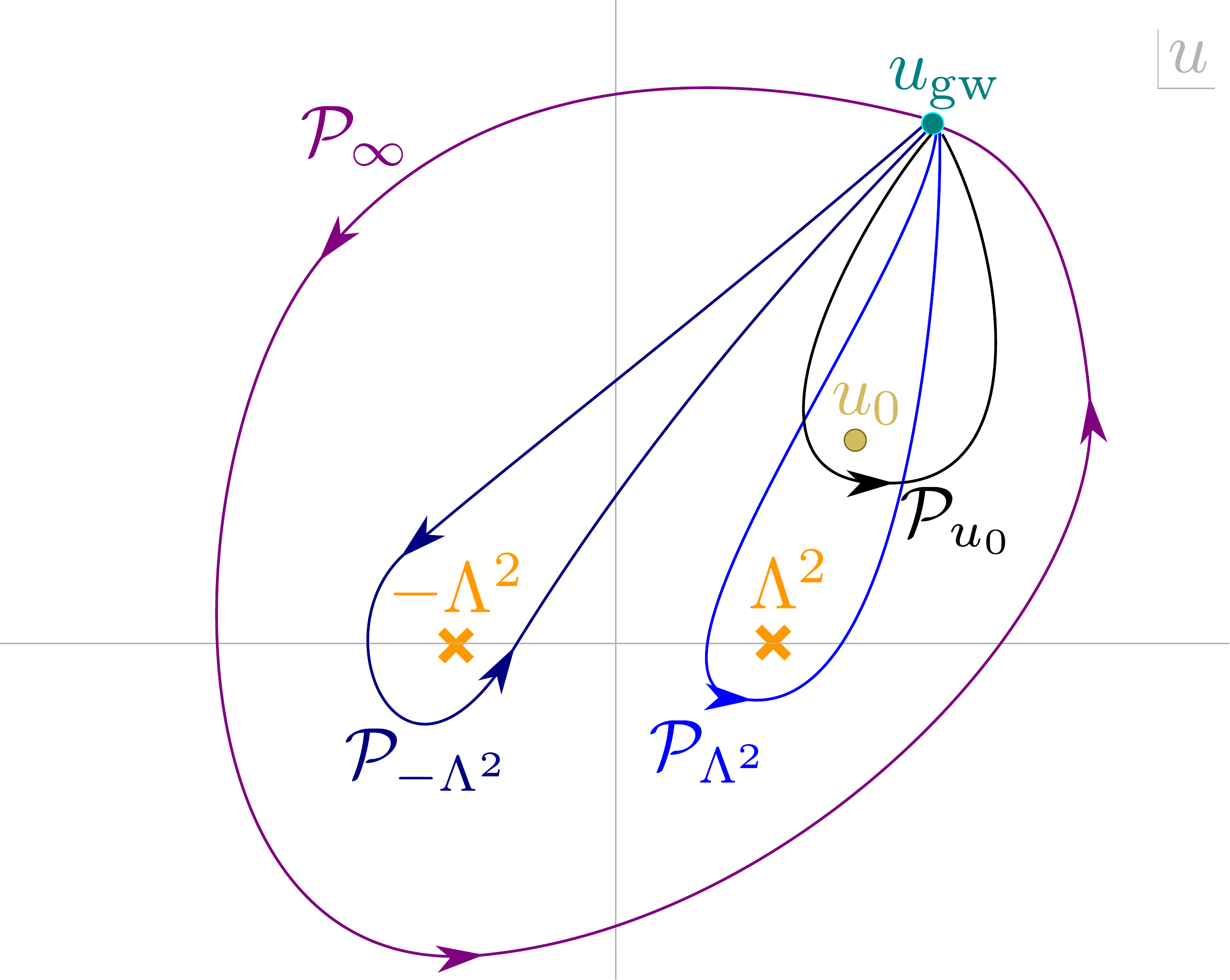}{Paths in the $u$-plane
defining the monodromy action on the fiber of $\Gamma_{12}$ above a point $u=u_{\rm gw}$
in the semiclassical GW regime.}

We now construct the torsors   $\Gamma_{ij} \to \CB C^*$ where $\CB C^*$ is
$\CB \times C$ with the divisors $\{ (\pm \Lambda^2, t) \}$ and \eqref{eq:branch-locus}
removed.  Since
$\Sigma \to C$ is two-sheeted there will only be $\Gamma_{12}$ and
its negative $\Gamma_{21}$.
We will fix $t_0\in C$ with large positive real part
 as above and just describe the torsor over the $u$-plane $\CB^*$
at fixed $t_0$.
As we just explained physically,
there are then three singular points: $u_\pm = \pm \Lambda^2$ and
$u_0$ defined in \eqref{eq:u0-def}. The vacua over $t_0$, defined by
$x_\pm = \pm \sqrt{2 (u_0- u)}$,
get permuted if $u$ follows the path $\CP_{u_0}$ around  $u_0$ shown in
Figure \ref{fig:gamma12-locsys-gw},  and hence the covering $\CB_{\bS} \to \CB$
is ramified at $u=u_0$ and $u=\infty$.

Now, to construct $\Gamma_{12}$ explicitly we first
consider the fiber above $u=u_{\rm gw}$ in the
semiclassical GW regime, where we can define
two natural elements $\gamma^+_{12}$ and $\gamma^-_{12}$ of
$\Gamma_{12}$ which correspond to open paths on $\Sigma$
whose projection to $C$ goes  from $t_0$ to the respective  branch point  $t_\pm$
and back as shown in Figure \ref{fig:gamma12-pm}.
The  difference is clearly $\gamma^+_{12} - \gamma^-_{12} = \gamma_m$.
We then have, up to instanton corrections,
\begin{equation}
\begin{split}
Z_{\gamma_{12}^+} & = \frac{2}{\pi} \int_{t_+}^{t_0} x \de t' \approx \frac{2\sqrt{-2u}}{\pi} (t_0-t_+) =
\frac{a t_0}{\I \pi}  +\frac{1}{2} a_D, \\
Z_{\gamma_{12}^-} & = \frac{2}{\pi} \int_{t_-}^{t_0} x \de t' \approx \frac{2\sqrt{-2u}}{\pi} (t_0-t_-) =
\frac{a t_0}{\I \pi}  - \frac{1}{2} a_D.
\end{split}
\end{equation}
From these equations it is easy to compute that the
monodromy around the path  $\CP_{\infty}$  shown in
Figure \ref{fig:gamma12-locsys-gw} shifts $Z_{\gamma^\pm_{12}}  \to \pm 2 a -Z_{\gamma^\pm_{12}}$.

Now let us consider the monodromy around the path $\CP_0$ in Figure \ref{fig:gamma12-locsys-gw}.
In order to compute the monodromy of $Z_{\gamma^+_{12}}  $
it is actually easier to fix $u=u_0$ and displace $t\to t_0+ \delta t$ and
let the phase of $\delta t$ increase from $0$ to $2\pi$. Since we are computing
monodromy around a single divisor \eqref{eq:branch-locus} in $\CB \times C$ the
result will be the same as holding $t=t_0$ and letting $u$ run around $\CP_0$.
When $u=u_0$ and $\delta t$ is small, we have
\begin{equation}
x^2 = 2 (\Lambda^2 \cosh (t_0 + \delta t) - u_0) \approx (2 \Lambda^2 \sinh t_0) \delta t
\end{equation}
and hence $Z_{\gamma^+_{12}} \sim (\delta t)^{3/2}$. It follows that the
monodromy around this locus sends $Z_{\gamma^+_{12}} \to -Z_{\gamma^+_{12}}$.

We are now ready to discuss the full local system above the $u$ plane.
We have given a basis for the fiber of  $\Gamma_{12}$
over  $u_{\rm gw}$ together with
the monodromy around $\CP_0$ and $\CP_{\infty}$, shown in
Figure \ref{fig:gamma12-locsys-gw}. Around $\CP_\infty$ we have
 $a \to -a$, $a_D \to 4 a - a_D$, $Z_{\gamma^\pm_{12}}  \to \pm 2 a -Z_{\gamma^\pm_{12}}$
and around $\CP_{u_0}$   we   have the $Z_{\gamma^+_{12}} \to -Z_{\gamma^+_{12}}$
while $a,a_D$ are unchanged.
Note that it follows that around the composition $\CP_{\Lambda^2}\CP_{-\Lambda^2}$
we have
\begin{equation}\label{eq:pmL-mono}
Z_{\gamma^+_{12}}  \to 2 a +Z_{\gamma^+_{12}}.
\end{equation}

\insfig{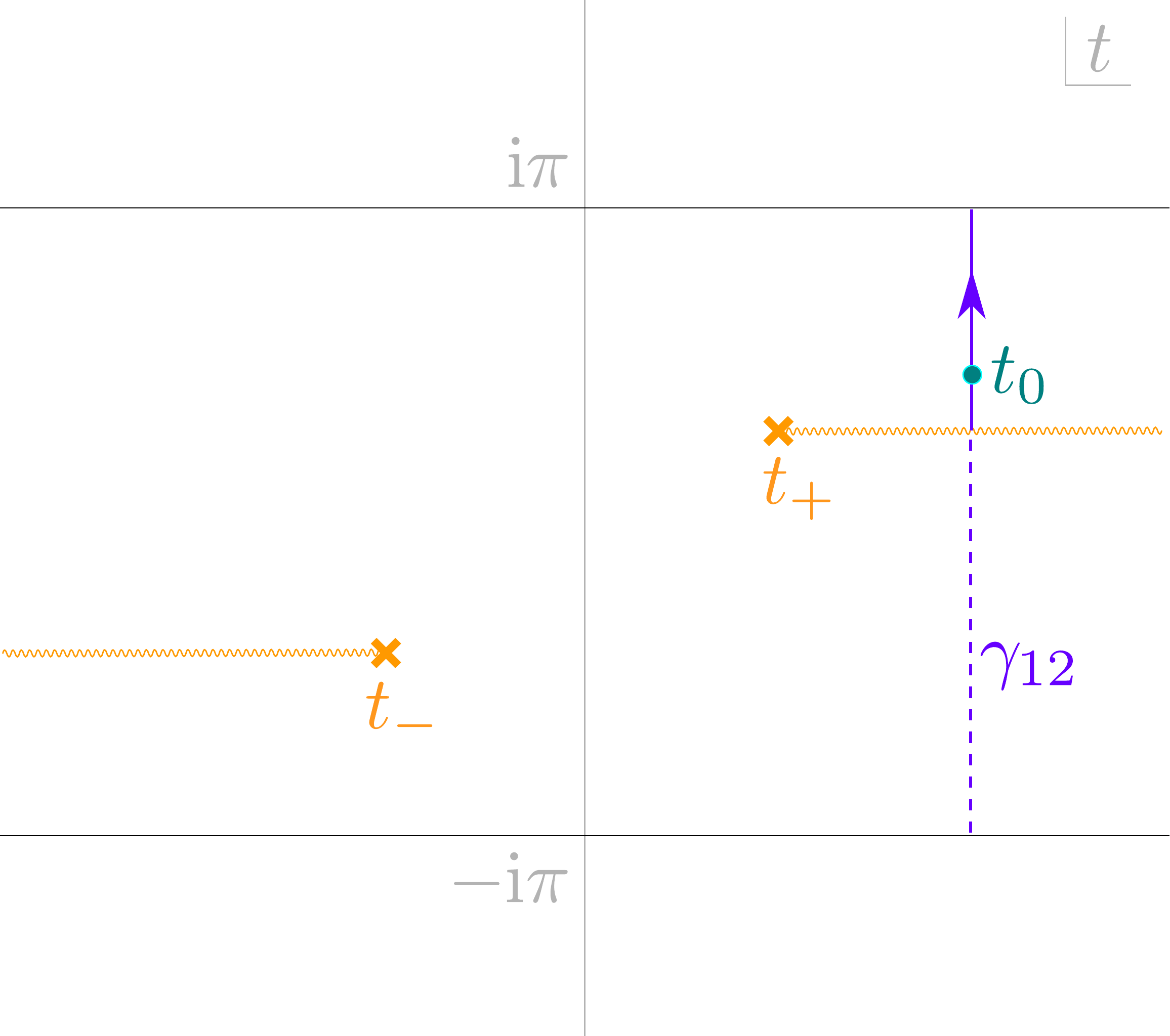}{When $u=u_{\sigma}$ is in the semiclassical sigma model regime, we
define an open path $\gamma_{12}$ on $\Sigma$, which gives a useful
element in the fiber of $\Gamma_{12}$ above $u_{\sigma}$.}

\insfig{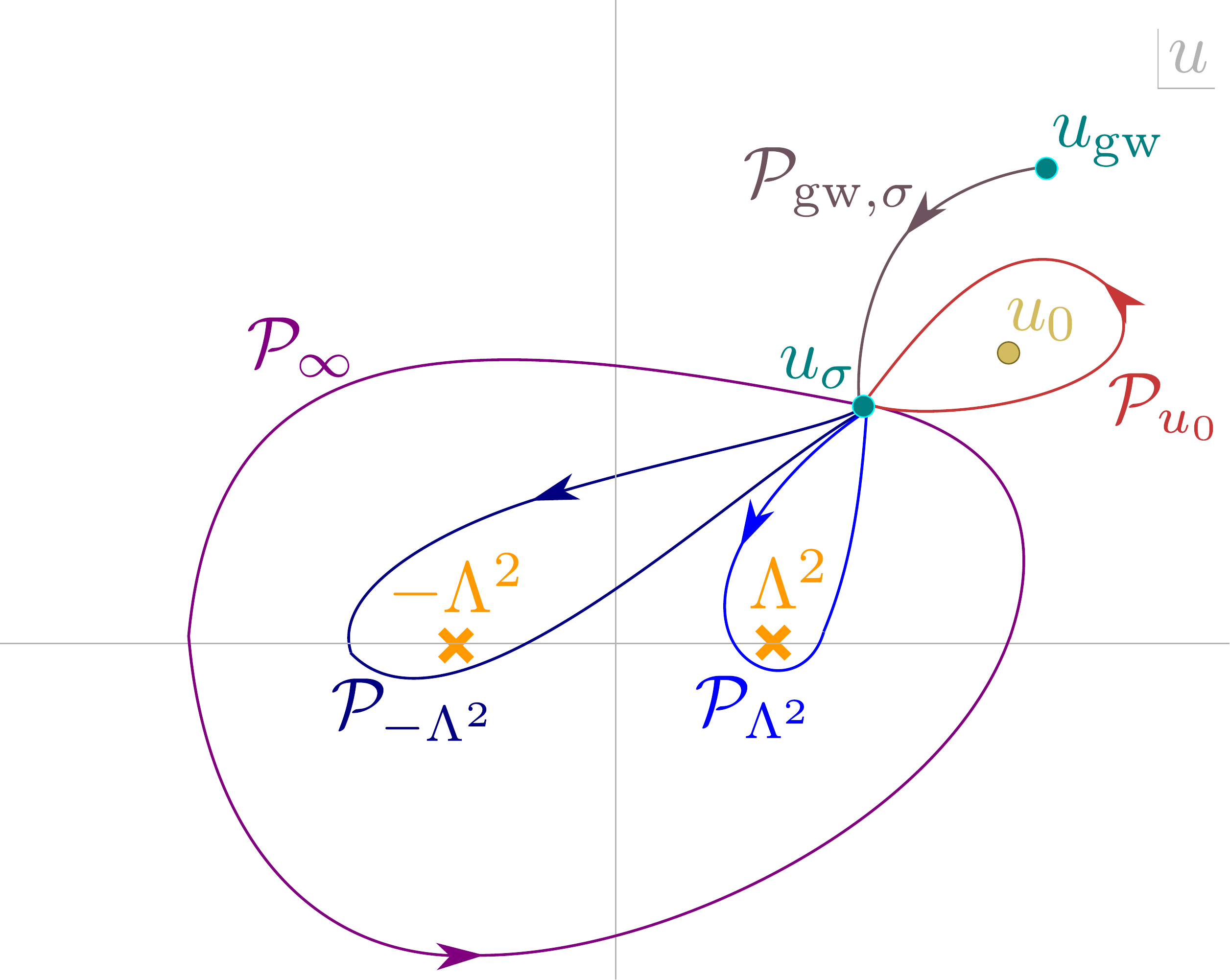}{Paths in the $u$-plane
defining the monodromy for $\Gamma_{12}$, acting on the fiber over
$u=u_{\sigma}$ in the semiclassical sigma model regime.}

In order to compute the monodromy around $\CP_{\pm \Lambda^2}$ separately we instead
move our fiber of $\Gamma_{12}$ to sit over $u=u_\sigma$ in the sigma-model regime,
where $\Lambda^2 \ll  |u_{\sigma} | \ll |u_0|$. (To be precise, we transport along
the path $\CP_{gw,\sigma}$ shown in Figure \ref{fig:gamma12-locsys-sig}.)  The cuts for the cover $\Sigma \to C$ are now
shown in Figure \ref{fig:gamma12}.
It is useful to consider a third element $\gamma_{12}$ in the fiber of $\Gamma_{12}$
over $u_{\sigma}$.
It corresponds to the open path on $\Sigma$ whose projection to $C$
is $t(s) = \left( t_0 + 2\pi \I s\right)  \mod~ 2\pi \I \IZ$, $0 \leq s \leq 1$.
The crucial point is that this element has no monodromy in the
 region $|u_\sigma|\ll |u_0|$, because the branch points always stay
away from this path.  Thus $\gamma_{12}$ is invariant under
transport around either of $\CP_{\pm \Lambda^2}$.  (As a simple check,
let us show that it is invariant around $\CP_{\Lambda^2}\CP_{-\Lambda^2}$. To do this  first
note that  $\gamma_{12} - \gamma^+_{12}$ is simply $\gamma_e$; see Figure \ref{fig:gamma12}. Next,
recall that we have \eqref{eq:pmL-mono}
around the composition $\CP_{\Lambda^2}\CP_{-\Lambda^2}$
and $a \to -a$ around this path, so
$Z_{\gamma_{12} } = Z_{\gamma_e} + Z_{\gamma_{12}^+} = a + Z_{\gamma_{12}^+}$
remains invariant.)  On the other hand, around $\CP_{u_0}$ we have
$\gamma_{12} \to 2\gamma_e - \gamma_{12}$. This completes our explicit description
of the local system $\Gamma_{12}$.

We are now ready to construct the local systems $\Gamma_1, \Gamma_2$ of $\Gamma$-torsors.
Note that if we define
\begin{equation}
Z_{\gamma_1} = - Z_{\gamma_2} = \half Z_{\gamma_{12}}= \frac{1}{2} Z_{\gamma_{12}^+}+ \frac{a}{2}
\end{equation}
then $Z_{\gamma_i}$ really do shift by elements $Z_{\gamma}$ for $\gamma \in \Gamma$
for all closed curves in $\CB^*$, thanks to the factor of $2$ in the
monodromy $\gamma_{12} \to 2\gamma_e - \gamma_{12}$ around $\CP_{u_0}$.
Thus, in this example, the   potential anomaly raised in
\S\ref{subsubsec:Potential-Anomaly} does not occur.

Now, with proper superpotentials in hand, we can fill in a gap in the above
discussion and explain the phrase ``semi-classical GW regime.''  The reason is
that we can use the weak coupling expressions to show that
the IR GW parameter is
\begin{equation}\label{eq:IR-UV-GW}
t_{IR} = \frac{\p Z_{\gamma_1}}{\p a} \approx \frac{t}{2\pi \I} + \frac{1}{4}\tau + \frac{1}{2}.
\end{equation}
Since the theory is weakly coupled we can identify $t$ itself with a GW parameter.

\subsubsection{An example of the anomaly}\label{subsubsec:Example-Anomaly}

One class of theories, discussed further in \S\S \ref{sec:Hitchin-Expl} and \ref{sec:Detailed-Expl}, are
the so-called Argyres-Douglas (``AD'') theories of type $N$. Here $C= \IC $ and the chiral ring/SW equation is of the
form:
\begin{equation}
\lambda^2 = P_N(z) (\de z)^2.
\end{equation}
The parameter $t$ of \S \ref{subsubsec:ExampleSU2-CP1} is here restored to
$z\in C$, and $P_N(z)$ is a polynomial of degree $N$.
See \cite{Gaiotto:2009hg} for background material. Crucial points for our discussion
here are that when $N$ is odd there are branch points at the $N$ roots of $P_N$ together
with a branch point at $z=\infty$, but when $N$ is even, $z=\infty$ is not a branch point.
Related to this, the local system $\Gamma$ has a one-dimensional flavor lattice when
$N$ is even, but no flavor lattice when $N$ is odd.

We claim that for $N$ odd it is possible to construct $\Gamma$-torsors $\Gamma_i$ compatible with
the $\IZ_2$ deck transformation so that
$\Gamma_{12} = \Gamma_1-\Gamma_2$, but for $N$ even this is not possible. Thus,
the potential anomaly of \S\ref{subsubsec:Potential-Anomaly} is realized when $N$ is even.
By ``compatible'' we mean the following. Since the chiral ring $x^2 = P_N(z)$
has a $\IZ_2$ symmetry exchanging the sheets it is natural to invoke
 the physical principle that there
should be symmetry under permutation of the two vacua so that $Z_{-\gamma_1}$
should have the same monodromy as $Z_{\gamma_2}$ in the
splitting $Z_{\gamma_{12}} = Z_{\gamma_1}-Z_{\gamma_2}$. But
$Z_{\gamma_1} - Z_{-\gamma_1}$ shifts by \emph{even} elements of $\Gamma$,
and hence so must $Z_{\gamma_{12}}$.

In order to justify this claim let us briefly discuss the case $N=3$ in
a way which will generalize to all odd $N$. The construction is very
closely analogous  to the example of \S  \ref{subsubsec:ExampleSU2-CP1} above.
In this case we may parameterize
\begin{equation}
P_3(z) = z^3 - 3 \Lambda^2 z + u.
\end{equation}
$\Lambda$ is a non-normalizable parameter and $\CB$ is the complex
$u$-plane.
We can choose cuts in the $z$-plane  going from each of the three roots
of $P_3$ to $z=\infty$. The local system $\Gamma \to \CB^*$ has monodromy
around $u= \pm 2 \Lambda^2$ (where a pair of roots collide) and is of rank $2$.

Now we add a surface defect at a point $z_0 \in C$. We assume
that $\vert z_0 \vert$ is very large. There is then one new singular
point in the $u$-plane, at $u_0$ such that $z_0^3 - 3 \Lambda^2 z_0 + u_0 =0$.
The analog here of the weakly coupled $\sigma$-model regime is
\begin{equation}
\vert z_0^3 \vert \gg \vert u\vert \gg \vert \Lambda^2 \vert.
\end{equation}
We can construct a special element $\gamma_{12}\in  \Gamma_{12}\vert_{u_\sigma}$,
where $u_\sigma$ is in the weakly coupled $\sigma$-model regime,
by choosing a path in $C$ beginning and ending at $z_0$ and circling around all
three branch points. Since it has crossed an odd number of cuts it lifts to
an open path in $\Sigma $ joining the vacuum $\lambda_0$ to $-\lambda_0$.
As in the $SU(2)$ example, this element has no monodromy around all the paths
in $\CB_*$ based at $u_\sigma$ which wind around zeros of the
discriminant of $P_3$ (namely $u= \pm 2 \Lambda^2$ ).  It \emph{does}
have monodromy around the path $\CP_{u_0}$ based at $u_\sigma$ which winds
once around $u_0$. To compute this monodromy, note that as $u_\sigma \to u_0$
(thus leaving the $\sigma$-model regime) some branch point, say $z_*(u_\sigma)$,
approaches $z_0$. We can then define an open path
$\gamma_{12}^*\in \Gamma_{12}\vert_{u_\sigma}$  (the analog of $\gamma_{12}^+$
above) whose image in $C$ begins at $z_0$, runs to $z_*$ passing through a cut and comes back
to $z_0$. Note that $Z_{\gamma_{12}} = Z_{\gamma_{12}^*} + Z_\gamma$ for some
$\gamma \in \Gamma$. On the other hand, by an argument similar to that of the
$SU(2)$ example, $Z_{\gamma_{12}^*} \to - Z_{\gamma_{12}^*}$ around $\CP_{u_0}$
and hence the only nontrivial monodromy of $Z_{\gamma_{12}}$ in $\CB^*$ is
around $\CP_{u_0}$ and given by $Z_{\gamma_{12}} \to 2 Z_\gamma - Z_{\gamma_{12}}$.
We can therefore define $Z_{\gamma_1} = - Z_{\gamma_2} = \frac{1}{2} Z_{\gamma_{12}}$,
and these periods will have monodromy shifts by proper periods associated to $\Gamma$.

The above construction generalizes to AD theories with odd $N$. On the other hand,
when $N$ is even the situation is very different. Consider, for example, the
first nontrivial case, $N=4$, with
\begin{equation}
P_4(z) = z^4 + 4 \Lambda^2 z^2 + 2m z + u .
\end{equation}
Here $\Lambda, m$ are fixed nonnormalizable parameters and $\CB$ is again the
complex $u$-plane. There are four branch points in $C$ given by the roots
of $P_4(z)$. The local system $\Gamma$ has rank $3$ with a one-dimensional
flavor sublattice and $\Gamma \to \CB^*$ has monodromy around the three
roots of the disciminant of $P_4$. When we add a surface defect at $z_0\in C$
there is a fourth singularity in the $u$-plane at $u_0$ such that $P_4(z_0)=0$.
Now, if $u=u_\sigma $ is in the weakly coupled sigma model regime
\begin{equation}
 \vert z_0^4 \vert \gg \vert u\vert  \gg \vert \Lambda^2\vert , \vert m \vert
\end{equation}
we \emph{cannot}  define the analog of the cycle  $\gamma_{12}\in \Gamma_{12}\vert_{u_\sigma}$
since it will cross an even number of cuts and hence will not connect the two sheets
above $z_0$. If, instead, we choose any other path
$\tilde \gamma_{12}\in \Gamma_{12}\vert_{u_\sigma}$
surrounding an odd number of branch points, then that path will pass through
a cut joining two branch points. Therefore, there will be a singularity in $\CB$
where those two branch points collide. The monodromy of $\tilde \gamma_{12}$
around such a point will be $\tilde \gamma_{12}\to \tilde \gamma_{12}+ \gamma_v$
where $\gamma_v$ is the vanishing cycle associated with the colliding branch points.
Therefore, if we attempt to introduce $Z_{\gamma_1} = - Z_{\gamma_2} = \half Z_{\tilde\gamma_{12}}$
then $\gamma_1,\gamma_2$ will have half-integral monodromy $\gamma_1 \to \gamma_1 +
\frac{1}{2} \gamma_v$. Thus our attempt has failed. Now, suppose there were some
splitting $Z_{\gamma_{12}} = Z_{\gamma_1} - Z_{\gamma_2}$ where $Z_{\gamma_1}, Z_{\gamma_2}$
have proper integral monodromy.
Then $-Z_{\gamma_1} = Z_{-\gamma_1}$ would have proper
monodromy, and $Z_{\gamma_1} - Z_{-\gamma_1}$ would have monodromy of the type we have
just proven cannot exist. Therefore, for $N=4$ AD theories there is no splitting
$\Gamma_{12} = \Gamma_1 - \Gamma_2$ which respects the symmetry between
the sheets. The argument we have just given extends to all
even values of $N$.

To use the language of \S \ref{subsubsec:Potential-Anomaly}
above, when $N$ is even we can construct $\Gamma_i$ as twisted torsors.
The three examples we have just discussed are examples drawn from a class of theories
described in \S \ref{sec:Hitchin-Expl} known as $A_1$ theories. As we
discuss in \S \ref{sec:Hitchin-Expl} the $A_1$ theories with regular singular
points all exhibit the anomaly we have discussed. Theories with irregular
singular points are limits of those with regular singular points. In such
limits particles can decouple and the nature of flavor symmetries can
change. It can be shown that theories with odd irregular singular points
(such as AD $N$ odd and $SU(2)$, $N_f=0$) have a natural candidate for the
analog of the cycle $\gamma_{12}^+$ used above and do not exhibit the
anomaly. Other theories with only even irregular singular points,
such as AD theories with $N$ even still exhibit the anomaly.

\subsection{BPS degeneracies and wall-crossing}\label{subsec:BPS-Deg-WC}

Now that we have given a physical description of the lattices $\Gamma$, torsors $\Gamma_i$, $\Gamma_{ij}$ and central charges $Z$ which appear in the wall-crossing formula,
we can move to the main actors, $\Omega$, $\mu$ and $\omega$.

First, recall that in the 4d wall-crossing
formula the integers $\Omega(\gamma)$ are interpreted as degeneracies of 4d BPS
particles \cite{Gaiotto:2008cd}.
This is their interpretation in the 2d-4d wall-crossing formula as well.

The integers $\mu(\gamma_{ij})$ in the 2d-4d wall-crossing formula
are a natural generalization of the index  $\mu_{ij}$
in the 2d case \cite{Cecotti:1992rm}.
$\mu(\gamma_{ij})$ is the degeneracy of 2d BPS solitons between vacuum $i$ and vacuum
$j$ of the surface defect, carrying total charge $\gamma_{ij}$. Using the
notation of Appendix \ref{app:d2d4-multiplets} it can be written:
\begin{equation}
\mu(\gamma_{ij}) := \Tr_{\tilde\CH_{\gamma_{ij}}^{\BPS} }  e^{\I \pi F} e^{- \beta H}.
\end{equation}
Here $\tilde\CH_{\gamma_{ij}}^{\BPS}$ is the space of BPS states with
center-of-mass degrees of freedom removed,
while   $F$ is some $u(1)_V$ charge. An important point is that there is
a \emph{choice} of $u(1)_V$ generator $F$. One natural choice is $F=so(12)$,
but there is  some  ambiguity in
picking the overall fermion number when removing the center of mass
degrees of freedom. We will take the $\mu(\gamma_{ij})$ to be integral
although $e^{\I \pi F}$ acting on the  full Hilbert space $\CH_{\gamma_{ij}}$   need not be
$\pm 1$.
It is standard in the 2d literature to choose $F$   so that
$\mu(\gamma_{ij}) = -\mu(-\gamma_{ij})$. The ambiguity in
$F$  will be relevant in \S \ref{sec:compactification}.

The affine-linear functions $\omega(\gamma, \cdot)$ require more discussion.
Roughly we want to interpret
$\omega(\gamma, \gamma_i)$ as the number of 2d BPS particles in vacuum $i$, carrying gauge charge
$\gamma$.  If this were literally the case, then $\omega(\gamma,\gamma_i)$
would actually depend only on $i$, not on the choice of a particular $\gamma_i$.
The reason for the dependence on $\gamma_i$ is a rather subtle difficulty.  In addition to 1-particle states
with charge $\gamma$ localized near the surface defect, the theory also contains 1-particle ``4d'' states with
charge $\gamma$.  When decomposed in terms of the unbroken 2d Poincare symmetry these states give a continuum of
representations: The momentum in the transverse space to the surface defect
is an \emph{internal quantum number} and hence the representation of the little group
has no gap from the BPS bound. The representation at the bottom of this continuum is
the same one where a genuine 2d particle
would be found.  It is therefore difficult to disentangle the 2d and 4d spectra.

To get a bit more insight, let us focus on a region of parameter space containing a point
where a specific $Z_{\gamma^e} \to 0$.  At this point the Lagrangian we have been discussing until now ---
which includes only the abelian gauge fields and their superpartners --- breaks down.
One gets a better description near this point by including fields
describing the light BPS particles with charge $\gamma^e$; the singularity of the original
theory would then be reproduced by integrating those fields out.
What kind of new fields do we need to add?
Light 4d particles should be represented by hypermultiplets; light 2d particles should be represented by
2d chiral multiplets.

The effect of integrating out the light hypermultiplets is very well known: A
simple 1-loop computation leads to a correction to the prepotential  which creates logarithmic divergences in the central charges:
\begin{equation} \label{eq:log}
 Z_\gamma = \frac{\I}{2 \pi} \Omega(\gamma^e) \inprod{\gamma, \gamma^e}Z_{\gamma^e} \log (Z_{\gamma^e})+ {\rm Regular}
\end{equation}
This logarithmic behavior implies the well-known Picard-Lefschetz-like monodromy:  as $Z_{\gamma^e}$ goes around $0$
we have $\gamma \to \gamma + \Omega(\gamma^e) \inprod{\gamma, \gamma^e} \gamma^e$.

It turns out that the effect of integrating out
a light 2d particle leads to a correction very similar to \eqref{eq:log}.
Indeed, we can invoke the computation of Witten in the two-dimensional
gauged linear sigma model (see \cite{Witten:1993yc}, eq. (3.9) et. seq. and
also \cite{Morrison:1994fr})
to conclude that integrating out $n$ two-dimensional chiral multiplets of
four-dimensional charge $\gamma^e$
induces a singular term in the two-dimensional \emph{superpotential}:
\begin{equation}\label{eq:chgesuper}
\CW = \frac{\I}{2 \pi} n Z_{\gamma^e} \log Z_{\gamma^e} + {\rm Regular}.
\end{equation}

For later reference (in \S \ref{subsec:One-d-Coul-Delta-Zero} below)
it will be useful to recall briefly the argument for \eqref{eq:chgesuper}.
The twisted chiral superfield is \eqref{eq:vm-to-tw-cm} $\Upsilon = a + \cdots + \vartheta^+ \vartheta^- (F_{03} + \I (D_{12}-F_{12}))$, where $Z_{\gamma_e} = q a$ vanishes at $a=0$ and the $n$ chiral multiplets $X_i$, $i=1,\dots, n$ have  $U(1)$ charge $q$ and hence have Lagrangian
\begin{equation}
\int \de x^0 \de x^3 \int \de^4 \vartheta \bar X_i e^{q V} X_i = - \int \vert \CD X_i\vert^2 + (m^2 +q (D_{12}-F_{12})) \vert X_i \vert^2  + \cdots
\end{equation}
where $m^2 = q^2 \vert a \vert^2$ and we have only written the terms relevant to our computation. Integrating out the chiral superfields
induces a one-loop determinant
$$\exp\left( - n \log \det[ - \CD^* \CD + m^2 + q ( D_{12}-F_{12}) ] \right) $$
which when expanded in the auxiliary field gives a coupling
\begin{equation}
\int \de x^0 \de x^3 (D_{12}-F_{12}) nq \int \frac{\de^2 k}{ (2\pi)^2} \frac{1}{k^2 + m^2}+ \cdots
\end{equation}
Supersymmetry now determines the rest. Comparison with the last term in
\eqref{eq:explct-action} reveals that we have induced a Gukov-Witten parameter
\begin{equation}\label{eq:alpha-correction}
(\Im \tau) \alpha = - n q \int \frac{d^2 k}{(2\pi)^2} \frac{1}{k^2 + q^2 \vert a \vert^2} = \frac{nq}{4\pi} \log \frac{\vert a\vert^2}{\vert \Lambda\vert^2}
\end{equation}
where $\Lambda$ is a UV cutoff, necessary in this IR free gauge theory. On the other
hand, a simple computation shows that such an $\alpha$ corresponds to
\begin{equation}\label{eq:alpha-Weff}
\CW = nq \frac{\I}{2\pi } \left( a \log \frac{a}{\Lambda} - a \right)
\end{equation}
thus establishing \eqref{eq:chgesuper}.

The above one-loop effect leads to the monodromy of the superpotential
discussed in equation \eqref{eq:w-shift}, and hence, as explained
there, if we wish to extend the period function and define
$\CW = Z_{\gamma_i^0}$ then $\gamma_i^0$ must shift by
$\gamma^0_i \to \gamma^0_i + n  \gamma^e$ under one turn around $a=0$.
In other words, the presence of light 2d particles leads
to affine-linear monodromies of the torsor $\Gamma_i$.  In particular, we cannot hope to find a single $\gamma^0_i$
which makes sense globally:  we will always have to make do with local descriptions, related to one
another by shifts.

Moreover, more general paths in moduli space force us to consider more general
monodromies, $\gamma^0_i \to \gamma_i = \gamma^0_i + \gamma$ for more general charges $\gamma$.
In this case we have an effective
superpotential  $Z_{\gamma_i} = Z_{\gamma^0_i}+ Z_{\gamma}$ which, by
\eqref{eq:log}, is just:
 \begin{equation} \label{eq:affine-log}
Z_{\gamma_i} =  \frac{\I}{2 \pi } (n + \Omega(\gamma^e) \inprod{\gamma_i - \gamma^0_i, \gamma^e}) Z_{\gamma^e} \log Z_{\gamma^e} + {\rm Regular}.
\end{equation}
Now, there is clearly something odd about this formula since it involves a the  choice of a specific $\gamma^0_i \in \Gamma_i$, which we used in writing our
effective Lagrangian.
 What would have happened if we had made a different choice, say replacing
$\gamma^0_i$ by $\gamma^0_i + \tilde\gamma$?  Perhaps surprisingly,
from \eqref{eq:affine-log} we see that this change must be compensated by a change of $n$ to $n + \Omega(\gamma^e) \langle \gamma^e,\tilde \gamma \rangle$ --- so a rather
innocuous-looking shift of the effective superpotential also requires a change in the number of 2d chiral multiplets!
Thus, the number of chiral multiplets depends on the flux and we should write $n(\gamma_i^0)$.
The most meaningful and invariant
quantity we can extract from our considerations is the coefficient
$n(\gamma_i^0) + \Omega(\gamma^e) \inprod{\gamma_i - \gamma^0_i, \gamma^e}$.
which appeared in \eqref{eq:affine-log}, giving the physical effect of integrating out both 2d and 4d
particles of charge $\gamma^e$.  We propose to identify this with $\omega(\gamma^e, \gamma_i)$.  In other
words, if $u$ is near a locus in $\CB$ where some $Z_\gamma \to 0$, we define $\omega(\gamma, \gamma_i;u)$ in terms of
the contribution from particles of charge $\gamma$ to $Z_{\gamma_i}$,
\begin{equation}
Z_{\gamma_i} =  \frac{\I}{2 \pi } \omega(\gamma, \gamma_i;u) Z_{\gamma} \log Z_{\gamma} + {\rm Regular}.
\end{equation}
Then the values of $\omega(\gamma,\gamma_i;u)$ at other vacua will be determined by wall-crossing.
Note that $\omega(-\gamma, \gamma_i;u) = - \omega(\gamma, \gamma_i;u)$.

\textbf{Example}: Let us return to Example \ref{sec:gen-example} above.
We discussed there the example of a singularity in $\CB^*$ which arises from the
presence of a single light hypermultiplet
whose charge is represented by a vanishing cycle $\tilde \gamma$ in $\Sigma$.
From the monodromy of $\Gamma_{ij}$, equation \eqref{eq:hyper-mono} we recognize immediately
$\omega(\tilde \gamma,\gamma_{ij}) = \langle \tilde \gamma', \gamma_{ij} \rangle$
in the neighbourhood of the singular locus.
Similarly, a
lot of simple wall-crossing examples can be derived from these statements,
by simply considering a locus where $z$ comes close to two branch points.
See \S \ref{sec:Hitchin-Expl} for several examples along these lines.

\textbf{Remark:}
The reader might well be distressed by the absence of a clear definition of $\omega(\gamma,\gamma_i)$.
Fortunately, the only quantities that really enter the 2d-4d wall-crossing formula are the differences
$\omega(\gamma,\gamma_{ij'})$.  These do have a clear definition as a trace in a Hilbert space, which will
be detailed in the next section \ref{sec:lineops}. These will also be the quantities which we will have direct control on in our examples.

\subsection{Cancellation of surface defect global anomalies by gauging flavor symmetries}
\label{subsec:Anomaly-Cancelation}

In \S \ref{sec:gen-example} we found that surface defects in \tf\ potentially
have global anomalies, and in \S\S \ref{subsubsec:Potential-Anomaly}
and  \ref{subsubsec:ExampleSU2-CP1}
we mentioned that there are simple examples which illustrate such anomalies.
In this section we demonstrate in somewhat general terms how such anomalies can
sometimes be cancelled by gauging an appropriate subgroup of the flavor group.
(Such a subgroup might or might not exist in any given theory.)

First, let us put the problem we are trying to solve in the proper context:

We are used to thinking about the spectrum of zero-dimensional defects, or local operators,
as an integral part of what a theory is, maybe as part of the very definition of the theory.
We also think of the spectrum of one-dimensional defects, or line defects, this way, although
this often involves a small refinement of our notion of ``theory.''
For example, in four dimensions, there is typically a discrete choice of which class of line defects
to allow:  in, say, a gauge theory with Lie algebra $su(N)$,
the fundamental Wilson loop and fundamental 't Hooft loop cannot appear simultaneously in correlation functions, because they
would not be mutually local.  The choice of which line defects to allow is part of the proper definition of the
theory and involves, for example, a choice of the actual compact gauge group. For example, no theory with
gauge algebra $su(N)$
can contain both the fundamental Wilson loop and the fundamental 't Hooft loop because they are mutually nonlocal.
The fundamental Wilson loop is allowed if the gauge group is $SU(N)$ while the fundamental 't Hooft loop is allowed
if the gauge group is $PSU(N)$.
By contrast, we think of a surface defect not as a part of the theory, but rather as a modification of the theory.
The spectrum of possible surface defects might be as intricate as the set of all possible 2d theories, or even more so.

In four dimensions, considering surface defects together with local operators might lead to problems with locality:
transporting a local operator around a surface defect might induce some unwanted monodromy phases.
As we usually consider the set of local operators as a given, we would normally just not allow such surface defects.
In \S \ref{sec:Hitchin-Expl} we will encounter situations, though, where these
surface defects (namely, the canonical surface defects $\bS_z$)
are too important simply to be  thrown away, so
it is the local operators which have to give. We will therefore change the theory
to accomodate these surface defects.
 The idea is as follows.
The effect of the monodromy around a surface defect
must be some symmetry of the bulk theory.  We will
assume that the symmetry transformations associated to the
surface defects we wish to retain  lies in a finite abelian group $\CD$
which is a subgroup of the group of flavor symmetries.
This is indeed what happens to the surface defects $\bS_z$ studied   in
\S\S \ref{sec:Hitchin-Expl},
\ref{sec:Detailed-Expl} and \ref{sec:Solve-Hitchin}. As with any
finite global symmetry group, we can gauge it.
In doing so we throw away the bothersome local operators, which are no longer
gauge invariant.  At the same time we introduce new surface defects corresponding
to nontrivial $\CD$-bundles in the space transverse to the surface
defect.\footnote{In a recent paper the effects of gauging finite global
symmetry groups in supergravity theories was studied \cite{Seiberg:2010qd,Banks:2010zn}.
Some of the considerations of that paper are similar to ours. We thank
N. Seiberg for discussions which considerably influenced our thinking
on this topic.}

Of course, we do not want to throw away the vanilla BPS states.
When the charge torsors $\Gamma_i$ are twisted and
have shifts by fractional elements of $\Gamma$, the AB phases of test particles of
charges $\gamma_g^{\rm test}$ will not be well-defined.
We will also see that by gauging a suitable finite flavor subgroup
this anomaly can in principle be cancelled.

In general the new surface defects are labeled by conjugacy classes  of the
gauged finite flavor symmetry group, $\CD$, and since we assume it is abelian
they will be labeled by elements of that group.
In what follows, one consequence of this gauging will be particularly important:
surface defects which carry nontrivial flavor
monodromy cannot have boundaries.  They can, however, have interfaces
to other surface defects carrying the same flavor monodromy.
Throughout this paper, we will see that gauging the flavor symmetry
has subtle and entertaining effects on the IR physics
of the surface defect.

\subsubsection{Flavor holonomies} \label{sec:flavor-phase}

In this section we assume that there is a
finite abelian flavor group $\CD$, trivially fibered over $\CB^*$,
so that the Pontryagin dual group of characters $\hat \CD$
fits in an exact sequence
\begin{equation}\label{eq:lattice-discrete-ext-seq}
0 \rightarrow \hat \CD \rightarrow \Gamma_g^{\rm ext} \rightarrow \Gamma_g\rightarrow 0.
\end{equation}
As usual, this extension is trivial locally but not globally.

A useful example to keep in mind is given by choosing a finite subgroup
of the fiber of $\Gamma_f^*\otimes \IR/(2\pi \IZ)$. Then
 there is a projection map from $\Gamma_f$ to  $\hat \CD$.  Applying this map
to \eqref{eq:lattice-ext-seq} we find that the action of the discrete flavor group on
charged particles is captured by an extension of local systems of abelian groups
\eqref{eq:lattice-discrete-ext-seq}.

What does the  exact sequence \eqref{eq:lattice-discrete-ext-seq}
mean?
A gauge-invariant state (one whose charge projects to zero in $\Gamma_g$) could carry some element of $\hat\CD$.
There is an ambiguity
when we try to lift this action to non-gauge-invariant states such as BPS particles:
in other words, the discrete flavor charge of a BPS particle is ambiguous.
To fix the flavor charges we have to choose a splitting of \eqref{eq:lattice-discrete-ext-seq}.
(This is very similar to the ambiguity we have in assigning the usual flavor charges to
BPS particles, i.e. in splitting the sequence \eqref{eq:lattice-ext-seq}.)
Any two different splittings differ by an element of $\Hom(\Gamma_g, \hat\CD)$.
If we gauge $\CD$, then $\Gamma_g^{\rm ext}$ will be the new extended group of all gauge charges.

The fiberwise Pontryagin
dual of \eqref{eq:lattice-discrete-ext-seq} is an extension of the algebraic integrable system $\CM$:
\begin{equation}\label{eq:torus-discrete-ext-seq}
0 \rightarrow \CM \rightarrow \CM^{\rm ext} \rightarrow \CD \rightarrow 0.
\end{equation}
The fiber of
$\CM^{\rm ext}$ has one connected component $\CM_d$ for each element $d \in \CD$; the identity component $\CM_1 = \CM$,
while the other $\CM_d$ are twisted versions of $\CM$.

Now we are ready to consider the surface defects which become available after
gauging $\CD$.  These surface defects carry
a flavor monodromy $d \in \CD$.  The Aharonov-Bohm phases of test particles
around such a surface defect give a section of $\CM^{\rm ext}$, or more precisely, of its connected component
$\CM_d$ along the lines of \eqref{eq:mult-sec-def}.
So globally the IR data of the surface defect is determined by a complex Lagrangian
multi-section of $\CM_d$.

We can describe the monodromy of $\CM_d$ more explicitly as follows.
Pick a splitting of \eqref{eq:lattice-discrete-ext-seq} into discrete flavor charges and gauge
charges, i.e. decompose charges $\gamma^{\rm ext}_g$ into $\xi \oplus \gamma_g$, where $\xi \in \hat \CD$ and $\gamma_g \in \Gamma_g$.
Under monodromy around a loop on the Coulomb branch $\CB^*$, these charges are transformed by
\begin{align}
\gamma_g &\to M(\gamma_g), \\
\xi &\to \xi + \delta \xi(\gamma_g) \label{eq:delta-xi}
\end{align}
where $\delta \xi$ is some homomorphism $\Gamma_g \to \hat \CD$.  Notice that we are using an additive notation
for the characters $\xi$ and $\delta \xi$. The multiplicative characters are $\exp 2 \pi \I \xi$, etc.

Then, we can attempt to define functions $\theta_{\gamma_g}$ on $\CM_d$
analogous to those on $\CM$: Choosing a splitting of \eqref{eq:lattice-discrete-ext-seq}
and a corresponding local trivialization of $\CM^{\rm ext}$, $\xi \oplus \gamma_g$ defines
a $U(1)$-valued function $\xi(d) e^{\I \theta_{\gamma_g}}$ on $\CM^{\rm ext}$.
On the other hand, monodromy takes $(0 \oplus \gamma_g)$   to $\delta \xi(\gamma_g) \oplus M(\gamma_g)$
and hence
$\delta \xi$ deforms the monodromy of the fiber coordinates $\theta_{\gamma_g}$ on $\CM_d$ from
linear to affine-linear:
\begin{equation} \label{eq:twisted-monodromy}
\theta_{\gamma_g} \to  \theta_{M(\gamma_g)} + 2 \pi [\delta \xi(\gamma_g)](d).
\end{equation}
Since $ [\delta \xi(\gamma_g)](d)$ is linear in both $\gamma_g$ and $d$ and
since $\Gamma_g$ is self-dual with respect to  $\langle \cdot, \cdot \rangle$,
we can define a useful quantity $\delta\xi(d) \in \Gamma_g \otimes \IR/\IZ$
by
\begin{equation}
[\delta \xi(\gamma_g)](d)  = \exp[ 2\pi \I \langle \delta \xi(d), \gamma_g \rangle] .
\end{equation}

Now let us explore the consequences for the Gukov-Witten parameters $\nu$ associated
to a surface defect with flavor monodromy. If we define a multisection by
\eqref{eq:mult-sec-def} then, after monodromy around a loop in $\CB^*$ we will find
\begin{equation}
\exp[ i \theta_{M(\gamma_g)} + 2\pi \I \langle \delta \xi(d), \gamma_g \rangle] =
\exp[2\pi \I \langle \nu' , \gamma_g \rangle] .
\end{equation}
It follows that, if we choose some lift $\widehat{ \delta \xi(d) } \in \Gamma_g \otimes \IR$
then the monodromy of the Gukov-Witten parameters is of the form
\begin{equation}
\nu' = M^{tr} \cdot \nu + \widehat{ \delta \xi(d) } + \gamma
\end{equation}
for some $\gamma \in \Gamma_g$. Therefore, by \eqref{eq:IR-GW} the superpotential
has a shift $\CW \to \CW + Z_{\gamma'}$
(where we lift $\gamma'$ from $\Gamma_g\otimes \IR$ to $\Gamma\otimes \IR$).

Now, we should stress that the shift by $ \widehat{ \delta \xi(d) }$ is \emph{not}
by an element of $\Gamma_g$ but rather by a fractional element of $\Gamma_g$.
But such shifts were precisely the sort we found in potential anomalies in
defining the torsors $\Gamma_i$ in \S \ref{subsubsec:Potential-Anomaly}!
This will be the key to the anomaly cancellation mechanism described in
\S \ref{subsubsec:Anomaly Cancelation} below.

It is useful to give a concrete example here. Consider a typical example of a singularity in $\CB^*$,
which arises from the presence of a single light hypermultiplet,
whose charge is $\tilde \gamma$.  The monodromy of the local system $\Gamma$
is of the Lefschetz type, $\gamma \to \gamma + \langle \tilde \gamma, \gamma \rangle \tilde \gamma$.
This is easily generalized to the case with discrete flavor symmetries. The flavor monodromy will be
$\xi \to \xi + \langle \tilde \gamma_g, \gamma_g \rangle \tilde \xi$ if the
discrete charge of the hypermultiplet is $\tilde\xi$.
But then $\delta \xi(\gamma_g) = \langle \tilde \gamma_g, \gamma_g \rangle \tilde \xi$ and thus
$\delta \xi(d) = \tilde \xi(d) \tilde \gamma_g$. In general $ \tilde \xi(d) $ will be
a nontrivial element of $\IR/\IZ$.

\subsubsection{Anomaly cancellation}\label{subsubsec:Anomaly Cancelation}

In the previous section we have learned that by gauging a finite abelian flavor
symmetry we can induce shifts of the GW parameters analogous to those forced on
us by the  cohomological obstruction discussed in
\S \ref{sec:gen-example}. Therefore, under suitable conditions,
gauging a finite flavor symmetry group can cancel the global anomalies
obstructing the existence of a surface defect associated with a
twisted torsor.

For example, consider the situation discussed in equations
\eqref{eq:tilde-Zi-def} to \eqref{eq:tilde-gammaij-shft}.
The shift \eqref{eq:tilde-gammaij-shft} means that
AB phases are ill-defined because   a monodromy around a loop in
 $\CB C^*$ changes the AB phase of a test particle by
$\exp[2\pi \I \langle \frac{1}{n}\sum_{j=1}^n\gamma^{(ij)}, \gamma_g^{\rm test} \rangle ]$.
On the other hand, if we can find a suitable finite flavor subgroup $\CD$
and an element $d$ such that under the same monodromy transformation
\begin{equation}\label{eq:ac-cond}
\widehat{\delta \xi(d)} = -  \frac{1}{n}\sum_{j=1}^n\gamma^{(ij)} \, \mod \, \Gamma
\end{equation}
then the AB phases of BPS test particles in the theory where the flavor group
is gauged will be well-defined.
In this case we can form a good theory of the surface defect, even
though the torsor $\Gamma_i$ is twisted.

\textbf{Example \ref{subsubsec:Anomaly Cancelation}}.
For an example of how we might find a suitable finite flavor
group $\CD$ in the IR, we return to the theories discussed in \S \ref{sec:gen-example}. We will show
how to construct an extension like \eqref{eq:lattice-discrete-ext-seq}
with $\CD \simeq \IZ_n$, which will serve for anomaly cancellation. We begin with the exact sequence
\begin{equation}\label{eq:a-secq}
0  \rightarrow \CP(z)  \rightarrow H_1(\bar\Sigma - \pi^{-1} \{z\},\IZ) \rightarrow H_1(\bar\Sigma,\IZ)\rightarrow 0.
\end{equation}
This sequence extends the charge lattice by a free abelian group
of rank $n$,
\begin{equation}
\CP(z) = \bigoplus_{i=1}^n P_i(z)\cdot \IZ \, \simeq \, \IZ^n,
\end{equation}
generated by the homology classes of small circles $P_i(z)$ around the $n$ preimages of $z$. Now consider the sublattice $\Lambda_n \subset \CP(z)$
given by the kernel of the homomorphism $\CP(z) \to \IZ_n$
defined by $ \sum a_i P_i(z)  \mapsto \sum a_i\,\mod\,n$. We can
take a quotient of \eqref{eq:a-secq} by this lattice and its image in $H_1(\bar\Sigma - \pi^{-1} \{z\},\IZ) $
 to produce
\begin{equation}\label{eq:a-secq-fin}
0  \rightarrow \hat \CD \rightarrow H_1(\bar\Sigma - \pi^{-1} \{z\},\IZ)/\Lambda_n \rightarrow H_1(\bar\Sigma,\IZ)\rightarrow 0.
\end{equation}
where $\hat\CD := \CP(z)/\Lambda_n \simeq \IZ_n$.
The actual charge lattice $\Gamma_g$ is not exactly $H_1(\bar\Sigma,\IZ)$ but is
closely related to it as a subquotient. The subgroup $\hat\CD$ survives taking this
subquotient, yielding a candidate for \eqref{eq:lattice-discrete-ext-seq}.

Now suppose we choose a system of charges $\gamma_{ij}\in \Gamma_{ij}$ as in
\S \ref{subsubsec:Potential-Anomaly}. Using this data we can split gauge
charges $\gamma_g^{\rm ext} \in H_1(\bar \Sigma - \pi^{-1} \{z\},\IZ)/\Lambda_n$
as $\xi \oplus \gamma_g$ where $\xi = [\sum a_i P_i] $ with
\begin{equation}
\sum_{i=1}^n a_i = \sum_{j=1}^n \langle  \gamma_g^{\rm ext}, \gamma_{ij} \rangle \quad \mathrm{mod}\,n.
\end{equation}
(One can check that the right hand side is indeed independent of $i$.)

Now let $d$ be a primitive generator of $\CD$. Let $\bS$ be a surface defect with
flavor holonomy $d$. Then, the flavor contribution to the AB phase of a particle
of charge $\gamma_g^{\rm test} \in H_1(\bar \Sigma - \pi^{-1} \{z\},\IZ)/\Lambda_n$
transported around such a defect $\bS$  will be
\begin{equation}\label{eq:flv-ab}
\xi(d) = \exp \left( - \frac{2\pi \I}{n} \left\langle \sum_{j} \gamma_{ij} , \gamma_g^{\rm test} \right\rangle \right).
\end{equation}
Monodromy of the local system $\Gamma_{ij}$ around cycles in $\CB C^*$ takes
$\gamma_{ij} \to \gamma_{ij} + \gamma^{(ij)}$, and hence \eqref{eq:flv-ab}
satisfies (trivially) the anomaly cancellation condition \eqref{eq:ac-cond}.

\bigskip
\textbf{Remarks}:

\begin{enumerate}

\item  We must stress that we have only given a potential mechanism for anomaly cancellation
and we have not given a  general proof that the problems of \eqref{eq:prop} can always be cured
by conjecturing an extension of the lattice of gauge charges by characters for a $\IZ_n$ flavor symmetry.
Moreover, it is far from obvious that such tentative finite
flavor charge groups $\CD$ would always arise from actual symmetries of the bulk UV 4d theory.
In \S \ref{subsec:Lag-Desc} we explain that for the $A_1$ theories with regular singular points
a suitable UV $\IZ_2$ flavor symmetry does indeed exist.
In addition, one can define a suitable flavor symmetry for the AD theories with $N$
even.

\item
Notice that if the surface defect carries a nontrivial flavor group label
the non-integral monodromies of the $\Gamma_i$ imply that the
$\omega(\gamma, \gamma_i)$ are nonintegral.  This is not so strange, considering that
we have twisted boundary conditions around the surface defect
for bulk particles. It is perfectly possible for bulk hypermultiplets with
twisted boundary conditions to contribute to the 2d effective superpotential as
a fraction of a 2d chiral multiplet.
It would be nice to compute such contributions explicitly.

\end{enumerate}

\section{Line defects, interfaces, framed BPS states, and a spin 2d-4d  wall-crossing formula} \label{sec:lineops}

\subsection{Interfaces}

Past experience with the pure 4d wall-crossing formula suggests that a useful way
of understanding it is to study supersymmetric \ti{line defects} $L$ \cite{Gaiotto:2010be}.  In the 2d-4d context,
in addition to the usual 4d line defects, we want to consider supersymmetric line defects which sit on
a surface defect, dividing it into two pieces --- or more generally, supersymmetric \ti{interfaces}
which have one surface defect $\bS$ on the left and another surface defect $\bS'$ on the right.

The supersymmetric interfaces we consider preserve $2$ out of the $4$ supercharges preserved by the
surface defect.  In the conventions of \cite{Gaiotto:2010be}, they are
\begin{equation}\label{eq:preserve-susy}
Q_1^1 -  \zeta^{-1} \bar Q_{\dot 2}^1,  \qquad Q_2^2 + \zeta^{-1} \bar Q_{\dot 1}^2
\end{equation}
where the lower index is a spin index and the upper index is an $su(2)_R$ symmetry index.
As in \cite{Gaiotto:2010be}, $\zeta$ is a phase, but later will be analytically continued to $\IC^{\times}$, and
the interfaces in general depend on a lift to the universal cover $\widetilde{\IC^\times}$.

Some simple examples of supersymmetric interfaces on surface defects can be constructed directly.  For example,
suppose we consider a supersymmetric gauge theory with gauge group $G$, with a Gukov-Witten-type surface
defect $\bS$ which breaks $G$ to a Levi subgroup $H$.  Then we can consider a supersymmetric Wilson line
in an irreducible representation of $H$ (which is \ti{not} a representation of $G$).  This gives a line
defect which is restricted to lie on $\bS$.  Another useful machine for producing such interfaces
is the ``Janus'' construction:  here we allow a twisted F-term coupling on the surface defect to jump
at some point $x^3$, and include an appropriate coupling at the interface to preserve supersymmetry
\cite{Gaiotto:2009fs}.

In the presence of the interface, the IR physics depends on the choice of a vacuum $i$ for $\bS$ and $j'$ for $\bS'$.
Having fixed that choice, the Hilbert space is graded by charges lying in $\Gamma_{i j'} = \Gamma_i - \Gamma_{j'}$.

\subsection{Review of the 4d case: Framed BPS states and halos} \label{subsec:halos-4d}

Let us recall briefly the case studied in \cite{Gaiotto:2010be}. This
is the case where  $\bS$ and $\bS'$ are both the null surface defect.
In this case we denoted a line defect of type $\zeta$ by $L_\zeta$.
Often $\zeta$ will be understood and we simply write $L$.
The  presence of the line defect modifies the Hilbert space of (one particle)
states which is denoted $\CH_{L,u}$, where $u\in \CB$. Often the $u$-dependence
will be suppressed in the notation. The Hilbert space $\CH_{L}$ is graded by
a $\Gamma$-torsor $\Gamma_{L}$ and we write  $\CH_{L}  = \oplus_{\gamma \in \Gamma_L}
\CH_{L, \gamma}$.  The preserved supersymmetries lead to a BPS bound
$E \geq - \Re(Z_\gamma/\zeta)$ in $\CH_{L, \gamma}$ and states satisfying this
bound are called \emph{framed BPS states}. We define a protected spin character for these
\begin{equation}\label{eq:Def-Framed-PSC}
\fro(u,L_\zeta, \gamma;y) := {\rm Tr}_{\CH_{u,L,\zeta,\gamma}^{\rm
BPS}} (-1)^{2J_3} (-y)^{2\CJ_3}
\end{equation}
where
\begin{equation}
\CJ_3 := J_3 + I_3,
\end{equation}
and $I_3$ is an R-symmetry
generator.
We will often lighten the notation and just write $\fro(L,\gamma;y)$. Specializing
to $y=-1$ defines the framed BPS index $\fro(L,\gamma)$.

When considered as functions of $(u,\zeta)$ the framed indices are piecewise constant
but undergo wall-crossing, just like ordinary BPS degeneracies. The main physical
justification for this is the ``halo picture'' described  in
\cite{Denef:2007vg,Gaiotto:2010be,Andriyash:2010qv,Andriyash:2010yf,PiTP}.
Near BPS walls, defined for populated charges $\gamma \in \Gamma$ by,
\begin{equation}\label{eq:4d-wall}
W_{\gamma} := \{ (u,\zeta) : Z_\gamma/\zeta < 0 \}
\end{equation}
some of the states in $\CH_{u,L,\zeta,\gamma}^{\rm BPS}$ can be described as
``halo configurations.''  They look like a ``core'' BPS state of the line defect
very weakly bound to some ``halo'' particles.
The distance between the line defect and the halo particles
has a universal form depending only on the IR data:  if $\gamma^c$, $\gamma^h$ are the core and halo charges
respectively, it is
\begin{equation}
 r_h = \frac{\inprod{\gamma^c, \gamma^h}}{2\,\im(Z_{\gamma^h} / \zeta)}.
\end{equation}
The bound state exists only when $r_h > 0$, i.e. only on one side of the wall; as we approach the wall
from that side  $r_h \to \infty$, so this bound state  disappears from the framed BPS spectrum.

The framed BPS states which appear/disappear upon crossing the wall $W_{\gamma^h}$ can be thought of as
states in a Fock space consisting of states containing a core charge and some number of
halo particles of charge $\gamma^h$
 orbiting around it. Since the halo particles are mutually BPS there are many such
states, and they can be enumerated with Fock space combinatorics. In order to describe this
more precisely
we first introduce a generating function
\begin{equation}\label{eq:gen}
F^{}(u, L_\zeta, \{x_\gamma\}; y) := \sum_{\gamma} \fro(u, L, \zeta,
\gamma; y) x_\gamma.
\end{equation}
where $x_{\gamma} x_{\gamma'} = x_{\gamma +
\gamma'}.$ In order to lighten the notation we will often drop various
arguments, and sometimes simply write this as $F^{}(L)$.
Now, one isolates the contributions from a core charge $\gamma^c$ surrounded
by halo configurations of particles with charge $\gamma^h$. The
$\IZ_2$-graded Fock space is built on $(J_{\gamma^c, \gamma^h}) \otimes
\tilde\CH^{\BPS}(\gamma^h;u)$, where $\tilde \CH^{\BPS}$ is the space of BPS states
with the center-of-mass degree of freedom factored out. Here $(J_{\gamma^c, \gamma^h})$
is the spin representation with maximal spin $2J_3 = \vert \langle \gamma^c,\gamma^h\rangle \vert$.
If we decompose
\begin{equation}
\Omega(u,\gamma^h;-z) =  {\Tr}_{\tilde \CH(\gamma^h;u)}(-z)^{2J_3} z^{2 I_3} =
\sum_{m\in \IZ}  a_{m, \gamma^h} z^m
\end{equation}
then $a_m\geq 0$ for $m$ even while  $a_m\leq 0$ for $m$ odd. There are creation
operators $A_{m',m,\alpha}^\dagger$  with
\begin{equation}\label{eq:enum-osc}
- \vert \langle \gamma^c,\gamma^h\rangle \vert\leq m'
\leq \vert \langle \gamma^c,\gamma^h\rangle \vert
\end{equation}
while $\alpha$ runs over $\vert a_{m,\gamma^h}\vert$ values for each $m$.
Somewhat counterintuitively, the oscillators with $m$ even correspond to fermionic
oscillators while those with $m$ odd correspond to bosonic oscillators.  The corresponding
Fock space factor when a halo is created is thus
\begin{equation}
(1  +   y^{m+m'} x_{\gamma^h} )^{  a_{m,\gamma^h}}
\end{equation}
when $m$ is even, and
\begin{equation}
\frac{1}{(1  -   y^{m+m'} x_{\gamma^h} )^{ \vert a_{m,\gamma^h}\vert }}
\end{equation}
when $m$ is odd.  These factors can be summarized as
\begin{equation}
(1  +(-1)^m   y^{m+m'} x_{\gamma^h} )^{ a_{m,\gamma^h}}.
\end{equation}

Reference \cite{Gaiotto:2010be} shows that the framed BPS indices (at $y=-1$)
can be described by replacing $x_\gamma$ by $X_\gamma$ satisfying the
twisted group law $X_\gamma X_{\gamma'} = (-1)^{\langle \gamma, \gamma'\rangle} X_{\gamma + \gamma'}$,
and making an $L$-independent ``coordinate'' transformation of the $X_\gamma$,
\begin{equation}
X_{\gamma'} \to (1-X_{\gamma})^{\langle \gamma, \gamma' \rangle\Omega(\gamma)} X_{\gamma'}.
\end{equation}
This transformation is nothing but (the inverse of) \eqref{eq:k-4d}!  This is the key
observation in one proof of the 4d wall-crossing formula for the $\Omega(\gamma)$ \cite{Gaiotto:2010be}.

If one wants a wall-crossing formula for the full $y$-dependent framed BPS indices, then the above
generating function must be split up into its contribution from various ``core'' charges.
The reason for this is that the halo Fock spaces (in particular the range $m'$
in \eqref{eq:enum-osc}) depend on the core charge.  This leads to somewhat
awkward wall-crossing formulae.  However, as  explained at
length in \cite{Gaiotto:2010be} (see also \cite{Andriyash:2010qv,PiTP})  one can write an
elegant formula by   replacing $F^{}$ by the same expression evaluated on
noncommuting variables,
\begin{equation}\label{eq:gen-2}
\widehat{F}(u, L_\zeta, \{\hat X_\gamma\}; y) := \sum_{\gamma} \fro(u, L, \zeta,
\gamma; y) \hat X_\gamma,
\end{equation}
where
\begin{equation}\label{eq:Heis-Alg}
\hat X_\gamma \hat X_{\gamma'} = y^{\inprod{\gamma, \gamma'}} \hat X_{\gamma +
\gamma'},
\end{equation}
Again, we will abbreviate this by $\widehat{F}(L)$.

The transformation of $\widehat{F}$ across walls $W_\gamma$ for populated charges
$\gamma$ can be expressed in terms of conjugation by products of quantum
dilogs evaluated on $\hat X_\gamma$, where the product is determined by
the data $a_{m,\gamma}$. See \cite{Dimofte:2009tm,Gaiotto:2010be} for details.

\subsection{Analog in two dimensions}

Let us now consider the special case where the four-dimensional theory is null but
there are nontrivial surface defects separated by an interface
$\bS L_\zeta \bS'$.
It is tempting to apply the approach of \S \ref{subsec:halos-4d}
to try to understand the purely 2d wall-crossing formula as well.
We do this in the present section,
although our arguments in this case will be somewhat less rigorous.
Just as in the 4d situation, we can define ``framed BPS states'' in the Hilbert space of the theory
with a supersymmetric interface $L_\zeta$ inserted between two surface defects.
Then we describe the wall-crossing of the framed BPS degeneracies.

At least in specific models such as  supersymmetric Landau-Ginzburg models, there is a picture
of this wall-crossing which is much like that in the 4d case:  there are ``halo'' states
consisting of a single 2d particle loosely bound to the line defect. Although there is
not a precise formula for this halo radius it can be shown that any reasonable measure
of such a radius has the crucial property that it diverges at walls $W_{ij}$ where
$Z_{ij}/\zeta \in \IR_-$.
See Appendix \ref{app:LG-Radius}.  At such walls, halo states appear or disappear from
the framed BPS spectrum.

As in the 4d case, after introducing appropriate generating functions $F^{}(L)$ for the framed BPS states, this
picture leads to a prediction for the transformation of $F(L)$ across the walls.
The transformation is
\begin{equation} \label{eq:tleft}
F^{}(L) \to \left(1 - \mu_{ij} X_{ij}\right) F^{}(L).
\end{equation}
For 2d particles on the right of the line defect, similarly
\begin{equation} \label{eq:tright}
F^{}(L) \to F^{}(L) \left(1 + \mu_{ij} X_{ij}\right).
\end{equation}
Note the change of sign between \eqref{eq:tleft} and \eqref{eq:tright}.  We will see later
that this is necessary for the wall-crossing invariance of the algebra of OPEs of line defects.

The transformation \eqref{eq:tleft} is precisely \eqref{eq:s-2d}.  Starting from this observation, and following again
the approach of \cite{Gaiotto:2010be}, we get a new way of understanding the 2d wall-crossing formula.

Our discussion of the ``halo states'' above invoked facts specific to the Landau-Ginzburg model.
It would be very natural to believe that these states in fact exist in any 2d theory and have the same
behavior we have described; if so, this would give a real derivation of the 2d wall-crossing formula using
only the halo picture.
Unfortunately (and in contrast to the 4d case) it seems difficult to argue directly that this behavior
is a universal feature depending only on the infrared physics.
The difficulty is that we are considering infrared theories which are massive:
the forces between the halo particle and the line defect
are generated by irrelevant operators which are, after all, irrelevant  at low energies.
In \cite{Cecotti:1992rm} a full proof of the 2d wall-crossing formula was given, independent of the details of
the theory, but it required the heavier machinery of $tt^*$ geometry.

\subsection{Framed indices in the 2d-4d case}\label{sec:2d-4dhalo}

Let us now consider the general 2d-4d case. There is a Hilbert space
of one-particle states in the  presence of surface defects. If we have vacua
$i \in \CV(\bS)$ and $j'\in \CV(\bS')$ then the Hilbert space is graded by
the $\Gamma$-torsor $\Gamma_{ij'}$,
\begin{equation}
\CH_{u, \bS L_\zeta\bS'}  = \bigoplus_{i\in \CV(\bS), j'\in \CV(\bS')}
\bigoplus_{a  \in \Gamma_{ij'} }
\CH_{u, \bS L_\zeta\bS', a}.
\end{equation}
As usual we will lighten the notation to $\CH_{L}$ and $\CH_{L,a}$.

The two supercharges \eqref{eq:preserve-susy} preserved by the line defect imply a BPS bound:  states of charge
$\gamma_{ij'}$ have
\begin{equation}
 E \ge - \re (Z_{\gamma_{ij'}} / \zeta).
\end{equation}
So just as reviewed above, we can consider \ti{framed BPS states} saturating this bound. Moreover, we can
introduce a framed BPS degeneracy:
\begin{equation}\label{eq:Interface-Framed-PSC}
\fro(u,\bS L_\zeta \bS', a;y) := {\rm Tr}_{\CH_{L,a }^{\rm
BPS}} e^{\I \pi F}  (-y)^{\CJ}
\end{equation}
As usual we abbreviate this by $\fro(L,a;y)$ and the value
at $y=-1$ by $\fro(L,a)$.
Here we have made a choice of $u(1)_V$ charge generator $F$. This generates a
one-dimensional abelian Lie algebra and hence its spectrum is some set of real
numbers, not necessarily integer-spaced. Moreover we have
introduced the operator $\CJ:= 2J_{12} + 2I_{12}$ which commutes with the preserved
supercharges. Again the spectrum is some set of real numbers, not necessarily
integer-spaced.

\insfig{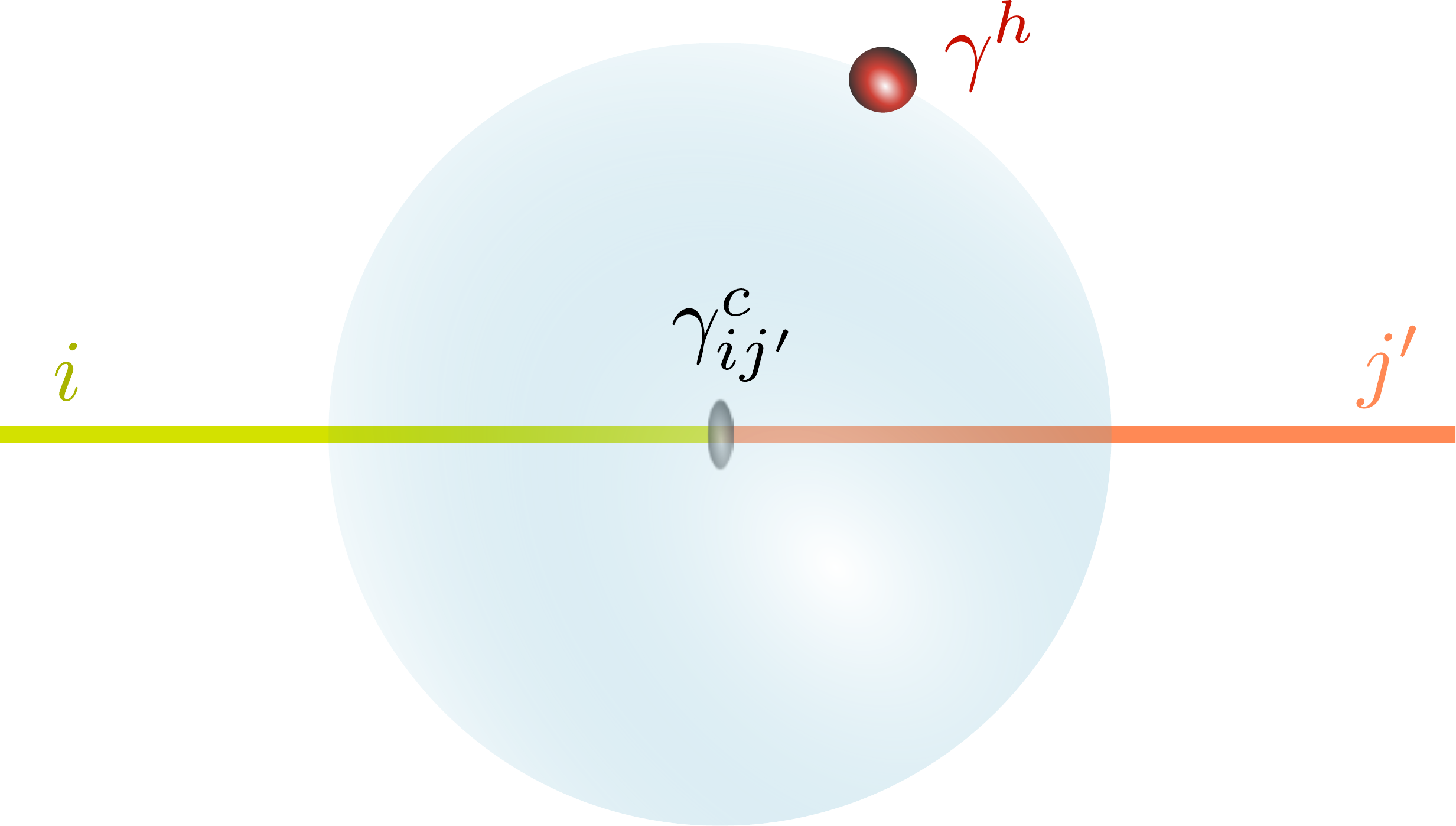}{A 4d  halo particle of charge $\gamma^{h}$
is bound to an interface in charge sector $\gamma_{ij'}^c$ (the ``core charge'')
to produce a framed
BPS state of charge $\gamma_{ij'} = \gamma^{h} + \gamma_{ij'}^c$.   }

\insfig{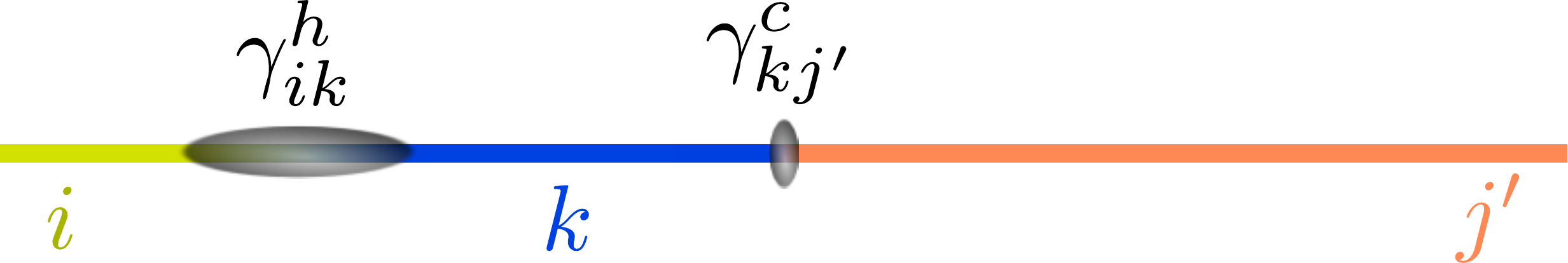}{A soliton of type $ik$ and 4d electromagnetic charge $\gamma_{ik}^h$
is bound to an interface in charge sector $\gamma_{kj'}^c$ to produce a framed
BPS state of charge $\gamma_{ij'} = \gamma_{ik}^h + \gamma_{kj'}^c$. A similar picture
applies to solitons of charge $\gamma_{k'j'}^h$ bound on the right. }

The framed BPS degeneracies will undergo wall-crossing across two
kinds of walls.  First,  halo states will be gained/lost
when crossing
walls of type $W_\gamma$ for populated charges $\gamma \in \Gamma$. Such states are
shown in Figure \ref{fig:halo-line-bound}.
Second, there will also be wall-crossing across \emph{soliton walls}  (again
for populated charges):
\begin{align}\label{eq:2d-wall}
W_{\gamma_{ik}} &:= \{ (u,\zeta) : Z_{\gamma_{ik}}/\zeta < 0 \} \qquad i,k\in \CV(\bS), \\
W_{\gamma_{ k'j'}} &:= \{ (u,\zeta) : Z_{\gamma_{k'j'}}/\zeta < 0 \} \qquad k',j'\in \CV(\bS'). \label{eq:2d-wall-p}
\end{align}
Across the walls \eqref{eq:2d-wall},
a bound state involving a 2d ``halo'' particle with charge $\gamma_{ik}$ and degeneracy $\mu(\gamma_{ik})$,
sitting to the left of a core state of the line defect, appears or disappears. Such states are illustrated in
Figure \ref{fig:soliton-line-bound}.  Similarly \eqref{eq:2d-wall-p} involves bound states where the halo
particle is on the right.

In \S \ref{subsec:Halo-Just}
we will interpret the transformations \eqref{eq:cv-like} and \eqref{eq:ks-like}
which enter the wall-crossing formula in terms of these  ``halo'' states.
The interpretation of \eqref{eq:cv-like} should be much like what we have just discussed in the pure 2d case.
Similarly, we should interpret the wall-crossing associated with $W_\gamma$
in terms of a halo of particles of charge $\gamma$, which surrounds a core state
of the line defect with charge $\gamma_{ij'} = \gamma_i - \gamma_{j'}$. However,
since there are a number of potentially subtle points in the halo discussion, we
first make some general remarks on the structure of the wall-crossing formula.

\subsection{Ring of line defects}\label{subsec:Line-Op-Ring}

In \cite{Gaiotto:2010be} an important role was played by an
algebra of line defects whose coefficients are vector spaces.
There is a natural generalization to the case of supersymmetric
interfaces.

Indeed, consider two interfaces  $\bS L_\zeta \bS'$ and $\bS' L'_\zeta \bS''$
preserving the same supersymmetry.
We can define a new interface $\bS \left(L \circ L'\right)_\zeta \bS''$
by a (nonsingular!) OPE, bringing them against each other along the surface defect.
It should be possible to compute the framed BPS generating function of the composition of the two defects
by counting the framed BPS states of the system of two well-separated line defects $L$ and $L'$:
the index should be independent of the distance between the line defects.
If the distance between the defects is sufficiently large, the Hilbert space of ground states of the
system should be the tensor product of the Hilbert spaces of the two subsystems.
We should allow the strip of surface defect between the two line defects to be in any of its possible vacua.
The only IR subtlety we should take care of
is the fermion number grading, which can be affected by the gauge charges carried by the
ground states: a system of two well-separated dyons might carry fractional amounts
of angular momentum hidden in the electro-magnetic fields.

The coefficients of this OPE are vector spaces
which are graded by $F$ and $\CJ$ \cite{Gaiotto:2010be}.
Replacing these vector spaces by their characters defines
a deformed product $L\circ_y L'$.
As in \cite{Gaiotto:2010be} it is useful to describe this
product by   introducing the ``quantum''
generating function
\begin{equation}\label{eq:q-F-gen}
\widehat{F}(L):= \sum_{i\in \CV(\bS), j'\in \CV(\bS'), \gamma_{ij'}\in \Gamma_{ij'}}
\fro(L_\zeta, \gamma_{ij'};y) \hat X_{\gamma_{ij'}}
\end{equation}
where, for the moment, $\hat X_{\gamma_{ij'}}$ are just placeholders.

We now claim that with a suitable multiplication law on $\hat X_a$ we have the
rule
\begin{equation}\label{eq:Fq-mult}
\widehat{F}(L \circ_y  L') = \widehat{F}(L) \widehat{F}(L')
\end{equation}
that is, $\widehat{F}$ is a homomorphism from the algebra of interfaces to the groupoid
algebra\footnote{The existence of supersymmetric interfaces extends the groupoid
algebra to include that for several surface defects. We understand $\IV$ to
mean this extended groupoid.}  $\IC[\IV]$.

In order to find the requisite multiplication law on $\hat X_a$   we use the relation
\begin{equation}\label{eq:Hilb-fact}
\CH_{L\circ_y L', c}   = \oplus_{a+b=c}
\CH_{ L,a} \otimes \CH_{ L',b} \otimes \CN_{a,b}
\end{equation}
where $\CN_{a,b}$ is a one-dimensional  representation of $so(2)_{12}\oplus u(1)_r$.
Let us define $\sigma(a,b)$ and $n(a,b)$ via:
\begin{equation}\label{eq:sigma-ab-def}
 e^{\I \pi F}\vert_{\CN_{a,b}} := \sigma(a,b)
\end{equation}
\begin{equation}\label{eq:n-ab-def}
 (-y)^{ \CJ}\vert_{\CN_{a,b}} := (-y)^{n(a,b)}
\end{equation}
Note that $\sigma(a,b)$ is in general a phase and $n(a,b)$ need not
be integer because of angular momentum in massless degrees of freedom.

Equation \eqref{eq:Hilb-fact} is reasonable because we are working with
indices and therefore -- by topological
invariance -- we can separate the line defects an arbitrary distance from
each other. The states can factorize but there can be shifts in the action of  various
$u(1)$ generators.

Now, equation \eqref{eq:Fq-mult} will hold
\emph{provided}
\begin{equation}\label{eq:2d-4d-heis}
\hat X_a \hat X_b =  \sigma(a,b) (-y)^{n(a,b)} \hat X_{a+b}
\end{equation}
for composable morphisms $a,b$. We continue to take $\hat X_a \hat X_b= 0$ if
$a,b$ are not composable.
Associativity is guaranteed if $n(a,b)$ and $\sigma(a,b)$ are 2-cocycles on the groupoid.
From the case with $\bS $ and $\bS'$ the null surface defects and the choice
$F = 2J_{12}$  we know from \cite{Gaiotto:2010be} that
$n(\gamma,\gamma') = \langle \gamma, \gamma' \rangle$  and
$\sigma(\gamma, \gamma') = (-1)^{ \langle \gamma, \gamma' \rangle}$.
Now, these cocycles can be shifted by coboundaries. Physically, these
correspond to shifts in the definitions of the generators $F$ and $\CJ$.
By an argument analogous to that at the end of \S \ref{subsubsec:2d-4d-GlobalPicture} we may
assume that $\sigma(a,b)$ are signs and $n(a,b)$ are integers, and we will
henceforth make that assumption.

\subsection{A 2d-4d spin wall-crossing formula}\label{subsec:2d-4d-motivic}

Now we come to a crucial point. Wall-crossing must be compatible with the multiplication of
line defects, since the latter is defined in the UV and hence independent of the choice
of the vacuum.  Therefore the wall-crossing transformation
must be an automorphism  of our groupoid algebra \eqref{eq:2d-4d-heis}. This is
a matrix algebra over the quantum torus, hence Morita equivalent to the
coordinate algebra of the quantum torus, and hence a simple algebra.
Therefore the automorphism must be inner.
Therefore, across a wall of type  $W_\gamma$, $\widehat{F}(L)$ should transform according to
\begin{equation}\label{eq:q-tmn}
\hat T_\gamma: \hat X_a \mapsto  \Phi_{\bS,\gamma} \hat X_a \Phi_{\bS',\gamma}^{-1}.
\end{equation}
As we will see below, the halo picture of wall-crossing suggests that $\Phi_{\bS,\gamma}$
must have the form
\begin{equation}\label{eq:phbs-def}
\Phi_{\bS,\gamma} =\prod_{i\in \CV(\bS)} \prod_{s\in \CS_{i,\gamma} }
( 1 + \varphi_{i,s}(y) \hat X_{\gamma_{ii}} )^{d(i,s)},
\end{equation}
with an analogous expression for $\Phi_{\bS',\gamma}$.
Here $\CS_{i,\gamma}$ is some countably infinite set of real numbers depending on $i,\gamma$
and $\varphi_{i,s}(y)$ is a ``monomial in $y$,'' that is, it is
a phase times a (possibly fractional) power of $y$. The important
thing is that  $\Phi_{\bS,\gamma}$ does not depend on the detailed core charge $a$
being conjugated.\footnote{In equation \eqref{eq:phbs-def} we have assumed, for simplicity, that on $W_\gamma$ the
only populated charges with central charge $Z$ parallel to $Z_\gamma$ are particles
that themselves have electromagnetic charge $\gamma$. We continue to make that assumption below.}

Similarly, across walls of the kind $W_{\gamma_{ij}}$ we have again a conjugation
\eqref{eq:q-tmn}, but now with
\begin{equation}\label{eq:Phib-sol}
\Phi_{\bS, \gamma_{ij}} = ( 1 + \varphi_{\gamma_{ij}}(y) \hat X_{\gamma_{ij}}).
\end{equation}
Here $\varphi_{\gamma_{ij}}(y)$ is a sum of monomials in $y$. (See \eqref{eq:phi-mu} below.)

Given the transformation \eqref{eq:q-tmn}, we can use
the same argument as in \cite{Gaiotto:2010be} to deduce a 2d-4d analog
of the Kontsevich-Soibelman motivic wall-crossing formula. Given a path
$\CP$ between two points
in $\CB \times \IC^\times$  the accumulated transformation of $\widehat{F}(L)$ is
\begin{equation}
\widehat{F}(L) \to  T^{\rm left}(\CP) \widehat{F}(L) T^{\rm right}(\CP)
\end{equation}
where
\begin{equation}
T^{\rm left}(\CP) = : \prod_{b}    \Phi_{\bS,b}  :
\end{equation}
  is the path-ordered product of the  transformations taken across
walls  $W_{b}$  of type \eqref{eq:4d-wall} or \eqref{eq:2d-wall}
crossed by the path $\CP$. Given two paths $\CP$ and $\CP'$ with
common endpoints such that the local system does not have
monodromy around the corresponding closed path,
we can state the  spin 2d-4d wall-crossing formula
in the form:
\begin{equation} \label{eq:spin-wcf}
T^{\rm left}(\CP) = T^{\rm left}(\CP').
\end{equation}
(An equivalent formula can be written with $T^{\rm right}$.)
A generalization can be stated which takes into account the monodromy
of an arbitrary path $\CP (\CP')^{-1}$.

\subsection{Halo picture}\label{subsec:Halo-Just}

We would now like to relate the quantities $\CS_{i, \gamma}$, $\varphi_{i,s}(y)$,
$d(i,s)$ and $\varphi_{\gamma_{ij}}(y)$ introduced in the wall-crossing
formula  above to physical quantities determined
by Hilbert spaces of BPS states.

\subsubsection{Crossing soliton walls}

The first main statement of framed wall-crossing is that across $W_{\gamma_{ik}}$ we have
\begin{equation}\label{eq:sol-wall-hilb}
\Delta \CH^{\BPS}_{L, \gamma_{ij'}} = \pm \tilde \CH^{\BPS}_{\gamma_{ik}} \otimes \CH^\BPS_{L, \gamma_{kj'}}
\otimes \CN_{\gamma_{ik}, \gamma_{kj'},L}
\end{equation}
as representations of the abelian Lie algebra $u(1) \oplus u(1)$ generated by $F$ and $\CJ$.
Here $\CN_{\gamma_{ik}, \gamma_{kj'},L}$ is a one-dimensional representation with
character $e^{\I \pi f_{\CN} } (-y)^{\CJ_{\CN} }$. The
overall sign of the right hand side depends on the direction the wall is crossed.

To express this in terms of generating functions, define
\begin{equation}
 \mu(\gamma_{ij};y) := \Tr_{\tilde\CH^{ \BPS}_{\gamma_{ik}} } e^{\I \pi F} (-y)^{ \CJ}.
\end{equation}

Comparing with the wall-crossing  from \eqref{eq:Phib-sol}  and taking traces we learn that
\begin{equation}
\varphi_{\gamma_{ik}}(y) = \pm \mu(\gamma_{ik};y)
\left( \frac{ \sigma(\gamma_{ik}, \gamma_{kj'})}{e^{\I \pi f_\CN}}\right)
y^{ \CJ_\CN - n(\gamma_{ik},\gamma_{kj'})}.
\end{equation}
Putting $y=-1$ we find agreement with the formal structure described in
equation \eqref{eq:cv-like} provided that $e^{\I \pi F}$ on the one-dimensional
space
$\CN_{\gamma_{ik}, \gamma_{kj'},L}$ has the value  $\sigma(\gamma_{ik}, \gamma_{kj'})$.
Similarly, it is natural to guess that  $n(a,b)$ defined in \eqref{eq:n-ab-def}
satisfies   $\CJ_\CN - n(\gamma_{ik},\gamma_{kj'})=0$ and hence we conjecture that
\begin{equation}\label{eq:phi-mu}
\varphi_{\gamma_{ik}}(y) = \pm \mu(\gamma_{ik};y).
\end{equation}

\subsubsection{Crossing 4d walls}\label{subsubsec:4d-halo}

Let us try to describe the halo particles relevant to wall-crossing
associated with $W_\gamma$.
We choose a specific IR Lagrangian description of the setup,
i.e. a particular $\gamma_i^0 \in \Gamma_i$ and $\gamma_{j'}^0 \in \Gamma_{j'}$.
As we have mentioned before, $\gamma_i^0$, $\gamma_{j'}^0$ measure the amount of
electromagnetic flux which is threaded through the surface defects.  The interface sources
a total flux $\gamma_{ij'}$, of which  $\gamma_i^0 - \gamma^0_{j'}$ is carried by the surface defects,
while the remaining
\begin{equation}
\gamma^c = \gamma_{ij'} - \gamma_i^0 + \gamma_{j'}^0 \in \Gamma
\end{equation}
emerges into space.
Thus, the duality invariant gauge field is
\begin{equation}\label{eq:F-decomp}
\IF = \gamma_{ij'} \omega_{S^2} + \gamma_{j'}^0 \delta_N - \gamma_i^0\delta_S
\end{equation}
See Figure \ref{fig:threaded-flux}.
 Here $\omega_{S^2}$ is the unit volume form of the
sphere and $\delta_N,\delta_S$ are unit weight delta functions at the north and south poles.
The decomposition \eqref{eq:F-decomp} is noncanonical, but $\IF$ is canonically defined
and satisfies Dirac quantization. (The projection of the
charges to $\Gamma_g$ is understood here.)

\insfigpng{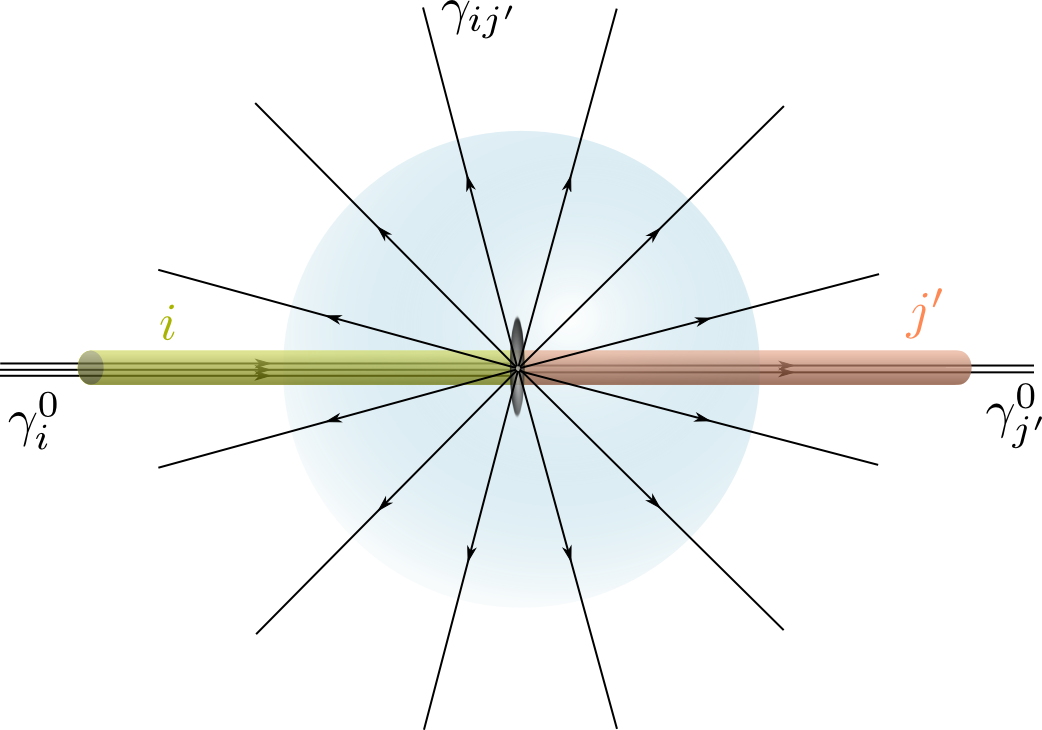}{The configuration of electromagnetic flux around an interface between two surface defects.
The total outgoing flux through the blue sphere is $\gamma_{ij'} + \gamma_{j'}^0 - \gamma_{i}^0 \in \Gamma$.}

The ``halo'' then consists of 4d particles which couple to this flux $\gamma_{ij'}$ as in the usual 4d case,
\ti{plus} the contribution from 2d particles on the two surface defects.
While these two contributions are not separately well-defined (they depend on our choice of Lagrangian),
their sum must make good physical sense.
As usual, the halo  particles generate a Fock space, and in order to discuss the
wall-crossing at the level of state spaces (i.e. at the ``categorified level'')
we need to describe the oscillators which generate the Fock space.
In  Appendix \ref{app:Landau-Levels} we analyze the Landau level problem for
halo particles of charge $\gamma$ in the presence of the field \eqref{eq:F-decomp}.
In a description where there are no chiral multiplets on the surface defect,
the halo particles may be described by fermionic and bosonic creation operators
$A_{m',m,\alpha}^\dagger$ where $\alpha$ again runs over $\vert a_{m,\gamma^h}\vert$ values
for every integer $m$, but now   $m'$ has a
fixed fractional part and satisfies the inequality \eqref{eq:Range-hatm}. The number of
such $m'$ values is the function $N_{\kappa_1, \kappa_2}$ given in \eqref{eq:Nkappa-def}. The signed
sum of the number of oscillators
is given by \eqref{eq:omega-Landau}.  We identify this quantity
with $\omega(\gamma,\gamma_{ij'})$.

The framed wall-crossing formula for these walls says that across $W_\gamma$
the Hilbert space of framed BPS states with core charge $\gamma_{ij'} + N \gamma$
for some $N$ gains or loses a Fock space factor generated by $A_{m',m,\alpha}^\dagger$.
To express this formula more concisely
we introduce a generating function $F^{}$, analogous to \eqref{eq:gen},
valued in the untwisted groupoid algebra:
\begin{equation}\label{eq:framed-decomp}
F^{}(u, L_\zeta, \{x_a\}; y) := \sum_{i\in \CV(\bS), j'\in \CV(\bS')}
\sum_{a\in \Gamma_{i,j'}} \fro(L,a; y) x_a,
\end{equation}
which will be abbreviated to $F^{}(L)$.  The $x_a$ satisfy
\eqref{eq:twisted-mult} with $\sigma(a,b)=1$.

As in the pure 4d case, it is useful to define the ``part of the
generating function with core charge $\gamma_{ij'}$'',
\begin{equation}
F^{}_{\bar \gamma_{ij'}}:= \sum_{n\in \IZ}
  \fro(L, \gamma_{ij'} + n \gamma ; y) x_{\gamma_{ij'} + n \gamma}.
\end{equation}
Now each halo particle   of type $A_{m',m,\alpha}^\dagger$ contributes
a Fock space factor
\begin{equation}\label{eq:fock-factor}
(1 + (-1)^m y^{m'+m}x_\gamma),
\end{equation}
and hence the transformation law across $W_\gamma$ is:
\begin{equation}\label{eq:class-tmn}
T_\gamma: F^{}_{\bar \gamma_{ij'}} \to
\prod_{m,m'} (1 + (-1)^m y^{m'+m}x_\gamma )^{a_{m,\gamma} }
F^{}_{\bar \gamma_{ij'}}.
\end{equation}
As before, we are assuming that the only 4d particles with central charge parallel
to $Z_\gamma$ along $W_\gamma$ in fact are those with charge $\gamma$.

Similarly, there is a transformation law for crossing
soliton walls $W_{\gamma_{ik}}$:
\begin{equation}\label{eq:class-tmn-sol}
T_{\gamma_{ik}} : F^{}(L)   \to (1 \pm \mu(\gamma_{ik} ) x_{\gamma_{ik}} ) F^{}(L) (1 \mp  \mu(\gamma_{ik}) x_{\gamma_{ik}} )
\end{equation}
where the sign is determined by the direction in which the wall is crossed.

We now face an issue analogous to that in the pure 4d story, discussed just above \eqref{eq:gen-2}.
The transformation \eqref{eq:class-tmn} is written directly in terms of the physical halo particles, but
has the awkward feature that the prefactor depends on the core charge $\gamma_{ij'}$,
through the range of $m'$.  On the other hand, in \S \ref{subsec:2d-4d-motivic}
we found that there should be a nicer way of expressing the framed wall-crossing:
indeed, it should be possible to write it as conjugation by an operator
$\Phi_{\bS,\gamma}$ depending only on the wall which is crossed, not on the core charge.
What we would really like is to determine $\Phi_{\bS,\gamma}$ physically,
in terms of the halo picture we have been discussing in this section.
We indicate briefly how to do this in \S \ref{subsubsec:consist-conds}.

While we will not carry out that analysis to the end, there is one specialization that is easier to deal with.
Beginning with \eqref{eq:class-tmn} and \eqref{eq:class-tmn-sol},
taking $y \to -1$, converting to twisted (but commuting) variables, and
making use of \eqref{eq:omega-Landau}, we recover the transformations \eqref{eq:ks-like} and \eqref{eq:cv-like}
which appeared in the 2d-4d wall-crossing formula of \S\ref{sec:Formal-Statements}.
In particular, after this specialization the spin 2d-4d wall-crossing formula \eqref{eq:spin-wcf} reduces to the ordinary
(non-spin) 2d-4d wall-crossing formula we use in the rest of this paper.

\subsubsection{Determining the generalized quantum dilogarithms $\Phi_{\bS,\gamma}$}\label{subsubsec:consist-conds}

In this section we finally explain how to relate the $\Phi_{\bS,\gamma}$ appearing in the framed wall-crossing
across $W_\gamma$ to physical data defined in terms of halo Fock spaces.

We define a linear map $\Psi$ between the quantum groupoid algebra generated by $\hat X_{a}$ satisfying
\eqref{eq:2d-4d-heis} and the (untwisted) groupoid algebra generated by $x_{a}$:
$\Psi(\widehat X_{a}) = x_{a}$.
Note that $\Psi$ is \emph{not} a ring homomorphism!

Compatibility of the framed wall-crossing formula with the physical halo picture is simply the statement that the following
diagram commutes:
\begin{equation}\label{eq:consistent-hwc}
 \xymatrix{ \widehat{F} \ar[d]^{\Psi} \ar[r]^{\hat T_\gamma} & \widehat{F}
 \ar[d]^{\Psi} \\      F^{} \ar[r]^{T_\gamma} & F^{} }.
 \end{equation}
This condition can be used to write recursion relations for the exponents $d(i,s)$
and $d(j',t)$
in terms of $a_{m,\gamma}$, $\gamma$ and $\gamma_{ij'}$. It can also be used to obtain
the monomials $\varphi_{i,s}(y)$.

As an illustration of how this works, let us consider the formula
\begin{equation}
\hat X_a \hat X_b = \frac{\sigma(a,b)}{\sigma(b,a)} (-y)^{n(a,b)-n(b,a)} \hat X_b \hat X_a,
\end{equation}
valid when both $a+b$ and $b+a$ are composable morphisms.  In particular this holds
for $a=\gamma$ and $b=\gamma_{ij'}$, and we can use this to rewrite \eqref{eq:q-tmn} as
\begin{equation}\label{eq:rewrite-tmn}
\hat X_a \to \prod_{i,s} ( 1 + \varphi_{i,s}(y) \hat X_{\gamma} )^{d(i,s)}
\prod_{j', t } ( 1 +  \tilde\varphi_{j',t}(y) \hat X_{\gamma} )^{-d(j',t)}\hat X_a,
\end{equation}
where $\tilde\varphi_{j',t}(y) = \frac{\sigma(a,\gamma)}{\sigma(\gamma,a)} (-y)^{n(a,\gamma)-n(\gamma,a)}
\varphi_{j',t}(y)$.

The products on the left only depend on the $\hat X_a$ through
the single variable $\hat X_\gamma$ and hence can be
treated as functions of a commutative variable.  In particular they can be expanded
as a series $(1 + \sum_{N\geq 0} \Phi_N \hat X_{N\gamma})$, which can be combined to
rewrite the right hand side of \eqref{eq:rewrite-tmn} in the form
\begin{equation}
\hat X_a + \sum_{N\geq 1} \Phi_N  \sigma(N\gamma, a) (-y)^{n(N \gamma, a)} \hat X_{N\gamma + a}.
\end{equation}
Now we can easily apply $\Psi$ since that map simply takes monomials to monomials.
Comparing the result with the transformation \eqref{eq:class-tmn} of $F$, we would find recursion
relations implying that $\Phi_{\bS,\gamma}$ is an infinite product, generalizing the
quantum dilogarithm which appeared in the 4d case.

It would be worthwhile working out further details and examples of the spin version of
the 2d-4d wall-crossing formula, but we leave this to another occasion.

\section{Compactification to 3d and hyperholomorphic bundles} \label{sec:compactification}

So far we have written the 2d-4d wall-crossing formula, explained in
detail  a physical context to which it is relevant,  and given a physical derivation
of it based on the halo phenomenon.  We now turn to the 2d-4d analog of $tt^*$ geometry.

\subsection{The dimensionally reduced action}

We consider the compactification of the combined 2d-4d theory on a circle of radius $R$.
A surface defect which wraps the compactification circle will give rise to
a line defect in three dimensions.  As such, at energies much lower
than $1/R$ we will obtain an effective 1d-3d theory, whose effective Lagrangian
we wish to study.  If $1/R$ is much smaller than the scale where
the 4d IR description breaks down, then a good approximation to the 1d-3d Lagrangian will be obtained
by naive dimensional reduction of the 2d-4d IR theory.  This gives the exact 1d-3d Lagrangian in the limit
$R \to \infty$.  For finite $R$ there are quantum corrections which can be interpreted as coming from BPS
particles of the 2d-4d theory going around the compactification circle; these corrections are crucial to our
story, but for now we are only discussing the limit of infinite $R$.
As for the metric on the target space, this was discussed
 in detail  in  \cite{Gaiotto:2008cd}.  After patching, the dimensional reduction of the
bulk 4d term in \eqref{eq:4d-action}
gives a sigma model into a manifold $\CM$,
a torus bundle over the 4d Coulomb branch $\CB$, with a simple explicit metric $g^\sf$.

To find the corresponding 1d action we turn to
 the reduction of the term integrated over the surface defect, namely,
the second line of \eqref{eq:explct-action}. Before reduction this term can be written as
\begin{equation}\label{eq:2d-W-F-action}
 \int \de x^0 \de x^3 \,  \left( \eta_I F_{03}^I - \alpha^I G_{03,I}\right)
\end{equation}
where the dual gauge field is given by \eqref{eq:G-with-source}. The auxiliary
field has been integrated out and the resulting infinite expression is cancelled by
taking into account the last term in the definition of $G_{03,I}$.
We can now   dualize all of the gauge fields to scalar fields in
3d according to
\begin{equation}
 \p_i \theta_\gamma = \oint \de x^3 \langle \gamma, \IF_{i3} \rangle.
\end{equation}
This defines a scalar field $\theta$ valued in $\Gamma^*\otimes \IR/(2\pi\IZ)$. \footnote{To be precise, there are some subtle shifts of $\pi$ when adding the scalars $\theta_\gamma$,
due to the supersymmetry-preserving boundary conditions.
The correct additivity statement is that
$e^{\I \theta_{\gamma}}e^{\I \theta_{\gamma'}} = (-1)^{\langle \gamma, \gamma' \rangle} e^{\I \theta_{\gamma+\gamma'}}$ }
Note that for flavor charges $\gamma_f \in \Gamma_f$ we learn that  $\theta_{\gamma_f}$ are constant.
After choosing a duality frame we define   $\theta_e^I :=  \theta_{e^I}$ and $\theta_{m,I} := \theta_{e_I}$.
Note that for a charge $\gamma = p^I e_I  + q_I e^I$ we have
$\de \theta_\gamma = \de \theta \cdot  \gamma   = q_I \de \theta_e^I + \de \theta_{m,I} p^I$.

Thus, \eqref{eq:2d-W-F-action} leads to
a very natural supersymmetric line defect in the sigma model.   Its bosonic part is the path-ordered exponential
of a simple $U(1)$ connection
\begin{equation} \label{eq:s1d}
\exp \int \de x^0 A^\sf
\end{equation}
where in this expression we mean the integral of the pullback of the locally-defined one-form on
$\CM$:
\begin{equation} \label{eq:a-sf}
A^\sf  = \I    \nu \cdot  \de \theta   = \I (  \eta_I \de\theta^I_e + \alpha^I \de \theta_{m,I} ).
\end{equation}
 When we wish to emphasize
the choice of the vacuum $i$ of $\IS$ we write $A^\sf_i$.

\subsection{Globalizing}\label{subsec:Globalizing}

\subsubsection{Line bundles}

As emphasized in \S \ref{subsubsec:2d-4d-GlobalPicture}, the 2d-4d action is only
 locally defined on $\CB$:  there is no single
Lagrangian description which encompasses the whole IR theory.
So strictly speaking, we must perform the dimensional reduction in each patch and then glue the
dimensionally reduced theories together, taking account of the necessary electromagnetic duality transformations
and shifts of $\CW_i$. The expression $A^\sf_i$ is duality invariant and will glue nicely as a one-form across
patches requiring an electric-magnetic duality transformation.
On the other hand,
 the superpotential $\CW_i$ is not globally defined. It determines, locally, a
choice of basepoint  $\gamma^0_i \in \Gamma_i$.  A change of superpotential
is equivalent to $\gamma_i^0 \to \gamma_i^0 + \tilde\gamma$, with $\tilde \gamma \in \Gamma$. This
leads to a shift of $\nu_i$ by $\nu_i \to \nu_i + \tilde\gamma$ which in turn induces a gauge transformation
\begin{equation}\label{eq:sf-gt}
A^\sf_i \to A^\sf_i + \I \de \theta_{\tilde\gamma}.
\end{equation}
Therefore, we can cover $\CB$ with patches $\CU_\alpha$ so that on patch
overlaps $\CU_{\alpha\beta}$ the semiflat connections differ by
gauge transformations of the form \eqref{eq:sf-gt} for some $\tilde\gamma_{i,\alpha\beta}\in \Gamma$
which satisfy the cocycle condition on triple overlaps.
 We can therefore interpret $A^\sf_i$ as
the connection on a line bundle $V_i$ over $\CM$, whose transition functions are
$e^{\I \theta_{\tilde \gamma_{i,\alpha\beta}}}$. (This statement will need to be
slightly amended. See the next section below.)
Changing the local description by taking $\gamma_i^0 \to \gamma_i^0 + \tilde\gamma_\alpha$
in patch $\CU_\alpha$ (with $\tilde \gamma_\alpha\in \Gamma$)
 changes the local framing of $V_i$ and modifies the transition functions
by a coboundary.  Below we will identify $V_i$ as the vacuum line bundle (for vacuum $i$)
of the reduced
one-dimensional quantum mechanical problem.

It is worth remarking that
 if we interpret $\nu$ as a section of the torus fibration for the algebraic integrable system
 $\widetilde{\CM}$ of the theory as in \S \ref{subsubsec:Alg-Int-Mirr} then the connection $A^\sf$
 naturally corresponds to a flat connection $A^\sf$ on the dual torus. Therefore, when restricted
to a torus fiber we can view $A^\sf$ as the restriction of the canonical connection on the
Poincar\'e line bundle to $\Gamma_g^*\otimes \IR/(2\pi \IZ)$.

\subsubsection{Vector bundles}\label{subsubsec:VB}

There is an important  subtlety we have suppressed above:
the index $i$ labeling the vacuum is not globally defined --- rather it labels a sheet
of the ramified cover $\CB_\bS\to \CB$.  This implies that the line bundles $V_i$ also do not exist globally
over $\CM$.
What does exist globally over $\CM$ is their direct sum, a rank $n$ vector bundle which we denote as
\begin{equation}\label{eq:V-directsum}
V_\bS := \bigoplus_i V_i.
\end{equation}
If we use trivializations given by local superpotentials as above then the transition functions
on patch overlaps $\CU_{\alpha\beta}$  will be either
diagonal matrices with $i^{th}$ element
$e^{\I   \theta_{\tilde \gamma_{i,\alpha\beta}}}$, or  (constant) permutation matrices.
$V_\bS$ carries a connection $A^\sf$ which is diagonal (in each patch) with respect
to the decomposition into the $V_i$. Given the above transition functions we
can define a Hermitian structure by declaring that the framings $s_{i,\alpha}$ on $\CU_\alpha$
implicit in choosing the gauge \eqref{eq:a-sf} form an orthonormal basis for $V_{\bS}$.
We will refer to this as the \emph{unitary framing}.

We hasten to add that equation \eqref{eq:V-directsum} is only a statement about $\CC^\infty$
vector bundles.  In particular, the
diagonal structure of $A^\sf$ is an artifact of the $R \to \infty$ limit.
At finite values of $R$  the quantum effects coming from
2d solitons (whose worldlines are reinterpreted as  1d instantons)
 imply that the connection $A$ on the bundle
$V = V_\bS$ will be corrected:  in particular, it is no longer
diagonal with respect to the decomposition of $V$ as a sum of line bundles
given in \eqref{eq:V-directsum}. In the framing we have described above,
the connection $A$ will be exponentially close to a diagonal connection
for large $R$.
The quantum-corrected connection is a rather nontrivial and interesting object:  it is the 2d analogue of the
quantum-corrected \hk metric on $\CM$ which played an important role in \cite{Gaiotto:2008cd}.
We will see below how to construct it more explicitly using results of
\S \ref{subsec:Integral-Equations} and Appendix \ref{app:Twistor-HH}.

\subsubsection{Twisted vector bundles and mirror symmetry}\label{subsubsec:Twsted-Mirror}

In \S \ref{subsubsec:Potential-Anomaly} we explained that the torsors $\Gamma_i$ might
not exist as $\Gamma$-torsors. In such cases there will be further complications
in the geometrical construction of $V_{\bS}$ since  the fractional shifts of $\nu$ are
 not good gauge transformations. As further explained in
\S \ref{subsubsec:Potential-Anomaly},
under some conditions one \emph{can} construct
the $\Gamma_i$ as \emph{twisted} torsors. In these cases the $\Gamma_i$ will have
have monodromy shifts by $\frac{1}{n}\Gamma$ for some integer $n$. In such cases the
overlap transition functions $e^{\I \theta_{\tilde \gamma_{i,\alpha\beta}}}$ will be
ambiguous by an $n^{th}$ root of unity. Even if we choose a good cover, so that
we can choose an unambiguous  $n^{th}$ root on each patch overlap $\CU_{\alpha\beta}$,
 the cocycle condition can fail on triple
overlaps by an $n^{th}$ root of unity. Such an object defines what is known as a
\emph{twisted bundle}. In principle, from the monodromies of the torsors $\Gamma_{ij}$
one can construct  the gerbe of the twisting. It would be desirable to have a
more streamlined version of this construction.

In string theory, twisted bundles are associated
with background $B$ fields  \cite{Witten:1998cd}. The role of $B$-fields in
three-dimensional sigma models is discussed further in Appendix \ref{app:Flavor-Twisted},
where they are related to global symmetries. Recall from \S \ref{subsubsec:Anomaly Cancelation}
that when there are obstructions to splitting $\Gamma_{ij}=\Gamma_i-\Gamma_j$ we can, in some
cases,   nevertheless modify the theory to incorporate the troublesome (but desirable)
surface defects in a well-defined
theory by gauging a suitable discrete flavor symmetry. As explained at the end of
 Appendix \ref{app:Flavor-Twisted} the $B$-field associated to the gauging of
 discrete flavor symmetry allows the definition of
 well-defined amplitudes in the 3d sigma model when the target space has
 a twisted bundle with (twisted) connection.

Another  useful viewpoint on the emergence of twisted bundles comes
from further reduction to two-dimensional sigma models, where we can use mirror
symmetry. It is well-known that reduction of four-dimensional gauge theory
on two circles produces a sigma model which has a self-mirror target. In
the 4d $\CN=2$ theories we are discussing, that target will be $\CM$ or $\widetilde{\CM}$.  Mirror
symmetry --- which amounts to $T$-duality on the fibers over $\CB$ ---
is induced by switching the compactification circles, and the theory
is self-mirror because the lattice of charges $\Gamma_g$ is self-dual.  If
we now consider \tf\ we can wrap the surface defect around one compactification
circle --- as we have been doing --- to produce a line defect in the three-dimensional
sigma model, and then we can take the second reduction circle to link the defect.  That is,
we reduce with respect to a $U(1)$ isometry, with a fixed point on
the defect.  This is often described as compactification on a cigar geometry.
As described in \cite{Nekrasov:2010ka}, the defect at the tip of the
cigar requires boundary conditions which describe branes in the sigma model.
As we have explained above, the resulting
brane has support on all of $\CM$ and carries (see \S \ref{subsec:HHVB} below)
a hyperholomorphic connection.
If, on the other hand, we first reduce  along the linking circle and then along the transverse
circle the resulting brane has fixed values of the scalar fields $\theta_\gamma$,
and hence is a Lagrangian brane, a section $\nu$ of $\widetilde\CM \to \CB$.
(See \S \ref{subsubsec:Alg-Int-Mirr}.) We conclude that these branes are mirror dual to
each other. \footnote{This description is incomplete. One should specify the bundle
and connection on the Lagrangian brane.}

Let us now consider this two-dimensional reduction in the case when the surface
defect has a set of superpotentials
leading to an anomaly in the splitting $\Gamma_{ij}= \Gamma_i-\Gamma_j$.  Then,
once again,  in good cases, we can
retain the surface defect by gauging a suitable discrete flavor symmetry.
In this case, the Lagrangian brane is a multisection of $\CM_d$ defined
below equation \eqref{eq:torus-discrete-ext-seq}, while  $V_{\bS}$ is a twisted
bundle. This picture fits in very well with the work of Hausel and
Thaddeus  \cite{mlh} who showed that for Hitchin fibrations the phenomenon
of having a disconnected fiber is mirror dual to having a discrete $B$-field.
(See also \S 7.2 of \cite{Kapustin:2006pk} for a discussion in the
physics literature.)

\subsection{Constraints from supersymmetry:  \hk and hyperholomorphic}\label{subsec:HHVB}

The compactified theory has $8$ supercharges ($\N=4$ supersymmetry in 3 dimensions).  This imposes strong constraints both on the 3d and 1d parts of the effective action.

The constraints on the 3d part are well known: they imply that the
metric $g$ on $\CM$ must be \hk.
The \hk property means $(\CM, g)$ is \kahler with respect to
a family of complex structures $J^{(\zeta)}$, parameterized by $\zeta \in \IC\IP^1$.

We now consider the analogous constraint on the 1d part of the action.
Thompson \cite{Thompson:2000pw} studied supersymmetric loop operators in (topologically twisted) $\CN=4$ sigma models,
with bosonic part of the form
\begin{equation} \label{eq:loopop}
\Tr \Pexp \oint \de x^0\,A
\end{equation}
where $A$ is a connection pulled back from some bundle $V$ over the target $\CM$.
(The full supersymmetric expression corrects $A$ by a fermion bilinear times
the curvature.)
He found a strong constraint on the connection $A$:
the field strength $F$ must be of type $(1,1)$ in all the complex structures $J^{(\zeta)}$ of the \hk manifold $\CM$.
Such connections are said to be \emph{hyperholomorphic}.
\footnote{Using the notation and conventions of Appendix B of \cite{Gaiotto:2010be} a simple explanation
of the result can be given as follows. The integral written out more fully is of the form
$\int \de x^0 \left( A_{ai} \dot \varphi^{ai}
+ \kappa F_{ai bj} \psi_\alpha^{A i} \psi_\beta^{Bj} \delta^{\alpha}_{A} \delta^{\beta}_B \epsilon^{ab}\right)$ for some constant $\kappa$. If $u^a$ is a vector in $S$ determining the complex structure then
$F_{ai bj}u^a u^b$ determines the $(2,0)$ part of the curvature in that complex structure. But this
symmetric combination does not appear in the fermion bilinear term, and hence for the supersymmetric
variation of $A_{ai} \dot \varphi^{ai}$ to cancel the fermion bilinear term the $(2,0)$ part of the
curvature must vanish in all complex structures.}

The reason for this moniker is that we can use the $(0,1)$ part of $A$ to
define a notion of holomorphic section of $V$, in any complex structure $J^{(\zeta)}$:
namely a holomorphic section $s$ is one for which $(\de + A) s$ is of type $(1,0)$.
So $A$ equips $V$ with the structure of a holomorphic vector bundle $V^{(\zeta)}$ over $\CM^{(\zeta)}$,
for all $\zeta \in \IC\IP^1$.

\subsection{The semiflat connection is hyperholomorphic in the semiflat geometry}

In this section we verify
 that the field strength of the connection we have found in the semiflat approximation,
\begin{equation}
F_i^\sf = \de A_i^\sf = \I \de \nu_i \cdot \de \theta,
\end{equation}
is indeed of type $(1,1)$ in all complex structures
on $(\CM, g^\sf)$.

To prove this, we first quickly review the \hk structure of $g^\sf$.
The complex structures $J^{(\zeta = 0)}$ and $J^{(\zeta = \infty)}$ are rather different from the rest:  in these structures
the torus fibers of $\CM^{(\zeta)} = (\CM, J^{(\zeta)})$ are compact complex submanifolds (in fact abelian varieties).
In particular, the fibers
admit no non-constant holomorphic functions.  In contrast, at other $\zeta$ there are plenty of holomorphic functions,
which we can even write explicitly:
\begin{equation} \label{eq:x-sf}
\CY^{\sf}_{\gamma} := \exp \left[\pi R Z_{\gamma}/\zeta + \I \theta_{\gamma} + \pi R\bar Z_{\gamma} \zeta \right].
\end{equation}
(Notice that $\CY^{\sf}_\gamma$ is only a \ti{locally} defined function on $\CM$.
If we try to analytically continue $\CY^{\sf}_\gamma$ to a global function it will typically
come out multivalued, because of the monodromies of $\Gamma$ around the singularities of $\CB$.)
It will be useful to introduce the symbol $\de \log \CY^\sf$ for a  1-form valued in $\Gamma^*$,
$\de \log \CY^{\sf}_\gamma = \de \log \CY^{\sf} \cdot \gamma$. Then the semiflat geometry
is summarized by the assertion that this form is of type $(1,0)$.
\footnote{It is sometimes useful to choose a duality frame and work with an
explicit basis of $(0,1)$ vector fields. One can check that the vector fields
\begin{equation}
\begin{split}
\frac{\p}{\p a^I } & + \frac{\I \pi R}{\zeta} \left( \frac{\p }{\p \theta_e^I} + \tau_{IJ} \frac{\p }{\p \theta_{m,J}} \right) \\
 \frac{\p}{\p \bar a^I } & + \I \pi R  \zeta \left( \frac{\p }{\p \theta_e^I} + \bar \tau_{IJ} \frac{\p }{\p \theta_{m,J}} \right) \\
\end{split}
\end{equation}
indeed annihilate $ \de \log \CY^{\sf} $. }

The relation \eqref{eq:IR-GW} between $\nu_i$ and the twisted superpotential $\CW_i$ may be rewritten
in the elegant form
\begin{equation}
\de \CW_i = \nu_i \cdot \de Z.
\end{equation}
It follows that $\de \nu_i \cdot \de Z=0$ and also (by complex conjugation) $\de
\nu_i \cdot \de \bar Z=0$.
Using these facts, and linearly extending $\de \log \CY^{\sf}$ to be valued in $\Gamma^*\otimes \IR$,
we can write
\begin{equation}
F_i^\sf = \I \, \de \nu_i \cdot \de \log \CY^{\sf}.
\end{equation}
Since $\de \log \CY^\sf$ is of type $(1,0)$
this proves that $F_i^\sf$
has no $(0,2)$ component in any complex structure, and, since it is pure imaginary, this
 implies it is $(1,1)$ in all complex structures.
We conclude that  $A_i^\sf$ is a hyperholomorphic connection on $V_i$.

We can actually write down holomorphic sections explicitly:  in the gauge \eqref{eq:a-sf} we
take
\begin{equation} \label{eq:xi-sf}
\CY^{\sf}_{i} := \exp \left[\pi R \CW_i /\zeta +  \I \theta_i + \pi R  \bar \CW_i \zeta \right].
\end{equation}
Here $\theta_i$ is a generic constant introduced for future convenience. To check that $\CY^{\sf}_{i}$ is a \ti{holomorphic} section, we calculate
\begin{equation}\label{eq:verify-holo}
\begin{split}
(\CY^{\sf}_{i})^{-1} (\de +  A^\sf_i) \CY^\sf_{_i} & =  \nu_i \cdot \left( \frac{\pi R}{\zeta} \de Z +
\pi R \zeta \de \bar Z \right) + \I \nu_i \cdot \de \theta \\
& = \nu_i \cdot \de \log \CY^\sf \\
\end{split}
\end{equation}
which is manifestly of type $(1,0)$.
If we take $Z_{\gamma_i^0}:=\CW_i$, then for $\gamma_i = \gamma_i^0 + \gamma$
we can define
\begin{equation} \label{eq:xi-sf-gi}
\CY^{\sf}_{\gamma_i} := \CY^{\sf}_i \CY_\gamma^{\sf} =
\exp \left[\frac{ \pi R  Z_{\gamma_i}}{\zeta} + \I \theta_{\gamma_i} +  \pi R\bar Z_{\gamma_i} \zeta \right]
\end{equation}
where for future convenience we set $e^{\I \theta_{\gamma_i}} = \sigma(\gamma_i^0, \gamma) e^{\I \theta_i}e^{\I \theta_{\gamma}}$.
The $\CY^{\sf}_{\gamma_i}$ provide a basis of holomorphic sections.

From now on, we will define angles $\theta_a$ and sections
\begin{equation} \label{eq:xi-sf-vev}
\CY^{\sf}_a=  \exp \left[\pi R Z_a /\zeta + \I \theta_a + \pi R \bar Z_a \zeta \right]
\end{equation}
in such a way that $\CY^{\sf}_a$ satisfy the same twisted multiplication
rules as the $X_a$ in \eqref{eq:twisted-mult} of \S\ref{sec:Formal-Statements}.
(Note that when $a,b$ are not composable it does not make geometric
sense to multiply $\CY^{\sf}_a$ by $\CY^{\sf}_b$, so  their
``product'' is zero in the sense that  we will never meet it.)
This choice is motivated in \S\ref{sec:x-gamma-interp}.

\textbf{Remark}:
As we have said, in the canonical framing implied by \eqref{eq:a-sf} we must make a gauge transformations on
patch overlaps $\CU_{\alpha\beta}$ so that
$A^\sf_i \to A^\sf_i + \I \de \theta_{\tilde \gamma_{i,\alpha\beta}}$ because $\CW_i \to \CW_i
+ Z_{\tilde \gamma_{i,\alpha\beta}}$.
From the description of the holomorphic sections in complex structure $\zeta$ which
we have just given we can see that we could take the  $\CY_i^\sf$ to be a local framing of $V_i$
and then across patches   the transition functions are
holomorphic:
$\CY_i^\sf \to \sigma(\gamma_i^0, \tilde \gamma_{i,\alpha\beta} ) \CY^\sf_{\tilde \gamma_{i,\alpha\beta} } \CY_i^\sf$, thus endowing $V_i$
with the structure of a holomorphic line bundle in complex structure $\zeta$.
Now, in \eqref{eq:V-directsum} we have defined $V$ as a $\CC^\infty$ bundle as a direct
sum of line bundles. However, the hyperholomorphic
connection we will construct is not diagonal with respect to this decomposition. Thus, although
both $V$ and the $V_i$ can be endowed with the structure of
holomorphic bundles in complex structure $\zeta$, \eqref{eq:V-directsum} is not an isomorphism of
holomorphic bundles.

\subsection{Physical interpretation of $V$: Expectation values} \label{sec:expectation-values}

Now let us consider the expectation values of supersymmetric interfaces.

In the uncompactified IR theory, such interfaces are easy to describe.
Indeed, to describe an interface in the IR between two surface
defects with twisted superpotentials $\CW_i$ and $\CW_{j'}$, we simply divide
$\IR^{1,1}$ into two half-spaces along some line, say $x_0=0$,
and integrate $\CW_i$ on one half-space and $\CW_{j'}$ on the other.
By itself this breaks supersymmetry, but we can make the interface
half-BPS by adding a simple boundary term to the action \cite{Gaiotto:2009fs}.
In other words, we insert at $x_0 = x_1 = x_2 = 0$ the line defect
\begin{equation}\label{eq:interfaceop}
L = L_{\gamma^0_{ij'}} = \exp \frac{1}{2} \int \de x_3\,[(\CW_i - \CW_{j'})/\zeta + (\bar \CW_i - \bar \CW_{j'}) \zeta].
\end{equation}
Here $\zeta$ is an arbitrary phase, so this is really a family of interfaces $L = L_\zeta$.
(See Appendix \ref{app:d2d4-multiplets} for further explanation of how \eqref{eq:interfaceop} restores
holomorphy.)

When we wrap the defect on a circle of radius $R$ we can speak of
expectation values $\inprod{L_{\gamma^0_{ij'}}}$.
In the $R \to \infty$ limit where naive dimensional reduction works,
we immediately see using \eqref{eq:interfaceop}
that the expectation value of this IR interface is simply
\begin{equation} \label{eq:xi-sf-vev-2}
 \inprod{L_{\gamma^0_{ij'}}} := \exp \left[\pi R \left(\CW_i- \CW_{j'}\right) / \zeta +   \pi R  \left( \bar \CW_i - \bar \CW_{j'} \right) \zeta \right] .
\end{equation}
The expectation value of UV-defined interfaces  will be a sum over sectors of
vevs of such IR interfaces. Combining \eqref{eq:lz-exp} with \eqref{eq:Interface-Framed-PSC}
leads then to a consistency condition on the choice of framings $e^{\I \theta_i} = e^{\I \theta_{\gamma_i^0}}$,
\begin{equation}
e^{\I \pi f_{ij'}} = \sigma(\gamma^0_{ij'}, \gamma^0_{j'}) e^{\I (\theta_i-\theta_{j'})}.
\end{equation}

$L_{\gamma^0_{ij'}}$ is not the only possible supersymmetric interface between these two surface defects.
We can get a whole family of such interfaces by bringing IR line defects with charge $\gamma \in \Gamma$
close to $L_{\gamma^0_{ij'}}$. The analog of \eqref{eq:interfaceop} for the IR line defect is
\begin{equation}
L_\gamma := \exp \frac{1}{2}\int \de x_3 \left[ \frac{Z_\gamma}{\zeta} + \I \theta\cdot \IA + \zeta \overline{Z}_{\gamma} \right].
\end{equation}
Thus, the vacuum expectation values of these other interfaces are
$\sigma(\gamma, \gamma^0_i - \gamma^0_{j'}) \CY^{\sf}_{\gamma^0_i - \gamma^0_{j'}} \CY^\sf_\gamma$.
What distinguishes $L_{\gamma^0_{ij'}}$ from the other members of this class is only that $L_{\gamma^0_{ij'}}$ looks particularly simple
with respect to the Lagrangian description we have chosen.

Now suppose we change our choice of Lagrangian description by
shifting the superpotential $\CW_i$ by $Z_{\tilde\gamma}$.
In the new picture the vacuum expectation values of
interfaces look slightly different.  There is again a ``simplest'' interface $L'$, whose vacuum expectation
value is given by \eqref{eq:xi-sf-vev-2} with the new $\CW_i$.  The interface $L'$ is not quite $L$, however;
it differs by the addition of an IR line operator with charge $\tilde\gamma$.  (One could see this, for
example, by comparing the central charges of the two interfaces.)  So $\inprod{L}$ in the new picture is
obtained by starting with \eqref{eq:xi-sf-vev-2}, shifting $\CW_i$ by $Z_{\tilde\gamma}$ and then
dividing by $\CY^\sf_{\tilde\gamma}$.  These two effects cancel each other out as far as the $\zeta$-dependent terms go,
but leave behind a $\zeta$-independent shift:  we find that if we change our Lagrangian
in this way, $\inprod{L}$ is multiplied by $e^{\I \theta_{\tilde\gamma}}$.
Similarly, if we shift $\CW_j$ by $Z_{\tilde\gamma}$
then $\inprod{L}$ is multiplied by $e^{- \I \theta_{\tilde\gamma}}$.
This fact can be summarized by saying that $\inprod{L}$ is a section of the bundle $\Hom(V_{j'}, V_i)$.

Moreover, since each $V_i$ has the connection $A_i^\sf$ making it into a hyperholomorphic bundle, we also get
a hyperholomorphic structure on $\Hom(V_{j'}, V_i)$, with respect to which $\inprod{L}$ is a holomorphic section.
This holomorphy reflects the fact that $L$ preserves the 2 supercharges \eqref{eq:preserve-susy}.

This interpretation of $\inprod{L}$ allows us to give a more physical understanding
of the line bundles $V_i$ and $V_{j'}$.  Indeed, on general grounds $\inprod{L}$ depends on
the choice of quantum vacua for the surface defects $\bS$ and $\bS'$;
it should really be thought of as a linear map from the vacuum ray in vacuum $i$   to that associated to vacuum $j'$.
This matches very well with the fact that $\inprod{L}$ is a section of $\Hom(V_{j'}, V_{i})$, if
we identify $V_i$ as the vacuum ray associated to the classical vacuum $i$ on $\bS$, and $V_{j'}$ as that associated to classical vacuum $j'$ on $\bS'$.

As we have mentioned above, at finite $R$ we expect that the vector bundles of vacua $V_\bS$ and $V_{\bS'}$ still make sense, but their
decomposition into individual line bundles $V_i$ and $V_{j'}$ is not very useful.
Nevertheless, the constraints from supersymmetry still operate, and imply in particular that
the vev $\inprod{L}$ of any supersymmetric line defect $L$ should be a holomorphic section of
the hyperholomorphic bundle $\Hom(V_{\bS'}, V_{\bS})$.

There is a useful observation concerning the twisted vector bundles
associated to surface defects carrying flavor monodromies.
Physically, interfaces are expected to exist only between surface defects which carry the same flavor monodromy.
This is consistent with the idea that the vector bundles associated
 to these surface defects are twisted by a $B$-field as in \S \ref{subsubsec:Twsted-Mirror}.  The bundle
  $\Hom(V_{\bS'}, V_{\bS})$ is a true vector bundle, with well-defined sections,
only if the two surface defects have the same twisting, which in turn means they must
carry the same flavor monodromy.

Finally, we remark that another physical interpretation of the connection $A$ is that it is
a Berry phase connection. This can be deduced from the considerations of Appendix A of
\cite{Gaiotto:2009fs}.

\subsection{Integral equations}\label{subsec:Integral-Equations}

In this section we introduce the main integral equations used to
construct the hyperholomorphic connection $A$ on
the bundle $V_{\bS}$. For brevity we simply denote the bundle by $V$.
The examples of \S \ref{sec:LocalModel} are meant to demonstrate that this is a
physically reasonable connection, and is indeed the exact result of the
sum of   the quantum corrections from 4d and 2d BPS states to the semiflat connection.

As we have mentioned, the construction of $A$ is closely analogous to the construction
in \cite{Gaiotto:2008cd}
of a \hk\ metric on $\CM$. A key step in reference \cite{Gaiotto:2008cd}  was the construction of
holomorphic functions $\CY_\gamma$, corrected versions of the $\CY_\gamma^\sf$ which appeared above.
The $\CY_\gamma$ were obtained as solutions of an appropriate integral equation, whose building blocks were
$\CY_\gamma^\sf$ and the BPS data $\Omega(\gamma)$:
\begin{equation} \label{eq:int-old}
 \CY_{\gamma}(\zeta) = \CY_{\gamma}^\sf(\zeta) \exp \left[  \sum_{\gamma'} \Omega(\gamma') \langle \gamma', \gamma \rangle \frac{1}{4 \pi \I} \int_{\ell_{\gamma'}} \frac{\de \zeta'}{\zeta'} \frac{\zeta' + \zeta}{\zeta' - \zeta} \log(1 - \CY_{\gamma'}(\zeta')) \right].
\end{equation}

In this section we present
a similar system of integral equations which determines certain sections $\CY_{\gamma_i}(\zeta)$ of $V$, likewise
built from $\CY_{\gamma_i}^\sf$ together with the BPS data
$\Omega(\gamma)$, $\mu(\gamma_{ij})$, and $\omega(\gamma, \gamma_a)$.
For any fixed $\zeta \in \IC^\times$, the $\CY_{\gamma_i}(\zeta)$ induce a holomorphic structure on $V$,
in a tautological way:  we take the holomorphic structure such that all $\CY_{\gamma_i}(\zeta)$
are holomorphic sections.  Moreover, all of these holomorphic structures come from a single hyperholomorphic
connection $A$ on $V$.  We explain the construction of $A$ from the $\CY_{\gamma_i}$ in Appendix \ref{app:Twistor-HH}.

The sections $\CY_{\gamma_i}$ are piecewise holomorphic in $\zeta$ and will undergo
morphisms corresponding to those of \S \ref{subsec:Formal-Statement} when $\zeta$ crosses
$\CK$-rays and  $\CS$-rays. The key to our system is to separate the Riemann-Hilbert problems
associated with $\CK$-rays and $\CS$-rays by
writing $\CY_{\gamma_i}$ as a product of two pieces:
\begin{equation} \label{eq:int-0}
\CY_{\gamma_i} = g_i x_{\gamma_i},
\end{equation}
where $x_{\gamma_i}(\zeta)$ is a section of $V_i$ and $g_i(\zeta)$ a linear map $V_i \to V$.

The $x_{\gamma_i}(\zeta)$ are defined by the equation
\begin{equation} \label{eq:int-1}
x_{\gamma_i}(\zeta) := \CY_{\gamma_i}^\sf(\zeta) \exp \left[
\sum_{\gamma'} \omega(\gamma', \gamma_i) \frac{1}{4 \pi \I}
\int_{\ell_{\gamma'}} \frac{\de \zeta'}{\zeta'} \frac{\zeta' + \zeta}{\zeta' - \zeta}
\log(1 - \CY_{\gamma'}(\zeta')) \right].
\end{equation}
For $i\not=j$, we also define sections $x_{\gamma_{ij}}(\zeta)$ of  $V_j^*\otimes V_i$ by
representing $\gamma_{ij}=\gamma_i - \gamma_j$ and
taking
\begin{equation}\label{eq:xgijdef}
x_{\gamma_{ij}} := \sigma(\gamma_{ij}, \gamma_j)^{-1}  x_{\gamma_i}x_{\gamma_j}^{-1} .
\end{equation}

Now the key integral equation for $g_k: V_k \to V$ is
\begin{equation} \label{eq:int-2}
g_k(\zeta) = g_k^\sf + \sum_{\ell \neq k, \gamma_{\ell k}} \mu(\gamma_{\ell k})
\frac{1}{4 \pi \I} \int_{\ell_{\gamma_{\ell k }}} \frac{\de \zeta'}{\zeta'}
\frac{\zeta' + \zeta}{\zeta' - \zeta} g_\ell (\zeta') x_{\gamma_{\ell k}}(\zeta'),
\end{equation}
where $g_k^\sf$ is a $\zeta$-independent and nowhere vanishing linear map $V_k \to V$.
(Recalling that as a $\CC^\infty$ bundle $V = \oplus V_i$, the obvious choice for $g_k^\sf$
would be just the inclusion map, but other choices could be convenient.)

In \cite{Gaiotto:2008cd,Gaiotto:2009hg} it was argued that the integral equation \eqref{eq:int-old} has
solutions when $R$
is taken large enough, and that in this case the solutions have no poles or zeroes, but the situation
might  become more complicated for smaller $R$.  We expect a similar story for \eqref{eq:int-2}.
In what follows we mostly restrict ourselves to large $R$ and
assume the solutions do exist. In particular we assume that
$R$ is sufficiently large that no zeros develop so that $g_k$ has no kernel.  In that case
$\CY_{\gamma_i}$ have well behaved asymptotics:  precisely, $\CY_{\gamma_i}(\zeta) (\CY^\sf_{\gamma_i}(\zeta))^{-1}$ is finite in the limit $\zeta \to 0$.  They are also multiplicative in the expected way:
\begin{equation}
\CY_{\gamma} \CY_{\gamma_i} = \sigma(\gamma, \gamma_i) \CY_{\gamma + \gamma_i}.
\end{equation}

It is also very convenient to consider $g_i(\zeta)$ to be column vectors in the unitary
framing of $V$ described in \S \ref{subsubsec:VB}. Choosing some ordering
of the vacua we can then assemble these column vectors into a matrix $g(\zeta):= \oplus g_i$.
We consider $g$ to be a linear transformation $g: \oplus V_i \to V$. When
$\zeta$ crosses a BPS ray $\ell_{\gamma_{ji}}$ in the counterclockwise direction we find that
\begin{equation}
g(\zeta) \to g(\zeta) (1 -  \mu(\gamma_{ji})e_{ji} x_{\gamma_{ji}})
\end{equation}
where $e_{ji}: \oplus V_k \to \oplus V_k $ is the matrix unit. (Thus, only
$e_{ji} x_{\gamma_{ji}}$ makes good geometric sense.)

Now, an important application of the above integral equations is that they allow us
to construct the expectation values of general supersymmetric interfaces between
$\bS$ and $\bS'$. Denote the corresponding vector bundles by $V,V'$ respectively.
Then we aim to construct locally defined
holomorphic sections $\CY_{\gamma_{ij'}}$ of ${\rm Hom}(V', V)$.
Recall that $\gamma_i$ is a morphism in ${\rm hom}(i,o)$ in the vacuum groupoid $\IV[\bS]$
so the $\CY_{\gamma_i}$ are geometrically sections of $V$ and physically
are  expectation values of interfaces with the
vacuum $i$ on the \emph{left} of the interface. Similarly we can define objects $\CY_{-\gamma_i}$
corresponding to having the vacuum $i$ on the right. Geometrically these are sections
of $V^*$. The analog of \eqref{eq:int-0} would   be $\CY_{-\gamma_i} = g_{-i} x_{-\gamma_i}$
where $g_{-i}: V_i^* \to V^*$. However, for our purposes it is more useful to invoke the
canonical isomorphism ${\rm Hom}(V_i^* , V^*) \simeq {\rm Hom}(V, V_i)$ and
regard $g_{-i}: V \to V_i$ and therefore
\begin{equation}\label{eq:int-0-right}
\CY_{-\gamma_i} = x_{-\gamma_i} g_{-i}
\end{equation}
as a section of $V^*$. The $g_{-i}(\zeta)$ are constructed from the integral
equation
\begin{equation} \label{eq:int-2-inv}
g_{-k}(\zeta) = g_{-k}^\sf - \sum_{\ell \neq k, \gamma_{  k \ell}} \mu(\gamma_{  k \ell} )
\frac{1}{4 \pi \I} \int_{\ell_{\gamma_{  k \ell} }} \frac{\de \zeta'}{\zeta'}
\frac{\zeta' + \zeta}{\zeta' - \zeta} x_{\gamma_{  k \ell}}(\zeta') g_{-\ell} (\zeta')  .
\end{equation}

In analogy to $g  = \oplus g_i$ we can define
$g_-:= \oplus_k g_{-k}$, and regard $g_-$ as a linear transformation $V \to \oplus_k V_k$.
Now, as $\zeta$ crosses a BPS ray $\ell_{\gamma_{ji}}$ in the counterclockwise direction we find that
\begin{equation}
g_-(\zeta) \to  (1 + \mu(\gamma_{ji})e_{ji} x_{\gamma_{ji}}) g_-(\zeta)
\end{equation}
It now follows that $g g_-: V\to V $ is free of singularities in the $\zeta$ plane, and hence must
be constant. Taking the $\zeta \to 0$ limit means that it is equal to its semiflat value,
which may be taken to be the identity transformation. Similarly, $g_-g: \oplus V_k \to \oplus V_k$
transforms by conjugation, but the identity matrix will satisfy the Riemann-Hilbert problem
and hence from either order we conclude that if we take $g_-^\sf = (g^{\sf})^{-1}$
then the quantum corrections respect  $g_- = g^{-1}$. In particular, if we regard
$g_k$ as column vectors then   $g_{-k}$
is the $k^{th}$ row of the inverse matrix $g^{-1}$.

Finally, with these remarks in hand we are ready to define, for $i\not=j$:
\begin{equation}\label{eq:off-ex}
\CY_{\gamma_{ij}} := g_i e_{ij} x_{\gamma_{ij}} g_{-j}
\end{equation}
and, more generally, with two different surface defects we have the obvious modification
giving:
\begin{equation}\label{eq:CX-ijprime}
\CY_{\gamma_{ij'}}(\zeta) := g_i e_{ij'} x_{\gamma_{ij'}} g_{-j'}
\end{equation}
as a locally defined section of ${\rm Hom}(V',V)$.

We conclude this section with a number of remarks:

\begin{enumerate}

\item The fact that $g_{-i}$ is the $i^{\rm {th}}$ row of the inverse matrix to $g$ is not obvious
from iterating the integral equations \eqref{eq:int-2} and \eqref{eq:int-2-inv}.

\item In general we expect the linear transformations $g_i$ and $g_{-i}$ to
become singular on the locus $\CB^{\rm sing}$.

\item If we define $x_{\gamma_{ii}} = \CY_{\gamma}$ then we can extend our definition
\eqref{eq:off-ex} to construct $\CY_{\gamma_{ii}} \in {\rm Hom}(V, V)$,
and $\sum_i \CY_{\gamma_{ii}} = \CY_{\gamma} 1_V$.

\item Again using the unitary framing, we
can also put the $x_{\gamma_i}$ together into a \emph{diagonal} matrix $x$. Then
the fact that $[ (\de + A) \CY_{\gamma_i}]^{0,1}=0$ (as explained in Appendix
\ref{app:Twistor-HH}) implies that
\begin{equation}\label{eq:diag-A-01}
( g^{-1} A g + g^{-1} \de g)^{0,1} = - (x^{-1} \de x )^{0,1}
\end{equation}
and hence $g$ can be regarded as the gauge transformation which diagonalizes $A$
to a direct sum of connections on $V_i$.

\item In \cite{Gaiotto:2008cd} a reality condition was imposed stating that
$\overline{\CY_{\gamma}(\zeta)} = \CY_{-\gamma}(-1/\bar \zeta)$.
This is compatible with the integral equation provided $\overline{\Omega(\gamma)} = \Omega(-\gamma)$.
(Of course, $\Omega(\gamma)$ is also real.) An analogous reality constraint can be
imposed on the quantities defined above. Recall these are expressed in the unitary
frame, allowing us to define an Hermitian structure on $V_k$ and $V$. First to $\gamma_i^0$
we associate a $-\gamma_i^0$ by requiring that $Z_{-\gamma_i^0} = - Z_{\gamma_i^0}$. Then
\begin{equation}
\overline{x_{\gamma_i}(\zeta)} = \frac{1}{\sigma(\gamma_i,-\gamma_i)} x_{-\gamma_i}(-1/\bar \zeta)
= x_{\gamma_i}^{-1}(-1/\bar \zeta)
\end{equation}
 provided
\begin{equation}\label{eq:unitarity-1}
\overline{\omega(\gamma, \gamma_i)} = \omega(-\gamma, - \gamma_i)
\end{equation}
Once again, we can also take $\omega(\gamma, \gamma_i)$ to be real.
Second, using the Hermitian structure on $V_k$ and $V$ (defined in
in \S \ref{subsec:Globalizing}) we can demand that
\begin{equation}\label{eq:unitarity-2}
(g_k(\zeta))^\dagger = g_{-k}(-1/\bar \zeta).
\end{equation}
One can check using the integral equations that \eqref{eq:unitarity-2} will hold provided that
it is true for the semiflat value and provided that
\begin{equation}\label{eq:unitarity-3}
\overline{\mu(\gamma_{\ell k} )x_{\gamma_{\ell k}}(\zeta)}
= - \mu(\gamma_{k\ell})x_{\gamma_{k\ell}}(-1/\bar \zeta)
\end{equation}
where $\gamma_{k\ell} = - \gamma_{\ell k}$. This can be written as
\begin{equation}\label{eq:unitarity-4}
\overline{\mu(\gamma_{\ell k} )} = - \mu(\gamma_{k\ell})\sigma(\gamma_{\ell k} ,\gamma_{k\ell}).
\end{equation}
(In deriving this one might
find it useful to note that if $\gamma_{ij}=\gamma_i - \gamma_j = - \gamma_{ji}$ then
\begin{equation}\label{eq:simp-cr}
\sigma(\gamma_{ij},\gamma_j) \sigma(\gamma_{ji},\gamma_i) = \sigma(\gamma_{ij},\gamma_{ji}),
\end{equation}
a relation which will be useful again later.)

\item Finally, regarding \eqref{eq:int-2} in its matrix form, a special case
reduces to the
important integral equation used by Cecotti and Vafa in their discussion of
$tt^*$ geometry \cite{Cecotti:1992rm}.

\end{enumerate}

\subsection{Wall-crossing formula as a consistency condition} \label{sec:discont}

We have made a proposal for a construction of the exact $A$ incorporating corrections both from 2d and 4d
BPS particles.  One of the main points of our story is that the 2d-4d WCF is the condition implying
that this $A$ does not jump at walls of marginal stability.  Let us now describe how this works.

Suppose $\CY_{\gamma_i}$ is constructed as above.  What is its analytic structure like?
From the integral equations we see that $\CY_{\gamma_i}(\zeta)$ depends holomorphically on $\zeta$, \ti{except}
when $\zeta$ hits one of the integration contours $\ell_\gamma$ or $\ell_{\gamma_{ij}}$.  $\CY_{\gamma_i}(\zeta)$ jumps when $\zeta$ crosses one of these contours.
The situation is quite parallel to that of \cite{Gaiotto:2008cd}.  As in that case, one can determine the jump of
$\CY_{\gamma_i}(\zeta)$ by Cauchy's theorem:
\begin{itemize}
\item As $\zeta$ crosses $\ell_\gamma$, we have
the discontinuity $x_{\gamma_i}(\zeta^+)= x_{\gamma_i}(\zeta^-)
(1 - \CY_\gamma)^{\omega(\gamma, \gamma_i)}$, where $\zeta^+$ is on
the counterclockwise side of $\ell_\gamma$, which gives
\begin{equation}
\CY_{\gamma_i}(\zeta^+)= \CY_{\gamma_i} (1 - \CY_\gamma)^{ \omega(\gamma, \gamma_i)}(\zeta^-).
\end{equation}
This is exactly the effect of the transformation $\CK_\gamma^{-\omega}$ from \eqref{eq:ks-like}.

\item As $\zeta$ crosses $\ell_{\gamma_{ji}}$, we have
$g_i(\zeta^+)=  g_i(\zeta^-) - \mu(\gamma_{ji}) g_j x_{\gamma_{ji}}(\zeta^-)$, which gives
\begin{equation}
 \CY_{\gamma_i}(\zeta^+)=    \CY_{\gamma_i}(\zeta^-)  + \mu(\gamma_{ji}) \sigma(\gamma_{ji},\gamma_i)
 \CY_{\gamma_j}(\zeta^-).
\end{equation}
This is exactly the effect of $\CS_{\gamma_{ji}}^{-\mu}$ as given in \eqref{eq:cv-like}.

\end{itemize}

These discontinuities are, fortunately, rather benign:  $\CY$ jumps, but (as explained in Appendix \ref{app:Twistor-HH})
this jump does not lead to a discontinuity of $A$.
However, in order for $A$ to be continuous, it is important that there should be no \ti{other} discontinuities
of $\CY$.  Imposing this requirement leads directly to the 2d-4d WCF.  The argument is basically the same
as one described in \cite{Gaiotto:2008cd,Gaiotto:2010be} for the 4d WCF,
and so we do not repeat it here.

\subsection{$\CY_{\gamma_i}$ and framed BPS states} \label{sec:x-gamma-interp}

So far we have constructed some holomorphic sections $\CY_{\gamma_i}(\zeta)$ of the holomorphic bundle $V^{(\zeta)}$ over
$\CM^{(\zeta)}$.
Let us now consider what
they mean physically.  It follows from the discussion
in  \S \ref{sec:expectation-values} that - heuristically - we can identify
$\CY_{\gamma_i}$ as the expectation value of an ``IR line defect'' with charge $\gamma_i$, which sits at
the boundary of the surface defect $\bS$.  This intuitive picture is not entirely sharp, because it is hard to give a
precise definition of an ``IR line defect.''

We can give a sharper interpretation of the $\CY_{\gamma_i}$ following our earlier work \cite{Gaiotto:2010be}, where we
addressed a similar issue for the functions $\CY_\gamma$.  There we argued that the vev of any UV line defect $L$
wrapped on the compactification $S^1$ decomposes as a sum of the $\CY_\gamma$:
\begin{equation}\label{eq:DarbouxExpansion}
 \inprod{L} = \sum_{\gamma} \fro(L, \gamma) \CY_{\gamma}(\zeta).
\end{equation}
This expansion identifies $\CY_\gamma$ as the contribution from a single framed BPS state of charge $\gamma$
going around the compactification circle.
Repeating the arguments of \cite{Gaiotto:2010be} we have a very similar story in our present 2d-4d context:
letting $L_\zeta$ be a UV line defect which is an interface between
$\bS$ and $\bS'$, we expect that there is a universal expansion
\begin{equation} \label{eq:lz-exp}
 \inprod{L} = \sum_{i, \gamma_{i}, {j'}, \gamma_{j'}} \fro(L, \gamma_{ij'}) \CY_{\gamma_{ij'}}(\zeta).
\end{equation}
(It is universal in the sense that the expansion functions $\CY_{\gamma_{ij'}}(\zeta)$ do not
depend on $L$.)
So we identify $\CY_{\gamma_{ij'}}$ similarly as the contribution from a single framed BPS state of charge $\gamma_{ij'}$ going around the compactification circle.
Of course in order for this expansion to be compatible with the OPE discussed
in \S \ref{sec:lineops} it is crucial that the
$\CY_a$ and $\hat X_a$ satisfy the same twisted multiplication rules.
This was the motivation for our introduction of the constant angles $\theta_a$ in \eqref{eq:x-sf}.
By analogy to \cite{Gaiotto:2008cd,Gaiotto:2009hg}, we expect that
the $\CY_{\gamma_{ij'}}$ could be characterized by a pair of requirements:
first, every $\inprod{L}$ admits a finite expansion of the form \eqref{eq:lz-exp};
second, the analytic continuation of $\CY_{\gamma_{ij'}}$ from a generic ray $\zeta \in e^{\I \vartheta} \IR_+$
to the half-plane $\IH_\vartheta$ centered on that ray behaves asymptotically like $\CY^\sf_{\gamma_{ij'}}$ as $\zeta \to 0$.

Note that while $\CY_{\gamma_{ij'}}$ and $\fro(L_\zeta, \gamma_{ij'})$ are both discontinuous as functions of $(u, \zeta)$
(the former as noted in \S \ref{sec:discont}, the latter because of the framed wall-crossing) these discontinuities
precisely cancel in \eqref{eq:lz-exp}, leaving $\inprod{L_\zeta}$ continuous, as it should be.

\section{Local models  and resolution of singularities}\label{sec:LocalModel}

In this section we consider two examples where the
quantum corrected connection $A$ can be described more explicitly.
These examples give a concrete check of our integral equations by studying models with various
simplified assumptions on the spectrum $\Omega, \omega, \mu$. These simplifications
typically arise when describing the light spectrum near certain local singularities
in $\CB$. These examples, which are the analog of \S 4 of \cite{Gaiotto:2008cd},
  also provide illustrations of one of the important features of the story:   the real-codimension-2
singularities in the connection $A^\sf$ are smoothed out (or at least improved)
by the quantum corrections,
in the sense that the corrected $A$ only has singularities in codimension greater than 2.

Recall that in the four-dimensional theory a typical singularity arises when a
 4d particle of charge $\gamma$ becomes massless, i.e. $Z_{\gamma}(u) \to 0$
as $u \to u_0$ but
with  $\Omega(\gamma;u_0)\not=0$. If --- as we assume --- the only particles
which become massless at $u_0$ have charges proportional to  $\gamma$
then near $u_0$, the space $\CM$ locally factors as a product of a smooth \hk\ space
times a four-dimensional piece, the latter fibered over a one-complex-dimensional Coulomb branch.
This four-dimensional geometry, which we call periodic Taub-NUT space,\footnote{It is also
known as the Ooguri-Vafa geometry in the math literature. We will use
the abbreviation ``PTN space.'' } can be
described rather explicitly by summing up a series of quantum corrections around
the semiflat geometry \cite{Ooguri:1996me,Seiberg:1996ns,Gaiotto:2008cd}.
These corrections smooth out the singularity of the semiflat geometry.

When generalizing this discussion to the 2d-4d case we can have $Z_\gamma(u)\to 0$
or $Z_{\gamma_{ij}}(u)\to 0$ for some $u\to u_0$. Generically we do not expect
these to happen at the same point $u_0$, so there are two cases to consider:

\begin{enumerate}

\item A 2d particle of 4d charge $\gamma$ becomes massless, i.e. again
$Z_{\gamma}(u) \to 0$ as $u\to u_0$,  but now with $\omega(\gamma, \cdot;u_0 ) \not=0 $.
Nevertheless, the 2d solitons measured by $\mu(\gamma_{ij};u_0)$ remain
heavy and can be ignored in a sufficiently small neighborhood of $u_0$.
In general there can be
several particles of different charges $\gamma$ becoming massless, but
we will make the key simplifying assumption that they are all collinear.
The case where mutually nonlocal particles become massless
is significantly more difficult.

\item A 2d soliton of 4d charge $\gamma_{ij}$ becomes massless, i.e.
$Z_{\gamma_{ij}}(u) \to 0$ for $u\to u_0$, with $\mu(\gamma_{ij};u_0) \not=0$.
Meanwhile all 2d particles with $\omega\not=0$ and 4d particles with $\Omega \not=0$
remain massive and can be ignored.

\end{enumerate}

In the following sections we will analyze the local 2d-4d geometry
in these two situations, and we will see how the singularity of the
semiflat connection becomes partially resolved.

As we have said, we do not expect to have $Z_\gamma\to 0$ and $Z_{\gamma_{ij}}\to 0$
simultaneously for populated charges. Nevertheless, this can of course happen,
and our equations \eqref{eq:int-1} and \eqref{eq:int-2} could be used
to describe hyperholomorphic vector bundles, i.e.
instantons, over the PTN geometry. The geometry of these instantons might
constitute an interesting generalization of the results of Kronheimer and Nakajima.
The end of \S \ref{subsubsec:ADN=2} describes a useful local model of this situation.

\subsection{Massless 4d and 2d particles on a one-dimensional Coulomb branch}\label{subsec:1d-Coul-Branch}

\subsubsection{Local system and BPS degeneracies}

We will describe the local system of (gauge) charges over a one-dimensional
Coulomb branch
\begin{equation}
D_\Lambda^* := \{ a \, \vert \,  0 < \vert a \vert < \vert \Lambda\vert \}
\end{equation}
where $\Lambda$ is a UV cutoff for the effective 4d IR free theory. We let
$D_\Lambda$ stand for the disk with $a=0$ restored.
We can trivialize the local system after pulling back
to the universal cover:
\begin{equation}
\IC_\Lambda := \{ z \, \vert \, {\rm Re}(z) < \log \vert \Lambda\vert  \},
\end{equation}
and we will use the projection $p(z) = e^z = a$, so we may
think of $z \sim \log a$ as a branch of the logarithm.
Our local system will be
\begin{equation}
\Gamma := \left( \IC_\Lambda \times \left( \IZ \gamma_e \oplus \IZ \gamma_m \right) \right)/\IZ
\end{equation}
where the generator  $1\in \IZ$ acts by taking
\begin{equation}
T\cdot (z;q \gamma_e + p \gamma_m ) := (z + 2\pi \I ; (q-\Delta p ) \gamma_e + p \gamma_m )
\end{equation}
where $q,p$ are integers, and   $\Delta$ is an integer characterizing the monodromy
of the local system. The symplectic structure is given by
\begin{equation}
\langle \gamma_e, \gamma_m \rangle = 1.
\end{equation}
The central charge function $Z \in {\rm Hom}(\Gamma, \IC)$ is defined by
\begin{equation}
Z(\gamma_e;z) = e^z = a,
\end{equation}
\begin{equation}\label{eq:zee-gamma-m}
Z(\gamma_m;z) = \tau_0 a + \Delta a_d,
\end{equation}
where $\tau_0$ is a constant with positive imaginary part and
\begin{equation}
a_d := \frac{1}{2\pi \I} e^z (z - (\log \Lambda +1 ) ) = \frac{1}{2\pi \I} (a \log \frac{a}{\Lambda} - a ),
\end{equation}
and finally we extend $Z$ by linearity.
To extend $Z$ to the torsor $\Gamma_i$  we   take  the low energy effective superpotential as
\begin{equation}\label{eq:bst-super}
\CW = \delta a_d +  \CW^{\an}
\end{equation}
where $\CW^{\an}= w_0 + w_1 a + w_2 a^2 + \cdots $ is some analytic superpotential.
The $w_i$ are complex numbers.
The constant $w_0$ does not affect the geometry, so we will take $w_0 = 0$.

The period matrix is
\begin{equation}
\begin{split}
\tau(a) &= \tau_0 + \frac{\Delta}{2\pi \I} \log   \frac{a}{\Lambda } \\
& = \frac{\Delta}{2\pi \I} \log   \frac{a}{\Lambda_4 },
\end{split}
\end{equation}
where the second equation is only valid if $\Delta\not=0$,
in which case we have $\Lambda_4 = \Lambda e^{-2\pi \I \tau_0/\Delta} $.
In this case we can also write
\begin{equation}
Z(\gamma_m) = \frac{\Delta}{2\pi \I} (a \log \frac{a}{\Lambda_4} - a ).
\end{equation}
If $\Delta =0$ then we cannot absorb $\tau_0$ into $\Lambda$.
In the remainder of \S\ref{subsec:1d-Coul-Branch}
we will assume that $\Delta\not=0$,
and thus, without loss of generality, we can assume $\Lambda=\Lambda_4$.
In \S \ref{subsec:One-d-Coul-Delta-Zero} we will consider the special
case where $\Delta=0$.
Similarly, if we take $\Lambda = \Lambda_4$ and $\delta \not=0$ then we can write
\begin{equation}
\CW = \frac{\delta}{2\pi \I}\left( a \log\frac{a}{\Lambda_2}- a\right) + w_2 a^2 + \cdots
\end{equation}
where
\begin{equation}
w_1 = \frac{\delta}{2\pi \I} \log \frac{\Lambda_4}{\Lambda_2}
\end{equation}
so we can interpret $w_1$ in terms of $\Lambda_2$.

We will assume that the light BPS degeneracies are of the form
\begin{equation}
\Omega( q \gamma_e + p \gamma_m; z) = \delta_{p,0}\Omega_q,
\end{equation}
\begin{equation}
\omega( q \gamma_e + p \gamma_m; \gamma_0; z) = \delta_{p,0} \omega_q.
\end{equation}
Here $q,p\in \IZ$ and $\Omega_q$ and $\omega_q$ are $z$-independent and satisfy the parity properties:
\begin{equation}
\Omega_{-q} = \Omega_q \qquad \qquad \omega_{-q} = - \omega_q
\end{equation}

Self-consistency of the above spectrum with the central charge and superpotential
require that
\begin{enumerate}

\item The integer $\Delta$ is
\begin{equation}
\Delta = \half \sum_{q\in \IZ} q^2 \Omega_q =\sum_{q>0} q^2 \Omega_q,
\end{equation}
because integrating  out light charge $q$ hypermultiplets with BPS degeneracy
$\Omega_q$ in a theory which originally has prepotential $\tau_0 a$ results in
the effective central charge $Z$ given above.

\item Similarly, as we discussed in equation \eqref{eq:alpha-Weff} above, integrating out
light 2d chiral multiplets results in a twisted chiral superpotential as above,
with
\begin{equation}\label{eq:delta-def}
\delta = \half \sum_{q\in \IZ} q \omega_q = \sum_{q>0}  q \omega_q.
\end{equation}

\item As we have emphasized, the description of the system depends on a choice of superpotential.
These are the elements denoted $\gamma_i$ in \S \ref{sec:Physical-Interpret}.
If we change $\gamma_i \to \gamma_i' = \gamma_i + (q_1 \gamma_e + p_1 \gamma_m)$,
then
\begin{equation}\label{eq:chg-omega}
\omega_q \to \omega_q + p_1 q \Omega_q
\end{equation}
and $\CW^\an \to \CW^\an + q_1 a$, or equivalently $\Lambda_2 \to \Lambda_2 e^{-2\pi \I q_1/\delta}$.

\end{enumerate}

\subsubsection{Review of the PTN geometry}

The dual of the local system over the cover $\IC_\Lambda$ is
\begin{equation}
\CN_\Lambda   :=
  \IC_\Lambda \times \left( (\IR/2\pi\IZ) \gamma_e^* \oplus
 (\IR/2\pi\IZ) \gamma_m^* \right)
 \end{equation}
We would like to define $\CM^\sf = \CN_\Lambda/\IZ$ where the generator
of $\IZ$ acts by
\begin{equation}
T\cdot (z; \theta_e \gamma_e^* + \theta_m \gamma_m^* )
= (z+2\pi \I ; \theta_e \gamma_e^* + (\theta_m + \Delta \theta_e)\gamma_m^* ).
\end{equation}
This gives the Pontryagin dual local system $\Gamma^*\otimes \IR/(2\pi \IZ)$ to $\Gamma$.
However, there is a subtle complication (related to self-duality) which
forces us to consider instead a local system $\widehat{\CM}^\sf = \CN_\Lambda/\IZ$
where the generator of $\IZ$ instead acts as
\begin{equation}\label{eq:thetam-shift}
\widehat T\cdot (z; \theta_e \gamma_e^* + \theta_m \gamma_m^* )
= (z+2\pi \I ; \theta_e \gamma_e^* + (\theta_m + \Delta \theta_e + \pi \Delta )\gamma_m^* ).
\end{equation}
The extra shift in $\theta_m$  will affect the signs of various quantities under the monodromy
transformation. Of course these two actions are related by $\theta_e \to \theta_e +\pi \Delta$
so the resulting spaces are isomorphic, but the quantum corrections are a bit more natural in
the coordinates appropriate to the $\widehat T$ quotient.

The semiflat metric is:
\begin{equation}\label{eq:sf-metric}
g^{\sf} =  R\,\Im\tau \vert \de a \vert^2 + \frac{1}{4\pi^2 R \, \Im \tau} \vert \de \theta_m - \tau \de \theta_e \vert^2
\end{equation}
This is invariant under $\widehat{T}$ and descends to a metric on $\widehat\CM^{\sf}$. It is positive definite for $a\in D^*_\Lambda$
provided $\Delta>0$.

Now recall the Gibbons-Hawking ansatz for a \hk\ four-dimensional metric
on a principal circle bundle over a region of $\IR^3$:
\begin{equation}\label{eq:gh-full}
U^{-1} \left( \frac{\de \chi}{2\pi} + A^\gh \right)^2 + U \de \vec x^2
\end{equation}
where
\begin{equation}
\de A^\gh = \star \de U
\end{equation}
where $\star$ is computed using the metric $\de \vec x^2 = (\de x^i)^2$ and orientation $\de x^1 \de x^2 \de x^3$.
Here $\Theta^{\gh}=\frac{\de \chi}{2\pi} + A^\gh$ is a
1-form on the total space of the circle bundle, normalized so that $\pi_*\Theta^\gh = 1$.

The semiflat metric $g^\sf$ can be put into the form \eqref{eq:gh-full} with
\begin{equation}
(x_1 + \I x_2, x_3)  = \left(a , \frac{\theta_e}{2\pi R} \right),\qquad\qquad  \chi = \theta_m,
\end{equation}
and
\begin{equation}
U^\sf = R \, \Im \tau,
\qquad
A^{\rm gh,sf} = - R \, \Re\tau \, \de x^3,
\qquad
F^{\rm gh,sf} = - \frac{\Delta}{(2\pi)^2} \de \phi \, \de \theta_e,
\end{equation}
where $a = \vert a \vert e^{\I \phi}$.

Note that since $\theta_e$ is invariant under the monodromy operator $\widehat{T}$
we have a projection  $\widehat\CM^\sf \to D^*_\Lambda \times S^1$ defining  a principal $U(1)$ bundle. The base
space contracts to $T^2$ and the first Chern class is $-\Delta$.

The semiflat metric is \hk, and we define twistor coordinates
\begin{equation}
\CX_e^\sf = \exp \left[ \frac{\pi R}{\zeta} a + \I \theta_e + \pi R \zeta \overline{a} \right],
\end{equation}
\begin{equation}
\CX_m^\sf = \exp \left[ \frac{\pi R}{\zeta} Z(\gamma_m;z) + \I \theta_m + \pi R \zeta \overline{Z(\gamma_m;z)} \right]
\end{equation}
with the untwisted group law $\CX^\sf_{\gamma}\CX^\sf_{\gamma'} = \CX^\sf_{\gamma + \gamma'}$.
(Had we worked on $\CM$ instead of $\widehat\CM$ we would have used a twisted group law.)
The coordinate $\CX_m^\sf$ is only a function on the covering space $\CN_\Lambda$ and under $\widehat{T}$ we have:
\begin{equation}
\widehat{T}^* \CX_m^\sf = (-1)^{\Delta} \CX_e^\Delta \CX_m^\sf.
\end{equation}

The quantum corrected metric is again of the form \eqref{eq:gh-full} but now with
\begin{equation}\label{eq:gh-full-U}
U =U^\Omega:= \sum_{q=1}^\infty q^2 \Omega_q  U_q,
\end{equation}
where
\begin{equation}
U_q := \frac{1}{4 \pi  } \sum_{n=-\infty}^\infty \left( \frac{1}{\sqrt{q^2   \abs{a}^2 + R^{-2}(q \frac{\theta_\elec}{2 \pi} + n)^2}} - \kappa_{n,4} \right).
\end{equation}
 The regularization $\kappa_{n,4}$
is chosen so that in the instanton expansion the semiflat term is $U^\sf$ given above with $\Lambda_4$.
 There is a corresponding gauge potential
such that $\de A^\Omega = \star \de U^\Omega$. Formulae for it, in one gauge, are in \cite{Gaiotto:2008cd},
eqs. (4.8)-(4.10).

For any integer $n$ we define the points $s_{n,q}:=(a=0, \theta_e = 2\pi n/q) \in    D_\Lambda \times S^1$,
together with the sets:
\begin{equation}
S_q := \{ s_{n,q} \vert n\in \IZ \}.
\end{equation}
Then $U^\Omega$ is nonsingular away from $S_\Omega = \cup_{q: \Omega_q \not=0} S_q$.
The  metric \eqref{eq:gh-full} with \eqref{eq:gh-full-U} defines a metric
on a  principal $U(1)$   bundle over $D_\Lambda \times S^1 - S_\Omega$.
We will not be extremely careful about specifying the different patches and
trivializations of this $U(1)$ bundle and will generally denote
the globally well-defined connection one-form on the
total space of the bundle by $\Theta^\Omega = \frac{\de \chi}{2\pi} + A^\Omega$.
The total space with its \hk\ metric will be
denoted $\CM(\Omega)$ in what follows and the projection is denoted
\begin{equation}\label{eq:PTN-fib}
\pi_{\Omega} : \CM(\Omega) \to D_\Lambda \times S^1 - S_\Omega
\end{equation}
If we restrict this principal bundle to a small sphere linking the point $s_{n,q}$,
where $(n,q)$ are relatively prime and $0\leq n \leq q-1$,
then
the function $U$ behaves like
\begin{equation}
U \sim \frac{ N_q  }{ 4 \pi \vert \vec x \vert }
\end{equation}
and hence the first Chern class on the linking sphere is just $c_1 = - N_q$.
Here $N_q$ is given by
\begin{equation}\label{eq:Nq-def}
N_q = \sum_{j=1}^\infty  \vert jq \vert \Omega_{jq}.
\end{equation}
Note that $N_q >0$ is required for a good metric.\footnote{In particular, massless vectormultiplets
lead to singularities at finite $a$.}

If we attempt to extend \eqref{eq:PTN-fib} over the points $S_\Omega$
then the fiber in the fibration collapses. We can complete $\CM(\Omega)$
to $\overline{\CM}(\Omega)$ by adding corresponding points $p_{n,q}\in\overline{\CM}(\Omega)$.
The metric near the point $p_{n,q}$ is locally
a $\IC^2/\IZ_{N_q} $ singularity.
Note that if $N_q$ is nonzero then there is such a singularity at all the
points $s_{n,q}$ for $n=0,1,\dots, q-1$.

Over the subspace
\begin{equation}
D_\Lambda^* \times S^1 \subset D_\Lambda \times S^1 - S_\Omega
\end{equation}
we can identify the $U(1)$ bundles and hence
compare the  metrics of the  PTN  and  semiflat spaces.  If
 we cut out a small tube around $a=0$ then the principal bundle of the
quantum corrected space has Chern class given by the sum over the Chern classes
for the  points $s_{n,q}$,
and this sum is just
$$
c_1 \vert_{D_\Lambda^* \times S^1} =
\sum_{n,q} (- N_q)  =  -\Delta,
 $$
where the sum is over $n,q$ such that $ q>0,$  $(n,q)=1,$ and $ 0\leq n\leq q-1$.
Since the bundles are isomorphic we can
 choose a ``common fiber coordinate $\chi=\theta_m$''
for both the semiflat bundle and the full PTN space.  When this is done the PTN metric
and the semiflat metric $g^\sf$ are exponentially close for $R\to \infty$ as can be
seen by rewriting $U$ as $U^{\sf}$ plus a series of instanton corrections.

To describe the quantum-corrected twistor coordinates we should introduce the functions
\begin{equation}\label{eq:Bsc-int}
F_q(z,\theta_e, \zeta) := \exp\left[ - \frac{1}{4\pi \I} \int_{ \ell_{q\gamma_e} } \frac{d \zeta'}{\zeta'}
\frac{\zeta' + \zeta}{\zeta' - \zeta} \log[1-\CX_e(\zeta')^q ] \right]
\end{equation}
defined for any $q\in \IZ$. Recall that
\begin{equation}
\ell_{q\gamma_e}:= \left\{ (z, \zeta) :  \frac{qa}{\zeta} < 0 \right\}.
\end{equation}
The functions $F_q$ live on the cover $\CN_{\Lambda}$  (and also on the universal cover
of the punctured twistor sphere).
There are in fact two possible definitions of $F_q$. We may define a
piecewise analytic function away from the BPS rays.
Suppose for fixed $\zeta$ that $z_0$ is a
value at which the BPS ray contains $\zeta$; then
\begin{equation}
F_q(z_0 + \I \epsilon , \theta_e, \zeta) = (1-\CX_e^q) F_q(z_0 - \I \epsilon, \theta_e, \zeta)
\end{equation}
The resulting function is \emph{periodic} in $z$ but only piecewise holomorphic as a function
of $z$, and in fact is not defined on the BPS rays. This is the definition
generally adopted in this paper. However, for our present purposes it
is more convenient to take
\eqref{eq:Bsc-int} to be the definition of $F_q$ only in the strip
\begin{equation}
\arg \zeta - \pi < \Im z < \arg \zeta + \pi
\end{equation}
for $q>0$, and for $q<0$ we shift the strip down by $\pi$. Then we
\emph{analytically continue} in $z$ to define a
function without discontinuities. This function satisfies
\begin{equation}\label{eq:period-F}
F_q(z+ 2 \pi \I,\theta_e,\zeta) = (1-\CX_e^q)^{-1} F_q(z,\theta_e,\zeta).
\end{equation}
It is the latter functions, entire in $z$ but
only quasi-periodic, which we will use in our discussion of the local PTN geometry.

The formula of \cite{Gaiotto:2008cd} for the quantum-corrected magnetic twistor
coordinate is
\begin{equation}\label{eq:q-CXm}
\CX_m = \CX_m^\sf \prod_{q\in \IZ, q\not=0} F_q^{q \Omega_q}.
\end{equation}
Now, under $z \to z+ 2\pi \I$ we have,\footnote{In deriving this the ``extra sign''
comes out more naturally in the form $(-1)^{n_{\Omega}}$ where
$n_\Omega :=  \sum_{q>0} q \Omega_q$. However $n_\Omega = \Delta\,\mod\,2$.}
according to \eqref{eq:period-F}
\begin{equation}
\prod_{q\in \IZ, q\not=0} F_q^{q \Omega_q} (z+2\pi \I) = (-1)^{\Delta} \CX_e^{-\Delta} \prod_{q\in \IZ, q\not=0} F_q^{q \Omega_q}(z)
\end{equation}
Thus, the functions $\CX_m$ are invariant under the deck transformation $\widehat T$, and
descend to $U(1)$-equivariant functions for the fundamental representation on the principal $U(1)$ fibration
$\widehat{\CM}(\Omega) \to D_\Lambda \times S^1 - S_\Omega$.

\subsubsection{The semiflat line bundle}

The semiflat connection is
\begin{equation}
A^\sf = \I (\eta \, \de \theta_e + \alpha \, \de \theta_m).
\end{equation}
If $t = \eta + \alpha \tau$ with $\eta, \alpha$ real,
\begin{equation}
\eta = \frac{ \im (\bar t \tau) }{\im \tau}, \qquad \qquad \alpha = \frac{\im t}{\im \tau}.
\end{equation}
Defining
\begin{equation}
 \frac{\p \CW^\an}{\p a}:= t^\an,
\end{equation}
then
\begin{equation}\label{eq:tee-tau}
 t = \frac{\delta}{\Delta} \tau(a) + t^\an
\end{equation}
and hence
\begin{equation}
\widehat{T}^*(\eta) = \eta - \Delta \alpha + \delta, \qquad \qquad \widehat{T}^*(\alpha) = \alpha.
\end{equation}
$A^\sf$  is globally defined on $\CN_\Lambda$
but does not descend to $\CM^\sf$.  Also note that the curvature of the
semiflat connection comes entirely from $\CW^{\an}$.

We should consider $A^\sf$ to define a connection on a principal $U(1)$ bundle
over $\widehat\CM^\sf$ defined by
\begin{equation}
\widehat P^\sf = (\CN_\Lambda \times U(1)) /\IZ,
\end{equation}
where the generator of $\IZ$ acts by
\begin{equation}
\widehat T\cdot (z; \theta_e \gamma_e^* + \theta_m \gamma_m^* ; e^{\I \psi} )
= (z+2\pi \I ; \theta_e \gamma_e^* + (\theta_m + \Delta \theta_e + \pi \Delta)\gamma_m^*;
e^{\I \psi } e^{-\I \delta \theta_e} (-1)^{n_\omega} ),
\end{equation}
so that $\Theta^\sf = \I ( \de \psi + \eta  \de \theta_e + \alpha  \de \theta_m) $ is well defined
on the total space of $\widehat P^\sf$:
\begin{equation}
\widehat{T}^* \Theta^\sf = \Theta^\sf
\end{equation}
Here $n_\omega = \sum_{q>0} \omega_q$ defines a shift of the fiber coordinate $\psi$,
analogous to the shift of $\theta_m$ in \eqref{eq:thetam-shift}.
Now we form the semiflat section:
\begin{equation}
\CX_{\CW}^\sf = \exp\left( \frac{\pi R}{\zeta} \CW - \I \psi +  \pi R \zeta \overline{\CW} \right),
\end{equation}
so that $(\de + \Theta^\sf)\CX_{\CW}^\sf$ is type $(1,0)$ in all complex structures.
Note that $\CX_{\CW}^\sf$ has the equivariance property
\begin{equation}
\widehat{T}^* (\CX_{\CW}^\sf) = (-1)^{n_\omega} \CX_{\CW}^\sf \CX_e^\delta,
\end{equation}
so, like $\CX_m^\sf$, it is well-defined on $\CN_\Lambda$ but does not descend to $\widehat\CM^\sf$.

\subsubsection{The quantum-corrected bundle $V$}

Since we are assuming $\mu(\gamma_{ij};a)=0$, it follows from
\eqref{eq:int-2} that there will be no quantum mixing
of line bundles and the quantum corrected connection is on the semiflat line bundle $V$.
Using the equation \eqref{eq:int-1} we see that the holomorphic section is given by
\begin{equation}\label{eq:q-corr-sec}
\CX_{\CW} =  \CX_{\CW}^\sf \prod_{q\in \IZ, q\not=0} F_q^{\omega_q},
\end{equation}
which transforms under $\widehat{T}$ as
\begin{equation}
\widehat{T}^* \CX_{\CW} = \CX_{\CW}.
\end{equation}
It thus descends to a $U(1)$-equivariant (under shifts $\psi \to \psi + \psi_0$)
function from $(\CN_\Lambda \times U(1)) /\IZ$ to $\IC$, and hence
defines a section of the associated line bundle $V \to \widehat\CM^\sf$.

There is a strong formal resemblance between the function $\CX_{\CW}$ and the
magnetic twistor function $\CX_{m}$ we used to describe the PTN geometry. To bring it out, define
\begin{equation}\label{eq:Omega-eff}
\Omega_q^{\eff} := \omega_q/q.
\end{equation}
Then if we also set $\theta_m^{\eff} := - \psi$ we can write:
\begin{equation}
\CX_{\CW} = \CX_{m}^{\omega} \CX^{\an}
\end{equation}
where $\CX_{m}^{\omega}$ is precisely the functions $ \CX_m $ computed with $\Omega^{\eff}$
(and let us stress that it is
independent of $\theta_m$, but does depend on $\theta_m^{\eff} = - \psi$).  Moreover
\begin{equation}
\CX^{\an} = \exp \left(  \frac{\pi R}{\zeta} \CW^{\an} + \pi R \zeta \overline{\CW}^{\an} \right).
\end{equation}

It follows from \eqref{eq:Omega-eff} that we should view $\CX_{\CW}$ as an equivariant
function on an \emph{effective} PTN space $\widehat\CM(\omega):= \widehat\CM(\Omega^{\eff})$.
It is equivariant under translation in the circle coordinate of the fibration by $\pi_\omega$
analogous to \eqref{eq:PTN-fib}.
Of course, we are supposed to be defining sections of a line bundle over the space $\widehat\CM(\Omega)$.
Since the sections $\CX_\CW$ are independent of $\theta_m$ we can say the following.
We consider the diagram:
\begin{equation}\label{eq:pllbckdiag}{
 \xymatrix{  & \widehat\CM(\omega) \ar[d]^{\pi_\omega}  \\   \widehat\CM(\Omega) \ar[r]^-{\pi_\Omega} & D_\Lambda\times S^1 - (S_\Omega \cup S_{\omega})  }  }
\end{equation}
For the moment we excise all the singular points in the base, and for simplicity we use the
same notation for the total space over this smaller base.
Finally, we can consider the pullback
\begin{equation}\label{eq:pull-fib}
\pi_\Omega^* ( \widehat\CM(\omega) ) \to \widehat\CM(\Omega)
\end{equation}
which is a $U(1)$ fibration over $\widehat\CM(\Omega)$.
The functions $\CX_m^\omega$ pull back to equivariant functions on the $U(1)$ fibration
\eqref{eq:pull-fib}.  These are equivalent to sections of a line bundle over
$\widehat\CM(\Omega)$.

\subsubsection{Deriving the connection}

Now we  use the differential equations (4.61)-(4.64) in \cite{Gaiotto:2008cd}
to give an explicit construction of the
hyperholomorphic connection.

We first note that since eqs. (4.61)-(4.64) of  \cite{Gaiotto:2008cd}
are supposed to be the Cauchy-Riemann
equations we can read off a basis for $T^{(0,1)}\CM(\Omega)$ in complex structure $\zeta$:
\begin{equation}
\begin{split}
V^{0,1}_1 & = \frac{\p}{\p a} + \frac{\I \pi R}{\zeta} \frac{\p}{\p \theta_e} - \left( \frac{\pi}{\zeta}(U^\Omega
+ 2\pi \I R A^\Omega_{\theta_e}) + 2\pi A_a^\Omega\right)\frac{\p}{\p \theta_m}\\
V^{0,1}_2 & = \frac{\p}{\p \bar a} +  \I \pi R \zeta \frac{\p}{\p \theta_e} + \left(  \pi \zeta(U^\Omega
- 2\pi \I R A^\Omega_{\theta_e}) - 2\pi A_{\bar a}^\Omega\right)\frac{\p}{\p \theta_m}\\
\end{split}
\end{equation}
where $U^\Omega$ denotes \eqref{eq:gh-full-U}, and so forth.
For later reference, in the semiflat limit
\begin{equation}
U^\Omega \to  R \frac{\Delta}{4\pi} \log \left\lvert \frac{\Lambda}{a} \right\rvert^2 = R\,\im \tau,
\end{equation}
\begin{equation}
A^\Omega \to \frac{\Delta}{4\pi^2 } \re\left(\I \log\frac{a}{\Lambda}\right) \de \theta_e = - \frac{1}{2\pi } \re(\tau) \de \theta_e
+ \CO(e^{- R \vert a \vert}).
\end{equation}
Now let $\Theta$ denote the full quantum-corrected connection.
We aim to extract $\Theta$ from the equations
\begin{equation}
\langle V_A^{0,1}, (\de + \Theta) \CX_{\CW} \rangle =0, \qquad \qquad A=1,2.
\end{equation}
First consider the case
where $\CW^{\an}=0$ and call the corresponding connection $\Theta^s$.
In the case $\CW^{\an}=0$ we have $\CX_{\CW} = \CX_m^\omega$ and hence we
can use the differential equations (4.61)-(4.64) of \cite{Gaiotto:2008cd} and then subtract
the terms corresponding to $\langle V^{0,1}_A, \de \CX_{\CW} \rangle$ to
get four equations for the components of $\Theta$. We find a somewhat
elegant result:
\begin{equation}\label{eq:thetam-s}
\Theta_{\theta_m}^s = \I \frac{U^\omega}{U^\Omega},
\end{equation}
\begin{equation}
\Theta_{i}^s = 2 \pi \I \frac{ (U^\omega A^\Omega_{i} - U^\Omega A^\omega_{i})}{U^\Omega},
\end{equation}
where $i=\theta_e, a, \bar a$. Here $U^\omega$ means the function
\eqref{eq:gh-full-U} computed with \eqref{eq:Omega-eff}, and $A^\omega$ is the
corresponding one-form. It is nice to check that the semiflat limit works out perfectly
and that there is then a series of instanton corrections to this limit.

Now when $\CW^\an$ is nonzero we can write
\begin{align}\label{eq:an-conn}
0 &= \langle V_A^{0,1}, \CX_{\CW}^{-1}(\de + \Theta) \CX_{\CW} \rangle \\
&= \langle V_A^{0,1}, (\CX_{m}^{\omega})^{-1}(\de + \Theta) \CX_{m}^{\omega} \rangle
+ \left\langle V_A^{0,1}, \de \left( \frac{\pi R}{\zeta} \CW^{\an} + \pi R \zeta \overline{\CW}^\an\right) \right\rangle.
\end{align}
Let us write the total connection as
\begin{equation}
\Theta  = \Theta^s + \Theta^{\an}.
\end{equation}
Then from \eqref{eq:an-conn} we derive
\begin{equation}\label{eq:thetm-an}
\Theta_{\theta_m}^\an = \I \frac{R (\im\tau)}{U^\Omega}\alpha^\an,
\end{equation}
\begin{equation}
\Theta_{\theta_e}^\an = \I ( \eta^\an + (\re \tau) \alpha^\an ) + 2\pi \I \frac{ R(\im \tau)\alpha^\an}{U^\Omega} A^\Omega_{\theta_e},
\end{equation}
\begin{equation}
\Theta_{i}^\an =  2\pi \I \frac{ R(\im \tau)\alpha^\an}{U^\Omega} A^\Omega_{i},
\end{equation}
where $i=a,\bar a$, and we have defined
$t^\an := \eta^\an + \tau(a) \alpha^\an$. Again, the semiflat limits reproduce the contribution of $\CW^{\an}$ to
the semiflat connection.

\subsubsection{Analyzing the singularities}

We consider the extension of  the pullback diagram \eqref{eq:pllbckdiag}
over the singular sets $S_\omega$ and $S_\Omega$. There are three different
cases to consider with a  different story for each case.
Denote the integer in equation \eqref{eq:Nq-def} computed for $\Omega_q^{\eff}$
of \eqref{eq:Omega-eff} by $N_q^{\eff}$.

\begin{enumerate}

\item $s_{n,q} \in S_\Omega -S_\omega \cap S_{\Omega}$: Above these points the metric is singular but
the line bundle is not. It pulls back to a locally trivial line bundle over the neighborhood
$\IC^2/\IZ_{N_q}$. The $\IZ_{N_q}$-action on the line over the origin is trivial.

\item $s_{n,q} \in S_\omega- S_\omega \cap S_{\Omega}$: Here the line bundle is singular but the metric is
not. The fiber above a point $s_{n,q}$ has $c_1 = - N_q^{\eff}$.
Above the point $s_{n,q}$ (in the metric fibration $\pi_\Omega$)
there is a whole fiber  in $\widehat\CM(\Omega)$. The line bundle is
singular all along that ring. Thus, in this case, the singularity is only reduced
from codimension two to codimension three.

\item $s_{n,q} \in  S_\omega \cap S_{\Omega}$.  Here we consider the pullback of $\IC^2/\IZ_{N_q^{\eff}}\to D^3_*$ to
a $U(1)$ fibration over $\IC^2/\IZ_{N_q}$ via  $\pi_\Omega: \IC^2/\IZ_{N_q} \to D^3_*$. This is easily
determined by the following simple remark. Consider the quotient  Hopf fibration
$\pi_{N_1}: S^3/\IZ_{N_1} \to S^2$. This is a principal $U(1)$ bundle over $S^2$ with first Chern class
$N_1$ (measured relative to a unit volume form generating $H^2(S^2;\IZ)$). On the other hand,
$\pi_{N_1}^*: H^2(S^2;\IZ) \to H^2(S^3/\IZ_{N_1};\IZ)$ is a homomorphism $\IZ \to \IZ_{N_1}$ and
this homomorphism is simply reduction modulo $N_1$.  Now therefore we can consider the principal
$U(1)$ fibration over $S^3/\IZ_{N_2}$ given by $\pi_{N_2}^*( S^3/\IZ_{N_1})$. This principal
$U(1)$ bundle over $S^3/\IZ_{N_2}$ has Chern class $N_1\ \mod\ N_2$. Now, there are tautological
line bundles $\CR_{\rho} \to \IC^2/\IZ_{N}$ labeled by $\rho$ in the Pontryagin dual $\widehat{\IZ}_N$.
They restrict to the linking $S^3/\IZ_N$ to be the associated bundle to the $U(1)$ principal bundle
with $c_1 = \rho$.  Thus in our pullback diagram \eqref{eq:pllbckdiag}, if we restrict to the
neighborhood of a singular point $s_{n,q} \in S_\Omega \cap S_{\omega}$ we get the tautological
line bundle over $\IC^2/\IZ_{N_q}$ given by $N_q^{\eff}\ \mod\ N_q$. Note this is invariant under
the transformations $\omega_q \to \omega_q + q \Omega_q$, as it must be.

\end{enumerate}

The connections we have derived are only singular on codimension three or codimension four singularities,
as opposed to the semiflat connections, where the line bundle is not defined on a
codimension two singularity. In this sense, the quantum corrections have smoothed out
the geometry.

\subsubsection{Mirror manifolds and mirror branes}\label{subsubsec:MirrorBranes}

We can illustrate the remarks we made above about mirror branes in the present example.
The PTN space is a torus fibration $\pi: \widehat\CM(\Omega) \to D_\Lambda$ with a
singular fiber over $a=0$. The fiber can be viewed as a necklace of intersecting ``spheres.''
More precisely, the circle fibers with coordinate $\theta_m$ collapse at the points $s_{n,q}\in S_\Omega$.
These fibers together with the intervals on the $\theta_e$ circle constitute a sphere with $\IZ_{N_q}$
orbifold singularities at the north and south poles.

The section $\tilde \nu = (\alpha \gamma_m + \eta \gamma_e)\ \mod\ \Gamma $ is a section of the
dual fibration, which should be interpreted as the mirror manifold. Indeed, the
T-dual of the semiflat metric $R\ g^\sf$ is the metric
\begin{equation}
\widetilde{Rg^\sf} = R^2\ \im\tau \vert \de a \vert^2 +    \frac{1}{4\pi^2\im\tau} \vert \de \tilde \theta_e - \tau  \de \tilde \theta_m \vert^2
\end{equation}
(where $\tilde\theta_i$ is T-dual to $\theta_i$) and hence isomorphic to the original
metric $R g^\sf$, with $\tilde \theta_e = \theta_m $ and $\tilde \theta_m = \theta_e$.
Hence the  mirror manifold to $\widehat{\CM}(\Omega)$ should be diffeomorphic to
$\widehat{\CM}(\Omega)$ itself. Note that it is the $\tilde \theta_e$ circle which shrinks
at the singular points $\tilde s_{n,q}$ defined by $\tilde \theta_m = 2\pi n/q$ with $\Omega_q\not=0$.

The section $\tilde \nu = (\alpha \gamma_m + \eta \gamma_e) \mod \Gamma $ has
coordinates $\tilde\theta_e = \eta$ and $\tilde \theta_m = \alpha$. On the other
hand, it follows from \eqref{eq:tee-tau} that
\begin{equation}\label{eq:eta}
\begin{split}
\eta
& = \re(w_1) - \im(w_1) \frac{ \re \tau(a)}{\im \tau(a) } + \CO( \frac{a}{\log a } ),\\
\end{split}
\end{equation}
\begin{equation}\label{eq:alpha}
\begin{split}
\alpha
& = \frac{\delta}{\Delta} + \frac{ \im w_1}{\im \tau(a) } +  \CO( \frac{a}{\log a } )\\
\end{split}
\end{equation}
(recall that $w_1$ is the linear term in $\CW^\an$). Hence their reductions modulo $1$
 have good limits at $a\to 0$. In particular, note that $\alpha \to \frac{\delta}{\Delta}$.

The section $\tilde \nu$ defines a Lagrangian cycle which is the support of an $A$-brane.
Sometimes this section goes through the singular points $\tilde s_{n,q}$. The $A$-brane
should be mirror to the brane described by the   line bundle with hyperholomorphic
connection over $\widehat{\CM}$ we have constructed above. However, as we have
seen, the latter brane depends on the details of $\Omega_q$ and $\omega_q$ whereas
the support of the $A$-brane is only sensitive to the data $w_1, \delta, \Delta$.
It must be that there is further data needed to specify this $A$-brane (in particular
its flat connection and possible binding to fractional branes).  There is undoubtedly an
interesting story here, related to \cite{Frenkel:2007tx}, but it lies beyond the scope of this
paper.

\subsection{A massless 2d particle with 4d gauge charges only}\label{subsec:One-d-Coul-Delta-Zero}

Our second example is slightly artificial, but illustrates nicely an important physical
point. We return to the analysis of \S \ref{subsec:1d-Coul-Branch}
but with $\Delta=0$, so that $\tau = \tau_0$ is constant. In particular,
we will assume that at some point $u_0$ of $\CB$
where is a charge $\gamma \in \Gamma$ with
$Z_\gamma(u_0) = 0$ and  $\omega(\gamma, \cdot; u_0) \neq 0$) but  $\Omega(\gamma;u_0) = 0$.
We will also assume that $\mu(\gamma_{ij};u_0)=0$ so the vacua don't mix and we
can take the case of a single vacuum.  Note that $\omega(\gamma,\gamma_i;u_0)$ is independent of
which $\gamma_i$ we choose in $\Gamma_i$.

We will take the simple effective superpotential
\begin{equation} \label{eq:4d-intout}
 \CW =\delta  \frac{\I}{2 \pi} \left(a \log \left( \frac{a}{\Lambda_2} \right) - a\right).
\end{equation}
The singularity of $\CW$ at $a = 0$ implies a singularity for $t = \partial_a \CW$ and hence for $A^\sf$ there.
Note that
\begin{equation}\label{eq:alphaeta-sf}
\eta = - \frac{\delta}{2\pi \Im \tau_0} \Re\left( \bar \tau_0 \log \frac{a}{\Lambda_2} \right),
\qquad \qquad \alpha =   \frac{\delta}{4\pi \Im \tau_0} \log \left\lvert \frac{a}{\Lambda_2} \right\rvert^2 .
\end{equation}
Now, we would like to study the IR Lagrangian obtained after compactifying the whole system on $S^1$.
The most straightforward way to proceed is to compactify the theory including the 2d chiral multiplet.
The KK mode expansion of the 2d chiral multiplet then gives an infinite set of 1d fields, each
charged under the 3d gauge field.  The $n$-th KK mode has
mass $m_n = \sqrt{\abs{qa}^2 + (n + \frac{q\theta_e}{2 \pi})^2 / R^2}$, where $\theta_e = \theta_{\gamma_e}$ is the
Wilson line of the $U(1)$ gauge field around $S^1$. Thus the compactified version
of \eqref{eq:alpha-correction} gives
\begin{align}
(\im \tau_0) \alpha_{1d} &= - \sum_{q>0} q\omega_q \frac{1}{4 \pi^2 R}
\sum_{n = -\infty}^\infty \int_{- \infty}^\infty \de k \frac{1}{m_n^2 + k^2}.
\end{align}
The integral on $k$ is elementary, but the sum requires regularization.
The regularized answer can be written
\begin{equation}
\begin{split}
(\im \tau_0) \alpha_{1d} & = - \sum_{q>0}
\frac{q\omega_q}{4 \pi R} \sum_{n = -\infty}^\infty \left( \frac{1}{m_n} - \kappa_{n,2}\right)\\
& = - \sum_{q>0} \frac{q\omega_q}{4 \pi R}
\sum_{n = -\infty}^\infty \left( \frac{1}{\sqrt{\abs{q a}^2
+ (n + q\frac{\theta_e}{2 \pi})^2 / R^2}} - \kappa_{n,2}\right)\\
\end{split}
\end{equation}
where $\kappa_{n,2}$ are some regularization constants, independent of $a, \theta_e$, chosen so that the sum converges
and the leading term in the large $R$ expansion reproduces $\alpha$.
Note that $\alpha_{1d} = - U^\omega/ ( R\,\im \tau_0)$,
where $U^\omega$ is the function  $U^\Omega$ of equation \eqref{eq:gh-full-U} computed with
$\Omega^{\eff}_q = q^{-1} \omega_q$.

The semiflat geometry is uncorrected, so we will be discussing connections on a line
bundle over $D^*_\Lambda \times T^2$.
The hyperholomorphic connection $\Theta$ has a component along $\theta_m$
determined by equation \eqref{eq:2d-W-F-action}
\begin{equation}
 \iota_{\partial_{\theta_m}} \Theta = - \I \alpha_{1d} .
\end{equation}
In fact, this together with the condition that $A$ is hyperholomorphic is almost enough to determine $A$ completely.
Hyperholomorphicity in the case of a 4-dimensional $\CM$ just means that $\de A$ is an anti-self-dual 2-form on $\CM$.
\footnote{We choose an orientation $e^1\wedge e^2 \wedge e^3 \wedge e^4$
where $e^1 + \I e^2 = \sqrt{R \im \tau_0} da$ and $e^3 + \I e^4 = \frac{\I}{2\pi \sqrt{R \im \tau_0}} (d \theta_m - \tau_0 d \theta_e)$.}
Writing
\begin{equation}
 \Theta = \Theta^{[3]} - \I \alpha_{1d} \left(\de \theta_m - (\re \tau_0) \de \theta_e \right)
\end{equation}
where $A^{[3]}$ is a $U(1)$ connection over the $\IR^2 \times S^1$ parameterized by $(a, \bar{a}, \theta_e)$, this amounts to
\begin{equation} \label{eq:a3}
\de \Theta^{[3]} =  (2 \pi \I \sqrt{R \, \im \tau_0}) \, \star_3 \de \alpha_{1d}.
\end{equation}
Here $\star_3$ denotes the 3-dimensional Hodge star with respect to the
metric $(e^1)^2 + (e^2)^3 + (e^3)^2$ with orientation $e^1 e^2 e^3$.  In particular
the existence of an $\Theta^{[3]}$ obeying \eqref{eq:a3} requires that $\alpha_{1d}$ is
a harmonic function on $\IR^2 \times S^1$, which is indeed the
case.  The general $\Theta^{[3]}$ obeying this equation was again written down in \S 4.1 of \cite{Gaiotto:2008cd}
(more precisely we have $\Theta^{[3]} = - \frac{2 \pi \I}{q} A^\there$ when $\omega$ has support at
one value of $q$ and value $1$ there.)

So we have determined the form of the hyperholomorphic connection $\Theta$.
The important new point here is that
 it agrees with our general expression \eqref{eq:thetam-s} with the
replacement $U^\Omega \to R \im \tau_0$. Therefore, in the general expression \eqref{eq:thetam-s} we
should view the instanton series for the numerator as due to instantons from worldlines of
2d BPS particles and the instanton expansion of the denominator as due to
instantons from worldlines of 4d BPS particles. We can therefore   interpret the result
\eqref{eq:thetam-s} as a combined 2d-4d instanton expansion.

The rest of the discussion of the topology of the quantum corrected line bundle
proceeds as in the previous section, with the replacement of $\widehat\CM(\Omega)$
by $  D^*_\Lambda \times T^2$.

\subsection{A massless 2d soliton}\label{subsubsec:Massless-2d}

Now let us turn to the other type of singularity, where $Z_{\gamma_{ij}}(u)\to 0$
for some 2d soliton with $\mu_{\gamma_{ij}}(u_0) \neq 0$
but all occupied 4d charges have nonvanishing $Z_{\gamma}$ at $u_0$.

There is a neat toy model for this situation:  take a free $U(1)$ gauge theory in 4 dimensions and
a surface defect supporting a single twisted chiral
field $X$, coupled together by a twisted superpotential $\CW = \frac{\Lambda_{\twod}}{3} X^3 - a X$,
where $a$ is the twisted chiral multiplet coming from the 4d theory.
(So this is essentially a 2d Landau-Ginzburg model whose superpotential depends on the Coulomb
branch modulus of the 4d theory.)

The 2d system has two vacua labeled by $i \in \{ +, -\}$.  These two vacua correspond to the two points
$X = \pm \sqrt{a / \Lambda_\twod}$, with
\begin{equation}
\CW_\pm = \mp \frac{2}{3\Lambda_{\twod}^{1/2}} a^{3/2}, \quad t_\pm = \partial_a \CW_\pm = \mp \left(\frac{a}{\Lambda_{\twod}}\right)^{1/2}.
\end{equation}
(For simplicity we will take $\Lambda_\twod$ positive and $\Lambda_{\twod}^{1/2}$ positive.)
There is a single BPS soliton interpolating between the two vacua, carrying charge $\gamma^0_{+-}$,
and a soliton interpolating the other way, with charge $\gamma^0_{-+}$.
We adopt the sign conventions for the $A_1$ theories in \S \ref{sec:2d-4ddata}, so
$\mu(\gamma^0_{+-}) = \mu(\gamma^0_{-+})=1$ while $\sigma(\gamma_{+-}^0,\gamma_{-+}^0)=-1$,
in accord with \eqref{eq:unitarity-4}.
There is an obvious singularity of $t_\pm$ at $a=0$ where the soliton becomes massless, which implies a singularity
of $A^\sf$.

In the exact compactified theory, $A^\sf$ should be corrected to some
smooth self-dual connection $A$ on a rank $2$ bundle
$V$ over $\CM = \IR^2 \times T^2$.  Unlike  the previous example, we will not compute this $A$ directly:
fortunately we will be able to determine it more indirectly.
We first describe $A$ in a bit more detail and then check that it indeed matches what comes
from our construction.

As in the previous example, the 4d theory is just the free $U(1)$ theory,
so the metric on $\CM$ is
given by \eqref{eq:sf-metric}.  Note that
the physics of the compactified system is independent of the electric and magnetic Wilson lines around $S^1$,
as all of the UV fields are uncharged.
Hence the exact corrected $A$ is invariant under translations on the $T^2$ factor of $\CM$.
  (Strictly speaking, the notion of
``translation invariant''
only makes sense once we fix a trivialization of $V$; here we are working with the trivialization determined by
the superpotentials $\CW_\pm$.)
Dividing such an $A$ up into
components as
\begin{equation}
 A = A^{[2]} + \varphi_a \frac{\de \theta_m - \bar\tau \de \theta_e}{2 \pi (\im \tau)} + \bar\varphi_{\bar a} \frac{\de \theta_m - \tau \de \theta_e}{2 \pi (\im \tau)}
\end{equation}
(where $A^{[2]}$ denotes a connection along the $\IR^2$ factor and $\bar\varphi_{\bar a}$ is the Hermitian adjoint of $\varphi_a$)
the self-dual Yang-Mills equations for $A$ become
\begin{align}
\partial_{\bar a} \varphi_a + [A^{[2]}_{\bar a}, \varphi_a] &= 0, \\
F^{[2]} + R^2 [\varphi_a, \bar\varphi_{\bar a}] &= 0,
\end{align}
i.e. the Hitchin equations for the pair $(A^{[2]}, \varphi_a \, \de a)$ on $\IR^2 = \IC$.
In the semiflat approximation \eqref{eq:a-sf}, $A^{[2]} = 0$ and $\varphi_a$
is diagonal, with eigenvalues $\pm \pi t $ with $\pi t= \pi \partial_a \CW = - \pi (\frac{a}{\Lambda_{\twod}})^{1/2}$.
This approximation is good at sufficiently large $a$, up to corrections
of order $e^{- 2 \pi R \abs{a}}$.  In particular, using Liouville's theorem these asymptotics imply
\begin{equation} \label{eq:trphi2}
 \Tr \varphi_a^2 = \pi^2 a / \Lambda_\twod.
\end{equation}
There is a unique smooth solution $(A^{[2]}, \varphi)$ of the Hitchin equations on $\IC$ obeying \eqref{eq:trphi2}.
It is radially symmetric on $\IC$, and determined by a solution of
Painleve III --- see for example \cite{Cecotti:1991me}, \S 8.1.
Fortunately, we
will not need the explicit form of $(A^{[2]}, \varphi)$.
Rather, we will use an indirect characterization, essentially the $tt^*$ technology of
\cite{Cecotti:1991me,Cecotti:1993rm} (see also \S 9.4.1 of \cite{Gaiotto:2009hg}.)

Let us briefly review how that technology works in this case.  We consider the flat connections
\begin{equation} \label{eq:hit-conn-pre}
\nabla(\zeta) = R \zeta^{-1} \varphi + A^{[2]} + R \zeta \bar \varphi.
\end{equation}
This family of connections can be completed to a single connection over the $\IC \times \IC\IP^1$
parameterized by $(a, \zeta)$.  This connection has irregular singularities
at $\zeta = 0$ and $\zeta = \infty$, exhibiting Stokes phenomena.
The formal asymptotics of solutions as $\zeta \to 0$ are
\begin{equation} \label{eq:formal}
\CY_\pm \sim \exp \left[ \mp \frac{\pi R}{\zeta}\frac{2}{3\Lambda_{\twod}^{1/2}} a^{3/2} \right] e_\pm
\end{equation}
where $e_\pm$ denote some $\zeta$-independent sections.
There are two anti-Stokes rays
 emerging from either $\zeta = 0$ or $\zeta = \infty$,
located where the exponent becomes real.  In the basis $(\CY_+, \CY_-)$
the corresponding Stokes factors are known to be simply
\begin{equation} \label{eq:okesokes}
\begin{pmatrix} 1 & 1 \\ 0 & 1 \end{pmatrix}, \qquad \begin{pmatrix} 1 & 0 \\ -1 & 1 \end{pmatrix}.
\end{equation}
See Appendix \ref{app:N1hitchin} for more details.
Moreover, these formal asymptotics and Stokes
data are enough to \ti{determine} the full connection, hence the solution $(A^{[2]}, \varphi)$ of
Hitchin's equations, and hence finally the hyperholomorphic connection $A$.

Now let us see how this description is reproduced by our construction.
In our present setup, where $\omega(\cdot, \cdot) = 0$,
\eqref{eq:int-0}, \eqref{eq:int-1} just say
\begin{equation} \label{eq:int-simp-0}
\CY_{\gamma_\pm} = g_\pm \CY^\sf_{\gamma_\pm},
\end{equation}
while $\mu(\gamma^0_{+-}) = 1$, $\mu(\gamma^0_{-+}) = 1$, with all other $\mu(\cdot) = 0$,
so \eqref{eq:int-2} says
\begin{align}
g_+(\zeta) &= g_+^\sf +  \frac{1}{4 \pi \I} \int_{\ell_{\gamma^0_{-+}}}
\frac{\de \zeta'}{\zeta'} \frac{\zeta' + \zeta}{\zeta' - \zeta} g_-(\zeta')
\CY^\sf_{\gamma^0_{-+}}(\zeta'), \label{eq:int-simp-1} \\
g_-(\zeta) &= g_-^\sf + \frac{1}{4 \pi \I} \int_{\ell_{\gamma^0_{+-}}}
\frac{\de \zeta'}{\zeta'} \frac{\zeta' + \zeta}{\zeta' - \zeta} g_+(\zeta')
\CY^\sf_{\gamma^0_{+-}}(\zeta'). \label{eq:int-simp-2}
\end{align}
(recall that $\CY^\sf_{\gamma^0_i}$ is defined in \eqref{eq:xi-sf}, \eqref{eq:xi-sf-gi}
but $\CY^\sf_{\gamma^0_{ij}}$ is defined by \eqref{eq:xgijdef}).
In particular, $\CY_{\gamma_{\pm}^0}$ are independent of base coordinates
 $\theta_e, \theta_m$ on $T^2$, i.e. they are invariant under
translations along $T^2$.
We may thus consider them as sections of a bundle over the base $\IC$.
As just explained, the translation invariance implies
this bundle supports a solution $(A^{[2]}, \varphi)$ of the Hitchin equations.
The Cauchy-Riemann equations for holomorphic
sections of $V^{(\zeta)}$, when restricted to translation invariant
sections, reduce to flatness under $\nabla(\zeta)$.
So $\CY_{\gamma_{\pm}^0}$ are $\nabla(\zeta)$-flat.  \eqref{eq:int-simp-0}-\eqref{eq:int-simp-2}
show that $\CY_{\gamma_{\pm}^0}$ have the asymptotics \eqref{eq:formal}.  Finally, we can read off
the Stokes factors from the discontinuities of $\CY_{\gamma_{\pm}^0}$ across the two rays
$\ell_{\gamma^0_{+-}}$ and $\ell_{\gamma^0_{-+}}$.  These discontinuities are
determined by \eqref{eq:int-simp-1} and \eqref{eq:int-simp-2}.
The net result is the following:

\begin{enumerate}

\item Across the wall $a^{3/2}/\zeta \in \IR_-$ we have
\begin{equation} \label{eq:a32jump1}
\begin{split}
\CY_+(+) & = \CY_+(-) + \sigma(\gamma_{-+},\gamma_+) \CY_- \\
\CY_-(+) & = \CY_-(-) = \CY_- \\
\end{split}
\end{equation}
where $\CY_\gamma(\pm)$ refers to the side of the
wall with $\Im (\pm a^{3/2}/\zeta) > 0$.

\item Across the wall $a^{3/2}/\zeta \in \IR_+$ we have
\begin{equation} \label{eq:a32jump2}
\begin{split}
\CY_+(+) & = \CY_+(-) =\CY_+  \\
\CY_-(+) & = \CY_-(-) + \sigma(\gamma_{+-},\gamma_-) \CY_+ \\
\end{split}
\end{equation}
where $\CY_\gamma(\pm)$ refers to the side of the
wall with $\Im (\mp a^{3/2}/\zeta) >0$.

\end{enumerate}

Using \eqref{eq:simp-cr} we know that $\sigma(\gamma_{+-},\gamma_-)\sigma(\gamma_{-+},\gamma_+)=-1$.
With an appropriate choice of $\sigma$ subject to this constraint,
\eqref{eq:a32jump1}, \eqref{eq:a32jump2} correspond to the desired Stokes factors \eqref{eq:okesokes}.
This is enough to show that our construction produces the $A$ we described above.

In particular, our construction has produced
a \ti{smooth} hyperholomorphic connection:  this is an example of the advertised resolution of
singularities induced by 2d BPS states.

\section{Example: $A$ type theories and Hitchin systems}\label{sec:Hitchin-Expl}

One important example of our construction is its application to the ``canonical surface defects''
in theories of class $\CS$.

Recall that theories of class $\CS$   are obtained from an $ADE$ group and a curve $C$
(with punctures) as described in \cite{Witten:1997sc,Gaiotto:2009hg,Gaiotto:2009we}.
For simplicity, throughout this section we will restrict our attention to the case where the $ADE$ group is of type $A_{n-1}$.
Then there are canonical surface defects $\bS_z$ corresponding to points $z \in C$
\cite{Alday:2009fs,Gaiotto:2009we}.  (If we think of the theory as obtained from a decoupling limit of M5-branes,
then $\bS_z$ can be constructed from an M2-brane ending on the M5-branes at the point $z$.)
In this class of theories and surface defects, we can describe the 2d-4d wall-crossing data of
\S\S \ref{subsec:Formal-Statement}
and \ref{sec:Physical-Interpret} explicitly in terms of the geometry of $C$ and its spectral cover.

We will need a bit of care in defining the canonical surface defect, due to some subtleties about locality of surface defects
in the 6d theory. It is not difficult to argue (looking for example at a KK reduction of the theory to 5d Yang-Mills or to the Abelian theory
on the Coulomb branch of the 6d theory) that the basic surface defects in 6d have a potential
mutual locality problem. If we keep a surface defect fixed, look at the $S^3$
which surrounds it, nucleate a second surface defect near the north pole of the
sphere, and annihilate it near the south (in analogy with what
is done in 4d with Wilson and 'tHooft operators) the partition function gains a factor of $e^{2 \pi \I/n}$.
\footnote{One way to see this is to work on the Coulomb branch. Then the worldsheet of the nucleated
surface defect has a topologically interesting term $\exp[2\pi \I \int_{\Sigma} (v_1, B) ]$ where
$B$ is a self-dual two-form valued in the Cartan subalgebra of $A_{n-1}$. Here $v_1$ is a vector in
the weight lattice of $A_{n-1}$. The other surface defect,
which we locate at the center of $\IR^4$ surrounded by the linking sphere $S^3$ is a magnetic
source of $B$ of strength $H = v_2 \omega_3$ where $\omega_3$ is the unit normalized volume
form on $S^3$ and $v_2$ is in the weight lattice of $A_{n-1}$. Here we have identitified the
Cartan subalgebra with its dual using a metric so that roots have length-squared two.
Then the topologically interesting phase for the process described above is
$\exp[2\pi \I (v_1, v_2)]$, and since both $v_1, v_2$ are in the weight lattice of $A_{n-1}$ this
is an $n^{th}$ root of unity.}
This factor plays an important role when one looks at mutual locality of line defects in 4d
corresponding to paths on $C$.
There is a second application, though, to mutual locality between surface
defects sitting at a point in $C$ and point operators in 4d engineered
by wrapping surface defects on $C$:  the point operator has $e^{2 \pi \I/n}$
monodromy around the surface defect.
The local operator which descends from a surface defect wrapping $C$ was identified in \cite{Gaiotto:2009gz}
with a specific chiral operator which receives vevs on the Higgs branch.
Hence in order to include the canonical surface defect $\bS_z$
in the 4d theory, strictly speaking one needs to gauge the $\IZ_n$ flavor
symmetry corresponding to the monodromy of these specific chiral operators
around the canonical surface defect.
At least in the $A_1$ case with regular singularities we will, in \S \ref{subsec:Lag-Desc},
identify the $\IZ_2$ flavor symmetry in the four dimensional Lagrangian
description of the theories, and verify that the $\Gamma_i$ fail to be $\Gamma$ torsors
in a way which is exactly controlled by the $\IZ_2$ flavor monodromy, leading
to the kind of anomaly cancellation discussed in  \S \ref{subsubsec:Anomaly Cancelation}.
The generalization to higher rank should not be hard,
though it might be hampered by the lack of a Lagrangian description.

The fact that $\bS_z$ carries flavor monodromies nicely resolves
 some puzzles rased in
\cite{Alday:2009fs} (in a way anticipated in \cite{Alday:2009fs}).
Whenever $C$ approaches a factorization limit and develops a long thin tube,
a weakly coupled gauge group appears in the four dimensional field theory.
If the point $z$ is in the tube region, then, as was observed in \cite{Alday:2009fs},
the surface defect $\bS_z$
 appears to be well described by a GW surface defect. However, a direct identification with a
GW surface defect leads to two puzzles: First, the GW surface defect alone does
 not break any flavor symmetry, but the canonical surface defect often appears to
break some flavor symmetry. The resolution of this puzzle is that the appropriate GW
defect must carry some flavor monodromy and this
 monodromy does indeed break the flavor group down to the commutant in the
 flavor group of the gauged $\IZ_n$ subgroup.
The second problem is that several class $\CS$ descriptions of the same 4d theory might exist,
and in general these will have sharply different canonical surface defects. How can
they all correspond to a single GW surface defect? The resolution is that there are
several GW defects, distinguished by having different flavor monodromies.

\subsection{Review of pure 4d data}\label{subsec:WKB-networks}

First we recall from \cite{Gaiotto:2009hg} some purely 4d aspects of the story.
Although the theory exists for any simply laced Lie algebra we will for simplicity
focus on the case of $A_{n-1}$. Later we will specialize to $n=2$.

A point $u$ of the Coulomb branch $\CB$ corresponds to
a set of meromorphic $k$-differentials $\phi_k$ ($k = 2, 3, \dots, n$)
on $C$.  The $\phi_k$ have poles only at the punctures of $C$, and part of the data defining the theory
is a set of linear conditions on the singular parts of the $\phi_k$ at the punctures; roughly these conditions identify the
residues of the $\phi_k$ with various combinations of the mass parameters of the theory.
$\CB$ is the space of all tuples $(\phi_2, \phi_3, \dots, \phi_n)$ of meromorphic differentials obeying these linear conditions.
$\CB$ is thus a finite-dimensional affine space modeled on $\oplus_{d=2}^{n} H^0(C,K^{\otimes d})$.

For every $u \in \CB$, there is a corresponding Seiberg-Witten curve,
\begin{equation}
 \Sigma_u = \{ \phi_n + \phi_{n-1} \lambda + \cdots + \phi_2 \lambda^{n-2} + \lambda^n = 0 \} \subset T^* C.
\end{equation}
$\lambda$ is a cotangent vector, so that each term in the above equation is an $n$-differential on $C$.
At a generic point of $C$ the equation has $n$ distinct solutions,
so $\Sigma_u \subset T^*C$ is an $n$-fold covering of $C$.  Locally the $n$ solutions give $n$ 1-forms $\lambda_i$ on $C$.  These 1-forms
pull back to a single globally defined 1-form on $\Sigma_u$ which we also call $\lambda$.

The local system of
lattices $\Gamma$ begins with a certain subquotient of $H_1(\Sigma_u;\IZ)$ and
then extends it so that $\Gamma_g$ is self-dual. This extension depends on a choice
of a set $\CL$ of maximal mutually local line defects and determines a
precise global structure of the gauge group. In the $A_1$ case we begin with
the sublattice $H_1(\Sigma_u;\IZ)^-$ odd under the deck transformation and then,
depending on the choice of $\CL$, we extend it to a self-dual lattice.
(For details see \cite{Gaiotto:2010be}.)
The central charge function $Z_{\gamma}$ is
\begin{equation}
 Z_{\gamma} = \frac{1}{\pi} \oint_{\gamma} \lambda.
\end{equation}

Finally we come to the most interesting part of the story:
the $\Omega(\gamma;u)$ are the 4d BPS degeneracies, described in \cite{Klemm:1996bj,Gaiotto:2009hg}.
They   count certain special networks on $C$, defined as follows:
First, define a \ti{WKB curve of type $ij$ and phase $\vartheta$} to be an oriented real curve on $C$,
along which   $e^{-\I \vartheta} \langle (\lambda_i - \lambda_j), \p_t \rangle$ is real
and positive for some pair of vacua $i,j$. Here $\p_t$ is a positively oriented tangent vector
along the path. \footnote{In  \cite{Gaiotto:2009hg} WKB curves for the $A_1$ case were defined to
be unoriented. This is compatible with   the present definition, because we only require orientation
with respect to \emph{some} pair of vacua $i,j$.}
A \ti{WKB network of phase $\vartheta$} is by definition
a network of curves on $C$ whose legs are WKB curves of phase $\vartheta$.
The network can have three-legged junctions where legs of types $ij$, $jk$, $ki$ come together.
A leg of type $ij$ is also allowed to end on a branch point where $\lambda_i - \lambda_j = 0$, i.e.
the $i$-th and $j$-th sheets of $\Sigma$ merge.

Following any leg of a WKB network in either direction, there are four possible outcomes:
either we reach a junction, we reach a branch point,
we return to where we started, or we spiral into one of the punctures of $C$.
We call the network \ti{finite} if this fourth possibility never occurs.
This last condition makes finite WKB networks rather special, since a generic WKB curve spirals into punctures in
both directions.  See Figure \ref{fig:finite-network}.
\insfig{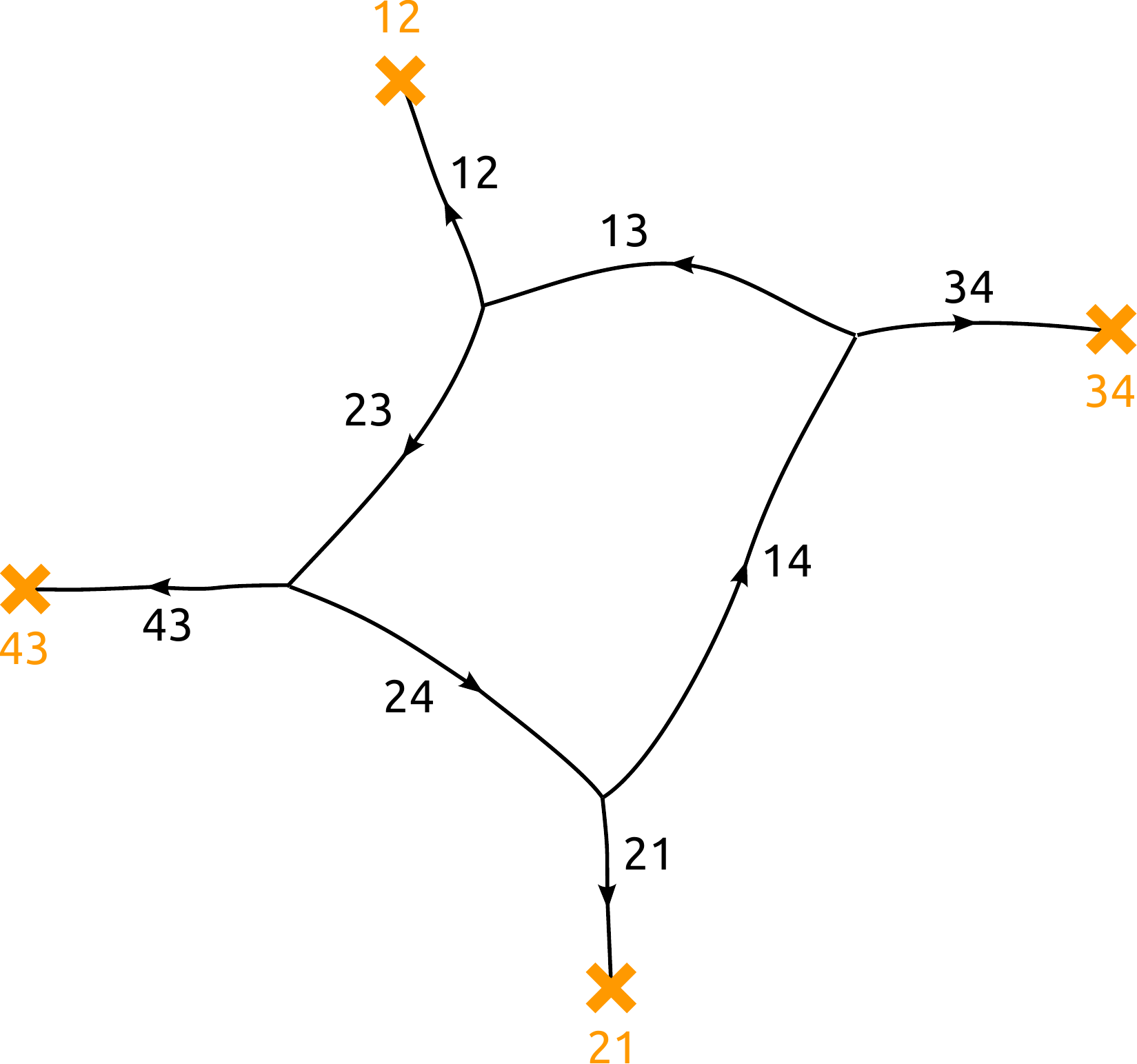}{A possible finite WKB network in the $A_3$ theory.  All legs are WKB $ij$-curves, with $ij$
as indicated.  Similarly the branch points (orange crosses) are $ij$-branch points as indicated.}

Each finite WKB network has a canonical lift to a union of closed curves on $\Sigma$, obtained by lifting
a leg of type $ij$ to a pair of curves on sheets $i$ and $j$ of $\Sigma$, with opposite orientations.
The $\Omega(\gamma;u)$ count finite WKB networks with phase $\vartheta = \arg Z_\gamma$,
for which the homology class of the lift is $\gamma$.\footnote{This is actually equivalent to
counting finite WKB networks with arbitrary phase --- indeed, if a finite WKB network has phase $\vartheta$
and lifts to a class $\gamma$, then $\vartheta = \arg Z_\gamma$.}
Isolated finite WKB networks
contribute $+1$ to $\Omega$.  The contributions from networks with moduli are more complicated
and have not been analyzed systematically, except in the case of the $A_1$ theory (see below).

\subsection{2d-4d data} \label{sec:2d-4ddata}

Now we describe the new data attached to the surface defect.
Most of what we say could be guessed just by looking for the most obvious generalizations
of the 4d data to incorporate the surface defect, consistent with the general structure we
have described in the rest of this paper.  A fuller justification of our claims here could
be given using a string theory realization of the $A_{n-1}$ theory.  Alternatively, in the $A_1$ case,
our claims are justified by the WKB analysis which we sketch later in this section.

\begin{enumerate}

\item {\bf Vacua}:  The finite set $\CV(\bS_z)$ of vacua of the surface defect is just the set of
preimages of $z$ in $\Sigma_u$; call them $x_i$, $i=1, \dots, n$. Often we will simply write $i$
for the vacuum $x_i$.
(This follows from the identification \cite{Gaiotto:2009fs} between the
twisted chiral ring of the 2d theory on the surface defect and the Seiberg-Witten curve of the
4d theory.)

\item {\bf Soliton charge torsors}:  The torsors $\Gamma_{ij}$ are taken to be
the $\Gamma$-torsor of (relative) homology classes of open paths from $x_i$ to $x_j$.
(Thus, we choose one connected path from $x_i$ to $x_j$ and then consider all
translates of the relative homology class in  $H_1(\Sigma_u, \{ x_i, x_j \} ;\IZ)$ by $\Gamma$.)
The central charge function $Z_{\gamma_{ij}}$ is then \footnote{In Seiberg-Witten theory it is
often said that the Seiberg-Witten differential is ambiguous by shifts $\lambda \to \lambda + \de f$.
This would affect the central charges here. In the theories of class $\CS$ there is a distinguished
Seiberg-Witten differential. This is the one which appears in the chiral ring, and is the one
we should employ here.}
\begin{equation}
 Z_{\gamma_{ij}} = \frac{1}{\pi} \int_{\gamma_{ij}} \lambda.
\end{equation}

\item {\bf 2d-4d BPS degeneracies $\omega$}:  The most naive guess for
$\omega(\gamma, \gamma_{ij})$ would be
\begin{equation} \label{eq:a1-omega-naive}
\omega(\gamma, \gamma_{ij};u) = \Omega(\gamma;u) \langle \gamma, \gamma_{ij} \rangle.
\end{equation}
This equation is indeed correct, but it needs some clarification:  choosing different representatives for the
class $\gamma$ would \ti{a priori} give different values for the intersection number
$\langle \gamma, \gamma_{ij} \rangle$, since representatives of $\gamma_{ij}$ are
open curves.  Which representative should we use?

First consider the special case where the only finite WKB networks in class $\gamma$ are isolated.
In this case, the lift of each such network
defines a canonical representative of the class $\gamma$.  In particular this representative defines
a definite class $\tilde\gamma$ in the   homology group $H_1(\Sigma_u-\{ x_i, x_j \} ;\IZ)$
(which pairs with $H_1(\Sigma_u,\{ x_i, x_j \} ;\IZ)$).
Having made this definition, we can say that this lifted network contributes
$\inprod{\tilde\gamma, \gamma_{ij}}$ to $\omega(\gamma, \gamma_{ij})$.
To get the full
$\omega(\gamma, \gamma_{ij})$ we sum over the networks contributing to $\Omega(\gamma)$.

If the networks contributing to $\Omega(\gamma)$
are not isolated, then the job of computing $\omega(\gamma,\gamma_{ij})$
is more complicated.  We do not give a general prescription here, but
we do indicate what should be done in the first nontrivial example, the $A_1$
theory.  In this case the networks we have been discussing in fact just consist of single WKB curves ---
there are no nontrivial junctions.  The new case we need to discuss is the case where the WKB curve
is closed:  in this case it is not isolated but rather lies in a 1-parameter family, sweeping out
an annulus on $C$.  In this case, we focus our attention
on the two closed WKB curves which make up the inner and outer boundaries of the annulus:
letting $\tilde \gamma, \tilde\gamma'$ denote the lifts of these two boundaries, we define
\begin{equation}
\omega(\gamma, \gamma_{ij} ) = - \inprod{\tilde \gamma, \gamma_{ij} } - \inprod{\tilde \gamma', \gamma_{ij} }.
\end{equation}
In the case where $z$ is not in the annulus both terms contribute equally
and this simplifies
$\omega(\gamma, \gamma_{ij}) = -2 \inprod{\tilde\gamma, \gamma_{ij}}$.
This is our original naive formula \eqref{eq:a1-omega-naive}, since in this case $\Omega(\gamma) = -2$.

\item {\bf 2d-4d BPS degeneracies $\mu$}:  Finally, the $\mu(\gamma_{ij};u)$
count \ti{finite open WKB networks of type $ij$} on $C$.  An open WKB network is just like the WKB networks we defined above,
except that it has one special leg which is of type $ij$ and  has one end at the point $z$.
The lift of any finite open WKB network
 to $\Sigma$ defines an element $\gamma_{ij} \in \Gamma_{ij}$.
When the moduli space of these networks is an isolated set of points, $\mu(\gamma_{ij})$ is a signed sum
over these points.  In general we do not have a prescription to fix the sign.  In the $A_1$ case
we do know the answer:  in this case there is at most a single finite open WKB network in each class,
and it contributes $\mu(\gamma_{ij}) = +1$.

\item {\bf Twisting function $\sigma$}:  We do not know the tricky $\pm 1$-valued twisting function
$\sigma$ in general.  In the $A_1$ case we do know it:  it is
\begin{align}\label{eq:Twisting-function}
 \sigma(\gamma, \gamma') &= (-1)^{\inprod{\gamma,\gamma'}}, \\
 \sigma(\gamma, \gamma_{ii}) &= (-1)^{\inprod{\gamma,\gamma_{ii}}}, \\
 \sigma(\gamma, \gamma_{ij}) &= \sigma(\gamma), \\
 \sigma(\gamma_{ij}, \gamma_{ji}) &= - \sigma(\gamma_{ij} + \gamma_{ji}),
\end{align}
where the $\sigma$ appearing on the \ti{right} side is the ``canonical quadratic refinement''
$\sigma: \Gamma \to \IZ_2$ which appeared in \S 7.7 of \cite{Gaiotto:2009hg}. In particular
recall that $\sigma(\gamma)=-1$ if $\gamma$ supports a hypermultiplet and $=+1$
if $\gamma$ supports a vectormultiplet. The above sign prescription is justified in
Appendix \ref{app:a1-signs}.

We will sometimes use the signs $\sigma(\gamma_{ij},\gamma_j)$ as well;
these involve gauge choices which we do not discuss here.

\item {\bf Supersymmetric interfaces}: In order to extend the wall-crossing
data to include interfaces we must describe a suitable class of such
interfaces. Suppose we pick two points $z$ and $z'$ of $C$.
We then have two surface defects $\bS = \bS_z$ and $\bS' = \bS_{z'}$
Interfaces $L$ between
$\bS$ and $\bS'$ correspond to paths $\wp$  in $C$ from $z$ to $z'$. One can justify this
by considering Janus configurations in the field theory, where $z$ is regarded as a
parameter of the surface defect. Alternatively, one can take a geometric engineering
viewpoint and construct the interface between $\bS_z$ and $\bS_{z'}$ by considering
three open M2 branes. The first ends on the surface $x^1=x^2=0$ with $x^0 \leq 0$
at $z\in C$. The third ends on the surface  $x^1=x^2=0$ with $x^0 \geq  0$ at $z'\in C$,
and the second ends at $x^1=x^2=x^0=0$ in $\IR^{1,3}$ times a path $\wp$ on $C$ from $z$
to $z'$. In a discussion analogous to that of \cite{Gaiotto:2010be}, \S 7.4,
the partial twisting of the theory along $C$ means that the defect only depends
on the homotopy class of $\wp$ (with fixed endpoints).  \footnote{One can also
introduce supersymmetric interfaces corresponding to the ``laminations'' discussed in
\cite{Gaiotto:2010be}, but we will leave that for another occasion.} The Hilbert
space $\CH_{\bS_z L_{\zeta,\wp}  \bS_{z'}}$ will be graded by a charge
lattice $\Gamma_{ij'}$, which will consist of a $\Gamma$ torsor of homology classes
of open paths from $x_i$ (above $z$)  to $x_{j'}$ (above $z'$).
Once again we take
\begin{equation}\label{eq:A1-Z-ijprime}
 Z_{\gamma_{ij'}} = \frac{1}{\pi} \int_{\gamma_{ij'}} \lambda.
\end{equation}

\item {\bf Framed BPS States}: These should be given by some analog of the
millipede construction of \S 10.8 of \cite{Gaiotto:2010be}, but we have
not developed the details here.
\end{enumerate}

Let us give one simple example of how the above geometric prescription is compatible
with the general wall-crossing formulae of \S \ref{subsec:Examples-2d-4d-wcf}.
In discussing wall-crossing we can vary both $u$ and $z$, the basepoint on $C$ of
the projection of the
open curves $\gamma_{ij}$ on $\Sigma$. When we vary $z$ we sometimes write $\gamma_{ij}(z)$
to emphasize the basepoint.  When we vary $z$, it can
cross finite WKB networks, and  this can change
$\langle \tilde\gamma, \gamma_{ij} \rangle$ thus inducing wall-crossing in
$\omega(\gamma, \gamma_{ij})$.  To illustrate this, consider a family of basepoints
$z$ following a path which crosses a finite WKB curve
connecting two branch points of type $(ij)$. This finite curve has
lift $\tilde \gamma$ with homology class $\gamma$.   By our rules $\Omega(\gamma)=1$ and $\sigma(\gamma)=-1$.
Suppose that the finite WKB curve
has phase $\vartheta_*$.  When $z=z_*$ is a point on the image of $\tilde \gamma$,
that is, when it is a point on the finite WKB network,  there are
two open finite WKB curves $\gamma_{ij}(z_*)$ and $\gamma_{ji}(z_*)$, such that
$\gamma_{ij}(z_*) + \gamma_{ji}(z_*)= \tilde\gamma$.
Now let $z_\pm$ be displaced above and below $\tilde\gamma$ as in
Figure \ref{fig:cross-finite-wkb}.
\insfig{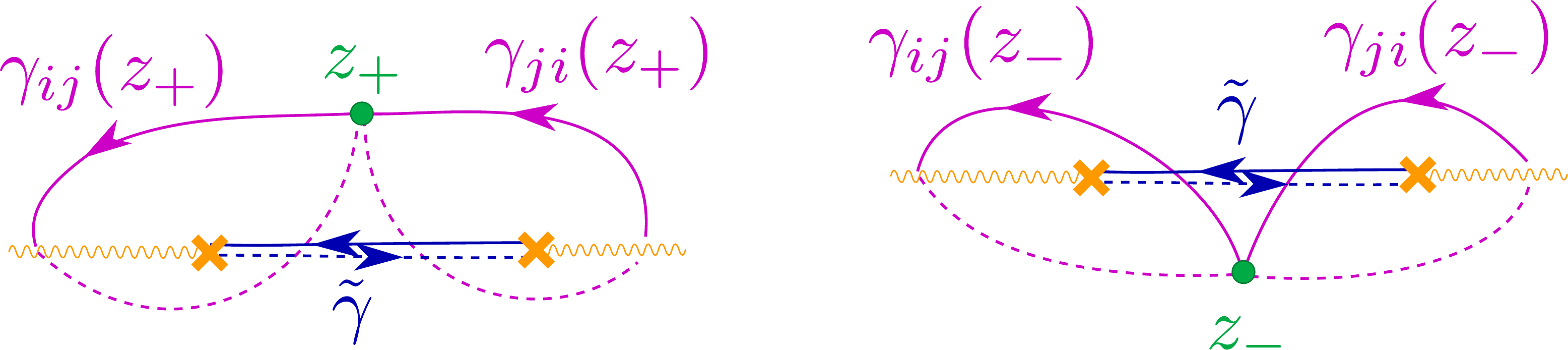}{$z$ crossing a finite WKB curve with lift $\tilde\gamma$.  When $z=z_+$,
the intersection $\langle \tilde\gamma,\gamma_{ij}(z_+) \rangle=+1$ (left).  When $z=z_-$,
$\langle \tilde\gamma,\gamma_{ij}(z_-)\rangle=-1$ (right).  This leads to a change of $\omega(\gamma, \gamma_{ij})$
as we pass from $z = z_+$ to $z = z_-$; this change is consistent with the 2d-4d wall-crossing formula,
as explained in the text.}
By continuity, the curves $\gamma_{ij}(z_\pm)$ and $\gamma_{ji}(z_\pm)$
continue to support finite open WKB curves with phase slightly displaced
from $\vartheta_*$. By our rules $\mu(\gamma_{ij}) = \mu(\gamma_{ji}) = 1$
and $\sigma(\gamma_{ij},\gamma_{ji})=1$, where $\gamma_{ij}$ applies to
any of the three values of $z=z_\pm, z_*$.
At $z=z_*$ the three central charges
$Z_{\gamma_{ij}(z_*)}$, $ Z_{\gamma_{ji}(z_*)}$, and $ Z_{\gamma}$ are aligned,
so as $z$ passes through the finite WKB network the BPS rays
of $\ell_{\gamma_{ij}}, \ell_\gamma, \ell_{\gamma_{ji}}$ change order,
exactly as in the wall-crossing formula \eqref{eq:example-4}.
Plugging $\Sigma_{ij}=\Sigma_{ji}=\Sigma_{ij}'=\Sigma_{ji}'=1$
into \eqref{eq:exple-4a} we find that the wall-crossing formula
demands that
\begin{equation}
\begin{split}
\Pi_i' & = \Pi_j \\
\Pi_j' & = \Pi_i \\
\Pi_i & = (1-X_\gamma) \Pi_j \\
\end{split}
\end{equation}
and hence, if primed quantities refer to the $z_+$ side
of the wall,
\begin{equation}
\begin{split}
\omega(\gamma, \gamma_{ij}(z_+) ) & = 1  \\
\omega(\gamma, \gamma_{ij}(z_-) )  & = -1  \\
\omega(m \gamma, \gamma_{ij}(z_\pm) ) & = 0 \qquad\qquad m>1 . \\
\end{split}
\end{equation}
These intersection numbers can be verified directly in Figure \ref{fig:cross-finite-wkb}.

One important general lesson we learn from the above example is that
\emph{the walls of marginal stability for the parameter $z$ of $\bS_z$
include the finite and separating WKB curves with phase $\vartheta  = \vartheta_*$,
where $\vartheta_*$ is the phase of some 4d
BPS state.}
 The reason  is that,
if $z$ lies on a finite or separating WKB curve of phase $\vartheta_*$,
there is at least one finite open network with phase $\vartheta_*$
(indeed this network just consists of a single finite open WKB curve from $z$ to a turning point).
Thus we have at least two BPS states with the same phase (one 2d, one 4d), which means $z$ is on a
wall of marginal stability. We will sometimes refer to the finite and separating WKB curves
at these special values $\vartheta_*$ as \emph{critical WKB curves}.

The details of what happens at the wall depend on whether $z$ is crossing a finite or a separating WKB
curve.  We showed above how $\omega$ can change as
$z$ crosses a finite WKB network, while $\mu$ is invariant.
On the other hand, as we will see in \S \ref{sec:Detailed-Expl},
when $z$ crosses a separating WKB curve $\mu$ can change,
while $\omega$ is invariant.  This corresponds to the
identity \eqref{eq:example-2} of type $\prod\CS\cdot \CK = \CK \cdot \prod\CS$.

\subsection{3d compactification and Hitchin's equations}\label{subsec:ClassS-3d}

Now, we consider the theory compactified on $S^1$ to 3 dimensions.

The moduli space $\CM$ of the theory on $\IR^{2,1} \times S^1_R$ is the moduli space of solutions
of Hitchin's equations \cite{Gaiotto:2009hg}.
Recall \cite{MR89a:32021} that these are equations for a
connection $A$ on a rank $n$ complex vector bundle $E \to C$
and a ``Higgs field'' $\varphi \in \Omega^{1,0}({\rm End}\ E)$:
\begin{equation}\label{eq:HitchinEqs-redux}
\begin{split}
F + R^2 [\varphi, \bar\varphi]  & = 0,\\
\bar\partial_A \varphi & = 0,\\
\partial_A \bar\varphi & = 0.
\end{split}
\end{equation}
Here the gauge group is $SU(n)$.\footnote{More
precisely $\CM$ is a certain finite quotient of the moduli space of solutions with gauge group $SU(n)$.  The precise quotient we pick depends
on some fine details of the 4d theory, like whether we take the gauge group to be simply connected or not; see  \S\S 6.3 and 7.3
of ref. \cite{Gaiotto:2010be} for some more discussion of this point.}
We impose the boundary conditions $(A, \varphi) = (A_0, \varphi_0) + ({\rm regular})$
near the punctures $z_a$ of $C$, where $(A_0, \varphi_0)$ is a suitable singular
solution near $z_a$.\footnote{See \S $4.1$ of \cite{Gaiotto:2009hg} for more details about the singular solution.}
These equations can be interpreted as the moment maps for a \hk quotient construction of $\CM$
as $\CM = \CN \hkq \CG$.  Here $\CN$ is the space of $(A,\varphi)$, not necessarily
satisfying \eqref{eq:HitchinEqs-redux}, but satisfying the boundary conditions.
$\CG$ is the group of gauge transformations which preserve the singular solutions
$(A_0, \varphi_0)$ near each point $z_a$.
Thus, at each puncture $z_a$ there is generically a reduction of structure group from $SU(n)$ to
the stabilizer $K_a\subset SU(n)$ of $(A_0,\varphi_0)$.

The map $\CM \to \CB$ is the ``Hitchin fibration,''
\begin{equation}
 \phi_k = \Tr \varphi^k.
\end{equation}
Each solution $(A, \varphi) \in \CM$
induces a family of flat connections parameterized by $\zeta \in \IC^\times$,
\begin{equation} \label{eq:hit-conn}
 \CA = R \frac{\varphi}{\zeta} + A + R \zeta \bar\varphi.
\end{equation}
Indeed Hitchin's equations can be interpreted as the statement that $\CA$ is flat for all $\zeta$.
For any fixed $\zeta \in \IC^\times$, \eqref{eq:hit-conn} gives an identification between
$\CM^\zeta$ and a moduli space of flat $SL(n,\IC)$ connections over $C$,
again with appropriate singularities at the punctures.

The family of solutions $(A, \varphi)$ can be packaged together into a
``universal'' solution to Hitchin's equations, living on a principal $PSU(n)$ bundle $\CU$ over $\CM \times C$.
Restricted to any $[(A,\varphi)]\times C$, $\CU$ is the principal bundle on which $(A, \varphi)$ is defined.
Restricting instead to the locus $\CM \times \{z\}\subset \CM \times C$ gives
a principal bundle $\CU_z$ over $\CM$, attached to the surface defect $\bS_z$.

A nice way to think about $\CU_z$ is to consider the ``pointed'' gauge group $\CG_z\subset \CG$ consisting of
gauge transformations which are the identity at $z$.  We can then split the \hk quotient construction of $\CM$
into two steps:  first take $\CN_z := \CN \hkq \CG_z$, and then
$\CM = \CN_z \hkq SU(n)$.  The center $\IZ_n \subset SU(n)$ acts trivially
on $\CN_z$, but the quotient $PSU(n) = SU(n)/ \IZ_n$ acts freely.  Thus
the zero locus of the \hk moment map for the $su(n)$-action on $\CN_z$
is a principal $PSU(n)$-bundle over $\CM$.  This principal bundle is $\CU_z$.  In particular, this way
of building $\CU_z$ makes it clear that it carries a canonical hyperholomorphic connection,
using general properties of the \hk quotient \cite{MR1139657}.

To define the ``vector bundle'' $V_{\bS_z}$, we would like to take the bundle associated
to $\CU_z$ by the $n$-dimensional representation of $SU(n)$.  There is a difficulty here:
$\CU_z$ is a $PSU(n)$ bundle, which does not have a canonical lift to an $SU(n)$ bundle.
(In fact, in similar situations it is known that it does not lift at all
\cite{MR2353678,Biswas1,Biswas2}.)
Therefore, $V_{\bS,z}$ only exists as a twisted bundle over $\CM$, twisted by a $B$-field
of order $n$. This obstruction is consistent with the fact that the surface defect
$\bS_z$ carries a $\IZ_n$ flavor monodromy, and hence, as we discuss in Appendix \ref{app:Flavor-Twisted},
$V_{\bS_z}$ should be twisted by a $\IZ_n$-valued $B$-field.

Because there is a reduction
of structure group of $E$ at $z_a$ to $K_a \subset SU(n)$, the fiber of $E$ must decompose into
irreducible representations of $K_a$. We write $E_a \simeq \oplus_s L^{(s)}_a$. In particular for $n=2$
we have
$E_a \simeq L_a \oplus L_a^{-1}$ for a complex line $L_a$. Now, the twisting of $V_{\bS_z}$
over $\CM$ is independent of $z$, and therefore ${\rm Hom}(V_{\bS_{z'}},V_{\bS_z})$ should
exist as an honest vector bundle over $\CM$. Therefore, again restricting to $n=2$,
the lines $L_a^{\pm 1}$ should define twisted line bundles over
$\CM$ with the same twisting as $V_{\bS_z}$.

\medskip

\textbf{Remarks}

\begin{enumerate}

\item The fact that
$\Hom(V_{\bS_z},V_{\bS_{z'}} )$ is a true bundle and not a twisted bundle has an
important physical interpretation. The
physical interpretation of sections of this bundle is related to
expectation values of supersymmetric interfaces. Indeed,
suppose we have a supersymmetric interface $L_{\wp}$ between surface defects
$\bS_z$ and $\bS_{z'}$. There will be
corresponding hyperholomorphic bundles $V$ and $V'$ over $\CM$.
In close analogy with \cite{Gaiotto:2010be}, equation $(7.13)$,  the expectation value
of the supersymmetric interface $\inprod{L_{\wp}} $ should be
identified with the parallel transport operator of $\CA$ along $\wp$ from   $z$ to $z'$
(or vice versa, depending on the orientation of $\wp$):
\begin{equation}\label{eq:Lvev-Ptrpt}
\inprod{L_{\wp}} = \Pexp \left(-\int_{\wp} \CA\right).
\end{equation}
Note that this equation only makes sense if we interpret $V \to \CM \times C$
as the universal bundle. Then both  the left hand side and the right hand side of this equation
can be understood as elements of the vector space ${\rm Hom}(V_{\CA,z}, V_{\CA,z'})$,
(when $\wp$ is oriented from $z$ to $z'$),
so it makes sense to equate them.

\item The (twisted) line bundles $L_a$ over $\CM$ also admit hyperholomorphic connections.
Indeed, we can take the limit $z\to z_a$ in the construction of \S \ref{subsec:Integral-Equations}.
Then $Z_{\gamma_{ij}(z)} \sim m_a \log(z-z_a) + {\rm regular}$ so that the semiflat coordinates
$\CY^\sf_{\gamma_i}$ will require renormalization.  These divergences suppress the
integral terms in \eqref{eq:int-2} (physically, the soliton masses go to infinity), and
hence there is no mixing between line bundles. The connection $\nabla$ on $L_a$ obtained
from the construction remains hyperholomorphic. We expect that the periods of $\frac{F(L_a, \nabla)}{2\pi}$
will be in $\frac{1}{n}\IZ$, reflecting the twisting, but we have not checked this.

\item The bundles $L_a$ should admit interpretations in terms of physical surface defects.
We leave this interesting question to the future.

\end{enumerate}

\subsection{Lagrangian descriptions of $A_1$ theories and their surface defects}\label{subsec:Lag-Desc}

At this point we restrict attention from the general $A_{n-1}$ theories to the $A_1$ case.
In this class of theories $\Sigma \to C$ is a $2$:$1$ covering. These theories have nice
Lagrangian descriptions and are amenable to the kind of WKB analysis
we discuss in \S \ref{subsec:WKB-2d-4d}.  We consider theories associated to
genus $g$ Riemann surfaces $C$, carrying $\ell$ regular punctures.
These theories have Lagrangian descriptions which
are in one-to-one correspondence with pair-of-pants decompositions of $C$; each Lagrangian description
involves $3g-3+\ell$ $SU(2)$ gauge groups corresponding to the tubes,
and $2g-2+\ell$ hypermultiplets in the $({\bf 2},{\bf 2},{\bf 2})$ associated to the trinions
\cite{Gaiotto:2009we}; call these hypermultiplets ``hypertrinions.''
By taking careful decoupling limits of these theories, one can produce other $A_1$ theories,
including asymptotically free Lagrangian theories but also non-Lagrangian, Argyres-Douglas-like theories.
These limiting theories correspond to Riemann surfaces $C$ carrying irregular punctures. See
\cite{Gaiotto:2009hg} for a detailed discussion.

Given the four-dimensional gauge theory description of the $A_1$ theories we can
define the surface defects $\bS_z$ as Gukov-Witten defects, at least in certain weak-coupling regimes.
An explicit example of this was discussed in \S \ref{subsubsec:ExampleSU2-CP1}.
More generally, $\bS_z$ can be defined as a Gukov-Witten defect for any of the $SU(2)$ factors
of the gauge group.
This description is valid when the $SU(2)$ in question is weakly coupled; in that case one should think of
$C$ as developing a long tube, and the point of $C$ labeling the surface defect as sitting somewhere
on that tube.  There is a natural coordinate $t$ on the tube; the position of the surface
defect in that coordinate system gives the (UV) Gukov-Witten parameter.
As the surface defect moves around $C$ from one long tube to another,
2d and 4d strong coupling effects mediate between different weakly coupled descriptions,
as illustrated in the discussion of \S \ref{subsubsec:ExampleSU2-CP1}. In that discussion
we noted that the 2d-4d instanton expansion breaks down due to light states when
$z$ hits a branch point. These light states are represented by the WKB curves from $z$ to the
nearby branch point.

We can now fill in a gap left open in our discussions of the twistings associated
with $\Gamma_i$ and $V_{\bS_z}$.  To do so we need to investigate the flavor symmetries
in this class of theories.  Associated to each puncture $z_a$ of $C$ is a flavor group $SU(2)_{a}$,
so the full flavor group is $\CF \simeq SU(2)^\ell$.  The mass parameter of each factor
$SU(2)_{a}$ is related to the residue $m_a$ of the Higgs field at $z_a$.
The flavor group $\CF$ does not act effectively on the hypertrinions modulo gauge
transformation.  Indeed, identifying the center $Z(\CF) \simeq \IZ_2^\ell$, any two of the $\IZ_2$
generators are equivalent up to a gauge transformation.\footnote{To
prove this, draw the trivalent graph representing the pants
decomposition of $C$.  The gauge transformation
by the nontrivial element of the center of an $SU(2)$ factor associated to an edge is equivalent
to an action by $-1$ on the two hypertrinions associated with the two vertices of that
edge. Now use the fact that the graph is connected.}
Therefore, letting $\CZ \subset \IZ_2^\ell$
be the index $2$ subgroup consisting of products of even numbers of generators, the
effective flavor group is $\bar \CF = \CF/\CZ$, which sits in a sequence
\begin{equation}
1 \rightarrow \IZ_2 \rightarrow \bar\CF  \rightarrow SO(3)^\ell \rightarrow 1.
\end{equation}
The distinguished $\IZ_2$ subgroup is the center $Z(\bar \CF)$, and can be
represented (in many ways) as acting by $-1$ on an odd number of
hypertrinions.  Now, the chiral operator in the 4d $\CN=2$ theory
which is not mutually local with respect
to $\bS_z$ is formed by taking the product of all the hypertrinions and contracting
gauge indices \cite{Gaiotto:2009gz}. Clearly this operator is projected out by
gauging the central
subgroup  $Z(\bar \CF)$  and hence we conclude that if we want the troublesome
(but desirable) surface defect $\bS_z$ then we must gauge $Z(\bar \CF)$
and the flavor monodromy carried by $\bS_z$ is the nontrivial element of
 this distinguished $\IZ_2$ subgroup.

Now, following the discussion in \S\ref{sec:flavor-phase},
if we gauge $Z(\bar \CF)$, then there will be
an extension of the group of gauge charges as in
\eqref{eq:lattice-discrete-ext-seq}, with $\hat\CD\simeq \IZ_2$.
We claim that this sequence, with $\CD=Z(\bar \CF)$, can be identified
with the sequence \eqref{eq:a-secq-fin} given in Example \ref{subsubsec:Anomaly Cancelation},
where $z$ can be taken to be any of the $z_a$.
Therefore, according to our discussion of that example, this UV flavor
symmetry allows us to cancel the anomalies in Aharonov-Bohm phases arising from
splitting the $\Gamma_{ij}$.

It remains to compare the twisting of $V_{\bS_z}$ arising from the fractional
shifts of $\Gamma_{ij}$, as described in \S \ref{subsubsec:Twsted-Mirror},
with that derived from the description in terms of \hk\ quotients in
\S \ref{subsec:ClassS-3d}.
Consistency of our description of \tf\ requires that they be the same, but
we lack a direct proof. Two ways to approach the problem would be

\begin{enumerate}

\item Consider the $R\to \infty$ limit. Then the connection becomes
diagonal, but a $\IZ_n$ gerbe cannot depend on $R$, so the identification
should be established in the semiflat geometry. It might be possible to
combine the $R\to \infty$ limit with the $z\to z_a$ limit and thereby
relate the $V_i$ (in the $z\to z_a$ limit) with the $L_a$.

\item One could also relate the curvature $F(L_a,\nabla)$ to the
$B$-field associated with the flavor symmetry $SU(2)_a$ as in
Appendix \ref{app:Flavor-Twisted}. Note that there are three
real mass parameters associated with $SU(2)_a$: these are the
complex residue $m_a$ of the Higgs field and the real flavor Wilson line
parameter $m_{a,3}$.  In complex structure $\zeta=0$, $m_{a,3}$ is
a K\"ahler parameter and $m_a$ a complex structure parameter, so
$\omega_I$ is affine-linear in $m_{a,3}$, while $\omega_J + \I \omega_K$
are affine-linear and holomorphic in $m_a$. In real coordinates,
with a proper normalization of $m_{a,i}$, we have
\begin{equation}
\frac{\p \omega_i}{\p m_{a,j}} = \delta_{ij} F(L_a, \nabla).
\end{equation}
On the other hand, according to Appendix \ref{app:Flavor-Twisted},
$\frac{\p \omega_I}{\p m_{a,3}}$ is the $B$-field associated with
the flavor symmetry $SU(2)_a$. With this link between the twisting of $L_a$
and the flavor symmetries established one might be able to
prove the desired identity of twistings.

\end{enumerate}

\subsection{A brief review of the WKB analysis and its extension to 2d-4d}\label{subsec:WKB-2d-4d}

According to our general discussion in \S \ref{sec:x-gamma-interp}, the parallel transport operators
$\inprod{L_\wp}$ of \eqref{eq:Lvev-Ptrpt} admit an expansion \eqref{eq:lz-exp}
in terms of canonically defined sections $\CY_{\gamma_{ij'}}$ of $\Hom(V, V')$.
The coefficients of this expansion are the framed 2d-4d BPS degeneracies.
The wall-crossing properties of these framed degeneracies
contain the combinatorial data of the BPS degeneracies $\mu$ and $\omega$.

For $A_1$ theories and their canonical surface defects, we can get our hands on the $\CY_{\gamma_{ij'}}$
very explicitly.  The main tool is a slight amplification of the
WKB analysis of \cite{Gaiotto:2009hg}, which we used in the pure 4d case to give a formula for the functions
$\CY_\gamma$.  Its extension similarly gives a formula for the sections
$\CY_{\gamma_{ij'}}$.  It also gives the expansion \eqref{eq:lz-exp} of the $\inprod{L}$
(at $y=-1$)
in terms of framed 2d-4d BPS degeneracies (generalizing \cite{Gaiotto:2010be}).
One can then verify directly that the wall-crossing of these framed degeneracies is indeed governed by
the bulk 2d-4d BPS degeneracies $\mu$ and $\omega$ which we described above.
In the rest of this section we will sketch how to construct the $\CY_{\gamma_{ij'}}$, assuming a certain degree of familiarity with the properties of $A_1$ Hitchin systems and with the results of \cite{Gaiotto:2009hg}.
We will leave a more detailed analysis, possible extension to other surface defects and to higher
rank theories to a future publication.

\subsubsection{4d review}

The main player in our WKB analysis is the family of flat connections $\CA$ on $C$
given in \eqref{eq:hit-conn}.
We have specialized to $A_1$ theories, so
$\CA$ is a flat $sl(2,\IC)$-valued connection on $C$, with singularities
at various points of $C$, which we call ``regular punctures'' or ``irregular punctures'' depending
on the nature of the singularities.  We assume that there is at least one puncture.

Given a point $u \in \CB$ and a choice of phase
$\vartheta \in \IR / 2 \pi \IZ$, in \cite{Gaiotto:2009hg} we defined the ``WKB foliation'' $F_\WKB(u, \vartheta)$.
This foliation in particular determines a decomposition of $C$ into a finite collection of simply
connected cells $C_{ab}$.  Here the indices $a$ and $b$ are labeling the possible asymptotic ends for
generic WKB curves:  these are either regular punctures
or Stokes sectors around irregular punctures.  Each cell $C_{ab}$ is a union of WKB curves of
phase $\vartheta$ which run from $a$ to $b$.  ($a=b$ is allowed, giving a
``degenerate cell.'') See Figure 25 in \cite{Gaiotto:2009hg}.

The WKB analysis of \cite{Gaiotto:2009hg} relied strongly on the concept of
a \emph{small flat section} associated with a singularity $z_a$ on $C$.
This will be important in the 2d-4d extension also, so let us review the
definition. Given a branch $\lambda$ of the covering $\Sigma \to C$,
WKB curves have a standard orientation, determined by $e^{-\I \vartheta } \langle \lambda, \p_t \rangle >0$
where $\p_t$ is a positively oriented tangent vector.  Near a singularity
$z_a$, choose the branch so that the WKB curve with its standard orientation flows
\emph{into} $z_a$. There then exists a flat section, $(\de + \CA) s_a^\vartheta(z)=0$,
such that $s_a^\vartheta(z(t))$ becomes exponentially small as $t\to \infty$ and $z(t) \to z_a$.
(Thus, $s_a^\vartheta$ also grows as $z$ moves away from $z_a$ along a WKB curve.)
Moreover, this condition determines the section $s_a^\vartheta$ uniquely up to
an overall scale. Such a small flat section will be variously denoted as
$s_a^\vartheta(  z;u, \zeta)$,  $s_a^\vartheta(z)$, or just $s_a^\vartheta$.
Our main constructions involving $s_a^\vartheta$ do not require us to choose
the scale. Now suppose that $\zeta$ is in the half-plane
 $\IH_\vartheta$ centered on the ray $\zeta \in e^{\I \vartheta} \IR_+$.
 Then, the WKB formula says that for the small flat section the expression
 \begin{equation}
 s_a^\vartheta(z) \exp \left( \frac{R}{\zeta} \int^z_{z_a(\epsilon)} \lambda\right),
 \end{equation}
 where the integral is along the curve with  $e^{-\I \vartheta } \langle \lambda, \p_t \rangle \in \IR $
 from $z_a(\epsilon)$ to $z$,
has a finite limit as $\zeta \to 0$ in the half-plane $\IH_\vartheta$.
Here $z_a(\epsilon)$ is infinitesimally close to the singular point $z_a$.

One of the main results of \cite{Gaiotto:2009hg} was a direct construction
of the ``Darboux functions'' $\CY_{\gamma}$ given the data of a WKB triangulation
and its small flat sections. A somewhat streamlined version of the construction
can be summarized as follows. Given $\vartheta$ we construct
antisymmetric inner products $T^\vartheta_{ab} := (s^\vartheta_a, s^\vartheta_b)$ of the sections
associated to a cell $C_{ab}$.  Here the inner product of two sections $(s,s')$ is defined
to be $\frac{s \wedge s'}{\vol}$
where $\vol$ is the $SL(2,\IC)$ invariant ``volume'' form. We stress that it is $z$-independent.
The asymptotics of $s_a^\vartheta$ imply similar asymptotics for $T^\vartheta_{ab}$,
controlled by $\int_{z_b(\epsilon)}^{z_a(\epsilon)} \lambda$. More precisely,
let $P_a^\pm$ be the lifts of the singular points $z_a$ on $\Sigma$.\footnote{The
notation presupposes regular singularities, but the
generalization to irregular singularities is clear.} Then,
for each ordered pair $a,b$ there is a unique oriented lift
$\hat E_{ab}\subset \Sigma - \{P_a^\pm \} $  of the unoriented edge $e_{ab}\subset C$
which is odd under the deck transformation and satisfies
\begin{equation}\label{eq:Tab-asym}
T_{ab}^\vartheta  e^{ - \frac{R}{\zeta} \oint_{\hat E_{ab}} \lambda} \sim {\rm finite}
\end{equation}
as $\zeta \to 0$ in the half-plane $\IH_{\vartheta}$.
(Once again we need to use the $\epsilon$ regularization.)    Note that since
$T_{ab}^\vartheta  = - T_{ba}^\vartheta $ we have $\hat E_{ab} = \hat E_{ba}$.

Now, let $n_{ab}$ be any
matrix of integers such that the diagonal elements vanish and for every $a$,
\begin{equation}\label{eq:zerosum}
\sum_{b} (n_{ab}+n_{ba})=0.
\end{equation}
To such a matrix we associate
\begin{equation}\label{eq:xenn-def}
\CX_n := \prod_{a,b} \left( T_{ab}^\vartheta\right)^{n_{ab}}.
\end{equation}
The condition \eqref{eq:zerosum} guarantees that each small flat section $s_a$
enters the product $\CX_n$ as many times in the numerator and denominator
and hence is independent
of the normalization of  $s_a$.\footnote{Combinations $\prod T_{ab}^{n_{ab}}$ have been considered many times in the literature --- in particular the ``Fock-Goncharov coordinates'' \cite{MR2233852}
are of the form $\frac{T_{ab} T_{cd}}{T_{bc} T_{da}}$ where $a,b,c,d$ are the
four vertices of some quadrilateral in an ideal triangulation of $C$.}
Thus $\CX_n$ are well-defined on $\CM$.
Moreover, by \eqref{eq:Tab-asym} they have $\zeta\to 0$ asymptotics
of the form
\begin{equation}
\CX_n \sim \CN_n \exp\left( \frac{R}{\zeta}\oint_{\gamma(n)}\lambda \right)
\end{equation}
where $\CN_{n}$ has a well-defined limit for $\zeta\to 0$
and   $\gamma(n) = \sum_{ab} n_{ab} \hat E_{ab}$. The expression $\gamma(n)$
defines, \emph{a priori},
a class in the relative homology $H_1(\Sigma, \{ P_a^\pm \};\IZ)^-$.
However, thanks to \eqref{eq:zerosum} that class is in the image of $H_1(\Sigma;\IZ)^-$.
Moreover the map
$n \to \gamma(n)$ is linear, onto, and has kernel consisting of antisymmetric matrices.\footnote{To
prove this we use the dual pairing of $\hat E_{ab}$ with
the cycles $\gamma_E$ associated to the edges of the WKB triangulation
$T_{\WKB}(\vartheta)$. See \S 7.1.1 of \cite{Gaiotto:2009hg}.}
Finally, $\CY_{\gamma}^\vartheta$ will be defined by
\begin{equation}\label{eq:cy-gamm-def}
\CY_{\gamma}^\vartheta := \varepsilon(\gamma,n) \CX_n,
\end{equation}
where $n$ on the right side obeys $\gamma=\gamma(n)$,
and $\varepsilon(\gamma,n) = \pm 1$.
Since the map $n\to \gamma(n)$ has a kernel, it is not immediately obvious that this ``definition'' is well
defined.  Fortunately, if
$k_{ab}=-k_{ba}$ is antisymmetric then $\CX_{n+k} = (-1)^{\sum_{a<b}k_{ab}} \CX_{n}$,
so \eqref{eq:cy-gamm-def} is indeed well defined provided we take $\varepsilon(\gamma,n+k) = (-1)^{\sum_{a<b}k_{ab}} \varepsilon(\gamma,k)$.
To fix $\varepsilon$ completely, we need a little more input.  In \cite{Gaiotto:2009hg} we fixed it
by determining it on a basis of $\Gamma$ and then using the multiplicative law of the $\CY_\gamma$.
This is briefly reviewed in Appendix \ref{app:a1-signs}.

Finally, the Darboux functions are defined by specializing $\vartheta$,
$\CY_\gamma := \CY_\gamma^{\vartheta = \arg \zeta}$.

There is a well-known algorithm \cite{dts,MR2233852,MR2349682,Teschner:2005bz},
reviewed in our context in
\cite{Gaiotto:2009hg,Gaiotto:2010be}, for decomposing the trace of the parallel transport
around a loop $\wp$ as a linear combination of the $\CY_\gamma$. Physically this gives
the ``Darboux expansion'' \eqref{eq:DarbouxExpansion} of the line defect vev $\langle L_{\wp} \rangle$
in terms of the ``IR line defect vevs,'' as explained at length in \cite{Gaiotto:2010be}.

As we have recalled in \S\S \ref{sec:lineops} and \ref{sec:discont},
the functions $\CY_\gamma$ are supposed to have discontinuities when $(u, \zeta)$ meet
any ``BPS wall'' $W_\gamma$, defined in \eqref{eq:4d-wall},
with the jump across $W_\gamma$ identified with a transformation $\CK_\gamma^{\Omega(\gamma)}$.
This jump reflects the effect of 4d BPS particles of charge $\gamma$.
In the $A_1$ theories, these discontinuities arise
from jumps in the topology of the WKB foliation $F_\WKB(u,\vartheta = \arg \zeta)$ as we vary $(u, \zeta)$.
The topological content of this foliation is captured by the ``decorated WKB triangulation''
$T_\WKB(u, \vartheta)$.  The vertices of $T_\WKB(u, \vartheta)$ are the asymptotic ends $a$, decorated by
the sections $s_a$.  The edges of $T_\WKB(u, \vartheta)$ are in 1-1 correspondence with the cells $C_{ab}$.
The 4d spectrum of the $A_1$ theory follows directly from the combinatorics of
how $T_\WKB(u, \vartheta)$ evolves as $\vartheta$ is varied.

More precisely, as $\zeta$ varies, $T_\WKB(u, \vartheta = \arg \zeta)$ may jump in three different ways,
which we call the
\ti{flip}, \ti{juggle} and \ti{pop}:

\begin{itemize}
 \item A flip is a very simple change: we remove one edge $E$, leaving behind a quadrilateral $Q_E$, and put back another edge along the
other diagonal of $Q_E$.  This jump of $T_\WKB$ leads to a corresponding transformation of the $\CY_\gamma$:
it is $\CK_{\gamma_E}$,
where $\gamma_E$
is the lift to $\Sigma$ of a cycle running around the pair of turning points inside the quadrilateral.
See \S 7.1 of \cite{Gaiotto:2009hg} for further details.
The occurrence of a flip is signaled by the presence
of a finite WKB curve (necessarily of phase $\vartheta = \arg -Z_{\gamma_E}$) joining the two turning points in $Q_E$;
if such a finite WKB curve exists, then $T_\WKB$ undergoes a flip of edge $E$, at the BPS ray $Z_{\gamma_E}/\zeta \in \IR_-$.
This flip corresponds to a BPS hypermultiplet of charge $\gamma_E$, i.e. to $\Omega(\gamma_E) = 1$.
\insfig{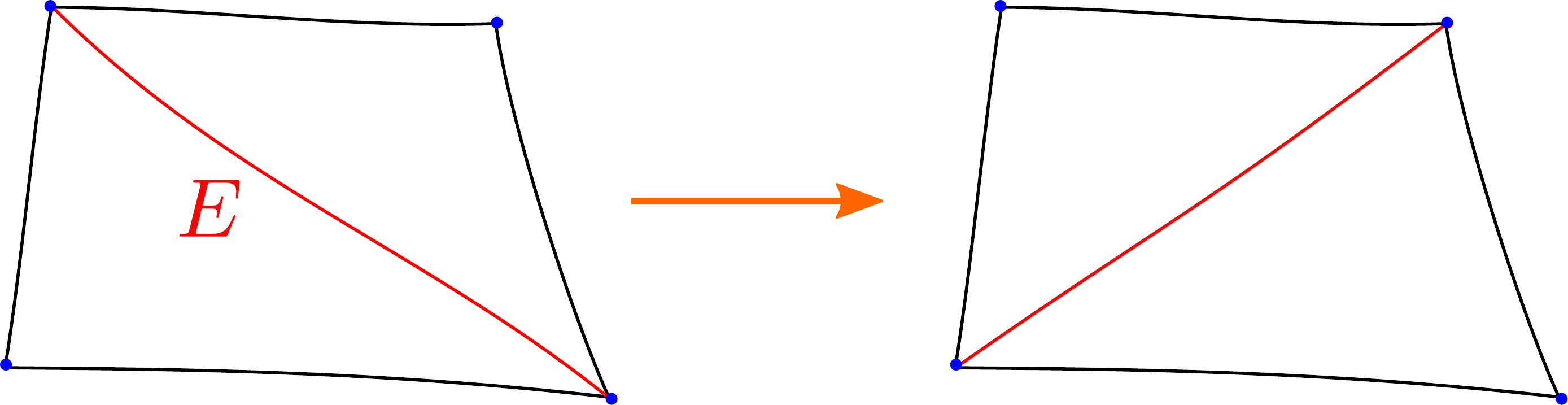}{The effect of a flip on $T_\WKB$.}

\item A pop replaces the chosen small flat section $s_a$ at some regular singularity $a$ with the opposite monodromy eigenvector
$\tilde s_a$. When $T_\WKB$ undergoes a pop, the functions $\CY_\gamma$ are not changed, but for a rather subtle reason:
the functions $T_{ab}$ change,
but the recipe for building $\CY_\gamma$ from the $T_{ab}$ also changes,
in just such a way that the $\CY_\gamma$ are left invariant.
(See \S 7.6.3 of  \cite{Gaiotto:2009hg}.)
A pop occurs at the BPS ray corresponding to a pure flavor
central charge $\gamma^f_a$, $Z_{\gamma^f_a}/\zeta \in \IR_-$.
The triangulation $T_\WKB(u, \zeta)$ for $(u, \zeta)$ near this BPS ray is always \ti{degenerate}:  it has only a single edge
coming out of the vertex $a$.
The pop is signaled by the appearance of a closed WKB curve surrounding the singularity $z_a$.
(The cycle $\gamma^f_a$ is an odd lift of this closed WKB curve.)
In the language of the 4d theory, this closed WKB curve corresponds to a BPS state carrying only flavor charges.

\item A juggle is a subtler transformation:  it does not relate two triangulations in the ordinary sense,
but rather two \ti{limits} of triangulations, each obtained from an infinite sequence of flips of the
common edges of two triangles which form an annulus on $C$.
The effect of the juggle on the limiting coordinates $\CY_\gamma$ is a $\CK$-transformation $\CK_{\gamma_0}^{-2}$,
where $\gamma_0$ is an odd lift of a closed loop going around the annulus once.
This transformation appears sandwiched between the two infinite sequences of transformations coming from the flips.
The juggle happens at the BPS ray $Z_{\gamma_0} / \zeta \in \IR_-$.
It is signaled by the appearance of a family of closed WKB curves running around the annulus, corresponding
to a BPS vector multiplet of charge $\gamma_0$.
\insfig{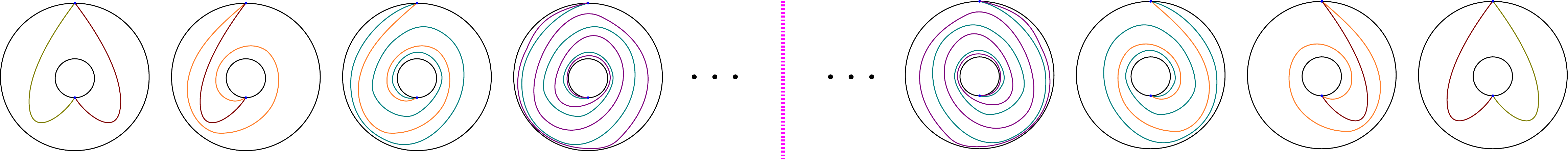}{A juggle arises as the limit of an infinite sequence of flips.}

\end{itemize}

\subsubsection{Extension to 2d-4d}

Now let us construct the Darboux sections $\CY_{\gamma_{ij'}}$ for the
2d-4d case. These have been defined in \S \ref{sec:compactification} in a way
analogous to the definition of the $\CY_\gamma$ in \cite{Gaiotto:2008cd}.
However, as in \cite{Gaiotto:2009hg}, for $A_1$ theories there should also be
a direct expression in terms of small flat sections, analogous to \eqref{eq:cy-gamm-def}.
Now we should have  $\CY^\vartheta_{\gamma_{ij'}} \in {\rm Hom}(V_{\bS_z}, V_{\bS_{z'}})$
where one end of $\gamma_{ij'}$, $x_i$, lies above $z$ and the other end, $x_{j'}$, lies above $z'$.
\footnote{Unfortunately, we have reversed conventions here relative to
\S \ref{subsec:Integral-Equations}, where we would have taken $\CY_{\gamma_{ij'}}$ to
be in $ {\rm Hom}(V_{\bS_{z'}},V_{\bS_z})$.}

The $\CY^\vartheta_{\gamma_{ij'}}$ should be piecewise holomorphic in $z,z',u,\zeta$
with wall-crossing discontinuities discussed above and moreover should satisfy
reality constraints and, importantly, the asymptotic condition that
\begin{equation}
\CY^\vartheta_{\gamma_{ij'}}   e^{-\frac{R}{\zeta} \int_{\gamma_{ij'}} \lambda }
\end{equation}
has a finite limit as $\zeta \to 0$ in $\IH_\vartheta$.
Finally, these sections should satisfy
\begin{equation}
\CY_{\gamma_{ij'}} \CY_{\gamma_{j'k''}} = \sigma(\gamma_{ij'}, \gamma_{j'k''}) \CY_{\gamma_{ij'}+
 \gamma_{j'k''}}
 \end{equation}
for the twisting function defined in \eqref{eq:Twisting-function}.

In analogy to the $\CX_n$, we again take $n_{ab}$ to be a matrix of integers with
zero on the diagonal. Then we define the
analog of \eqref{eq:xenn-def}:
\begin{equation}\label{eq:Chi-ij-ess}
\CX_{n,c,d}^{\vartheta}(z ,z') := \prod_{a,b} (T^\vartheta_{ab})^{n_{ab}} s^\vartheta_c(z) \otimes s^\vartheta_d(z').
\end{equation}
Here $z$ lies in some cell $C_{a_0,b_0}$ and $c$ is one of $a_0$ or $ b_0$, while $z'$ lies in
some cell $C_{a_0',b_0'}$ and $d$ is one of $a_0'$ or $b_0'$.
Here and in several places below, we
use the $SL(2,\IC)$-invariant volume form to identify $V$ with $V^*$.
To be explicit, $s_a^\vartheta(z) \otimes s_b^\vartheta(z')  \in {\rm Hom}(V_{\bS_z},V_{\bS_{z'}})$ is
defined by
\begin{equation}\label{eq:complaw}
s_a^\vartheta(z) \otimes s_b^\vartheta(z'): v \mapsto (v,s_a^\vartheta) s_b^\vartheta(z')
\end{equation}
and thus \eqref{eq:Chi-ij-ess} indeed determines a section of $\Hom(V_{\bS_z}, V_{\bS_{z'}} )$.
It will be independent of the choice of normalization of the $s_a^\vartheta$
provided that $n_{ab}$ satisfies the analog
of \eqref{eq:zerosum},
\begin{equation}
\sum_{b} (n_{ab} + n_{ba} ) = \begin{cases} -1 & a=c,d, \\ 0 & {\rm otherwise.} \\ \end{cases}
\end{equation}
Now we need the analog of the map $n \to \gamma(n)$ we found in the case
when $\gamma$ is closed.  As before
this will be determined by $\zeta \to 0$ asymptotics. As we have explained, to
a singularity $z_a$ we can associate a branch $\lambda$, such that WKB curves
with their standard orientation flow into $z_a$.  Given $z \in C_{a,b}$, let $x_{a,z} \in \Sigma$
be the lift of $z$ to this branch.  Then there is a lift
of the WKB curve between $z_a$ and $z$ to an oriented path $\hat E_{a, x_{a,z}}$ between a lift of $z_a$ and $x_{a,z}$,
such that
\begin{equation}
s_a^\vartheta(z) \exp\left( -\frac{R}{\zeta} \int_{\hat E_{a, x_{a,z}}} \lambda \right)
\end{equation}
is finite as $\zeta \to 0$ in $\IH_\vartheta$.  It follows that
\begin{equation}
\CX_{n,c,d}^{\vartheta}(z ,z') \exp\left( -\frac{R}{\zeta} \int_{\gamma_{ij'}(n,c,d)} \lambda \right)
\end{equation}
is finite for $\zeta \to 0$ in $\IH_\vartheta$, where we have defined\footnote{We use $\sigma$ for both the deck transformation and the twisting function.  We hope these will not be confused.}
\begin{equation}
\gamma_{ij'}(n,c,d):= -\sigma^*(\hat E_{c,x_{c,z}}) + \sum_{a,b} n_{ab} \hat E_{ab} + \hat E_{d,x_{d,z'}}.
\end{equation}
As with the closed curves, we can now define $\CY^\vartheta_{\gamma_{ij'}}$ by
\begin{equation}
\CY^\vartheta_{\gamma_{ij'}} := \varepsilon(\gamma_{ij'}, n) \CX_{n,c,d}^\vartheta(z ,z')
\end{equation}
where $\gamma_{ij'} = \gamma_{ij'}(n,c,d)$.  As before, $\varepsilon(\gamma_{ij'}, n)$ is a slightly
delicate sign; a systematic way of fixing it is given in Appendix \ref{app:a1-signs}.

Finally, the $\CY_{\gamma_{ij'}}^{\vartheta = \arg \zeta}$ are to be identified with the
$\CY_{\gamma_{ij'}} \in \Hom(V_{\bS_z}, V_{\bS_{z'}})$ which we are after.

\insfig{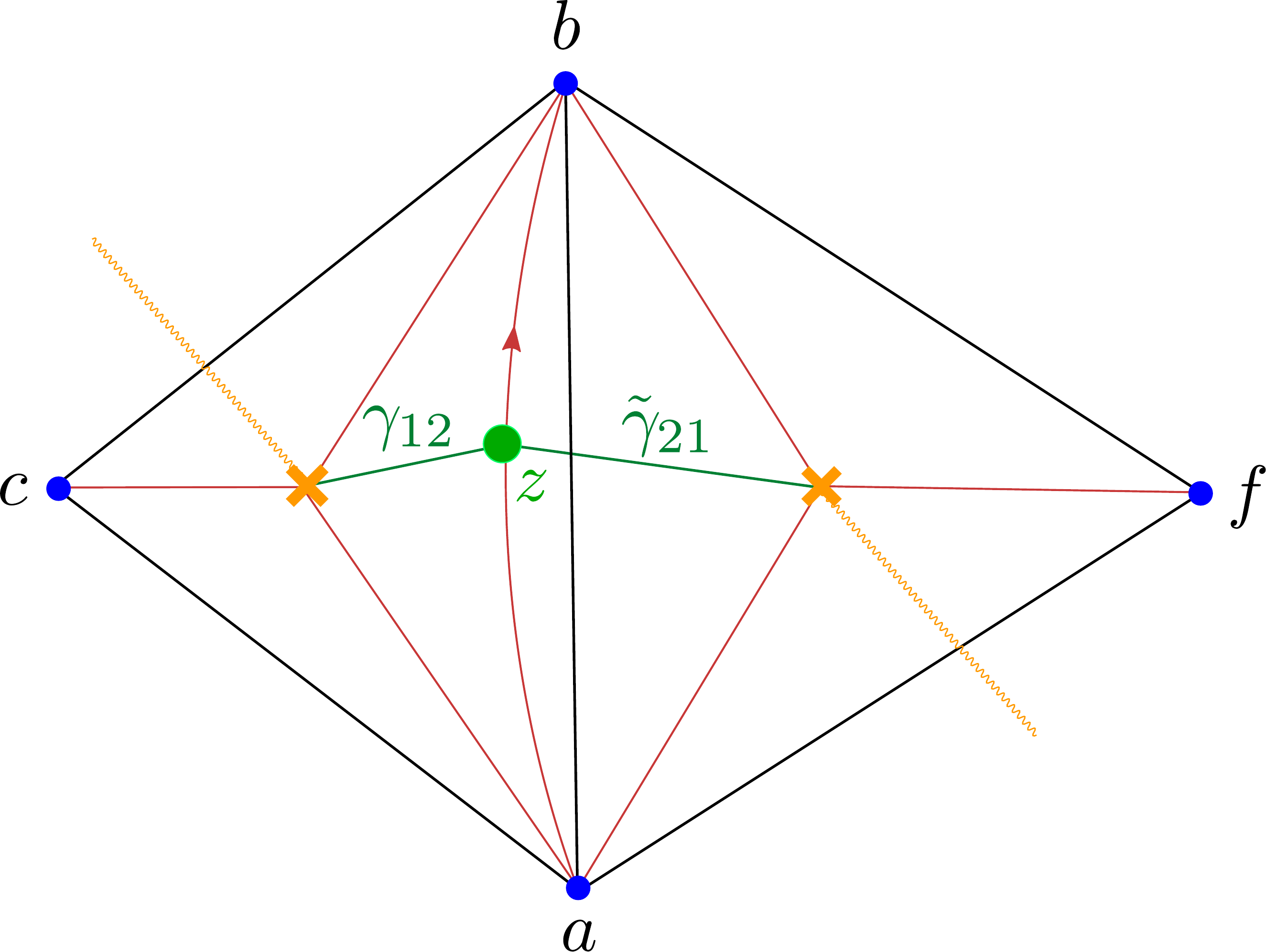}{$z$ is in the cell $C_{ab}$ and also in the triangle $abc$, with $abc$ in
counterclockwise order. The first sheet is such that with orientation
$e^{-\I \vartheta} \langle \lambda, \p_t \rangle >0$ the WKB curves flow from $a$ to
$b$, counterclockwise on the triangle.
The green curves are the projections of open paths $\gamma_{12}$
and $\tilde \gamma_{21}$ connecting sheets $1$
and $2$ above $z$ so that $\gamma_{12}+\tilde\gamma_{21}=-\gamma_E$.}

\textbf{Example}.  Suppose $z$ is in the cell $C_{ab}$ as in Figure \ref{fig:simple-y},
and we define sheet $1$ to be the root $\lambda$ so that the WKB curve with
$e^{-\I \vartheta} \langle \lambda, \p_t \rangle >0$
through $z$ is oriented to flow counterclockwise around the triangle, i.e. from $a$
to $b$. Then $x_{b,z}$ lies on sheet $1$ and $x_{a,z}$ lies on sheet $2$.
Now consider $\gamma_{11} = 0 \in \Gamma_{11}$. This is associated with
\begin{equation}\label{eq:cy11}
\CY_{\gamma_{11}=0} = \frac{ s_{a}(z) \otimes s_b(z) }{(s_b,s_a)},
\end{equation}
while  $\gamma_{22} = 0 \in \Gamma_{22}$ is given by
\begin{equation}\label{eq:cy22}
\CY_{\gamma_{22}=0} =  \frac{s_b(z) \otimes s_{a}(z)}{(s_a,s_b)}.
\end{equation}
Now consider the simple path $\gamma_{12}(z)$ from sheet $1$ to sheet $2$ and
projecting to a simple curve from $z$ to the turning point in the triangle $abc$, as in Figure \ref{fig:simple-y}.
Then
\begin{equation}\label{eq:cy12}
\CY_{\gamma_{12}} = \frac{(s_a,s_c)}{(s_b,s_a)(s_b,s_c)} s_{b}(z) \otimes s_{b}(z).
\end{equation}
As a check, if we compose with a similar expression for the open curve $\tilde\gamma_{21}$
whose projection is shown in Figure \ref{fig:simple-y}, so that $\gamma_{12}+ \tilde\gamma_{21} = - \gamma_E$,
we find indeed that
\begin{equation}
\begin{split}
\CY_{\gamma_{12}} \CY_{\tilde\gamma_{21}} & = \frac{(s_f,s_b)(s_a,s_c)}{(s_a,s_f)(s_b,s_c)} \cdot
\frac{s_{a}(z)\otimes s_b(z)}{(s_b,s_a)} \\
& = - \CY_{-\gamma} \CY_{\gamma_{11} }, \\
\end{split}
\end{equation}
where $\gamma_{11}=0\in \Gamma_{1}$. This is summarized in the equation
$\CY_{\gamma_{12}} \CY_{\tilde\gamma_{21}}= \sigma(\gamma_{12},\tilde\gamma_{21}) \CY_{\gamma_{12}+\tilde
\gamma_{21}}$.

\insfig{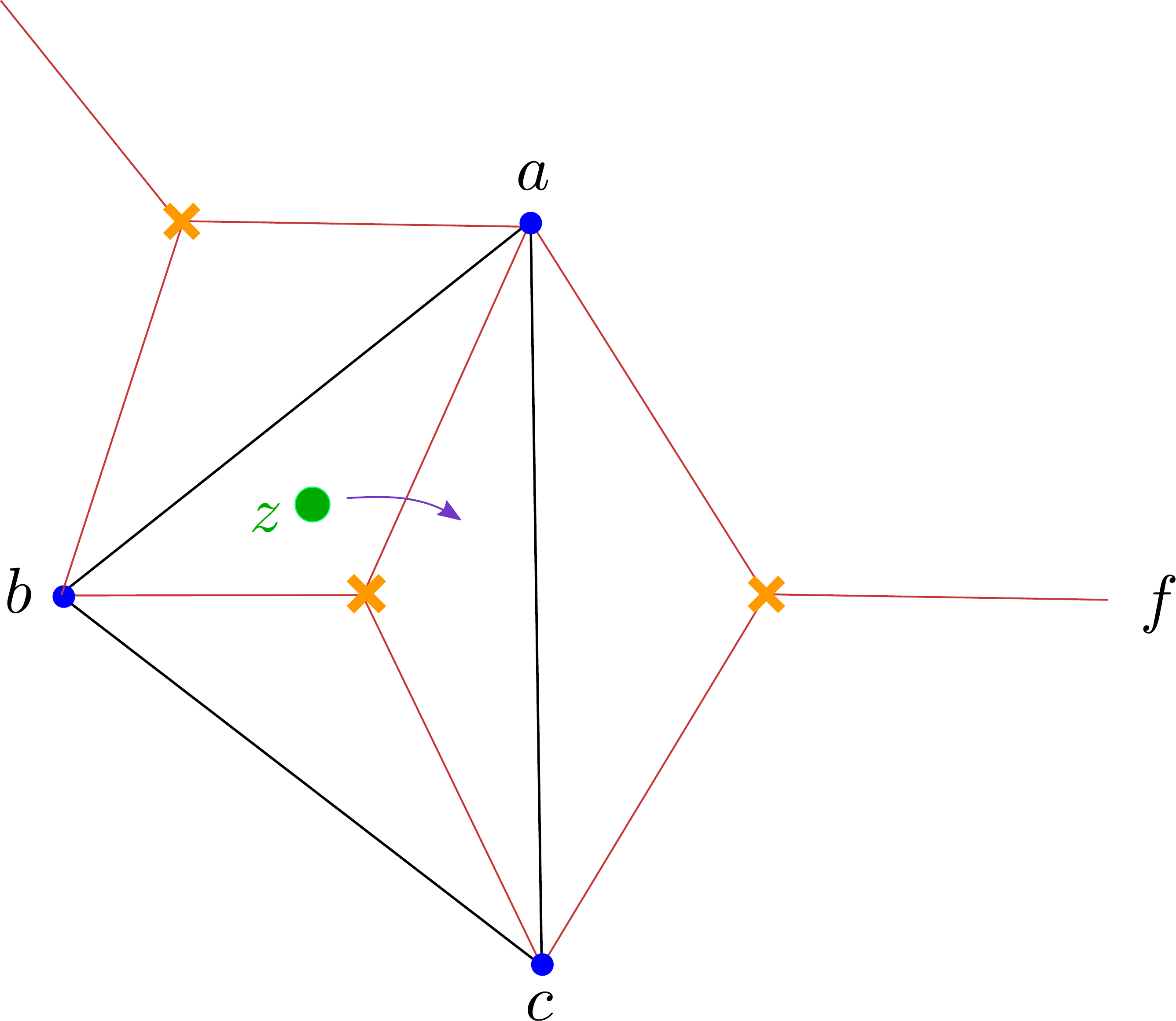}{The parameter $z$ of a surface defect is changed so that it moves
from a cell $C_{ab}$ to a cell $C_{ac}$ across a separating WKB curve. Blue dots are singularities,
and orange crosses are branch points, aka turning points.
A similar picture applies if $z$ is held fixed but $(u,\zeta)$ are changed so that a separating
WKB curve sweeps across $z$.}

To complete the demonstration that the $\CY_{\gamma_{ij'}}$ are correctly defined, we
must demonstrate how the full 2d-4d BPS spectrum is captured by the discontinuities of
$\CY_{\gamma_{ij'}}$.  In this case the combinatorial data that goes into $\CY_{\gamma_{ij'}}$ is
slightly more than what we had in the pure 4d case:  we need the WKB triangulation $T_\WKB$ \ti{plus} the
information of to which cell $C_{ab}$ the points $z$, $z'$ belong.  As we vary $\vartheta = \arg \zeta$ these combinatorial data
might jump, and we ask how these jumps induce jumps in $\CY_{\gamma_{ij'}}$.  We have to consider flips, pops and juggles
of $T_\WKB$ as before, as well as the new possibility that $z$ or $z'$ moves from one cell to another:

\begin{itemize}
\item Suppose $z$ moves from a cell $C_{ab}$ to a cell $C_{ac}$ as in Figure \ref{fig:z-crossing-cell}.  In this case
we can use the simple identity
$s_a T_{bc} + s_b T_{ca} + s_c T_{ab} = 0$ to relate the $\CY^{\vartheta}_{\gamma_{ij'}}$
before and after the jump.  The resulting transformation turns out to be
an $\CS$-factor, with $\mu_{\gamma_{ij}} =  1$,
where $i$ and $j$ refer to the two preimages of $z$,
and $\gamma_{ij}$ is the path which starts from one preimage of $z$, runs to the branch
point in the middle of the triangle $(abc)$, and comes back to the other preimage of $z$.
The detailed demonstration of this can be found in Appendix \ref{app:a1-signs}.
If we hold $z$ fixed and let $\vartheta$ vary, such a jump will occur
whenever there is a WKB curve of phase $\vartheta$
which connects $z$ to the branch point.  Such a finite WKB curve represents
a 2d soliton with charge $\gamma_{ij}$.
A very similar story
also holds with $z$ replaced by $z'$.

\item A flip of $T_\WKB$ occurs when, as $\vartheta$ is varied, one of the cells of $C$ shrinks to zero size.
It follows that at the $\vartheta$ where the flip occurs, $z$ and $z'$ are (at least generically) not contained in this collapsing cell:
even if they were in that cell for some value of $\vartheta$, they will fall out before the flip is reached.
From this it follows that the flip affects $\CY_{\gamma_{ij'}}$ only through its action on the functions $T_{ab}$.
Indeed, the action turns out to be precisely given by a transformation $\CK^\omega_{\gamma_{ij'}}$, where
$\omega(\gamma_E, \gamma_{ij'}) = \langle \gamma_E, \gamma_{ij'} \rangle$ (in the sense we defined in
\S \ref{sec:2d-4ddata} above.)

\item A pop of $T_\WKB$ occurs when the WKB curves going into a regular singularity $a$ start winding around $a$
more and more and finally become closed curves.  In particular, near the $\vartheta$ where the pop occurs, $T_\WKB$
contains a degenerate cell, which contains $a$ in its interior and also contains
a WKB curve which runs from a turning point to $a$.  As $\vartheta$ approaches the critical value, this WKB curve
winds more and more around $a$.  In particular,
if $z$ lies in this degenerate cell, there are infinitely many values of $\vartheta$ for which this WKB curve
meets $z$.  These values accumulate at the critical $\vartheta$ where the pop occurs.  As we have explained above, each time
the WKB curve meets $z$, the $\CY_{\gamma_{ij'}}$ undergo an $\CS$ transformation:
so we get an infinite sequence of such transformations, corresponding to solitons with charges
$\gamma_{ij} + n \gamma^f_a$, as $n \to \infty$.  On the other side of the critical $\zeta$ we have a second
infinite sequence of
$\CS$ transformations corresponding to solitons with charges $\gamma_{ji} + n \gamma^f_a$ as $n \to \infty$.
We may then ask whether there is a further transformation sandwiched between these two infinite sequences,
exactly at the value of $\vartheta$ where the pop occurs.  It turns out that the answer is yes:
there is a single $\CK_{\gamma^f_a}^\omega$ transformation, with $\omega(\gamma^f_a, \cdot)$ corresponding
to a 2d particle carrying
only 4d flavor charges.  One could in principle prove this
directly by studying the limits of $\CY_{\gamma_i}(\zeta)$ as $\zeta$
approaches the critical value from either side, along the lines of what was done for the juggle transformation of
the $\CY_\gamma$ in \cite{Gaiotto:2009hg}.  In practice, it is faster to work indirectly --- move the point $z$ just
outside the degenerate cell, so that the infinite sequences of transformations disappear, and then use the
wall-crossing formula \eqref{eq:wallcp1} to deduce what happens when $z$ is in the degenerate cell. See \S
\ref{subsec:cp1-sigma} below for a concrete illustration of these remarks.

\item A juggle of $T_\WKB$ occurs when closed WKB curves form an annular region.
If $z$ and $z'$ are outside the annular region where the juggle
is taking place, then the effect of the juggle on the $\CY_{\gamma_{ij'}}$ is simply a $\CK_{\gamma_0}^\omega$ transformation with
$\omega(\gamma_0, \gamma_{ij'}) = -2 \langle \gamma_0, \gamma_{ij'} \rangle$.  This is a straightforward extension of
our discussion above about the juggle in the pure 4d case.  Also as in the pure 4d case,
this transformation is surrounded by two infinite sequences of $\CK$ corresponding to flips.
The more interesting case is when
either $z$ or $z'$ lies in the annulus.  Then one finds an infinite sequence of $\CS$-factors interspersed
with these infinite sequences of $\CK$,
as the WKB curves wind more and more around the annulus.  At the critical $\zeta$
where the juggle occurs, there is a $\CK$-factor with appropriate $\omega(\gamma_0, \cdot)$.
Much as we just discussed for the pop, the simplest approach to determining exactly what transformation
occurs at the critical $\zeta$ is to use a wall-crossing formula. We give an example in
\S \ref{subsec:WC-exple-su2}.

\end{itemize}

\bigskip
\bigskip
\textbf{Remarks}

\begin{enumerate}

\item The analog of the Darboux expansion for interfaces $L_{\wp}$ associated
with open curves $\wp$ on $C$ from $z$ to $z'$ is given in \eqref{eq:lz-exp}.
As in the case of closed curves, for $A_1$ theories we can compute
the framed BPS degeneracies quite explicitly, using
a standard algorithm for writing the parallel transport from $z$ to $z'$
as a linear combination of  $\CY^\vartheta_{\gamma_{ij'}} \in \Hom(V, V')$.
For details see, for example, \cite{Gaiotto:2009hg}, Appendix A, equation $(A.8)$.

\item It should be possible to define the more subtle quantities $\CY_{\gamma_i}$
in terms of expressions of the form $\prod_{a,b} T_{ab}^{n_{ab}} s_c(z)$
where the $n_{ab}$ are now half-integers, raising issues with the choices
of squareroots.  We will not attempt to define them
here, but we note that doing so would amount to an explicit
construction of the twisting $B$ field of the universal bundle, so it might be
worthwhile to work this out carefully.

\item
Finally let us remark that there exists a 2d-4d analog of the ``spectrum generating transformation''
written down in \S 11 of \cite{Gaiotto:2009hg}.  There we wrote an
explicit formula for the $\CY^\vartheta_{\gamma}$
in terms of the opposite monodromy eigenvectors $\tilde s_{a,b,c,\cdots}$, which is equivalent
to working out the transformation relating $\CY_\gamma^\vartheta$ to $\CY_\gamma^{\vartheta+\pi}$.
This transformation turned out to be given by a combinatorial recipe depending only on the triangulation
$T_\WKB(\vartheta)$.  On the other hand this transformation
is the composition of all the factors $\CK_\gamma^{\Omega(\gamma)}$ attached to the BPS rays
which one encounters
in varying $\arg \zeta$ from $\vartheta$ to $\vartheta + \pi$, and hence contains
complete information about the 4d BPS spectrum.
All of that discussion can be generalized to the 2d-4d setting:  one can give a combinatorial recipe for
the transformation
relating $\CY_{\gamma_{ij'}}^\vartheta$ to $\CY_{\gamma_{ij'}}^{\vartheta+\pi}$, as an automorphism ${\bf S}$
of the vacuum groupoid algebra. This ${\bf S}$ has a unique ordered decomposition of the form
\begin{equation}\label{spectrumgen}
{\bf S} = \quad : \prod_{\vartheta< \arg -Z_{\gamma}, \arg -Z_{\gamma_{ij'}} < \vartheta+\pi} \CK_{\gamma}^\omega
\CS_{\gamma_{ij'}}^\mu :
\end{equation}
which determines the
complete 2d-4d BPS spectrum.  We do not write the details
here. The main new ingredient is a formula for the
transformation taking $s_a \to \tilde s_a$ for a
pop at a single vertex. It involves a rational expression of the
Fock-Goncharov coordinates on all the edges of a star neighborhood of the   vertex $a$.
We will defer the details to another occasion.

\end{enumerate}

\section{Examples of 2d-4d wall-crossing in some $A_1$ theories}\label{sec:Detailed-Expl}

In this section we illustrate the general remarks of \S \ref{sec:Hitchin-Expl} with a few
simple examples.  Many of the more intricate examples of \cite{Gaiotto:2009fs} could be
reconsidered along the lines below, but we have not done this yet.  We expect that
some rich phenomena remain to be discovered in those examples.

In the following sections we will be writing formulas for $A(\sphericalangle)$ in various different theories and regions of
moduli space.  To lighten the notation, we take advantage of the fact that,
with our sign convention for $\sigma$ understood,
$\CS^\mu_{\gamma_{ij}}$ only depends on $\mu$ through the value $\mu(\gamma_{ij})$;
this value is always $+1$ or zero in our examples, so
we drop the exponent and just write $\CS_{\gamma_{ij}}$ when
$\mu(\gamma_{ij})$ is nonzero.   We also sometimes drop the explicit $\omega$
in $\CK_\gamma^\omega$ when it is unchanged in a wall-crossing formula.

\subsection{Simple Argyres-Douglas-type theories}\label{subsec:AD-EXPL}

A particularly simple class of theories for studying wall-crossing are the Argyres-Douglas-type superconformal theories of $A_1$
type.  These theories are obtained by compactifying the
$A_1$ $(2,0)$ theory from six to four dimensions on $C = \IC \IP^1$, with an irregular singularity at infinity, whose strength is specified
by an integer $N \ge 1$.
(Usually one only considers $N \ge 3$, since the theories with $N=1,2$ have rather trivial
4d dynamics.  As we will see below, though, even the case $N=1$ is interesting when coupled to a surface defect.)

Following the discussion of \S \ref{sec:Hitchin-Expl}, these theories correspond to the Hitchin system on $\IC\IP^1$ with
gauge group $SU(2)$ and
a single irregular puncture at infinity.  The Coulomb branch $\CB$
is thus parameterized by meromorphic quadratic differentials $\phi_2(z)$ on $\IC\IP^1$
of the form
\begin{equation}
\phi_2(z) = - P_N(z) \, \de z^{2},
\end{equation}
where $P_N(z)$ is a polynomial of degree $N$.
The low-order coefficients of $P_N(z)$ are coordinates on the Coulomb branch, while the higher-order
ones are parameters of the theory (see e.g. \cite{Gaiotto:2009hg}, \S 9.2).

The BPS spectrum of these theories has been investigated in \cite{Shapere:1999xr} and revisited from our current perspective in \cite{Gaiotto:2009hg}.
At any point of $\CB$, the 4d BPS spectrum consists of a \ti{finite} set of hypermultiplets.
The wall-crossing phenomena which occur when we move in $\CB$
can always be understood using only the basic identity ``$\CK \CK = \CK \CK \CK$'' \eqref{eq:pentagon}.

Below we consider the 2d-4d BPS spectrum that we get by coupling these theories to their canonical surface defects $\bS_z$.
As we will see, the answer is again very simple:  we will find a finite set of 2d solitons between the two vacua.
Studying these solitons will provide examples of the various phenomena we have discussed in this paper.
For $N=1$ there is no wall-crossing for the solitons, but the solitons do
induce wall-crossing for framed 2d BPS states.  For $N>1$
the soliton spectrum undergoes wall-crossing, which provide nice examples of the
finite 2d-4d wall-crossing formulae. The essential formulae we need are summarized by special   cases of \eqref{eq:example-2} and
\eqref{eq:example-4}.  In the $A_1$ theories
the only identities we will need are:

\begin{enumerate}

\item
First, the result when $\bS_z$ crosses a separating WKB curve:
\begin{subequations}\label{eq:ks-ssk-gen}
\begin{align}
\CK_a  \CS_b & = \CS_{b} \CS_{b+a}\CK_a  \qquad \omega(a,b)=-1 \label{eq:ks-ssk}\\
\CS_b \CK_a  & = \CK_{a}  \CS_{b+a}\CS_b \qquad \omega(a,b)=+1\label{eq:sk-kss}
\end{align}
\end{subequations}
Note that this does not change the $\omega$ degeneracy, so we have suppressed the
superscript.

\item Second, the result when $\bS_z$ crosses a finite WKB curve:
\begin{equation}\label{eq:sks-sks}
\CS_b \CK_{a}^\omega \CS_{a-b} = \CS_{a-b} \CK_{a}^{\omega'} \CS_{b}
\end{equation}
which leaves the soliton spectrum unchanged but reverses the
sign $\omega(a,b)=-1$ to $\omega'(a,b)=+1$.
As discussed in  the example at the end of \S \ref{sec:2d-4ddata} above,
what has happened is that the meaning of the charge ``$b$''  has changed
since the point $z$ has moved.  The prescription for computing $\omega$ in
terms of intersection numbers remains in force.
\end{enumerate}

\subsubsection{$N=1$}\label{subsubsec:AD-EXPL-N1}

The $N=1$ theory is obtained by taking $P(z) = z$.
In this theory the 4-dimensional IR physics is trivial, so there are no 4d gauge fields;
there are also no flavor symmetries, so the charge lattice $\Gamma$ is trivial,
and there are no 4d BPS states. Although the four-dimensional theory is trivial there can still
be a nontrivial two-dimensional theory. Indeed  we will see that
this theory corresponds to a Landau-Ginzburg
theory with $W= \frac{1}{3} x^3 - z x$, which we have already studied in \S \ref{subsubsec:Massless-2d}
and Appendix \ref{app:N1hitchin}.\footnote{Although the mathematics is the same,
there is an important conceptual difference.
The Hitchin system in \S \ref{subsubsec:Massless-2d} is on the $u$-plane $\CB=\IC$ and there is
a 4d $U(1)$ gauge theory. Here there is no 4d gauge theory, $\CB$ is a point, and the Hitchin system is
on the Riemann surface $C$.}

When we include the canonical surface defect $\bS_z$ associated
to a point $z \in C$, the theory does have 2d BPS states. To see them explicitly
we can use our description from \S \ref{sec:2d-4ddata}:
a 2d BPS state corresponds to a finite WKB curve running from the turning point at $z = 0$ to $z$.
The finite WKB curves emerging from the turning point in this case are just three straight rays from $0$ to $\infty$,
with inclinations $\frac23 \vartheta, \frac23 (\vartheta+2\pi), \frac23 (\vartheta+4\pi)$.  As we vary $\vartheta$ through a phase
$\pi$ (say from $0$ to $\pi$), the three rays rotate, and one of them sweeps across the point $z$.  Hence there is exactly one
2d BPS state with phase between $0$ and $\pi$.  The full BPS spectrum consists of this particle plus its antiparticle (which
has phase between $\pi$ and $2\pi$).  These BPS states persist for all values of $z$, which is the only modulus around,
so there is no wall-crossing for the ordinary 2d BPS spectrum.  The spectrum of vacua and BPS states coincide with those of a
Landau-Ginzburg model with superpotential $\frac{1}{3} x^3 - z x$.  From this point of view,
our statements will just be a rewriting of standard statements from $tt^*$ lore \cite{Cecotti:1991me,Cecotti:1993rm}.

\insfig{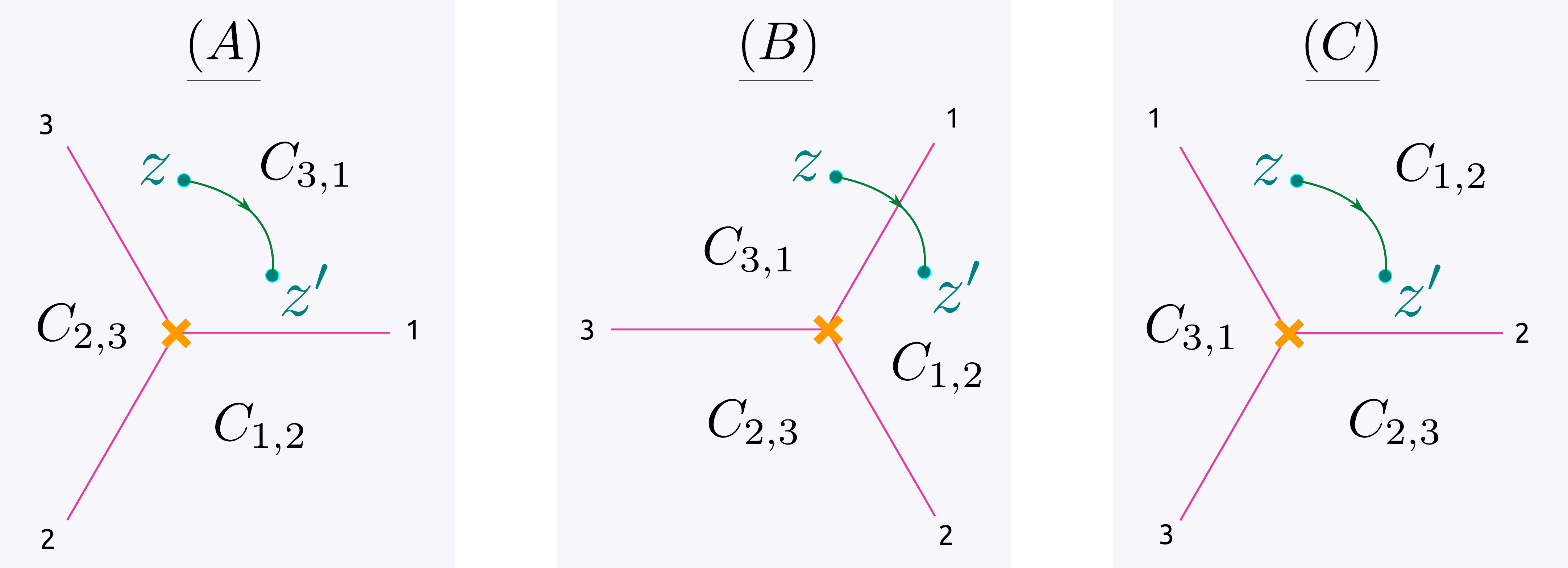}{WKB sectors and interface paths for three values of $\zeta$.
The separating WKB curves emerging from the turning point are
illustrated for three values of $\vartheta=\arg\zeta$.  In (A) $\zeta$ is a
positive real number. As the phase of $\zeta$ is increased the picture rotates
counterclockwise to produce figures (B) and (C).}

\insfig{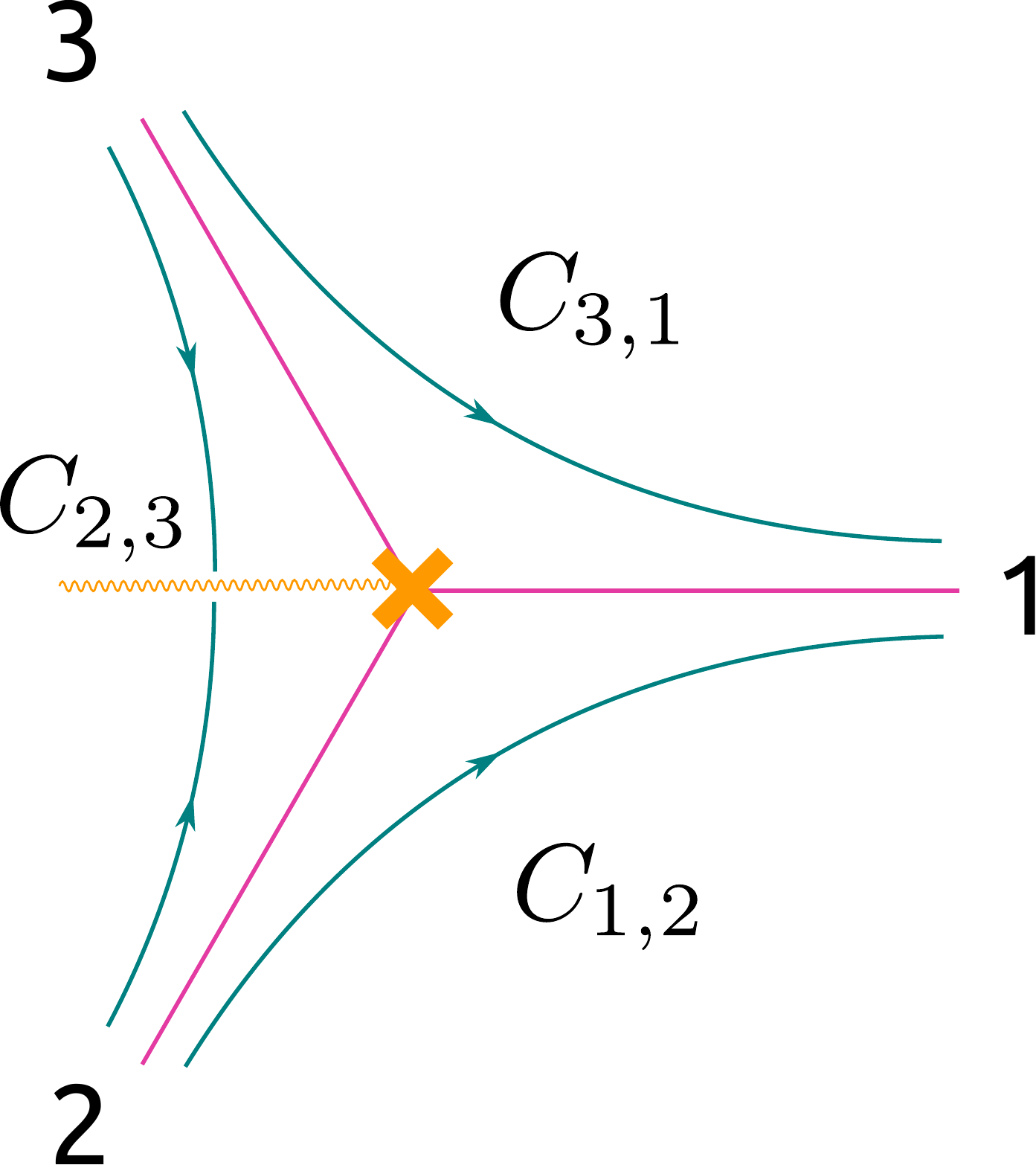}{Using the principal branch of the logarithm to define $z^{1/2}$ determines a sheet
of the covering $\Sigma \to C$, which we denote $+$.  The standard orientation of WKB curves is
shown here for WKB curves on sheet $+$, at $\vartheta=0$.  This orientation
allows us to fix a small flat section given a point $z \in C$ and a choice of sheet above $z$.}

Although there is no wall-crossing for the ordinary BPS states,
there is still some wall-crossing in the story:  indeed the presence of 2d BPS states implies that
there is wall-crossing for 2d \ti{framed} BPS states.  Let us explore this a bit.
We consider an interface $L_{\wp,\zeta}$ between two canonical surface defects $\bS_z$, $\bS_{z'}$
associated to a path $\wp$ from $z$ to $z'$ as shown in Figure \ref{fig:adn1-line}.
We initially assume that $z$ and $z'$ are ``close'' in the sense that the angle between them
is less than $2\pi/3$, and discuss the case when the angle is between $2\pi/3$ and $\pi$ later.

First of all, it is straightforward to express $\inprod{L_{\wp,\zeta}}$. The small flat
sections are defined first in the neighborhood of singular points and then defined
for other values of $z$ by parallel transport.  Using the convention
\eqref{eq:complaw} we can write
\begin{subequations}\label{eq:Lss}
\begin{align}
\inprod{L_{\wp,\zeta}}
& =  \frac{s_1(z)\otimes s_3(z')}{(s_3,s_1)} + \frac{s_3(z)\otimes s_1(z')}{(s_1,s_3)}\label{eq:Lss-13}\\
& =  \frac{s_1(z)\otimes s_2(z')}{(s_2,s_1)} + \frac{s_2(z)\otimes s_1(z')}{(s_1,s_2)}\label{eq:Lss-12}\\
& =  \frac{s_2(z)\otimes s_3(z')}{(s_3,s_2)} + \frac{s_3(z)\otimes s_2(z')}{(s_2,s_3)}\label{eq:Lss-23}.
\end{align}
\end{subequations}
All three of these expressions are valid for any $z,z'$.
Which one of them is \emph{useful}
for extracting framed BPS states depends on the sectors to which $z,z'$ belong.
Recall that the expressions $\CX_{n,c,d}(z,z')$ with good asymptotics
are only defined for $c\in \{ a, b\}$ when $z\in C_{ab}$
and similarly for $d$.  Therefore in case (A) of Figure \ref{fig:adn1-line} we should use
\eqref{eq:Lss-13}. In accordance with our discussion of equations \eqref{eq:cy11}
and \eqref{eq:cy22} we define
\begin{equation}\label{eq:cy++--}
\begin{split}
\CY_{\gamma_{++'}} & := \frac{s_3(z)\otimes s_{1}(z')}{(s_1,s_3)} \\
\CY_{\gamma_{--'}} & := \frac{s_1(z)\otimes s_{3}(z')}{(s_3,s_1)} \\
\end{split}
\end{equation}
because for $z,z'\in C_{3,1}$ the $+$ branch of $\lambda$ is associated
with $z_1$ and the $-$ branch with $z_3$. (See Figure
\ref{fig:adn1-or-wkb}.)  Therefore in case (A) of Figure \ref{fig:adn1-line}
we can write
\begin{equation}\label{eq:Lcycy}
\inprod{L_{\wp,\zeta}} = \CY_{\gamma_{++'}} + \CY_{\gamma_{--'}}.
\end{equation}
There are therefore two framed BPS states associated to the open paths indicated above.

Now let us increase the phase of $\zeta$ so that the WKB curves rotate to
case (B) of Figure \ref{fig:adn1-line}. Now $z$ remains in $C_{3,1}$, but $z'\in C_{1,2}$,
and given our rules for defining $\CX_{n,c,d}$ we must use $c\in \{ 1,3 \}$
and $d\in \{ 1,2\}$. Thus, none of the expressions \eqref{eq:Lss} is directly
useful. However, we can use the relation
\begin{equation}
s_1 (s_2,s_3) + s_2 (s_3,s_1) + s_3 (s_1,s_2)=0
\end{equation}
to eliminate $s_3(z')$  in  \eqref{eq:Lss-13} in favor of $s_1(z'),s_2(z')$ to get
\footnote{One can show that
consistency of the wall-crossing formulae given below fixes
all signs up to the choices of $\sigma(+-,-+') $ and $\sigma(+-,--') $.
We will make the choice
$\sigma(+-,-+')=+1$ and $\sigma(+-,--')=-1$, which simplifies the
formulae.}
\begin{equation}
\begin{split}
\inprod{L_{\wp,\zeta}} &= \frac{ s_3(z) \otimes s_1(z')}{(s_1,s_3)} +
\frac{(s_2,s_3)}{(s_1,s_2)(s_1,s_3)} s_1(z)\otimes s_1(z') +
 \frac{s_1(z)\otimes s_2(z')}{(s_2,s_1)}\\
& = \CY_{\gamma_{++'}}  - \CY_{\gamma_{-+'}} + \CY_{\gamma_{--'}} \\
\end{split}
\end{equation}
The path $\gamma_{--'}$ is the same as
that used in \eqref{eq:cy++--}, although its expression in terms
of the $s_a$'s has changed. Also note that a third term, similar to
\eqref{eq:cy12}, has emerged, so we now have three framed BPS states.
This is an example of the wall-crossing phenomenon for
framed BPS states.

Now we can check the framed wall-crossing formula. We have
\begin{equation}
 \CY_{\gamma_{++'}}  - \CY_{\gamma_{-+'}} + \CY_{\gamma_{--'}} =
 ( \CY_{\gamma_{++'}}    + \CY_{\gamma_{--'}}) (1 + \CY_{\gamma_{-'+'}})
 \end{equation}
which is the expected wall-crossing transformation by $\CS^\mu$.

If we further increase the phase of $\zeta$ so that we come to
case (C) of Figure \ref{fig:adn1-line} then we use \eqref{eq:Lss-12}
and recover once again \eqref{eq:Lcycy}. Now the wall-crossing is
given by
\begin{equation}
(1-\CY_{\gamma_{-+}})( \CY_{\gamma_{++'}}  - \CY_{\gamma_{-+'}} + \CY_{\gamma_{--'}}) =
   \CY_{\gamma_{++'}}    + \CY_{\gamma_{--'}}
 \end{equation}
which again is the expected transformation by $\CS^\mu$.

\insfig{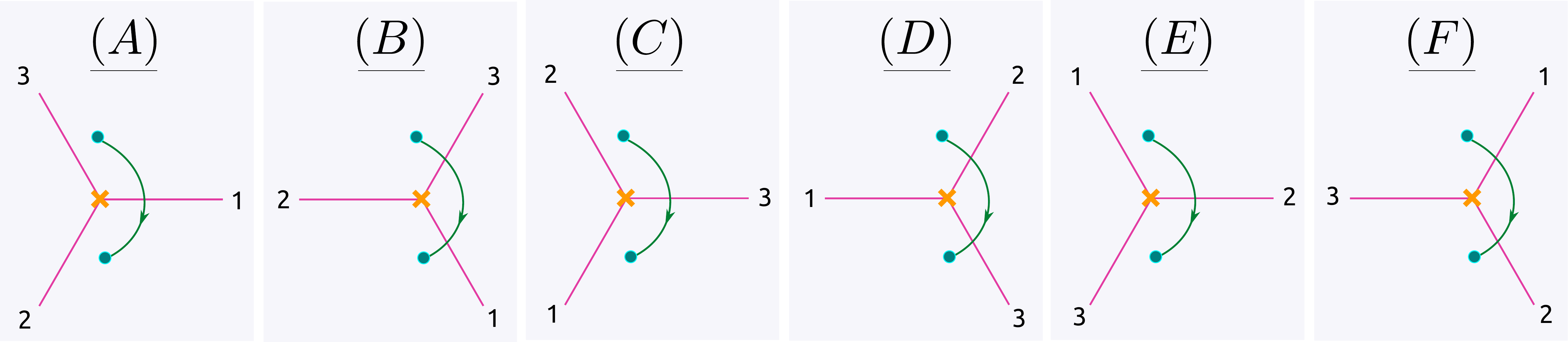}{When $z$ and $z'$ are separated by an angle between $2\pi/3$ and $\pi$ there
are six different Darboux expansions (up to monodromy).}

Let us now briefly indicate what happens when $z$ and $z'$ are separated by an angle between $2\pi/3$ and $\pi$,
as in Figure \ref{fig:wideangle}.
The Darboux expansion for case (A) is
\begin{equation}
\begin{split}
\inprod{L_{\wp,\zeta}} &= \frac{ s_1(z) \otimes s_2(z')}{(s_2,s_1)} +
\frac{(s_2,s_3)}{(s_1,s_2)(s_1,s_3)} s_1(z)\otimes s_1(z') +
 \frac{s_3(z)\otimes s_1(z')}{(s_1,s_3)}\\
& = \CY_{\gamma_{--'}}  + \CY_{\gamma_{-+'}} + \CY_{\gamma_{++'}} \\
\end{split}
\end{equation}
Denoting this expansion by $\CE_A$, etc. the wall-crossings are
\begin{equation}
\begin{split}
(1-\CY_{+-})\CE_A  & = \CE_B = \CY_{--'}+\CY_{+-'} + \CY_{-+'} \\
\CE_B(1+\CY_{-'+'}) & = \CE_C = \CY_{--'} + \CY_{+-'} + \CY_{++'} \\
(1-\CY_{-+})\CE_C & = \CE_D = \CY_{+-'} + \CY_{++'} + \CY_{-+'} \\
\CE_D (1+ \CY_{+'-'}) & = \CE_E = \CY_{++'} + \CY_{-+'} + \CY_{--'} \\
(1-\CY_{+-})\CE_{E} & = \CE_F = \CY_{-+'} + \CY_{--'} + \CY_{+-'} \\
\CE_F (1+ \CY_{-'+'}) & = M\cdot \CE_A = \CY_{--'} + \CY_{+-'} + \CY_{++'} \\
\end{split}
\end{equation}
In the last line $M$ is the monodromy operator and for brevity we have
shortened the notation to $\CY_{\gamma_{++'}} = \CY_{++'}$, etc.

As a final remark note that since the bulk theory is so trivial, and in particular since
there is no degree of freedom
charged under the $\IZ_2$ flavor monodromy for $\bS_z$,
we should be able to consider boundaries for the surface defect. The small flat sections
 $\pm s_a(z)$ are perfectly good candidates for the vevs of
such boundary line defects. Because they can be canonically normalized (up to sign) by the condition $(s_a,s_{a+1})=1$,
they are actual holomorphic sections of $V_\bS$.
When $z$ is in an appropriate cell, they will define $\CY_+$, $\CY_-$, $- \CY_+ - \CY_-$.

\subsubsection{$N=2$}\label{subsubsec:ADN=2}

We next consider the case $N=2$, where the polynomial $P$ is of the form
$P_2(z) = z^2 + 2 m$.  The IR dynamics are still trivial, so there are no gauge fields and no gauge charges, $\Gamma_g = 0$.
The Coulomb branch $\CB$ is just a single point, so there is no possibility of wall-crossing for the 4d BPS spectrum.
However, there is a 1-dimensional flavor charge lattice $\Gamma_f = \Gamma$, and the 4d BPS spectrum is nonempty --- there is a single BPS
particle, carrying a flavor charge $\gamma$, corresponding to a finite WKB curve which connects the two zeroes of $P$.
So $\Omega(\pm \gamma) = 1$, while all $\Omega(n \gamma) = 0$ for $n \neq \pm 1$.

The story becomes more interesting when we introduce the canonical surface defect $\bS_z$.
Then we get a local model for the general behavior of a 2d-4d theory with two vacua, a light bulk particle and
light 2d particles. (See the end of this section for further remarks on how to interpret the model.)

Unlike the $N=1$ example where
the 2d BPS spectrum was independent of $z$, we now find four distinct
regions in the $z$-plane, with different 2d BPS spectra.
Indeed, we can use  the general principle mentioned at the end of \S \ref{sec:2d-4ddata}.
The regions are cut out by the finite and separating WKB curves emanating from the two turning points,
at the value of $\vartheta=\vartheta_*$
for which the finite WKB curve is present; see Figure \ref{fig:N2-cells-critical}.
Recall that when $z$ lies on one of these curves,
there is an obvious finite open WKB network from $z$ to the turning point, with phase $\vartheta_*$.
This finite WKB network
corresponds to a 2d BPS particle whose phase is $\vartheta_*$.  But $\vartheta_*$ is also the
phase of the 4d BPS particle; so for this value of $z$ the central charges of the 2d and the 4d particle are aligned.
Thus, the finite and separating WKB curves of phase $\vartheta_*$ shown in  Figure \ref{fig:N2-cells-critical}
are the walls of marginal stability in the parameter $z$.

\insfig{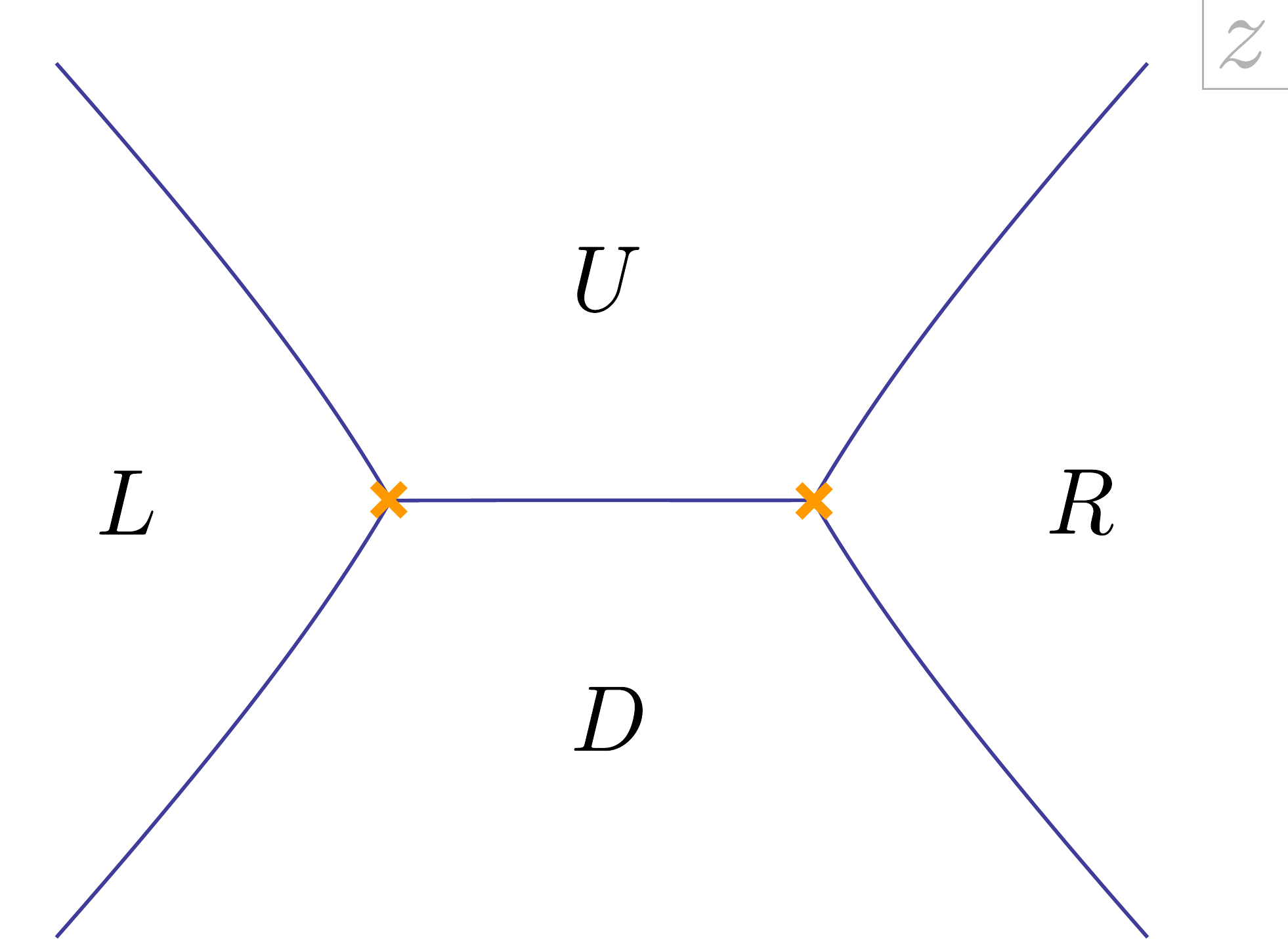}{The cell decomposition induced by the WKB foliation $F_\WKB(\vartheta)$ in the $N=2$ Argyres-Douglas-type theory, when $\vartheta$ is set to the critical value $\vartheta_*=\arg Z_\gamma$.
The WKB curves shown here are also the walls of marginal stability for
the parameter $z$ of the canonical surface defect.  We label the regions $R$, $U$, $L$, $D$. Here $m$ is real
and negative.}

\insfig{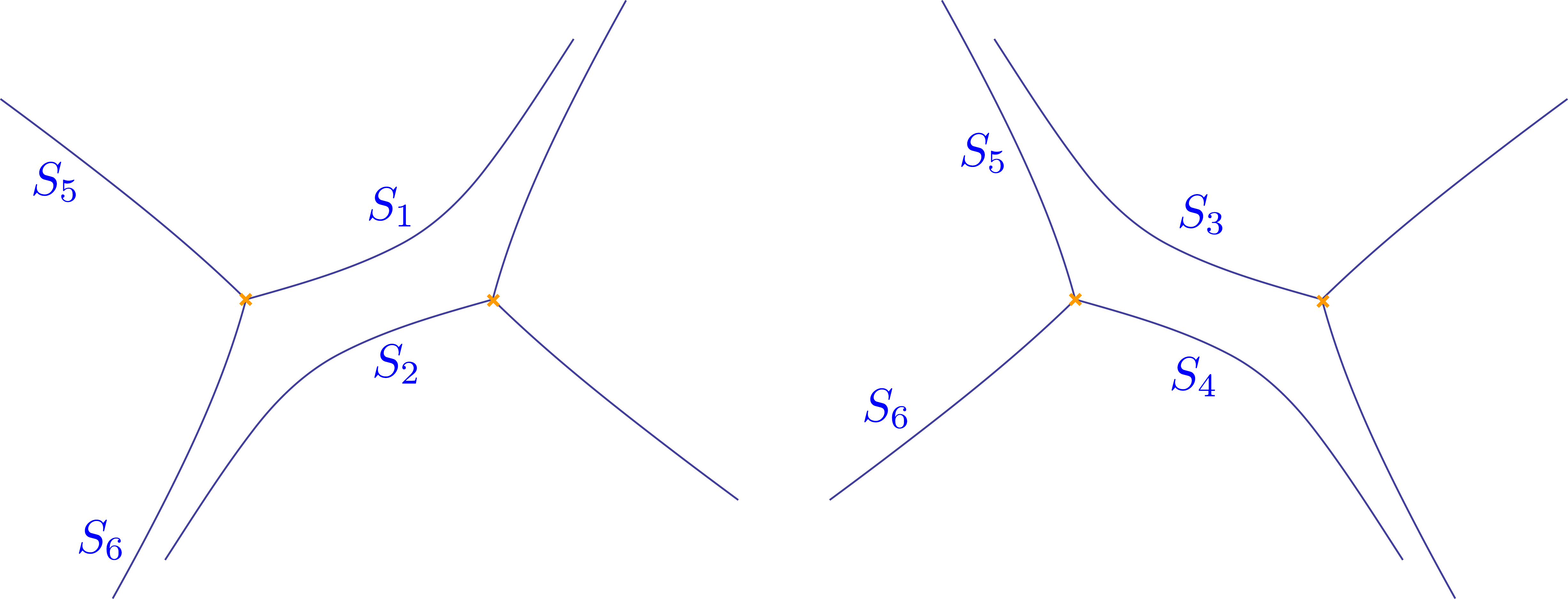}{When $\vartheta \not= \vartheta_*, \vartheta_*+\pi$ the separating WKB curves
look as shown. As $\vartheta$ increases the three prongs around each turning point
both rotate with the same sense. As $\vartheta$ crosses $\vartheta_*$ the curve of
type $S_1$ jumps to a curve of type $S_4$ and the curve of type $S_2$ jumps to a
curve of type $S_3$.}

The spectrum $\mu$ of 2d solitons can be determined geometrically as follows.
In the   regions $L$ and $R$ which have a single turning point on their boundary,
the 2d spectrum consists of a single soliton (and its antiparticle)
 between the two vacua: it corresponds to a single finite WKB curve running
from $z$ to the turning point at the boundary of the region.
 In the  regions $U$ and $D$ which have
two turning points on the boundary, the 2d spectrum consists of two 2d particles, corresponding to two finite
WKB curves from $z$ to the two turning points on the boundary. This can be proved as follows.
For $\vartheta \not=\vartheta_*, \vartheta_*+\pi$ the separating WKB curves look like those
shown in Figure \ref{fig:N2-cells-noncritical}. As $\vartheta$ rotates the curves rotate around the turning points.
The region between $S_1$ and $S_2$ or between $S_3$ and $S_4$ rotates and fills the
upper and lower regions of Figure \ref{fig:N2-cells-critical}. For any point in these regions
there will be values of $\vartheta$ such that that point lies on each of the curves of type $S_1$ and $S_3$
or $S_2$ and    $S_4$. Because the critical separating curves (i.e. those at $\vartheta=\vartheta_*$)
are walls of marginal stability
we know that the curves $S_i$ cannot enter into the left and right regions. Similarly,
for any point in the left region there will be some value of $\vartheta$ such that this point
lies on a separating curve of type $S_5$ or $S_6$.

\insfig{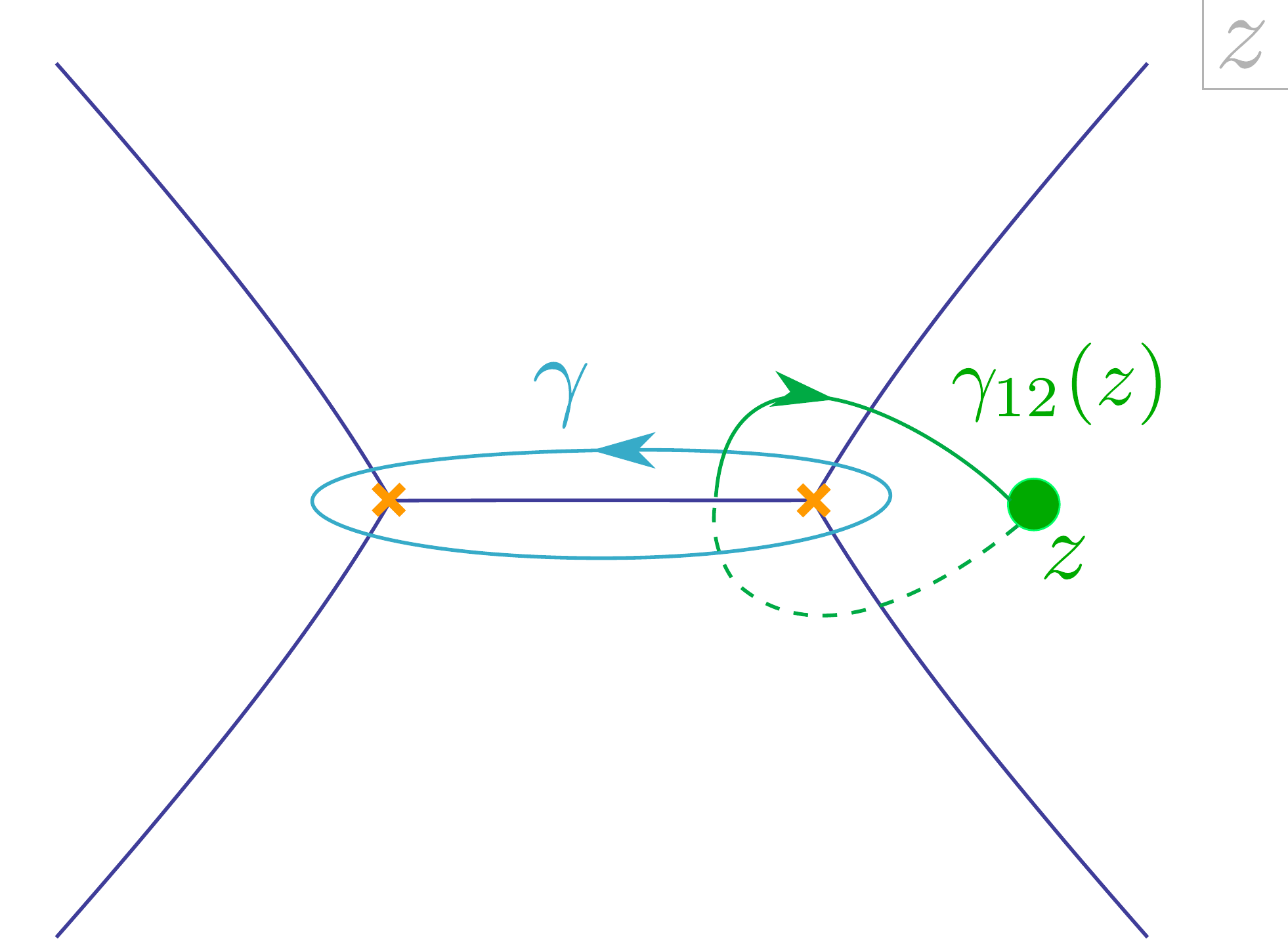}{A choice of cycles so that $\omega(\gamma,\gamma_{ij}(z))=-1$.
Here we have taken $P_2= z^2 - \alpha^2$ where $\alpha$ is a positive real number,
and we have placed the canonical surface operator $\bS_z$ on the ray
$(\alpha, \infty)$.}

In order to illustrate the wall-crossing formulae explicitly
we now describe the 2d soliton spectrum in some detail. Our technique will be
to determine the periods in some easily accessible region and then derive the
spectrum in the remaining regions using wall-crossing. Because the
local system $\Gamma_{12}$ has monodromy around the two roots of $P_2$
we will find different spectra when comparing using homotopically different
paths.

We begin with $z$ in the region $R$ and call the charge of the single 2d particle $\gamma_{12}(z)$,
or sometimes just $\gamma_{12}$ with $z$ understood. In order for this to be unambiguous we should
choose cuts for the double cover $\Sigma \to C$. It is convenient to define
$2m=-\alpha^2$ and write $\sqrt{P} = \sqrt{z+\alpha} \sqrt{z-\alpha}$, choosing the principal branch
of the logarithm to define each factor. When $\alpha$ is real $\sqrt{P}\sim z$ at large $z$
and has a cut running along the segment $[-\alpha,\alpha]$. For definiteness we take $\alpha>0$.
 We choose orientations as in
Figure \ref{fig:N2-cycles} so that  $\omega(\gamma, \gamma_{12}) = -1$. With these choices
we can determine the local system $\Gamma_{12}$.
A simple path taking $z$ around the root $\alpha$ leads to monodromy $\gamma_{12}(z) \to - \gamma_{12}(z) =: \gamma_{21}(z)$,
and a simple path taking $z$ around the root $-\alpha$ leads to monodromy $\gamma_{12}(z) \to - \gamma_{12}(z) + 2 \gamma$.
This follows most easily by noting that $Z_{\gamma_{12}(z)-\gamma}$ has a zero at $z=-\alpha$ and has a square-root
branch cut there.

It is also useful to note that for $z$ large and positive the WKB
curve clearly runs along the positive axis so $Z_{\gamma_{12}(z)}$ is positive. For large $\vert z \vert$
it behaves like $Z_{\gamma_{12}(z)}\sim \pi^{-1} z^2$. (In fact, in this extremely simple model we
can write the periods explicitly:
\begin{equation}
\begin{split}
Z_{\gamma} & = -\I \alpha^2, \\
Z_{\gamma_{12}(z)} & = \frac{1}{\pi} \left( z \sqrt{z^2 - \alpha^2} - \alpha^2 \log \left( z + \sqrt{z^2 - \alpha^2}\right) + \alpha^2 \log \alpha\right) \\
\end{split}
\end{equation}
but we will phrase our arguments so that such explicit formulae are not needed, since in general explicit formulae for
periods are not available.)

Using the above description of the periods, we can form
the product $A(\sphericalangle)$, taking
$\sphericalangle$ a sector of opening angle $2\pi$, i.e. the whole plane:
\begin{equation}\label{eq:A-R}
A(\sphericalangle;R) =\CK_\gamma \CS_{\gamma_{12} } \CK_{-\gamma} \CS_{-\gamma_{12}}
\end{equation}
Although we have derived it for $m$ real and negative and for $z$ large and positive,
this will hold for all $z$ in the region $R$. Moreover, changing the phase of $m$ merely
rotates the whole picture. (We will return to the wall-crossing in $m$ at the end of this
section.)

\insfig{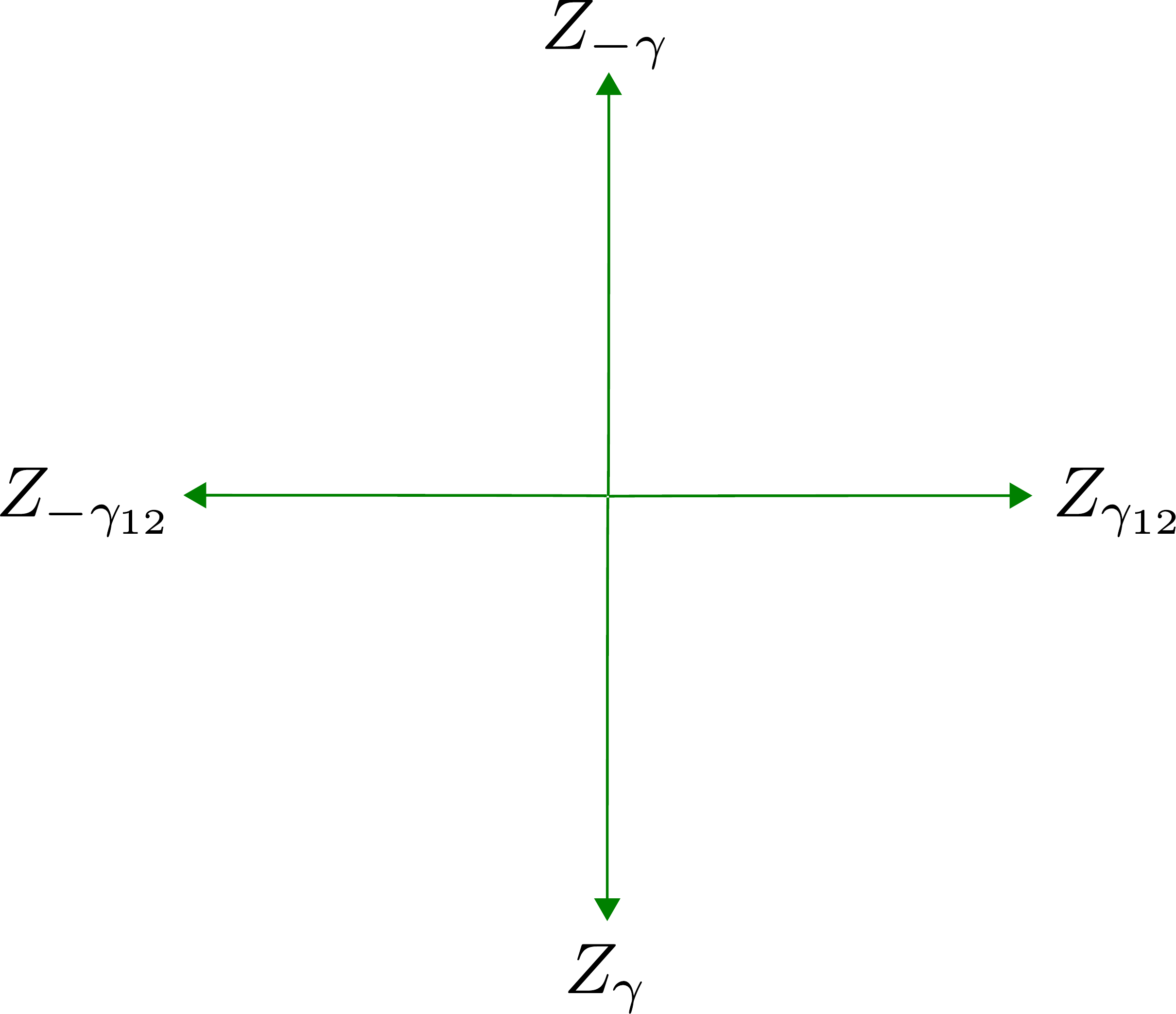}{The central charges for $z$ in the region $R$ on the real axis, and $m \in \IR_-$.
If $m$ is fixed and $z$ is continued into the region $U$, then $Z_{\gamma_{12}(z)}$
rotates counterclockwise, and aligns with $Z_{-\gamma}$ at the $R\to U$ wall of marginal stability.
If $z$ is continued into the region $D$ then the central charge $Z_{\gamma_{12}(z)}$
rotates clockwise, and aligns with $Z_{\gamma}$ at the $R\to D$ wall of marginal stability.}

Before proceeding let us make a few comments about the notation \eqref{eq:A-R}.
First, in general, products of wall-crossing factors are only expected to make sense for $\sphericalangle$ of
opening less than $\pi$. However, in the AD examples the product is always finite, so the
full $2\pi$ product is sensible.  Moreover, it is sometimes useful
to write the product for the entire range $2\pi$ since one might
wish to split it into different half-planes.  Of course, one can cycle the factors in
these expressions.  Second, the spectrum does not depend just on the
region $R,U,L,D$ but rather on a homotopy class of paths used to continue from $R$ to the region.
We will suppress this in our simple discussion below. Third, the expression \eqref{eq:A-R}
summarizes the configuration of central charges of occupied charges shown in
Figure \ref{fig:adn2-central} above. We will not draw the analogous figures for the
other regions; rather, the reader is urged to draw pictures of the central charges in order to
make the subsequent formulae comprehensible.

We are now ready to proceed with the wall-crossing analysis. We will first take $\vert z\vert$ large
and move counterclockwise through regions $R\to U \to L \to D$. Then we will compare with moving
clockwise from $R\to D$, as well as moving $z$ through the cut $[-\alpha, \alpha]$ directly
from $U \to D$.

If we first move $z$ from $R$ to $U$ then $Z_{\gamma_{12}(z)}$ becomes parallel to $Z_{-\gamma}$.
(This can be easily seen by making $\vert z \vert $ large. Then the marginal stability wall
is at $\arg z \approx \frac{\pi}{4}$ and $Z_{\gamma_{12}} \sim \pi^{-1} z^2 $ has argument approximately
$\pi/2$.) We can therefore apply \eqref{eq:sk-kss} with $a=-\gamma$ and $b=\gamma_{12}(z)$ to produce
the spectrum in region $R$:
\begin{equation}\label{eq:A-UP}
A(\sphericalangle;U) = \CK_\gamma \CS_{\gamma_{21}+\gamma } \CS_{\gamma_{21}}\CK_{-\gamma} \CS_{\gamma_{12}-\gamma}\CS_{\gamma_{12}}
\end{equation}
If we continue to increase the phase of $z$ so that we encounter the wall-crossing $U \to L$ then
the periods continue rotating and at the $U\to L$ wall $Z_{\gamma_{12}}$ becomes parallel to $Z_{\gamma}$.
We now apply \eqref{eq:ks-ssk} with $a=\gamma$ and $b= \gamma_{12}-\gamma$ to produce the spectrum in $L$:
\begin{equation}
A(\sphericalangle;L) =\CK_\gamma \CS_{\gamma_{12}-\gamma }  \CK_{-\gamma} \CS_{\gamma_{21}+\gamma}.
\end{equation}
Continuing to increase the phase of $z$ at the $L\to D$ wall $\arg z \approx \frac{5 \pi}{4}$ and
$Z_{\gamma_{12}-\gamma}$ becomes parallel to $Z_{-\gamma}$. We now apply \eqref{eq:sk-kss} to produce
\begin{equation}
A(\sphericalangle;Dl) =\CK_\gamma \CS_{\gamma_{21}+2\gamma } \CS_{\gamma_{21}+\gamma}
\CK_{-\gamma} \CS_{\gamma_{12}-2\gamma}\CS_{\gamma_{12}-\gamma}
\end{equation}
We write $Dl$ to indicate that this is the spectrum obtained from continuation from the region $L$. (Recall that
$A(\sphericalangle)$ actually depends on a homotopy class of paths from region $R$.)

Going back to the region $R$ we could instead have continued $z$ into the $D$ region by moving $z$ clockwise. In
this case $Z_{\gamma_{12}(z)}$ rotates clockwise and at the marginal stability wall becomes parallel to $Z_{\gamma}$
and \eqref{eq:ks-ssk} produces
\begin{equation}
A(\sphericalangle;Dr) =\CK_\gamma \CS_{\gamma_{21}  } \CS_{\gamma_{21}-\gamma}
\CK_{-\gamma} \CS_{\gamma_{12} }\CS_{\gamma_{12}+\gamma}
\end{equation}

\insfig{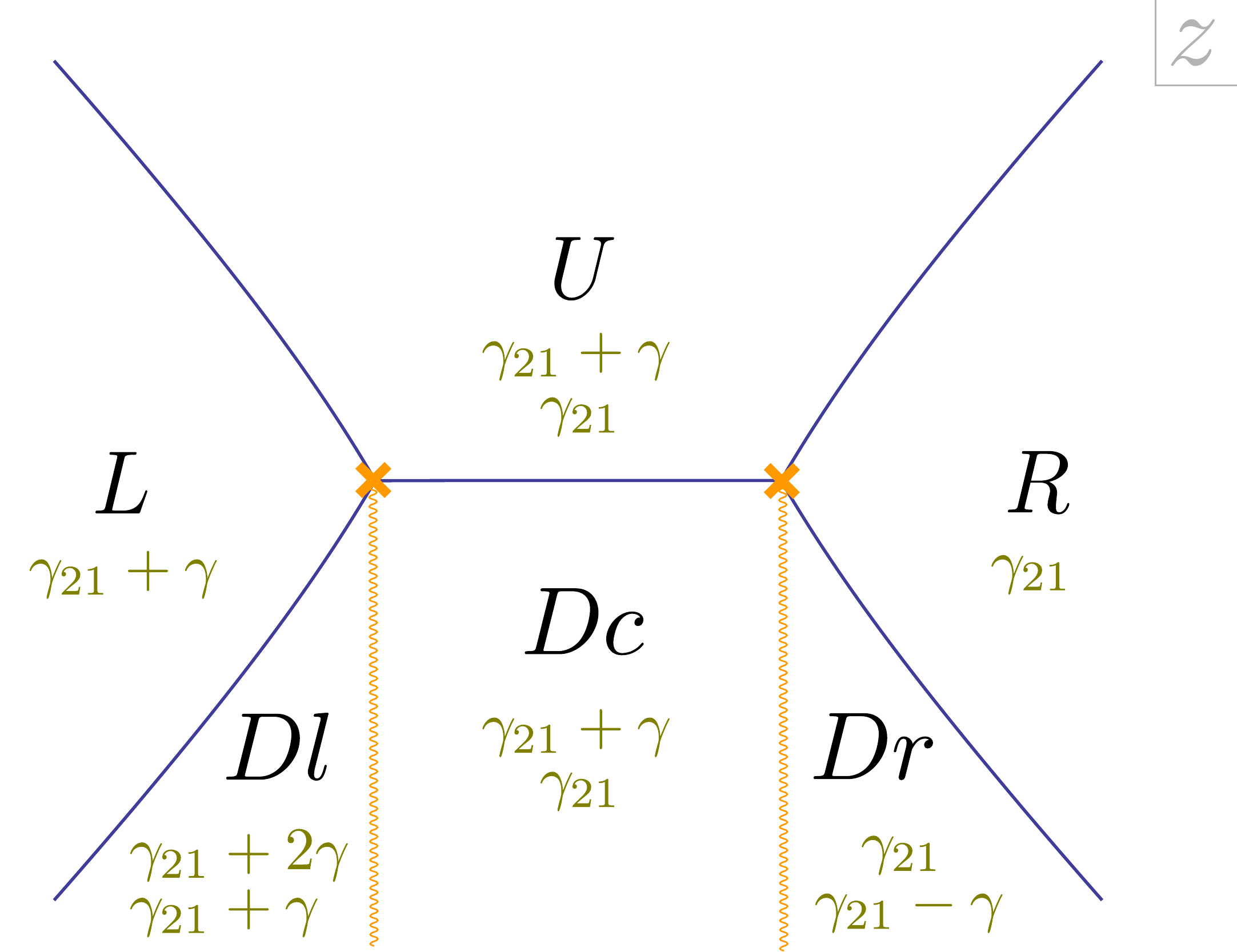}{The soliton spectrum for the $N=2$ AD theory. We have chosen two cuts, on the
complement of which the torsor $\Gamma_{12}$ can be trivialized.  These cuts divide the
``down region'' $D$ into three subregions.  The spectra in these three regions are
related by the monodromy transformations associated with the two turning points.
We have only given half the spectrum; the other half consists of the antiparticles.}

Finally, we could start with the spectrum summarized by $A(\sphericalangle;U)$ in \eqref{eq:A-UP} and continue $z$ directly
into the region $D$. Note that when $z\in (-\alpha, \alpha)$ the period $Z_{\gamma_{12}}$ is pure imaginary.
Moreover, expanding the period around $z=\alpha$ say, $z=\alpha + w$, with $w$ small
 we find $Z_{\gamma_{12}}\sim w \sqrt{2\alpha w}$. From this it follows that as $z$ goes through the
 marginal stability line $U \to D$ the central charges $Z_{\gamma_{12}}$ and $Z_{\gamma_{21}+\gamma}$
 simultaneously align with $Z_{\gamma}$. Thus   we should apply the wall-crossing formula \eqref{eq:sks-sks} to obtain:
\begin{equation}
A(\sphericalangle;Dc) =\CK_\gamma \CS_{\gamma_{12}} \CS_{\gamma_{12}-\gamma}
\CK_{-\gamma} \CS_{\gamma_{21} }\CS_{\gamma_{21}+\gamma}.
\end{equation}
In this way we arrive at the full spectrum shown in Figure \ref{fig:adn2-spectrum}.
The fact that we found three different spectra in the region $D$ is the result
of the monodromy of the local system $\Gamma_{12}$ around the two zeroes of $P$. As we have seen, when
comparing $Dl \to Dc$ we should take $\gamma_{12} \to \gamma_{21} + 2\gamma$
(since $Z_{\gamma_{12} - \gamma}$ vanishes at the left turning point), while
to compare $Dr \to Dc$ we should take $\gamma_{12} \to \gamma_{21}$.
It is easy to check that with these monodromy transformations
the spectrum is nicely consistent.

\insfig{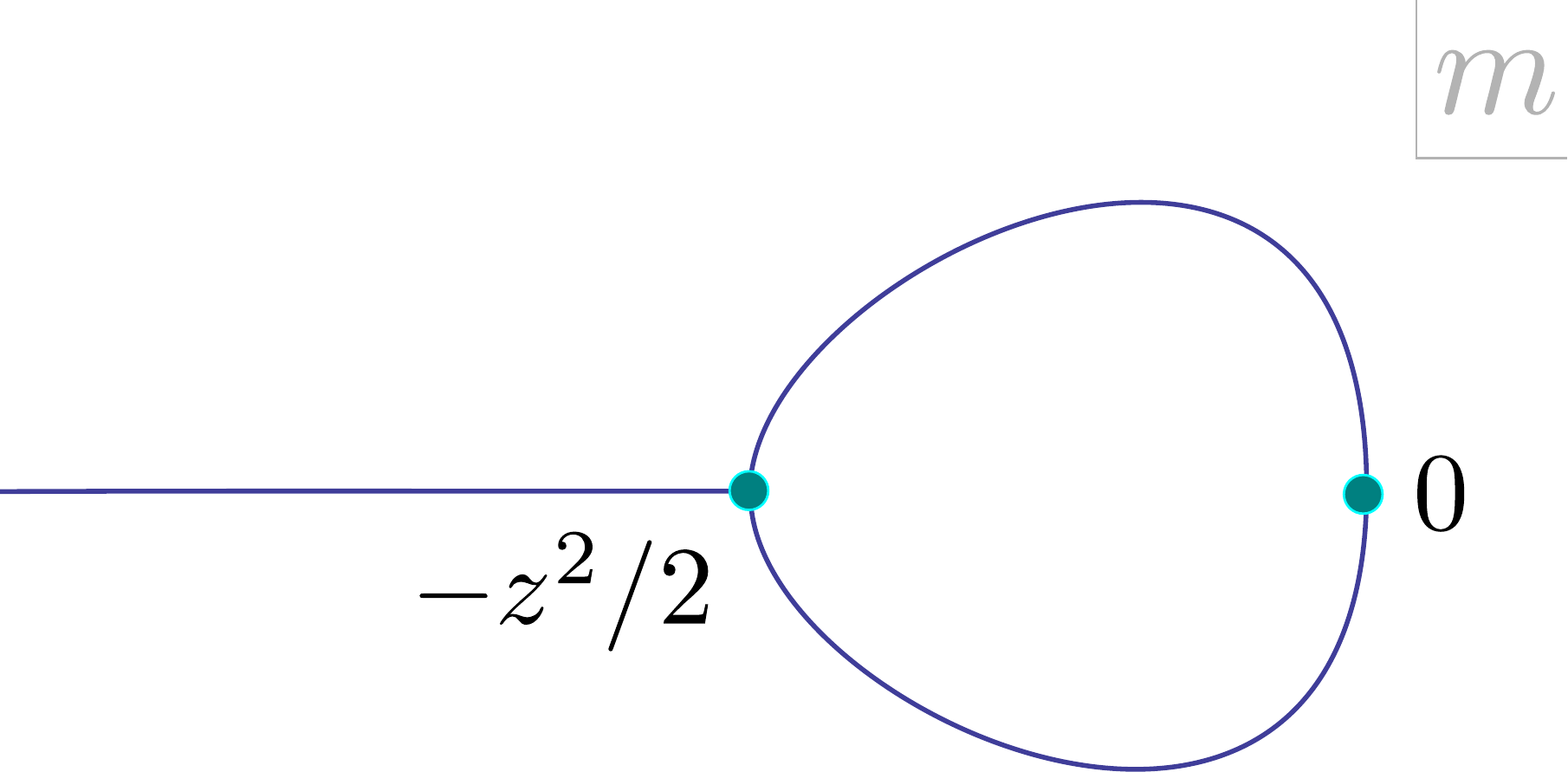}{Walls of marginal stability in the $m$-plane for fixed $z$.
The wall emanating to the left is of type $\CS\CK\CS = \CS \CK\CS$, while the
two walls emanating from $m=0$ are of type $\CK \CS = \CS \CS \CK$ and
$\CS \CK = \CK \CS \CS$ respectively.  There is a branch cut
emanating from $m=0$.  When $z\to 0$ the 4d and 2d
periods $Z_{\gamma}$ and $Z_{\gamma_{12}}$ simultaneously vanish.
(This corresponds to a singularity in the semiflat geometry, which
is resolved by quantum effects; the resolved structure looks like
a hyperholomorphic bundle on periodic NUT space.)}

Finally, it is interesting to study the behavior of the model in the $m$ plane,
since it serves as a useful local model for the general 2d-4d wall-crossing structure
near a locus of $\CB$ where $Z_\gamma = Z_{\gamma_{ij}} = 0$. Working at fixed $z$,
there are two singularities in the $m$ plane:
$m=0$, where $Z_\gamma = 0$, and $m = m_0 = - \frac{z^2}{2}$, where $Z_{\gamma_{12}}=0$.
Rotating the phase of $m$, the pattern of walls in Figure \ref{fig:N2-cells-critical}
rotates, so that if $m$ rotates by $2\pi$ then the figure rotates in the same sense by $\pi$.
When $\vert m \vert > \frac{1}{2} \vert z^2 \vert$ the segment between the two roots of $P_2$
rotates through $z$ and we apply the wall-crossing formula \eqref{eq:sks-sks}. When
$\vert m \vert < \frac{1}{2} \vert z^2 \vert$  as the phase of $m$ increases by $2\pi$
two of the separating WKB curves pass through $z$.  As we have seen, these correspond to
wall-crossings of type \eqref{eq:ks-ssk} and \eqref{eq:sk-kss}.  Hence at fixed $z$
we will find two walls of marginal stability
emanating from $m=0$, and the three types of walls meet at $m_0$.  The $m$-plane is thus
divided into two regions, an inner region with a single 2d soliton, and an outer region
with two 2d solitons. The two regions are separated by
a closed wall of marginal stability running through both singularities.
Another wall, of the $\CS\CK\CS \to \CS \CK\CS$ type
goes from $m_0$ all the way to infinity.  See Figure \ref{fig:adn2-m-walls}.

\bigskip
\textbf{Remarks}

\begin{enumerate}

\item The example of this section is actually a member of a larger family of local models,
\begin{equation}
P_N(x) = z Q_{N_f}(x)
\end{equation}
with $N>N_f$, which correspond to theories with $N_f$ flavor charges only, no gauge charges, and describe the general behavior of
2d-4d $A_n$ theories near loci where a certain number of mutually local bulk particles and domain walls between $N$ vacua
become simultaneously light. Indeed $x^2=z^2 + 2 m$ is equivalent to $x^2 - 2 m = 2 z x$. It would be interesting to study this general class, but
we will leave it for future work.

\item Although  we said that a bulk particle of flavor charge $\gamma$ is present,
 this is rather immaterial, unless one gauges that flavor symmetry,
as would happen
if this were a local model for a larger theory. Rather, what matters is that there
will be non-zero $\omega(\gamma,\gamma_{ij})$.
It is perfectly fine to interpret the results of this subsection as concerning
 the behavior of a certain 2d theory with 2 vacua, with 2d particles carrying a flavor charge $\gamma$.
It should not be difficult to engineer a theory, with the correct twisted
chiral ring relations as to reproduce the spectral curve of this Hitchin system.
We can get pretty close if we consider a $U(1)$ gauged linear sigma model,
with a chiral multiplet of charge $1$ and a chiral multiplet of charge $-2$.
Integrating away the massive chiral multiplets gives us a twisted superpotential
\begin{equation}
2 \sigma \log \sigma -(\sigma+m) \log (\sigma +m) - 2 \I \pi t \sigma
\end{equation}
 which is minimized if $\sigma^2 = e^{2 \pi \I t}(\sigma+m)$: if we were to
 set $z = e^{\I \pi t}$ and $\sigma = x z$, so that $2 \I \pi \sigma \de t \sim x \de z$
 then we get the desired curve. It would be interesting to compare the BPS
 spectrum of this theory with the one we discuss below.

\item We have not systematically investigated the framed BPS degeneracies in
this model, but we have given all the techniques needed to do so.

\end{enumerate}

\subsubsection{Larger $N$}\label{subsubsec:AD-LargerN}

At larger $N$, the picture becomes more intricate, but no essentially new wall-crossing phenomena happen:
all wall-crossings involving 2d particles
can be understood using the basic identities \eqref{eq:ks-ssk}, \eqref{eq:sk-kss}, and
\eqref{eq:sks-sks} we used above.

Let us describe a sample of the behavior one finds.
We begin from a
region of the Coulomb branch $\CB$ where all roots of $P(z)$ are real.  Here the 4d spectrum is easily described
\cite{Shapere:1999xr}:
finite WKB curves appear only between adjacent roots.  We denote their charges as
$\gamma^{s}$, $s = 1, \dots, N-1$, where  $s=1$ corresponds to the
\emph{rightmost} root. The intersection products are
  $\langle \gamma^s, \gamma^{s-1} \rangle = 1$ and $\langle \gamma^s, \gamma^{t} \rangle = 0$
  if $\vert s-t\vert > 1$.  The corresponding BPS rays $\ell_{\gamma^s}$
all lie either along the real or the imaginary $\zeta$ axis, for even and odd $s$ respectively. See,
for example, Figure \ref{fig:adn6-walls} for the case $N=6$.
(We have here a rather degenerate situation, where several BPS rays of different nonintersecting
charges coincide. Nevertheless,
this is not really a wall of marginal stability since the
corresponding charges have zero symplectic product with one another.)

\insfig{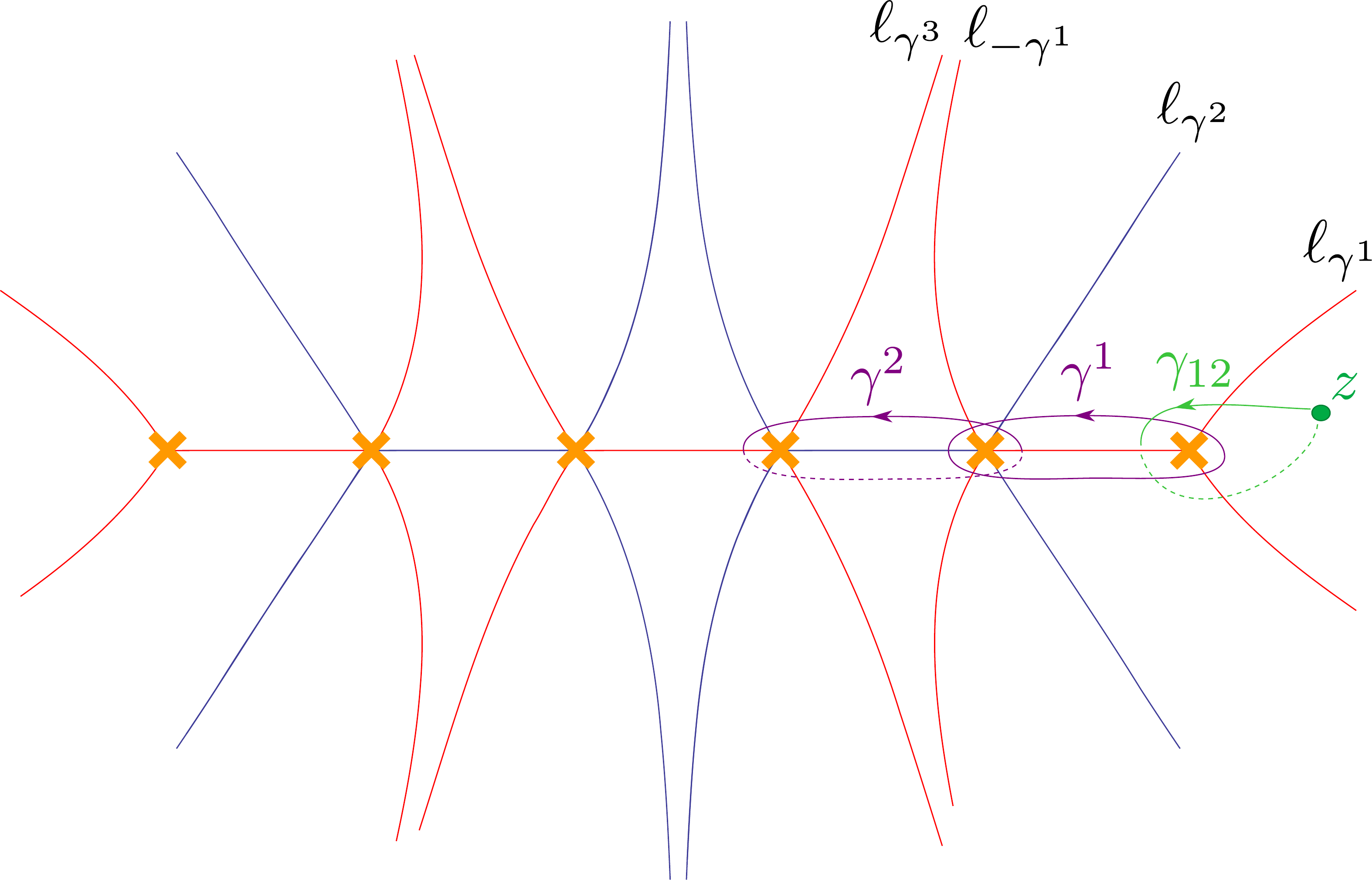}{The walls of marginal stability in $C$ for 2d-4d BPS states in the $N=6$ Argyres-Douglas theory,
when $u$ is in the region of the Coulomb branch where all roots of $P_6$ are real.  Finite WKB curves
join successive roots. The separating WKB curves asymptote to the fourteen lines of phase
$2\pi n/16$ not parallel to the $x$-axis. The red walls correspond to wall-crossings
involving BPS rays with phase $\vartheta = \pm \pi/2$ and the blue walls correspond to those rays with
$\vartheta=0, \pi$. We have shown $\gamma_{12}(z)$, $\gamma^1$ and $\gamma^2$ in some cut system,
and indicated the $\CK$-rays which become parallel with the $\CS$-ray $\ell_{\gamma_{12}(z)}$
for the first few walls.}

Now how about the spectrum with the surface defect $\bS_z$ included?
If $z$ is well to the right of all the roots, this spectrum just consists of
a single 2d particle, corresponding to
a finite WKB curve running from $z$
to the rightmost root.  Call its charge $\gamma_{12}$.  So we have an additional
BPS ray $\ell_{\gamma_{12}}$ carrying the transformation $\CS_{\gamma_{12}}$.
Furthermore there are no other open finite WKB networks starting from $z$.
In order to see this we use reasoning similar to that used near
Figure \ref{fig:N2-cells-noncritical}. At $\vartheta=\vartheta_*$ which supports a
4d BPS state the walls will look like those in Figure \ref{fig:adn6-walls}. For $\vartheta\not= \vartheta_*,
\vartheta_*+\pi$, the separating WKB curves will rotate but there will always be a
region far to the right containing the large positive real axis and sitting between
the rightmost separating WKB curves. Now, WKB rays can never cross, and
therefore,  if $z$  is a large positive number then for any
$\vartheta$ there can be at most one open finite  WKB network. On the other hand,
the walls of marginal stability are determined by the pattern at $\vartheta_*$.
Therefore, if $z$ is any point in the rightmost region of  Figure \ref{fig:adn6-walls}
 there is one and only one
open finite  WKB network connecting $z$ to the turning point $z_1$.
Moreover, $\omega(\gamma^s, \gamma_{12}) = 0$ for all $s$, except for $s=1$, for which
$\omega(\gamma^1, \gamma_{12}) = 1$.

\insfig{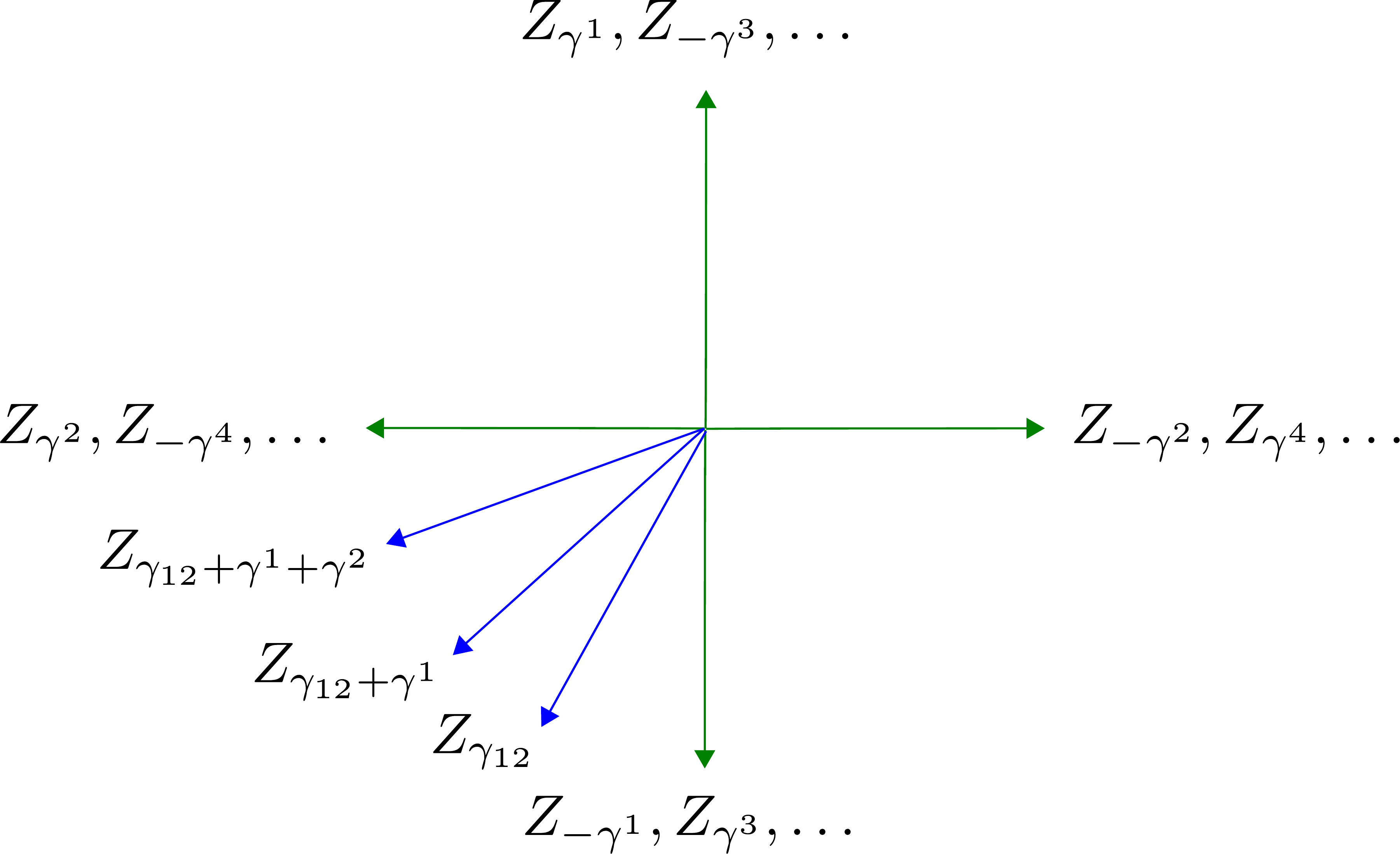}{When $z$ is large and lies in the region in the upper half plane between
the separating WKB curves associated to $\gamma^2$ and $-\gamma^1$, the soliton spectrum is that
shown here.}

Let us consider how the configuration of BPS rays evolves as $z$ is varied.
As we move $z$ counterclockwise around the configuration of roots of $P$, staying a large distance away,
the single $\CS_{\gamma_{12}}$ ray first meets a BPS ray carrying the transformation
$\CK_{\gamma^1}$ (and also carrying various other $\CK$ transformations corresponding to other 4d BPS states,
but those transformations commute with $\CS_{\gamma_{12}}$, so they play no role at this moment.)
Applying the WCF \eqref{eq:sk-kss}, one sees that after these two rays cross,
the $\CS_{\gamma_{12}}$ ray is replaced by two rays, carrying $\CS_{\gamma_{12}}$ and $\CS_{\gamma_{12}+ \gamma^1}$.
As we continue to move $z$, these two rays next cross $\CK_{\gamma^2}$. Since $\gamma_{12}$ has zero
intersection with $\tilde \gamma_2$ there is no wall-crossing as the ray for $\gamma_{12}$
passes through, but this is not so for $\gamma_{12}+\gamma^1$.
Indeed, using again \eqref{eq:sk-kss} with $a=\gamma^2$ and $b=\gamma_{12}+\gamma^1$
wall-crossing produces  three rays,
$\CS_{\gamma_{12}}$, $\CS_{\gamma_{12}+ \gamma^1}$, $\CS_{\gamma_{12}+ \gamma^1+\gamma^2}$, leading to the
configuration of rays shown in Figure \ref{fig:adn6-sample}. Next, $z$ crosses a ray for $-\gamma^1$ and $\gamma^3$.
In our degenerate situation with large $z$ these happen almost simultaneously. First the rays for
$\gamma_{12}(z)+\gamma^1$ and $\gamma_{12}(z)$ cross the ray with factor $\CK_{-\gamma^1}$ and we apply the wall-crossing formula \eqref{eq:ks-ssk}:
\begin{equation}
\CS_{\gamma_{12}+\gamma^1 + \gamma^2} \CS_{\gamma_{12}+\gamma^1} \CS_{\gamma_{12}} \CK_{-\gamma^1}\CK_{\gamma^3}
= \CS_{\gamma_{12}+\gamma^1 + \gamma^2} \CK_{-\gamma^1} \CS_{\gamma_{12}+\gamma^1}  \CK_{\gamma^3}.
\end{equation}
Next, since $\omega(-\gamma^1, \gamma_{12}(z)+\gamma^1 + \gamma^2)=0$, when the ray with charge
$\gamma_{12}+\gamma^1 + \gamma^2$ passes through $\CK_{-\gamma^1}$ we simply commute factors:
\begin{equation}
  \CS_{\gamma_{12}+\gamma^1 + \gamma^2} \CK_{-\gamma^1} \CS_{\gamma_{12}+\gamma^1}  \CK_{\gamma^3}
  = \CK_{-\gamma^1} \CS_{\gamma_{12}+\gamma^1 + \gamma^2} \CS_{\gamma_{12}+\gamma^1}   \CK_{\gamma^3}.
\end{equation}
Next as $z$ crosses the separating line for $\gamma^3$, $\CS_{\gamma_{12}+\gamma^1}$ commutes through
(again because intersection products vanish) and we apply \eqref{eq:sk-kss} to produce
\begin{equation}
\CK_{-\gamma^1} \CS_{\gamma_{12}+\gamma^1 + \gamma^2} \CS_{\gamma_{12}+\gamma^1}   \CK_{\gamma^3}
=\CK_{-\gamma^1}  \CK_{\gamma^3}  \CS_{\gamma_{12}+\gamma^1 + \gamma^2+ \gamma^3} \CS_{\gamma_{12}+\gamma^1 + \gamma^2} \CS_{\gamma_{12}+\gamma^1}.
\end{equation}
Thus the 2d particle of charge $\gamma_{12}$ decays into a 2d particle of charge
$\gamma_{12} + \gamma^1$ plus a 4d particle of charge $-\gamma^1$,
and almost simultaneously a new 2d particle of charge $\gamma_{12}+\gamma^1+\gamma^2+\gamma^3$ is formed as a bound state of a 2d $\gamma_{12}+\gamma^1+\gamma^2$
and a 4d $\gamma^3$.   The pattern continues as $z$ moves across the upper half of the plane:  we encounter a series of pairs of walls at which
one state decays and a new one is born.  At the last two walls we have only decays with no new states created, so
when $z$ reaches a point far to the left of all of the roots of $P$, we find just a single BPS state.
We can then bring $z$ back around the circle to its original position:  we encounter a sequence of walls very similar
to what we just described.
See Figure \ref{fig:adn6-walls} for a picture of the walls in the $z$-plane in the case $N=6$.
As in the discussion of $N=2$ above, when $z$ comes back to the original point we must
have only the original soliton, but to see this explicitly as the outcome of a composition of wall-crossings
requires us to take account of the monodromy of $\Gamma_{12}$ around the loop.
\insfig{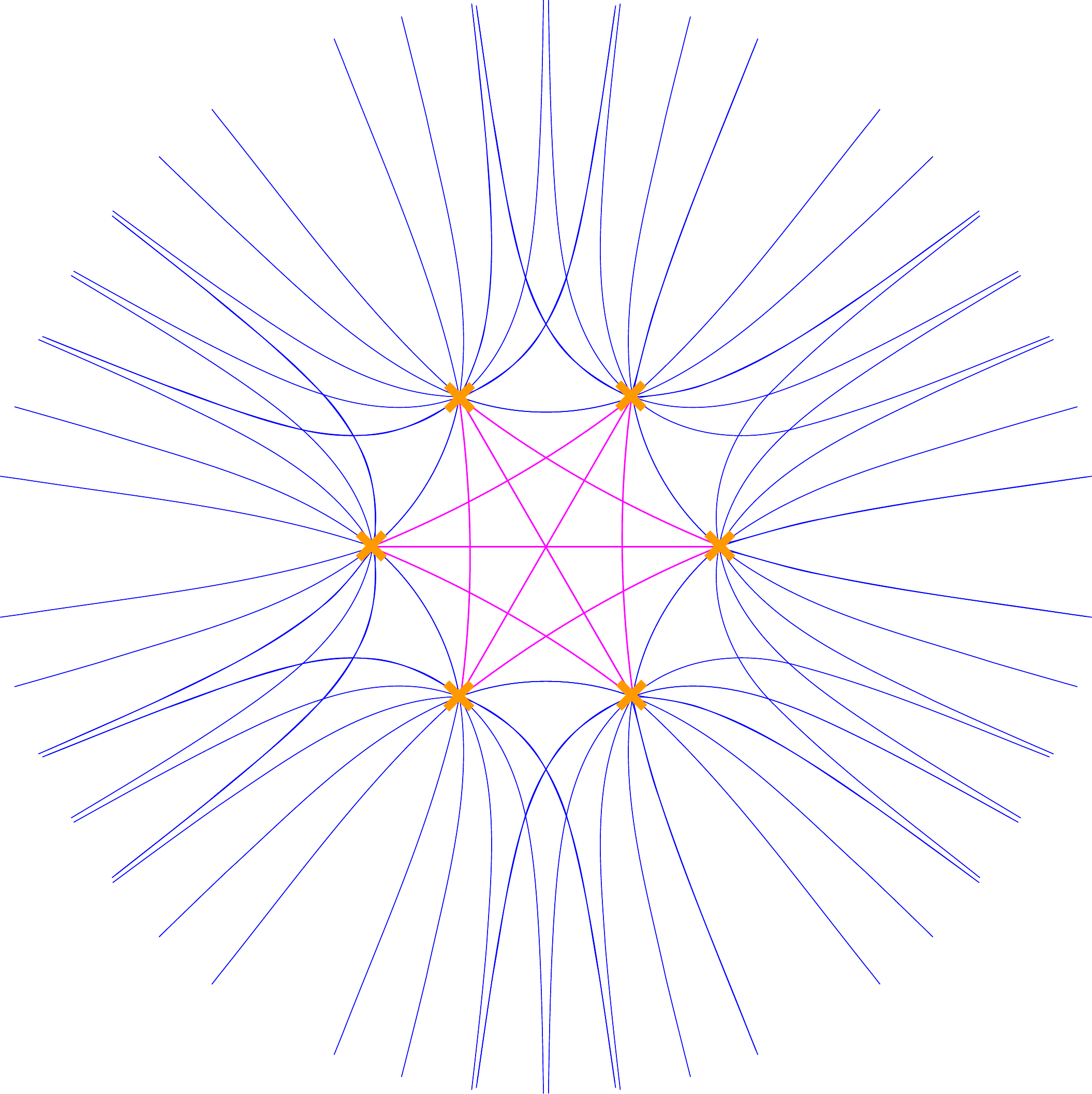}{Walls of marginal stability for the $N=6$ AD theory
at a particular point $u$ in $\CB$ as described in the text.}
We emphasize that this is only the picture at one specific locus in the Coulomb branch; as we vary $u \in \CB$
the walls would deform and even change topology.
One could encounter more complicated patterns of wall-crossing in these cases. As a simple
illustration of this we plot the walls of marginal stability for the $N=6$ theory
for $\lambda^2 = (z^6-1)(\de z)^2$ in Figure \ref{fig:adn6-wkb-symmetric}.   Still, we believe that
one could study all
the wall-crossing phenomena systematically using the same basic wall-crossing formulas we used above.

It is worthwhile discussing the case $N=3$ a bit further.  In this example the
Coulomb branch $\CB$ is one-dimensional and has a simple singularity structure.
We take the polynomial $P_3(z) = z^3 - 3 \Lambda^2 z + u$.  The Coulomb branch for the
bulk theory has two singular loci $u_\pm$, where an ``electron'' and a ``monopole'', of
charges $\gamma_e$ and $\gamma_m$, $\langle \gamma_e, \gamma_m\rangle =1$,
are respectively massless. There is a single wall of marginal stability,
separating an inner region where the spectrum consists of the two particles only,
and an outer region where the spectrum consists of three particles, of
charges $\gamma_e$, $\gamma_m$, $\gamma_e + \gamma_m$.
Notice that the monodromy at infinity which arises from the combination
of the monodromies around the two singular points
is $(\gamma_e, \gamma_m,\gamma_e+\gamma_m) \to (\gamma_e - \gamma_m, \gamma_m, \gamma_e)
\to (- \gamma_m, \gamma_m + \gamma_e, \gamma_e)$,
a cyclic permutation of the three particles.

If we set $\Lambda=0$ the two singular loci coalesce into a AD point, the inner region disappears, and the setup
has an exact $\IZ_3$ symmetry acting on the $z$ plane which coincides with the cyclic permutation of the populated
charges $\gamma_e$, $\gamma_m$, $\gamma_e + \gamma_m$. If we set the surface defect parameter at the origin, $z=0$,
the spectrum of 2d solitons must also enjoy this symmetry. Indeed, it is not difficult to argue
that there are three 2d solitons, whose charges correspond to three straight WKB segments from $z=0$ to the three turning points $z=u^{1/3}$. Even if $\Lambda \neq 0$, this is the general spectrum
for sufficiently large $u$. It is not difficult to fill in the full spectrum for large $|z|$ as a function of $u$, interpolating from very large to small $|u|$, but we will not describe the details here.
We simply note that the three 2d solitons at very large $|u|$ can be continuously connected to the soliton of charges
$\gamma_{ij}$, $\gamma_{ij} + \gamma_e$, $\gamma_{ij} + \gamma_e + \gamma_m$ which we encountered at the third step of the general $N$ analysis.

\subsection{The $\IC \IP^1$ sigma model}\label{subsec:cp1-sigma}

Now let us consider another example where the 4d dynamics is trivial, namely the 2d $\IC\IP^1$ sigma model.
This model nicely illustrates the examples related to degenerate cells and pop transitions
discussed in \S \ref{subsec:WKB-2d-4d}. It is also useful preparation for the
rather more challenging example of the four-dimensional $SU(2)$ $N_f=0$ theory
coupled to the  the $\IC \IP^1$ sigma model, to be discussed in \S \ref{subsec:WC-exple-su2}.

The BPS spectrum in this model has been studied by Dorey \cite{Dorey:1998yh}. Let us
briefly recall the salient results. The chiral ring with
twisted mass parameter $m$ is
\begin{equation}\label{eq:chir-ring}
x^2 = \Lambda^2 e^t + m^2,
\end{equation}
where $t$ is the K\"ahler parameter of the target $\IC \IP^1$.
The twisted mass defines two vacua which we can take to be the north and south poles of the
target space $\IC \IP^1$. This mass breaks the $SU(2)$ global symmetry (coming from the
isometries of the target space) to $U(1)$. We write the root lattice of
$SU(2)$ as $\IZ \alpha$ and measure $U(1)$ charges in terms of $\alpha$.
In the strong coupling region,
defined by  $|m^2| \ll |\Lambda^2 e^t|$, the effect of the twisted mass is negligible
and there are two solitons interpolating between the vacua.
At $m=0$ they form a doublet and hence have $U(1)$ charges
$\pm \frac{1}{2} \alpha$. On the other hand, in the weak coupling
region, defined by
$\lvert m^2 \rvert \gg  \lvert \Lambda^2  e^t \rvert$,
the twisted mass localizes the dynamics to one of the two vacua at the
north or south poles of $\IC \IP^1$.
The massive sigma-model fluctuations around those vacua give 2 BPS particles of central charge $Z = \I m$
together with their antiparticles. There is also an infinite tower of
semi-classical solitons interpolating between the two vacua.  These solitons
(and their antiparticles)
carry global $U(1)$ charge $(n+\frac{1}{2})\alpha$ for all integers $n\in \IZ$,
with degeneracy $\mu=1$ for each $n$.  Evidently, there is a marginal stability transformation
in the BPS spectrum, very reminiscent of the strong-weak coupling transition in the four-dimensional
$SU(2)$ theory \cite{Dorey:1998yh,Tong:2008qd}.

It is instructive to see how these results are reproduced in the geometric formulation
of this paper. We take $C=\IC \IP^1 -\{0,\infty\}$. ($C$ is not related
to the $\IC \IP^1$ target of the sigma model!)
The chiral ring corresponds to the spectral cover equation
\begin{equation}
\lambda^2 = \left( \frac{\Lambda^2}{z} + \frac{m^2}{z^2} \right) \de z^2
\end{equation}
where $z=e^t$ and $\lambda = x\,\de t$.
This covering is the spectral cover of
an $A_1$ Hitchin system on $C $
with the weakest possible irregular singularity at infinity, and a regular singularity with residue
$m \sigma^3$ at the origin.
The Hitchin equations in this system coincide with the $tt^*$ equations for the $\IC \IP^1$ sigma model.
The double covering has two branch points: one  at $z=z_0=-\frac{m^2}{\Lambda^2}$ and one at $z=\infty$, so
$\Sigma$ is a thrice-punctured sphere. Let $P^\pm$ be the two lifts of $z=0$.
The homology of $\Sigma$ is generated by $\gamma$, a small curve winding once
around $P^+$, and $\sigma^*(\gamma)$, a small curve winding once around $P^-$.
The class $\gamma_f:=\gamma - \sigma^*(\gamma)$ generates the
odd homology lattice $\Gamma = H_1(\Sigma;\IZ)^-$ and corresponds to
a closed curve around $z=\infty$. We will identify $\Gamma$ with the
flavor lattice $\alpha \IZ$ and hence $\alpha$ with $\gamma_f$.

\insfig{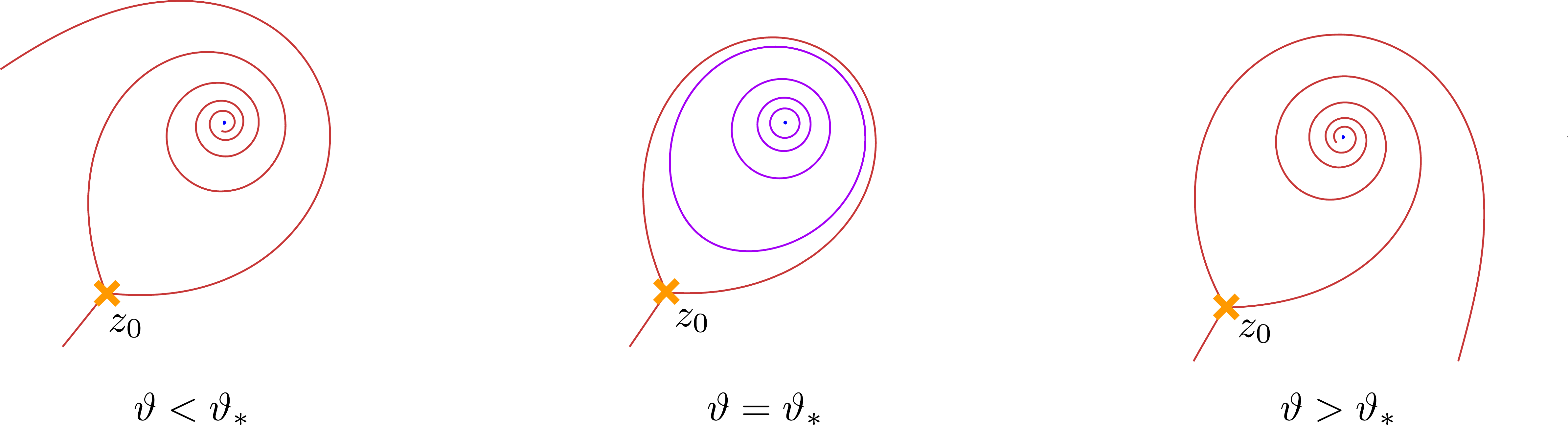}{At $\vartheta=\vartheta_*$ there is a closed WKB curve
connecting the turning point $z_0$ to itself. For $\vartheta\not=\vartheta_*$
the curves in red are separating WKB curves connecting $z_0$ to a singularity.}
Let us now derive the BPS spectrum geometrically. As in the previous examples this can
be inferred from the structure of the WKB foliations, shown in this case in
Figure \ref{fig:spiral-jump}. At the critical values $\vartheta = \vartheta_*, \vartheta_*+\pi$
given by $e^{\I \vartheta_*} = \pm \I \frac{m}{\vert m\vert}$
there is a finite WKB curve joining $z_0$ to itself, as in the
middle of Figure \ref{fig:spiral-jump}. As we have explained above the finite and
separating WKB curves at this critical value of $\vartheta$ are the marginal stability
lines in $z$.
The finite WKB curve at $\vartheta_*$ separates $C$ into two regions,
with the inner region near $z=0$ corresponding to weak coupling and the outer
region corresponding to strong coupling.

If $z$ is any point in the strong coupling
region there are precisely two special values of $\vartheta$ for which there is an open finite
WKB curve connecting $z$ to $z_0$.
Moreover, the open paths $\gamma_{12}$ and $\tilde\gamma_{21} $
corresponding to these two curves glue together to form $\gamma_{12}+ \tilde\gamma_{21}=\gamma_f$.
These BPS states, with $\mu(\gamma_{12})=\mu(\tilde \gamma_{21})=1$, correspond to the two strong-coupling
solitons with flavor charges $ \half \alpha$. Together with their antiparticles we
have two $SU(2)$ doublets. Note that $\tilde\gamma_{21} = \gamma_{21}+\gamma_f = -\gamma_{12} + \gamma_f$.
Thus we can identify both $\Gamma_{12}$ and $\Gamma_{21}$ with   the torsor $\half \alpha + \Gamma$.

Now suppose $z$ is any point in the weak coupling region.  In this case the open path $\gamma_{12}$
has intersection $2$ with $\gamma_f$ (where we choose a specific representative for $\gamma_f$,
namely the difference of lifts of the closed WKB curve forming the boundary of the punctured disc.)  We thus
have $\omega(\gamma_f, \gamma_{12}) = 2$.
Dividing this up symmetrically requires us to
assign $\omega(\gamma_f, \gamma_1) =  1$ and $\omega(\gamma_f,\gamma_2)=-1$.
So we find one particle of charge $\gamma_f$ in each vacuum.  Similarly we find one particle of charge $-\gamma_f$
in each vacuum.
This fits with the field-theoretic expectation:  each of the two vacua supports a particle of charge $\gamma_f$,
as well as its antiparticle of charge $-\gamma_f$.
Incidentally, there is another way of picturing these BPS states:  rather than thinking about the intersection between
$\gamma_{12}$ and a closed WKB curve at the boundary of the disc, which seems a bit indirect, we observe
that $z$ itself is lying on a closed WKB curve.  The two BPS particles in vacuum $1$ can be naturally identified
with the two possible orientations of the lift of this closed WKB curve to sheet $1$ of $\Sigma$, and similarly
for vacuum $2$.

Moreover, for any point $z$ in the weak coupling region there will be infinitely many
values of $\vartheta$, accumulating at the critical value $\vartheta_*$ from above and from
below, such that the spiraling WKB curves join $z$ to $z_0$.  These correspond to the
infinite tower of particles with flavor charges
$\gamma_{12}+n\gamma_f$ and $\gamma_{21}+n \gamma_f$ where $n\in \IZ$ corresponds to the winding number around
the singularity at  $z=0$.

Exactly at $\vartheta = \vartheta_*$ there is a separating WKB curve joining $z_0$ to $z=\infty$.
However, the relevant intersection product is zero, so this WKB curve does not contribute a BPS state.

We have thus reproduced the BPS spectrum of the model. Moreover, the wall-crossing
in the $z$ parameter
takes place along the finite WKB curve with $e^{\I \vartheta}= \pm \I m/\vert m\vert$.
The   corresponding wall-crossing formula is in perfect correspondence
with  \eqref{eq:wallcp1} if we translate $\gamma_{ij} \to \gamma_{12}$,
$\gamma_{ji}\to \tilde\gamma_{21}$ and $\gamma \to \gamma_f$.

With a bit more patience, we could also explore the spectrum of framed BPS states
for simple line defects in the model. We will only sketch the analysis.
As we have an irregular singularity of odd degree,
we can consider the associted small flat section $s(z)$, canonically normalized
by the requirement $(s, M s)=1$. Here $M$ is the monodromy matrix which represents parallel transport
once around the cylinder.  To be precise, we need to pick a path from the irregular singularity to $z$ in order to define $s(z)$.  Other choices of paths
winding $n$ more times around the cylinder will give sections $s_n(z) = M^n s(z)$ for all integer $n$. Notice that $(s_n,s_{n+1})=1$.

Standard Stokes theory always implies linear relations of the form
\begin{equation}
s_{n-1} + x_n s_n + s_{n+1} = 0
\end{equation}
for some scalar $x_n$. In this simple setup, all the $x_n$ are equal, and
\begin{equation}
x_0 s = - \left(M + M^{-1} \right) s = - \left( {\mathrm{Tr}} M \right) s.
\end{equation}
But the matrix $M$ is just the monodromy matrix around the regular singularity, and has
eigenvalues $\CY_{\pm \frac{1}{2} \gamma_f}$. Hence
\begin{equation}
x_n = x_0 = -\CY_{\frac{1}{2} \gamma_f} - \CY_{-\frac{1}{2} \gamma_f}.
\end{equation}
Finally, notice that
\begin{equation}
M\left(M s - \CY_{\pm \frac{1}{2} \gamma_f} s \right) = \CY_{\mp \frac{1}{2} \gamma_f}  M s - s.
\end{equation}
Hence the vectors $s_\mp = M s - \CY_{\pm \frac{1}{2} \gamma_f} s$ are monodromy eigenvectors
with eigenvalue $\CY_{\mp \frac{1}{2} \gamma_f} $.

We can naturally conjecture that the $s_n(z,\zeta)$ are vevs of a boundary line defect $L_n$ in the
$\IC \IP^1$ sigma model, i.e. a brane in the 2d sigma model. We can make an educated guess on the nature of the $L_n$
brane: a space-filling brane (Neumann boundary conditions for the sigma model) with a Chan-Paton $U(1)$ bundle with $n$ units of
flux. A first check is the observation that a shift of the sigma model $B$-field $z \to e^{2 \pi \I} z$ sends $s_n \to s_{n+1}$, and hence it should
induce a monodromy $L_n \to L_{n+1}$. But a $B$-field shift is indeed gauge equivalent to a shift of one unit in the Chan-Paton $U(1)$ bundle flux
of all branes.

A more refined check is to derive the spectrum of framed BPS degeneracies associated to
the $\IC \IP^1$ sigma model on a segment, with boundaries $L_0$ and $L_n$, by expanding
$(s_0,s_n)$ recursively:
\begin{equation}
(s_0,s_1)=1, \qquad (s_0,s_2) = \CY_{\frac{1}{2} \gamma_f} + \CY_{-\frac{1}{2} \gamma_f}, \qquad (s_0,s_3) = \CY_{ \gamma_f} +1+\CY_{- \gamma_f}.
\end{equation}
Generally
\begin{equation}
(s_0,s_n) = \frac{\CY_{ \frac{n}{2} \gamma_f} -\CY_{- \frac{n}{2}  \gamma_f}}{ \CY_{\frac{1}{2} \gamma_f} - \CY_{-\frac{1}{2} \gamma_f}}.
\end{equation}
Hence the framed BPS Hilbert space contains $n$ states, with the quantum numbers of a irreducible
representation of the $SU(2)$ flavor symmetry of the model. We can give a physical interpretation of this calculation.
We can reduce the $\IC \IP^1$ sigma model on the segment
to the supersymmetric quantum mechanics of the zeromode, in the presence of the $n$ units of Chan-Paton $U(1)$ bundle flux
on the $\IC \IP^1$ target space. It is natural for the ground states of this problem,
the framed BPS states, to form an irreducible $SU(2)$ multiplet of $n$ lowest Landau levels.

Finally, the cells of the WKB triangulation are associated to pairs of sections with good asymptotics, either of the form
$(s_\pm; s_n)$ or of the form $(s_n,s_{n+1})$, which should be identified with the $\CY_{\gamma_i}$ for appropriate cycles $\gamma_i$.
One can easily expand $s_{n'}(z)$ in any of those bases, and derive the framed BPS spectrum for the theory on the half line,
with $L_{n'}$ boundary conditions. It would be interesting to match it to an explicit calculation.
Also, it is conceivable that $s_\pm$  could themselves be identified with the vevs of certain boundary line defects
$L_\pm$. The relation $s_\mp = M s - \CY_{\pm \frac{1}{2} \gamma_f} s$ would seem to relate $L_\pm$ to the ``difference''
of space-filling brane branes  with one or zero units of flux. This suggests the identification of the tentative
$L_\pm$ with point-like branes  (Dirichlet boundary conditions for the sigma model), possibly located at the North and South poles
of $\IC \IP^1$ because of the twisted mass in the Lagrangian.

\subsection{The canonical surface defect in pure $SU(2)$ gauge theory}\label{subsec:WC-exple-su2}

We are now ready to deal with the most complicated example in this paper, the canonical surface defect $\bS_z$
in the pure $SU(2)$ gauge theory.  The spectral curve in this case is
\begin{equation} \label{eq:spectral-su2}
\lambda^2 = \left( \frac{\Lambda^2}{z^3} + \frac{2u}{z^2} + \frac{\Lambda^2}{z} \right) \de z^2.
\end{equation}

Since the analysis to follow is long and technical let us summarize the basic points
and lessons first here.
We will first analyze the soliton spectrum at strong four-dimensional coupling. We will find that
in the $z$-plane there are a finite number of domains in each of which there is a finite
soliton spectrum. We will derive the spectrum of the   BPS degeneracies $\mu$ and $\omega$
at fixed $u$ as functions of $z$.  We will   check that the 2d-4d wall-crossing
formulae are consistent with the monodromy of the local system $\Gamma_{12}$ over the $z$ plane.
 Then we will find a
convenient regime in which to fix $z$ and continue $u$ from the strong coupling
domain to the weak coupling domain. Since the 4d spectrum changes (dramatically)
the walls of marginal stability in the $z$ plane also change (dramatically).
There is now an extremely complicated pattern of walls of marginal stability in the
$z$ plane (because there are infinitely many 4d bulk particles). We will describe some
aspects of the resulting chambers in the $z$-plane qualitatively and check several
nontrivial aspects of the wall-crossing phenomenon and soliton spectrum without
giving a full description of $\mu$ and $\omega$ for $u$ in this weak-coupling domain.
In particular, at weak four-dimensional coupling there are two very different regimes.
The weak two-dimensional coupling regime is defined by taking  $z$ to be in the ``vectormultiplet
annulus,'' that is, the annulus foliated by the finite WKB curves arising when $\vartheta$ is the
phase of the vectormultiplet central charge. The strong two-dimensional coupling regime
is then the complementary region in the $z$-plane. In the strong two-dimensional
regime the soliton spectrum is finite in any chamber, but there are a countably infinite number of
chambers and the spectrum is unbounded. In the weak two-dimensional regime the
soliton spectrum is infinite and chamber-independent. Nevertheless, there is an uncountable number
of chambers, and the BPS degeneracies $\omega$ depend on these chambers.

\subsubsection{Preliminaries on local systems}

\insfig{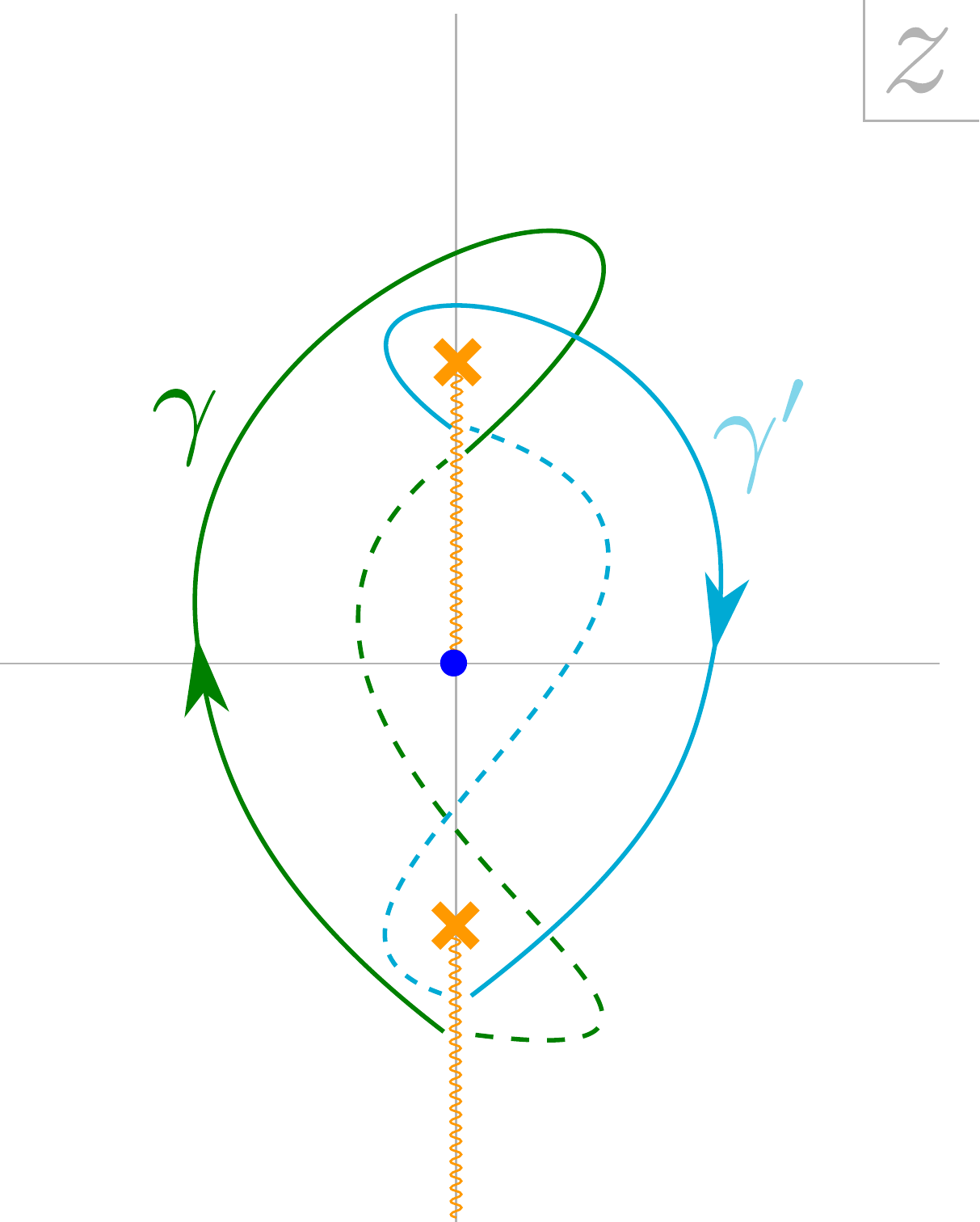}{A basis for $H_1(\bar\Sigma;\IZ)$ at $u=0$. We choose branch cuts
emanating from $z=\I$, $z=0$ and $z=-\I$ and running along the imaginary axis.
$Z_{\gamma} = K$ and $Z_{\gamma'} = -\I K$ for $K$ a positive constant.  The cycles
$\gamma$ and $\gamma'$ have $\inprod{\gamma, \gamma'} = 2$.}

Let us begin with some technical preliminaries.
Some aspects of the theory have already been discussed in \S \ref{subsubsec:ExampleSU2-CP1}.
In comparing with
\S \ref{subsubsec:ExampleSU2-CP1} one should take $u\to - u$ and $z=e^{t}$.
Above, we described the
local system $\Gamma_{12}$ over the $u$-plane for fixed $u$.  We also
described the local system $\Gamma$ over the $u$-plane $\CB^*$.
However, in this section it will be convenient to describe the local system $\Gamma \to \CB^*$ in a slightly
different way.  We begin with $u=0$ (strong 4d coupling) and
choose cuts for the covering \eqref{eq:spectral-su2} as in Figure \ref{fig:uzerocuts}. As shown
in Figure \ref{fig:uzerocuts} we have chosen cycles $\gamma$ and $\gamma'$ with $\langle \gamma, \gamma'\rangle=2$.
(We can map to \S \ref{subsubsec:ExampleSU2-CP1} by identifying $\gamma_m = \gamma'$
and $\gamma_e -\sigma^*(\gamma_e) = (\gamma \pm \gamma')$.
The choice of sign depends on how we continue $t_\pm$ to the imaginary axis;
the fact that there is a choice reflects the fact that monodromy of the local system on the $u$-plane can
take one form into the other.)  The 4d spectrum for $u$ in the
strong coupling regime has $\Omega(\pm \gamma) = \Omega(\pm \gamma') = 1$,
and all other degeneracies vanish. We will continue into the weak coupling
regime; the weak coupling spectrum consists of a
vector multiplet of charge $\pm(\gamma+\gamma')$ and
$\Omega = -2$ and two infinite towers of hypermultiplets of charge
$\gamma_n  :=\gamma + n (\gamma+\gamma')$ and $\gamma'_n := \gamma' + n(\gamma+\gamma')$,
$n\in \IZ$.  Note that $-\gamma_n = \gamma'_{-(n+1)}$.

\insfig{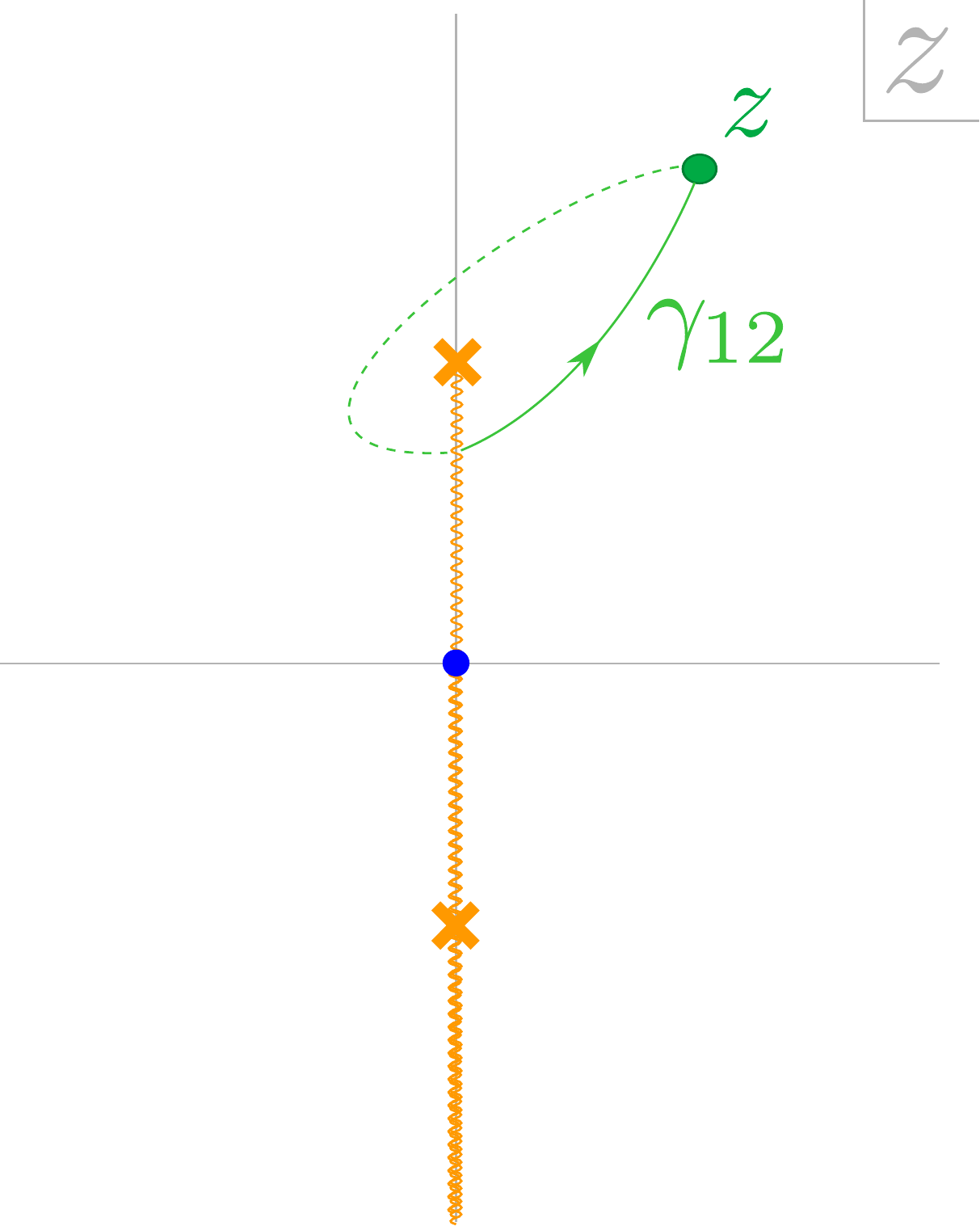}{We choose $\gamma_{12}(z)$ as shown for $z$ in the region $U$. The local system is
trivialized on the complement of the cuts running down the imaginary axis, starting from $z=\I, 0, -\I$.}

\insfig{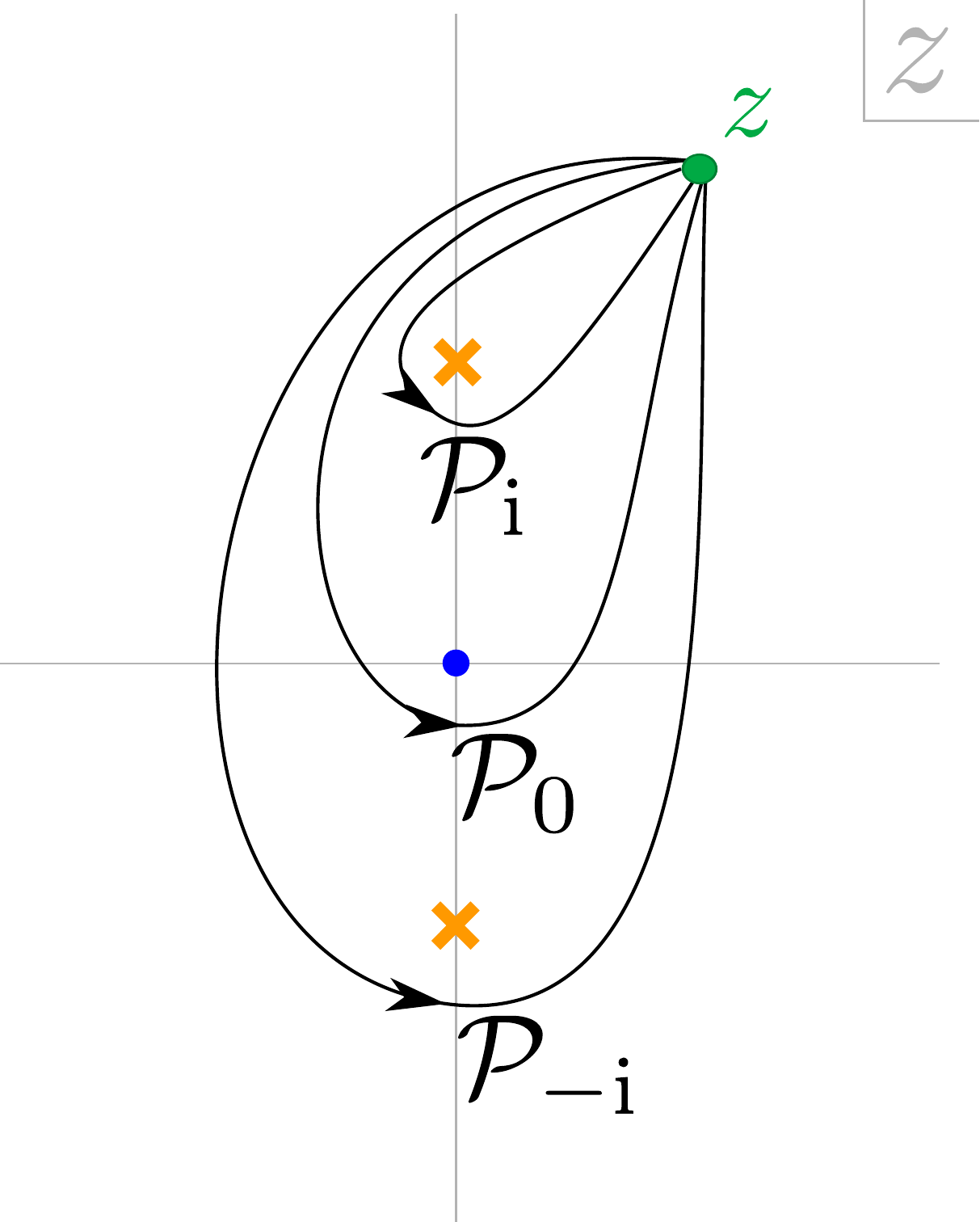}{The local system $\Gamma_{12}$ in the $z$-plane is determined by the monodromy around the
three paths $\CP_{\I}$, $\CP_0$, $\CP_{-\I}$ shown.}

We will also need some facts about the local system $\Gamma_{12}$ in the $z$-plane. We
define a class $\gamma_{12}(z)$ as in Figure \ref{fig:gamma12def}. We can
trivialize $\Gamma_{12}$ on the complement of the cuts shown there.
Using these pictures one can easily check that
for $u=0$ and $\im z$ sufficiently large and positive we have
$\omega(\gamma, \gamma_{12}(z)) = \omega(\gamma', \gamma_{12}(z))=+1$.

The monodromy of $\Gamma_{12}$ around the three paths shown in Figure \ref{fig:gamma12locsys} is
\begin{equation}\label{eq:PPP-mon}
\begin{split}
\CP_\I: \gamma_{12} & \to  \gamma_{21}, \\
\CP_0: \gamma_{12} & \to \gamma_{12} - \gamma - \gamma', \\
\CP_{-\I}: \gamma_{12} & \to \gamma_{21} + \gamma' - \gamma.
\end{split}
\end{equation}
The most straightforward way to prove these equations is simply to transport $\gamma_{12}(z)$
around the respective paths and
compute intersection products with $\gamma, \gamma'$. Alternatively, if we continue
$z$ along $\CP_0$ to $z\approx 0$, we will find that  $Z_{\gamma_{12}(z)+ \half (\gamma+\gamma')} \to 0$
as $z\to 0$ with a $\IZ_2$ branch cut. Therefore, a cycle that only encircles $z=0$
leads to a reflection
\begin{equation}
\left( \gamma_{12}(z) + \half (\gamma+\gamma')\right) \to - \left( \gamma_{12}(z) + \half (\gamma+\gamma')\right).
\end{equation}
In a similar way we find that continuation along a path $\CP_{-\I}$
gives $Z_{\gamma_{12}(z)+\gamma}\to 0$ for $z\to -\I$.

\subsubsection{The soliton spectrum at strong coupling}

\insfig{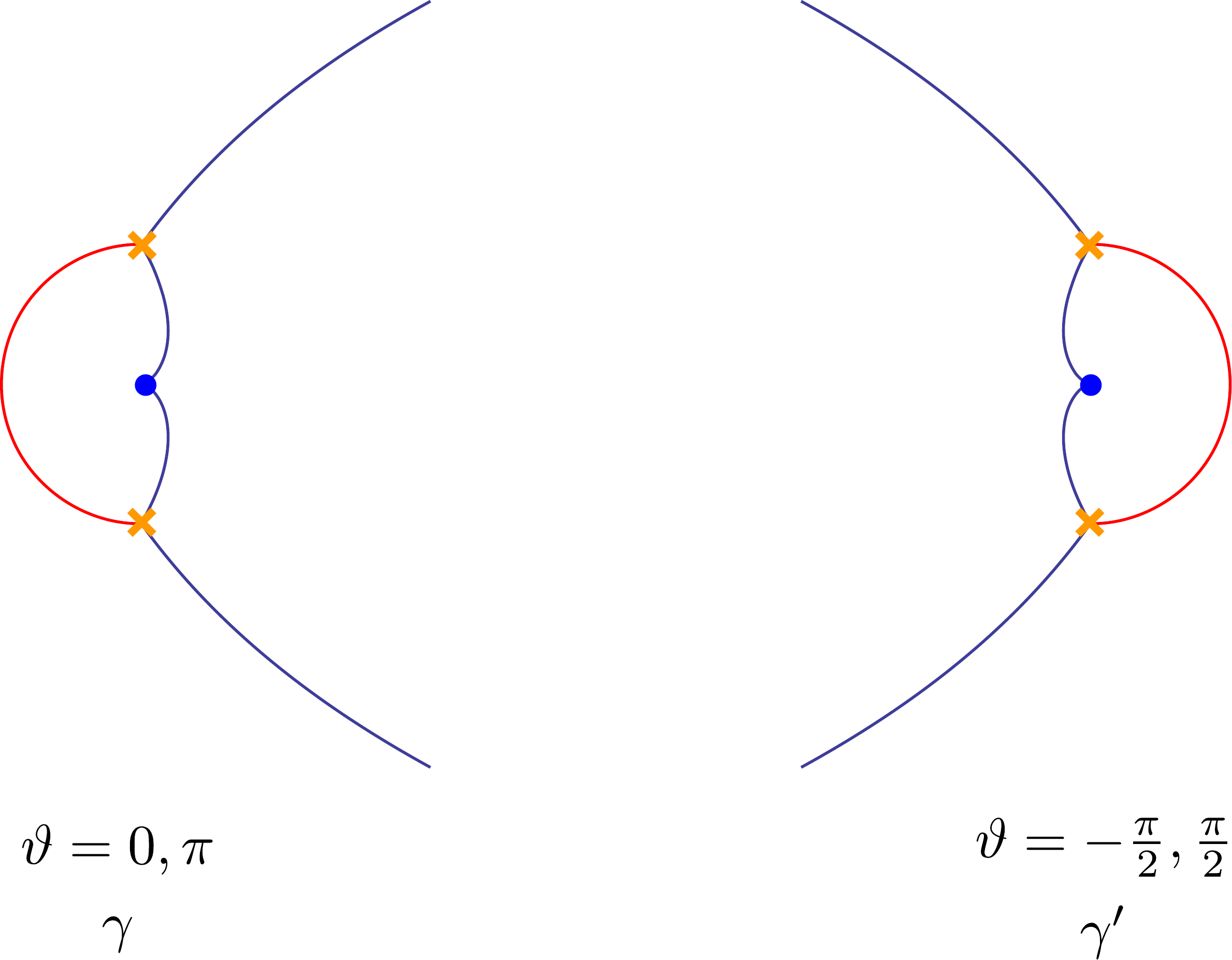}{The critical WKB curves for the monopole and dyon, of charges $\pm \gamma$ and $\pm \gamma'$
respectively. The turning points are at $z=\pm \I$. The separating WKB curves asymptote to horizontal lines.}

\insfig{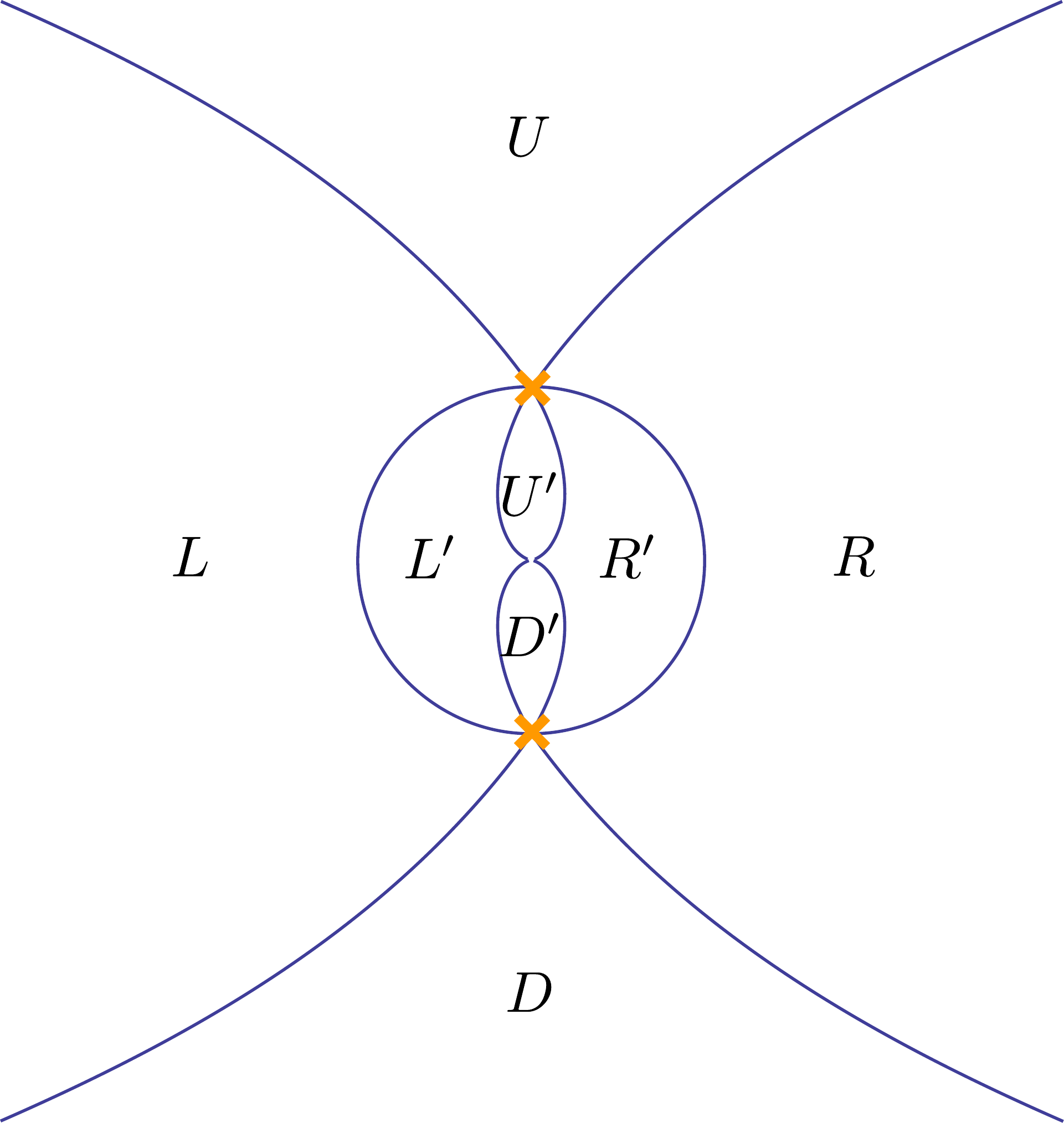}{Lines of marginal stability in the $z$-plane, for $u$ in the
strong coupling regime ($u=0$ in this figure). The region $R'$ is the image of
$R$ under $z \to 1/z$.}

Given the strong coupling 4d spectrum, the walls of marginal stability in the $z$-plane are obtained from the
finite and separating WKB curves at the critical values $\vartheta = \arg \pm Z_{\gamma}$
and $\vartheta = \arg \pm Z_{\gamma'}$.  For the case $u=0$
these are shown in Figure \ref{fig:mdwkbcrit}. The complement of the
union of these critical curves is the union of connected
regions $R$, $U$, $L$, $D$, and their respective reflected images $R'$, $U'$, $L'$, $D'$.
The wall-crossing analysis will be very similar to that in
\S \ref{subsubsec:ADN=2}, but now with \emph{two} flavor charges $\gamma$ and $\gamma'$.

\insfig{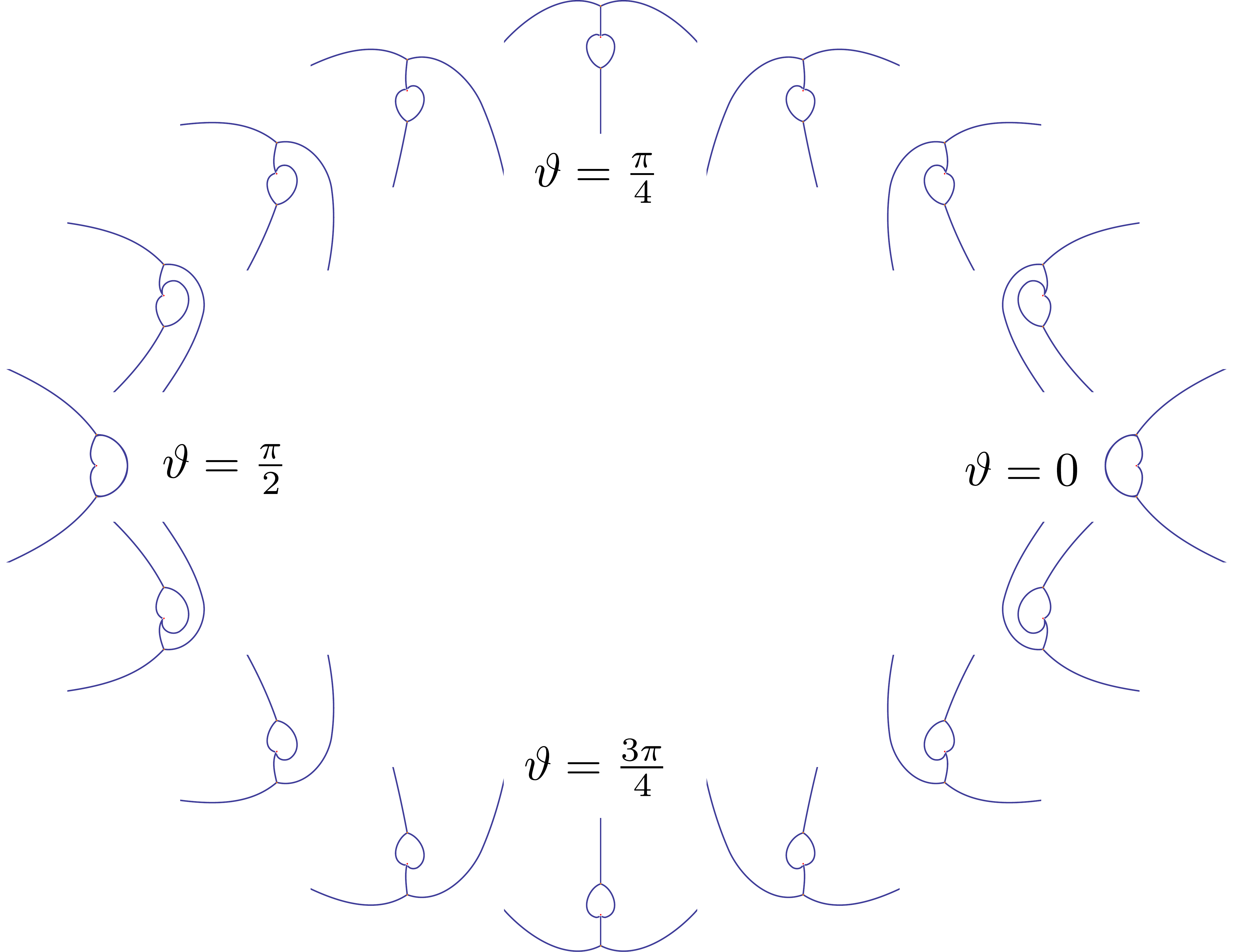}{The separating WKB curves as a function of $\vartheta$ at $u=0$. The critical
WKB curves for 4d particles are shown at $\vartheta=0$ and $\vartheta=\frac{\pi}{2}$. The unoriented WKB curves are
the same for $\vartheta$ and $\vartheta+\pi$. Therefore, the figure only shows
the curves as $\vartheta$ varies from $0$ to $\pi$. The separating WKB lines going to infinity eventually become
parallel. Therefore, from the above configurations of these separating WKB lines we can
deduce that when  $z$ is in the region $U$ or $D$ there are precisely three finite open WKB curves connecting $z$ to the turning point.
When $\vert z \vert$ is large, the corresponding phases $\vartheta$ of the three finite open WKB curves are close together.}

\insfig{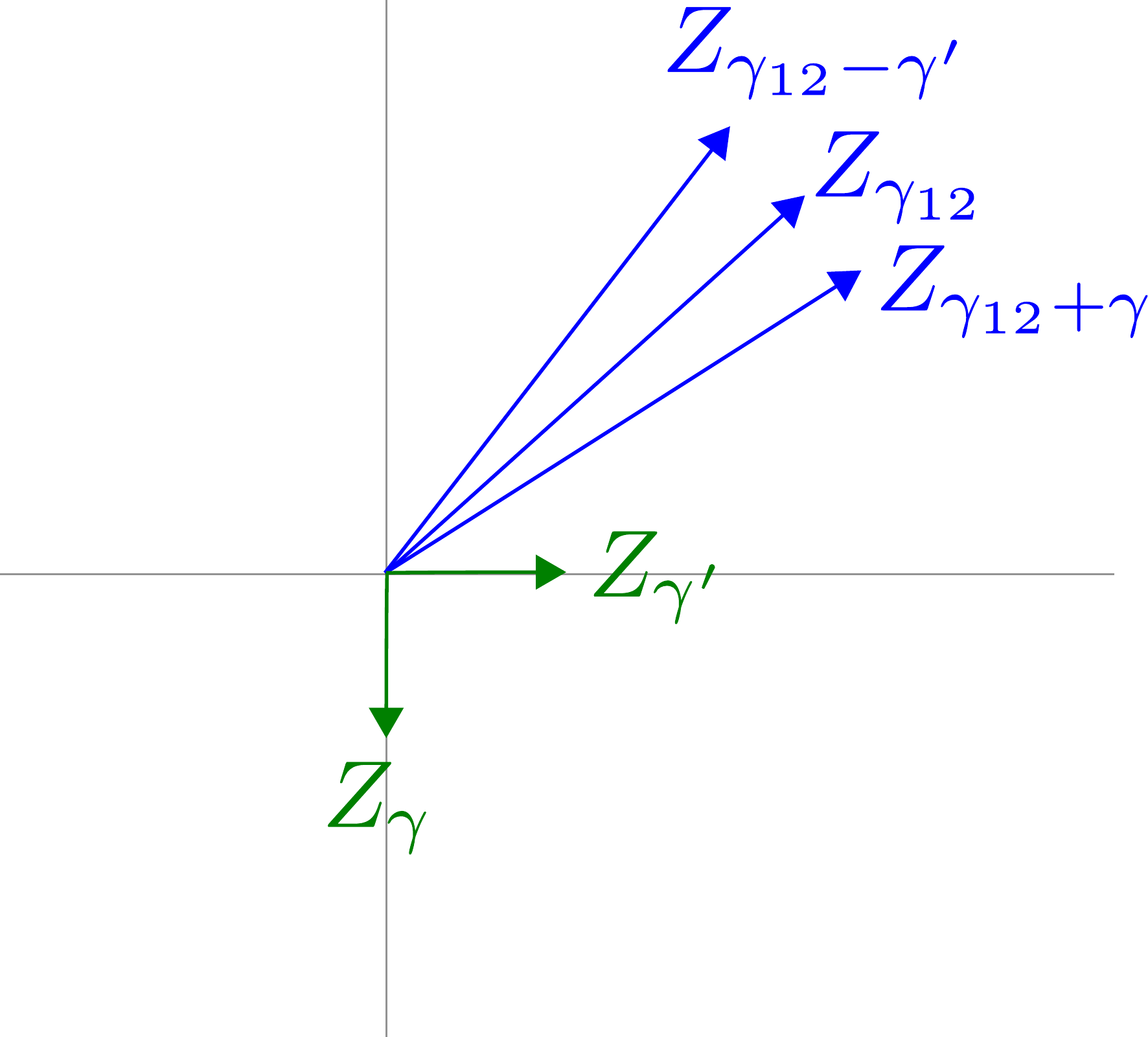}{Arrangement of the central charges for $z\in U$ in the strong coupling domain.
Here we take $u=0$ and $\vert z \vert$ large.}

It is easiest to begin the analysis of the soliton spectrum for $z$ in regions $U$, $D$.
As shown in Figure \ref{fig:wkb-movie}, in this case there are 3 distinct finite open WKB curves
ending at $z$, implying the existence of 3 solitons (plus their antiparticles) in both these regions.
We begin our analysis with $z$ in the region $U$.  In this case one of the open finite WKB curves is
the projection of a path in the class $\gamma_{12}(z)$ defined in Figure \ref{fig:gamma12def}.
By carefully computing the intersection numbers of the three finite open curves one can establish
that the charges of the three solitons in region $U$ are
$\gamma_{12}(z), \gamma_{12}(z)-\gamma', \gamma_{12}(z)+\gamma$ (together with their
antiparticles).  To find the proper ordering of the corresponding BPS rays, we must compute
periods.  One finds by direct computation that
$Z_{\gamma} = K$ and $Z_{\gamma'} = -\I K$ where $K>0$.
\footnote{In fact $K= 8 \sqrt{2\pi} \frac{\Gamma(3/4)}{\Gamma(1/4)}$.}
Now, for $Z_{\gamma_{12}(z)}$ it is useful to take $\vert z \vert$ large.
Then when $z\in U$ one finds $Z_{\gamma_{12}(z)} \approx \frac{2}{\pi} z^{1/2}$ where we
use the principal branch of the log.  Thus ordering of central charges
is as shown in Figure \ref{fig:regionuspec}.
As in \S \ref{subsubsec:ADN=2}, we will not give the analogous figures
for the other seven strong coupling regions, but will simply write the
corresponding products:
\begin{equation}\label{eq:strong-sol-u}
A(\sphericalangle;U) = \CK_{\gamma'} \CK_{\gamma} \CS_{\gamma_{12}+\gamma} \CS_{\gamma_{12}} \CS_{\gamma_{12}-\gamma'}.
\end{equation}
Here (unlike \S \ref{subsubsec:ADN=2}) we will choose $\sphericalangle$ to be an appropriate
sector of width $\pi$ (rather than $2\pi$).
We take it to be the right half-plane, in this and all subsequent expressions
(for the strong coupling region).

Now we consider the wall-crossings as $z$ moves along a path $U \to L \to D$.
As $z$ crosses from $U$ to $L$ the phase of $Z_{\gamma_{12}(z)}$ becomes close to $\pi/2$
for $\vert z \vert$ large, so the two rays $\ell_{\gamma_{12}(z)}$ and $\ell_{\gamma_{12}(z)- \gamma'}$
in Figure \ref{fig:regionuspec}
become parallel to $\ell_{-\gamma'}$ and we can apply the formula \eqref{eq:ks-ssk}.
(Note that $\omega(-\gamma', \gamma_{12}(z)) = -1$.)
This leads to a spectrum with only two solitons:
\begin{equation}\label{eq:strong-sol-l}
A(\sphericalangle;L) = \CK_{\gamma'}\CS_{\gamma_{21} } \CK_{\gamma} \CS_{\gamma_{12}+\gamma}.
\end{equation}
If we continue to increase the phase of $z$ and cross from $L$ into $D$ then
$\ell_{\gamma_{12}+\gamma}$ becomes parallel to $\ell_{-\gamma'}$ and \eqref{eq:sk-kss}
leads to
\begin{equation}\label{eq:strong-sol-dl}
A(\sphericalangle;Dl) = \CK_{\gamma'}\CS_{\gamma_{21}-\gamma+\gamma'} \CS_{\gamma_{21}-\gamma} \CS_{\gamma_{21}}
\CK_{\gamma}.
\end{equation}
Now let us compare moving $z$ along a path $U\to R \to D$. Crossing from $U \to R$
at large $\vert z\vert$ the wall is at $\vartheta\approx 0$, and $Z_{\gamma_{12}(z)}$
and $Z_{\gamma_{12}(z)+\gamma}$ become parallel to $Z_{\gamma}$. We can
then apply \eqref{eq:sk-kss} with $a=\gamma$ and $b=\gamma_{12}$ to get
\begin{equation}
A(\sphericalangle;R) = \CK_{\gamma'}\CS_{\gamma_{12}} \CK_{\gamma} \CS_{\gamma_{12}-\gamma'}.
\end{equation}
Then, moving from $R\to D$, $Z_{\gamma_{12}-\gamma'}$ aligns with $Z_\gamma$ and we get
\begin{equation}\label{eq:strong-sol-dr}
A(\sphericalangle;Dr) = \CK_{\gamma'}\CS_{\gamma_{12}}\CS_{\gamma_{12}-\gamma'}\CS_{\gamma_{12}+\gamma-\gamma'} \CK_{\gamma}.
\end{equation}
Now \eqref{eq:strong-sol-dl} differs from \eqref{eq:strong-sol-dr}, but we must take monodromy
into account. The cuts in Figure \ref{fig:gamma12locsys} divide $D$ into two regions
$Dl$ and $Dr$ and, by \eqref{eq:PPP-mon},  there is a discontinuity taking
$\gamma_{21}\to \gamma_{12}+\gamma - \gamma'$ when passing from $Dl$ to $Dr$.
Thus, the wall-crossing formula is consistent with the monodromy of the local system around $\CP_{-\I}$, as expected.

Now let us continue the spectrum from region $L\to L'$ in Figure \ref{fig:ruldregions}. In this case $z$ crosses the
finite WKB curve for the charge $\gamma$, and this curve has phase $\vartheta =0,\pi$.
Consistency requires that the BPS rays $\ell_{\gamma_{12}+\gamma}$ and $\ell_{\gamma_{21}}$ move toward
$\ell_{\gamma}$ and pass through each other when $z$ is on the marginal stability wall.
(This can be checked by noting that $Z_{\gamma_{12}(z)} \to 0$ as
$z \to \I$.)  Thus, the situation is very analogous to the $D\to U$ transition in the
$N=2$ AD example of \S \ref{subsubsec:ADN=2}. In particular we should apply
the wall-crossing formula \eqref{eq:sks-sks}. Note that
$\omega(\gamma, \gamma_{21}(z))=-1$ for $z\in L$ (as needed to apply the formula)
so that $\omega'(\gamma, \gamma_{21}(z))=+1$ for $z \in L'$. The spectrum is
now
\begin{equation}\label{eq:strong-sol-lp}
A(\sphericalangle;L') = \CK_{\gamma'}\CS_{\gamma_{12}+\gamma } \CK_{\gamma} \CS_{\gamma_{21} }.
\end{equation}
Similarly, when continuing from $z\in R$ to $z\in R'$, $z$ crosses the
finite WKB curve corresponding to $\gamma'$, with phase $\vartheta=\pm \pi/2$,
and we apply \eqref{eq:sks-sks} with the rays $\ell_{\gamma_{12}}$
and $\ell_{\gamma_{21}+\gamma'}$ sweeping through $\ell_{\gamma'}$ to discover that
$\omega'(\gamma', \gamma_{12}(z)) = -1$ for $z\in R'$, and the spectrum is
\begin{equation}
A(\sphericalangle;R') = \CK_{\gamma'}\CS_{\gamma_{21}+\gamma'} \CK_{\gamma} \CS_{\gamma_{21} }.
\end{equation}
Now we would like to continue from $L' \to U'$ and from $R' \to U'$ and compare.  To do this
$z$ must cross a separating WKB curve.  The separating WKB curve between $L'$ and $U'$
has phase $\vartheta = \pm \pi/2$, so the central charge $Z_{\gamma_{12}}$ lines up with
$Z_{\gamma'}$. In order to apply the wall-crossing formula \eqref{eq:sk-kss} with $b=\gamma_{12}(z)$
and $a=\gamma'$ we need to know
that $\omega(\gamma', \gamma_{12}(z)) =+1$ for $z \in L'$. This is indeed the
case, and is compatible with $\omega'(\gamma', \gamma_{12}(z)) = -1$ for $z\in R'$
thanks to the monodromy of the local system $\Gamma_{12}$ around $z=\I$.
In this way we get
\begin{equation}\label{eq:strong-sol-ulp}
A(\sphericalangle;U'l) = \CK_{\gamma'}\CS_{\gamma_{12}+\gamma'}\CS_{\gamma_{12}} \CS_{\gamma_{12}+\gamma} \CK_{\gamma}.
\end{equation}
In an analogous fashion we find that $\omega(\gamma, \gamma_{12}(z))=+1$ for $z\in R'$ and hence
\begin{equation}\label{eq:strong-sol-urp}
A(\sphericalangle;U'r) = \CK_{\gamma'}\CS_{\gamma_{21}+\gamma'}\CS_{\gamma_{21}} \CS_{\gamma_{21}+\gamma} \CK_{\gamma}.
\end{equation}
As in our comparison of $Dl$ with $Dr$, the cuts in Figure \ref{fig:gamma12def} divide $U'$ into
two regions $U'l$ and $U'r$, and \eqref{eq:strong-sol-ulp} agrees with \eqref{eq:strong-sol-urp} once
we take into account the discontinuity across the cut summarizing the monodromy
around $\CP_{\I}$.

To complete the analysis we consider the wall-crossing from $L' \to D'$ and $R' \to D'$. Since
$Z_{\gamma_{12}(z)+\gamma} \to 0$ as $z\to -\I$ from the region $L'$ we can see that
this central charge must line up with $Z_{\gamma'}$ and we apply \eqref{eq:ks-ssk}
with $a=\gamma' $ and $b= \gamma_{12}(z)+\gamma$ to obtain
\begin{equation}\label{eq:strong-sol-dlp}
A(\sphericalangle;D'l) = \CK_{\gamma'}\CK_{\gamma} \CS_{\gamma_{21} }\CS_{\gamma_{21}-\gamma} \CS_{\gamma_{21}-\gamma-\gamma'}.
\end{equation}
Finally, to do the wall-crossing from $R'$ to $D'$,
 since $Z_{\gamma_{12}(z)+\gamma'} $ aligns with $Z_{\gamma}$ passing from $R'$ to $D'$
  we apply \eqref{eq:sk-kss} to get
\begin{equation}\label{eq:strong-sol-drp}
A(\sphericalangle;D'r) = \CK_{\gamma'}\CK_{\gamma} \CS_{\gamma_{21}+ \gamma + \gamma' }\CS_{\gamma_{21}+\gamma'}
\CS_{\gamma_{21} }.
\end{equation}
In an analogous way to the previous cases \eqref{eq:strong-sol-dlp} is compatible
with \eqref{eq:strong-sol-drp} once we take into account the monodromy around $\CP_{0}$.
This completes our analysis of the soliton spectrum and its compatibility with the wall-crossing
formula.

\subsubsection{The soliton spectrum for weak 4d coupling and strong 2d coupling}

\insfig{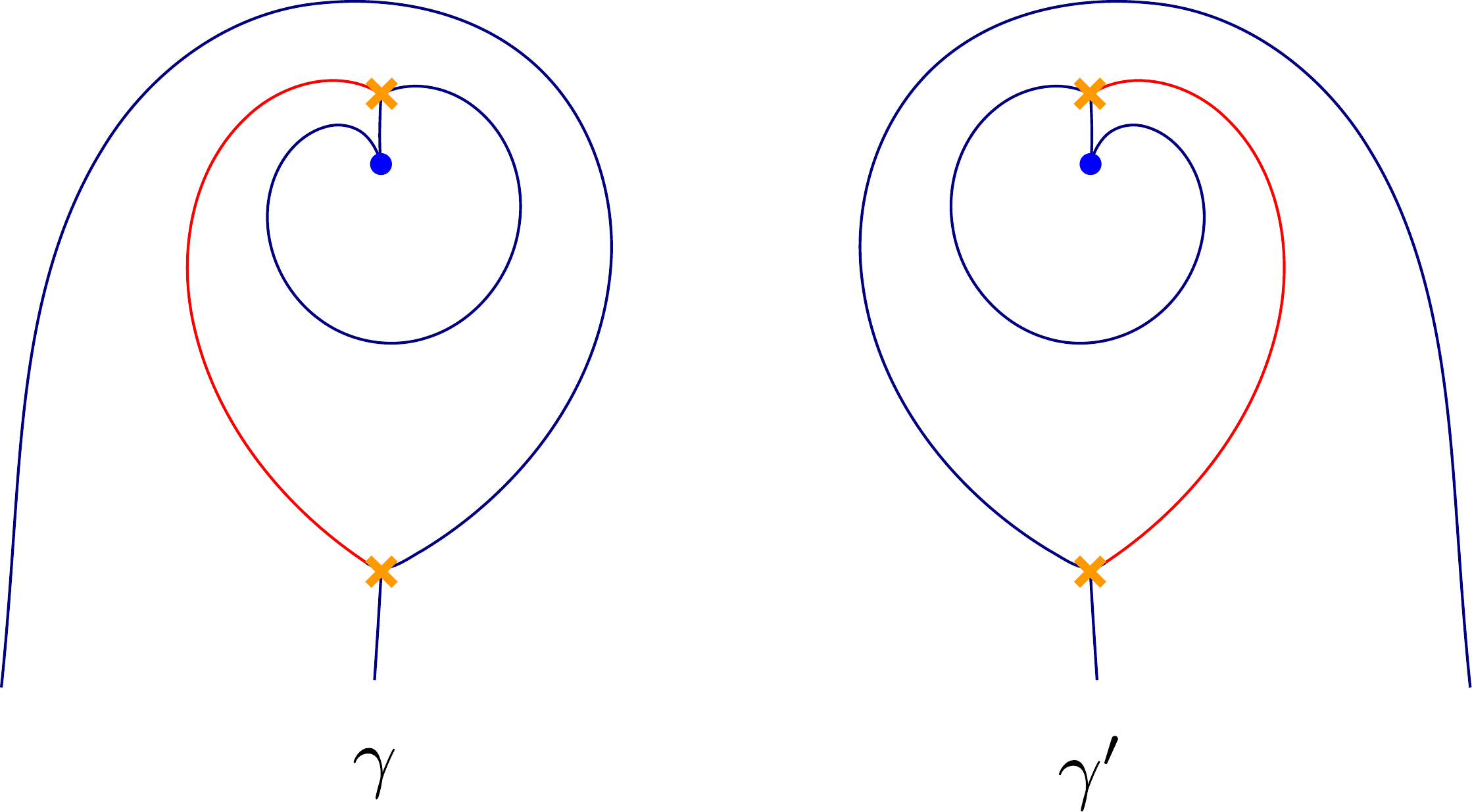}{Critical WKB curves associated to $\gamma$ and $\gamma'$, when $u$ is positive imaginary in the weak coupling domain.}

\insfig{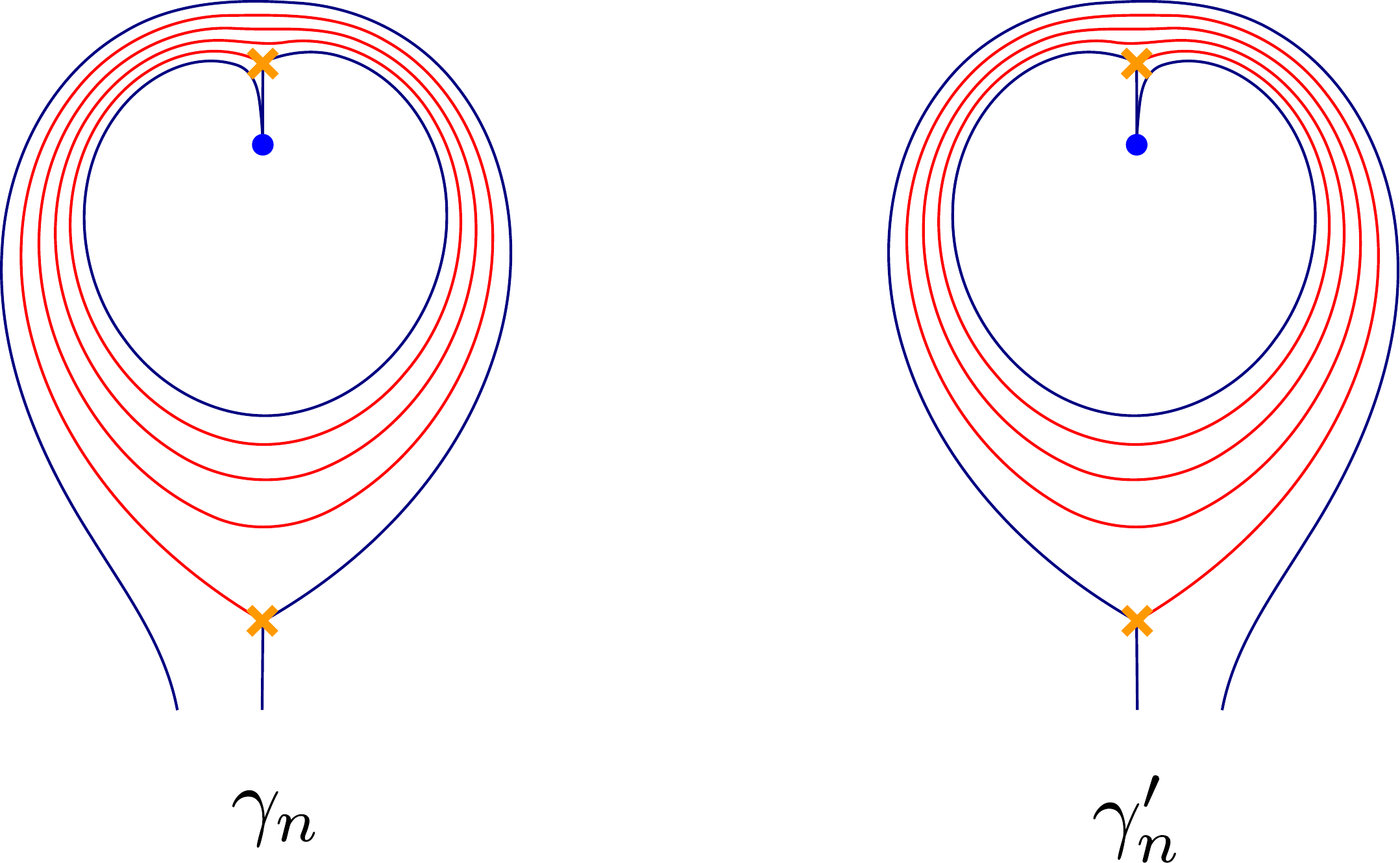}{Critical WKB curves associated to $\gamma_n$ and $\gamma'_n$, with some large $n>0$, when $u$ is positive imaginary in the weak coupling domain.
In each figure the finite WKB curve shown winds $n$ times around the origin.}

\insfig{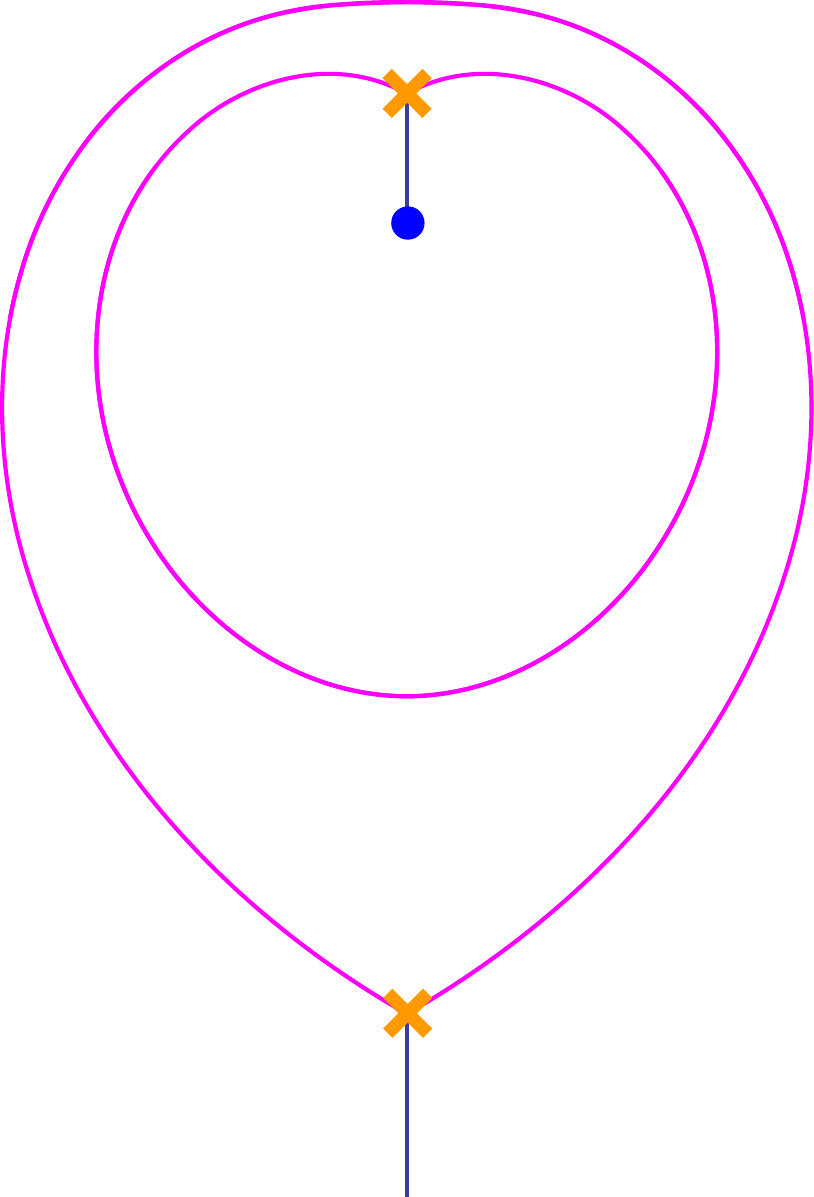}{Critical WKB curves associated to the vectormultiplet of charge $\gamma+\gamma'$. The two
purple finite WKB curves are the ``outer'' and ``inner' vectormultiplet curves.  The annulus between these
two curves is foliated by closed WKB curves and is referred to as the ``vectormultiplet annulus.'' This region corresponds to strong 2d coupling.  The complement of the annulus in the $z$ plane
consists of an ``outer region'' connected to $z=\infty$ and an ``inner region'' connected to $z=0$.
These are both weak coupling domains for the 2d coupling. }

\insfig{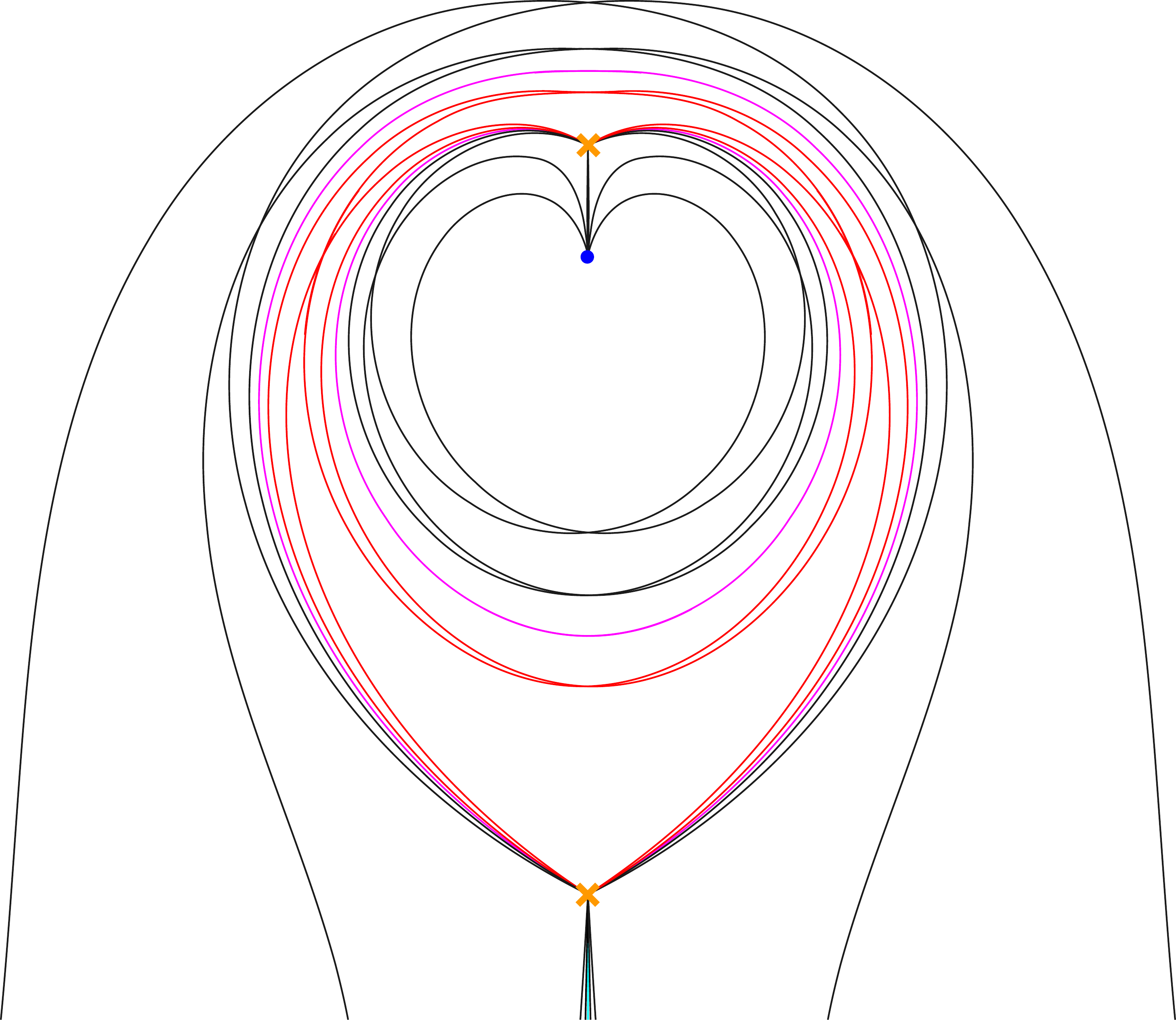}{Taking the union of the critical hypermultiplet WKB curves with the
inner and outer vectormultiplet curves produces an intricate pattern of walls of marginal stability.
There are an infinite number of chambers accumulating from the outer and inner regions complementary to the
annulus onto the outer and inner vectormultiplet curves, respectively.
In the interior of the vectormultiplet annulus the walls are dense and there is an uncountable
number of ``chambers.''}

Let us now consider what happens when $z$ is fixed (in an appropriate region discussed below) and
$u$ moves along a path from strong coupling to weak coupling.
To be specific, we consider a path so that the central charges $Z_{\gamma}$ and $Z_{\gamma'}$ align.
Then, the Kontsevich-Soibelman wall-crossing formula produces the weak coupling spectrum,
\begin{equation}
\left( \prod_{n\nearrow 0}^\infty \CK_{\gamma_n} \right) \CK_{\gamma+\gamma'} \left( \prod_{n\searrow \infty}^0 \CK_{\gamma_n'} \right),
\end{equation}
in agreement with the standard result \cite{Seiberg:1994rs,Ferrari:1996sv,Bilal:1996sk,Gaiotto:2008cd,Gaiotto:2009hg}.
In this notation $ \prod_{n\nearrow 0}^\infty $ means that the product is taken so that as one reads from
left to right $n$ increases $0,1,\dots, \infty$, while $\prod_{n\searrow \infty}^0$ similarly means that
as one reads from left to right $n$ decreases from $\infty$ to $0$.
We have suppressed the $\omega$ superscript but we remind the reader that $\Omega(\gamma+\gamma')=-2$.
Because there are many new 4d particles, the walls of marginal stability in the $z$-plane change dramatically.
If, for example, $u$ moves along the positive imaginary axis into the weak coupling region
then the critical WKB curves for $\gamma$ and $\gamma'$ evolve into those shown in Figure \ref{fig:crwkggp}.
In addition, there is an infinite tower of hypermultiplets with charges $\gamma_n, \gamma'_n$.
These lead to critical WKB curves such as those shown in Figure \ref{fig:crwkgngnp}. Finally a vectormultiplet emerges
with charge $\gamma+\gamma'$ and critical WKB curves shown in Figure \ref{fig:crwkvm}. The finite WKB curves
for the vectormultiplet foliate an annulus. The  finite WKB
lines associated to the hypermultiplets all wind around the origin and reside inside this annulus.
In addition each hypermultiplet contributes two separating WKB curves both outside and inside the annulus,
with each pair of lines differing by one unit of winding around the origin.
As $n\nearrow \infty$ these separating WKB curves accumulate
from the outside on the outer boundary of the annulus and from the inside on the inner boundary
of the annulus. The resulting pattern of marginal stability lines, taking into account the first
few hypermultiplets, is illustrated in
Figure \ref{fig:wkbsu2weak}.

\insfig{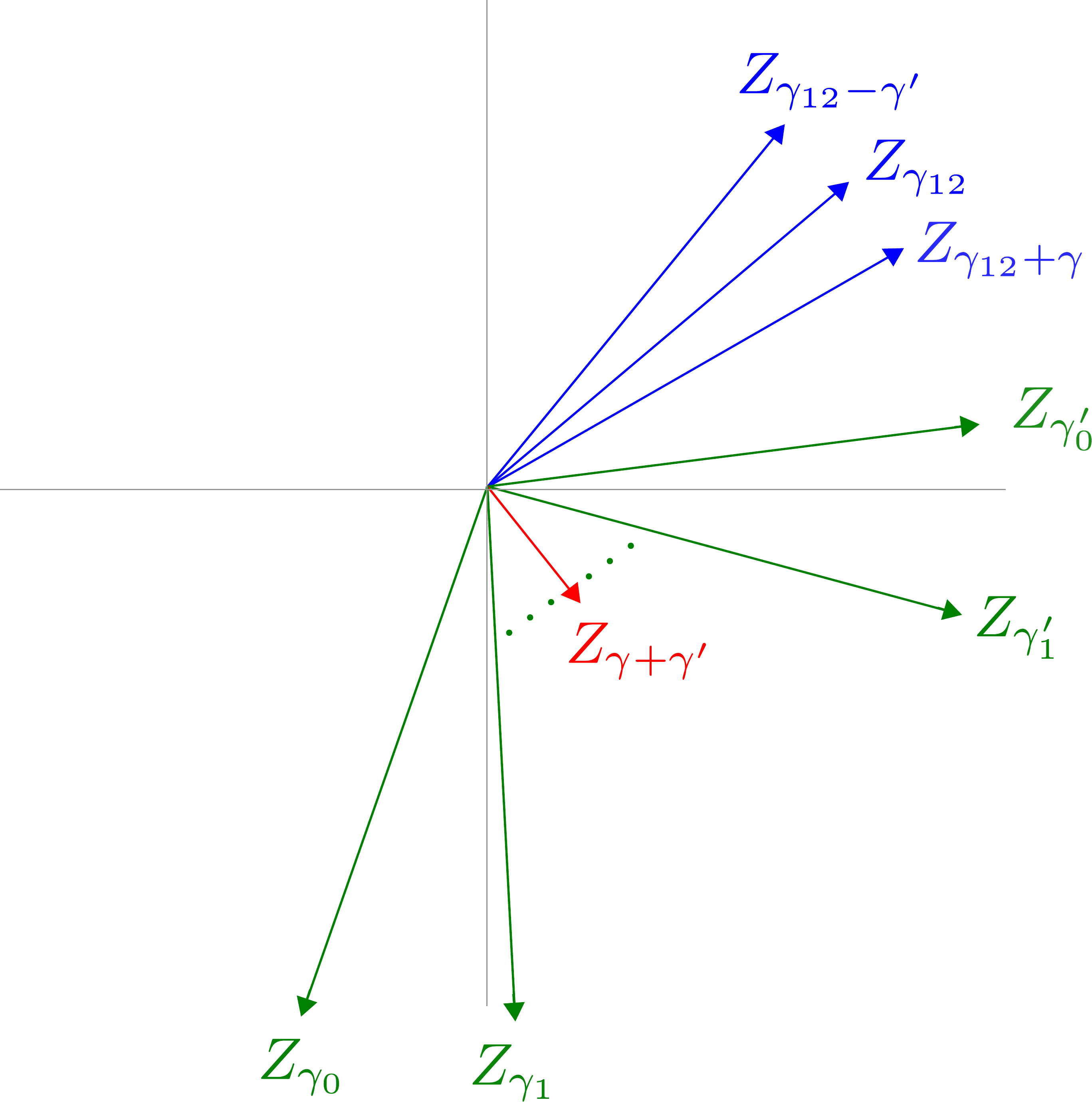}{The configuration of central charges after moving $u$ along an appropriate path into
the weak coupling domain, while holding $z$ fixed with large imaginary part, corresponding to the
upper region in Figure \protect\ref{fig:wkbsu2weak}.  The $\CK$-rays have experienced wall-crossing while staying
far from the $\CS$-rays.  The central charges corresponding to infinitely many more $\CK$-rays, with
phases accumulating to that of $Z_{\gamma + \gamma'}$, are suppressed.}

Now, to derive the soliton spectrum we first note that we can follow a path for $u \in \CB^*$ so that
the lines $\ell_\gamma$ and $\ell_{\gamma'}$ in Figure \ref{fig:regionuspec} sweep through each other without
passing through the soliton lines, thus producing the configuration of central
charges shown in Figure \ref{fig:regionuzerowk}. This is clearly true if we take $\vert z \vert $ to be sufficiently large
with large imaginary part.  We will call this region $U_0$.  Thus, for $z$ in $U_0$, the spectrum is given by
\begin{equation}\label{eq:wk4d-st2d}
A(\sphericalangle;U_0) = \left( \prod_{n\nearrow 0}^\infty \CK_{\gamma_n} \right) \CK_{\gamma+\gamma'} \left( \prod_{n\searrow \infty}^0 \CK_{\gamma_n'} \right)\CS_{\gamma_{12}(z)+\gamma} \CS_{\gamma_{12}(z)} \CS_{\gamma_{12}(z) - \gamma'}
\end{equation}
We have suppressed the superscript $\omega$, but we must
remember to extend our discussion slightly to find the $\omega$.
We consider the element $\gamma^0_{12} = \gamma_{12} + \frac{1}{2}(\gamma - \gamma')$, for which
$\omega(\gamma, \gamma^0_{12})=\omega(\gamma', \gamma^0_{12})=0$ before the wall-crossing.
It follows that also $\omega(\cdot, \gamma^0_{12}) = 0$ after the wall-crossing, just by acting
with the BPS product on $X_{\gamma^0_{12}}$
(cf. a similar computation in \S \ref{subsec:Examples-2d-4d-wcf}).
Then the affine-linearity of $\omega$ shows
$\omega(m \gamma +n \gamma', \gamma_{12}) = (n+m) \Omega(m \gamma +n \gamma')$.

We now comment on the wall-crossing of the soliton spectrum, without doing a
complete analysis. As we have noted, there are an infinite number of chambers outside the
annulus bounded by separating WKB curves for hypermultiplets of charges $\gamma_n$ and $\gamma'_n$.
If $z$ moves across such lines then there will be a wall-crossing formula of type \eqref{eq:ks-ssk}
or \eqref{eq:sk-kss}. Let us choose one particular path where $z$ begins on the imaginary axis
at large positive imaginary part and moves downwards towards the vectormultiplet annulus.
Referring to Figure \ref{fig:wkbsu2weak} it is clear that there will be an infinite sequence of
wall-crossings as $z$ moves across pairs of walls which intersect along the imaginary axis. We will call the
resulting chambers $U_0, U_1, U_2, \dots$ where $U_0$ is the noncompact chamber at large
$\vert z \vert$. These chambers get smaller and accumulate at the intersection
of the  outer ring of the
vectormultiplet annulus with the positive imaginary axis. Now, there is a canonical half-plane
we can use to describe the spectrum whose boundaries are the $W$-boson lines.
We will denote this sector as  $\sphericalangle_W$. Thus, we can rewrite the
spectrum \eqref{eq:wk4d-st2d} in the equivalent form:
\begin{equation}
A(\sphericalangle_W;U_0) =   \CK_{\gamma_W} \left(\prod_{n\searrow \infty}^0 \CK_{\gamma_n'} \right)
\CS_{\gamma_{12}(z)+\gamma} \CS_{\gamma_{12}(z)} \CS_{\gamma_{12}(z) - \gamma'}\left( \prod_{n\nearrow 0}^\infty \CK_{-\gamma_n} \right)
\end{equation}
Now, as $z$ crosses from the noncompact region $U_0$ in Figure \ref{fig:wkbsu2weak} across the topmost wall
the line for $ \gamma_{12}(z) - \gamma' $ sweeps counterclockwise across that for $-\gamma$ and simultaneously
the line for $\gamma_{12}(z)+\gamma$ sweeps across that for $\gamma'$.
 We therefore have simultaneous
wall-crossings of types   \eqref{eq:sk-kss} and \eqref{eq:ks-ssk} respectively, producing
\begin{multline}
A(\sphericalangle_W;U_1) = \CK_{\gamma_W}
 \left( \prod_{n\searrow \infty}^1 \CK_{\gamma_n'} \right) \CS_{\gamma_{12}+\gamma}    \\
   \cdot  (\CS_{\gamma_{12}+\gamma_W} \CK_{\gamma'}) \CS_{\gamma_{12} }  (\CK_{-\gamma} \CS_{\gamma_{12}-\gamma_W} )\cdot  \\
      \CS_{ \gamma_{12}-\gamma'}   \left( \prod_{n\nearrow 1}^\infty \CK_{-\gamma_n} \right).
\end{multline}
We are now set up for an inductive process, since once again on the left side of the product the ray
with charge $\gamma_{12} +\gamma $ can sweep clockwise across
the next $\CK$-ray of charge $\gamma_1'$, while on the right the
ray for $\gamma_{12}-\gamma'$ can sweep counter-clockwise across $-\gamma_1$.
After $N$ such steps we find
\begin{multline}\label{eq:wk4d-st2d-N}
A(\sphericalangle_W;U_N) = \CK_{\gamma_W} \left( \prod_{n\searrow \infty}^N \CK_{\gamma_n'} \right) \CS_{\gamma_{12}+\gamma}    \\
 \left(\prod_{n\searrow N-1}^{0} (\CS_{\gamma_{12}+ (n+1)\gamma_W} \CK_{\gamma_n'}) \right) \CS_{\gamma_{12}} \left( \prod_{n\nearrow 0}^{N-1}
 (\CK_{-\gamma_n } \CS_{\gamma_{12}- (n+1)\gamma_W} ) \right)\cdot  \\
\CS_{\gamma_{12}-\gamma'} \left( \prod_{n\nearrow N}^{\infty} (\CK_{-\gamma_n} ) \right).
\end{multline}

\bigskip
\textbf{Remarks}

\begin{enumerate}

\item One might wonder whether the path we have described meets
other wall-crossings, involving exchanges of
other $\CK$ and $\CS$ factors in \eqref{eq:wk4d-st2d-N}. The easiest way to see
that this cannot happen is
to note that the $\omega(a,b)$ associated to these other potential crossings would not
be compatible with \eqref{eq:ks-ssk} or \eqref{eq:sk-kss}.

\item Other paths from $U_0$ into the vectormultiplet annulus will produce interesting
variants of \eqref{eq:wk4d-st2d-N} which are not symmetrical between the products
in the middle line of this equation. We have not attempted to investigate the full
chamber structure and soliton spectrum in detail.

\item Using the $z \to 1/z$ symmetry we can also conclude that there is a similar spectrum of solitons
in the regions inside the vectormultiplet annulus.

\item There is a nice physical interpretation of the spectrum we have found. As we have discussed in
\S\ref{subsec:cp1-sigma}, the strongly coupled $\IC \IP^1$ sigma model has only two solitons
whereas \eqref{eq:wk4d-st2d} predicts 3 solitons. In order to compare these spectra we should take
the limit as the 4d $SU(2)$ gauge coupling becomes infinitely weak. One way to do this is to
set $\tilde z = \Lambda^2 z$ and take $\Lambda \to 0$ holding $\tilde z$ fixed. In this limit
the chiral ring equation \eqref{eq:spectral-su2}
degenerates to \eqref{eq:chir-ring}. Moreover, the period $Z_{\gamma_{12}(z)}$ diverges
as $\sim - \sqrt{2u} \log \frac{\Lambda^4}{2u}$ while $Z_{\gamma_{12}-\gamma'}$ and
$Z_{\gamma_{12}+\gamma}$ have finite limits.
\footnote{To prove these statements note that in this limit one turning point approaches $\tilde z=0$
like $\tilde z \approx -\Lambda^4/(2u)$ while the other approaches $\tilde z \approx -2u$. The WKB curve for $\gamma_{12}$ goes to the turning point near the origin,
and the term $\sqrt{2u} \frac{\de \tilde z}{\tilde z}$ dominates
the line integral of $\lambda$. On the other hand for the charges $\gamma_{12}+\gamma$ and
$\gamma_{12}-\gamma'$ the WKB curve goes to the other turning point $-2u$ and the
line integral does not diverge.} It  follows that the soliton with
charge $\gamma_{12}$ becomes large. Indeed, identifying $\sqrt{2\vert u\vert}= v$, where
$v$ is the scale of the vev of the 4d vacuum,  this soliton has a mass $\sim \frac{32\pi v}{g^2}$, and in the
weak coupling limit $g\to 0$ it decouples from the spectrum. By contrast,
the solitons with charges
$\gamma_{12}+\gamma$ and $\gamma_{12}-\gamma'$ remain in the spectrum. Similarly, in
\eqref{eq:wk4d-st2d-N} we find two towers of length $N$ of solitons
with charges $\gamma_{21}+ n(\gamma+\gamma')$ and
$\gamma_{12}+n(\gamma+\gamma')$, $n=1,\dots, N$. These too have divergent central charges
going like  $\sim - \sqrt{2u} \log \frac{\Lambda^4}{2u}$ and hence should be
viewed as bound states of the hypermultiplet dyons with the surface defect. All of these heavy particles
should correspond to solitonic field configurations visible in a semiclassical analysis
of the $\IC \IP^1$ sigma model coupled to the weakly coupled $SU(2)$ theory. It would be
an interesting check to produce these solitons directly. It would also be interesting
to test directly the predictions $\omega(\gamma_n, \gamma_{12}) = \omega(\gamma_n', \gamma_{12})= 2n+1$.
In any case, we conclude that the spectrum \eqref{eq:wk4d-st2d}
is physically sensible.

\item As $z$ moves towards the vectormultiplet annulus the product approaches the expression
\begin{equation}\label{eq:wk4d-st2d-inf}
\begin{split}
A(\sphericalangle_W;U_\infty) &  =K_{\gamma_W} S_{\gamma_{12}+\gamma}  \left( \prod_{n\searrow \infty}^{-\infty} (
\CS_{\gamma_{12}+(n+1)\gamma_W} \CK_{\gamma_n'} ) \right) \cdot \CS_{\gamma_{12}-\gamma'}\\
& = K_{\gamma_W} S_{\gamma_{12}+\gamma}  \left( \prod_{n\nearrow -\infty}^{\infty} (\CK_{-\gamma_n}
\CS_{\gamma_{12}-(n+1)\gamma_W}   ) \right) \cdot \CS_{\gamma_{12}-\gamma'}\\
\end{split}
\end{equation}

\end{enumerate}

\subsubsection{The soliton spectrum for weak 4d coupling and weak 2d coupling}

When $z$ moves into the vectormultiplet annulus both the 4d and 2d theories are weakly coupled. There
is an infinite spectrum of 4d hypermultiplets leading to a very intricate pattern of marginal stability
lines inside the annulus, as we have already mentioned, and in addition, experience with the $\IC \IP^1$
sigma model in \S \ref{subsec:cp1-sigma} suggests that there should also be an infinite spectrum of
weakly coupled 2d $\IC \IP^1$ solitons, in addition to those we have already discovered.
Indeed, let us return to the limiting expression \eqref{eq:wk4d-st2d-inf}. By slightly shifting the
half-plane we can write an equivalent product, cycling the factor $\CS_{\gamma_{12}-\gamma'}$ on the
right to its anti-particle factor $\CS_{\gamma_{21}+\gamma'}$ on the left. Then, the product
 $\CS_{\gamma_{21}(z)+\gamma'}    \CK_{\gamma+\gamma'}
\CS_{\gamma_{12}(z)+\gamma}$ has the property that $\omega(\gamma+\gamma', \gamma_{21}(z)+\gamma') = 0$
and $\omega(\gamma+\gamma', \gamma_{12}(z)+\gamma)=0$, as can be seen by computing intersection
products. Therefore, we can apply the $\IC \IP^1$ wall-crossing formula \eqref{eq:wallcp1} to this
product to produce a product on the left of the form
\begin{equation}
\left(\prod_{n\nearrow 0}^\infty \CS_{\gamma_{12}+\gamma_n} \right) \CK_{\gamma_W}^\omega
\left( \prod_{n\searrow \infty}^0 \CS_{\gamma_{21} + \gamma'_n} \right)
\end{equation}
where $\omega(\gamma+\gamma', \gamma_{12}+\gamma)=2$.
Using this one can derive a heuristic formula for the spectrum   when $z$ lies \emph{exactly on} the outer boundary
of the  annulus for the vectormultiplet. Written back in the standard half-plane $\sphericalangle_W$
this can be written in several forms, of which two of the more suggestive ones are:
\begin{equation}\label{eq:wk4d-st2d-ANN}
\begin{split}
A(\sphericalangle_W;Ann)  & =  \CK_{\gamma_W} \left( \prod_{n\searrow \infty}^0 \CS_{\gamma_{21}+\gamma_n'} \right)
\cdot \left( \prod_{n\searrow \infty}^{-\infty} (\CS_{\gamma_{12}+(n+1)\gamma_W} \CK_{\gamma_n'}) \right) \cdot
\left(\prod_{n\searrow -1}^{-\infty} \CS_{\gamma_{21}+\gamma_n'} \right) \\
& =  \CK_{\gamma_W} \left( \prod_{n\nearrow -\infty}^{-1} \CS_{\gamma_{21}-\gamma_n} \right)
\cdot \left( \prod_{n\nearrow -\infty}^{\infty} ( \CK_{-\gamma_n}\CS_{\gamma_{12}-(n+1)\gamma_W})  \right) \cdot
\left(\prod_{n\nearrow 0}^{\infty} \CS_{\gamma_{21}-\gamma_n} \right) \\
\end{split}
\end{equation}
Of course these are simply related by using the identity $\gamma_{-(n+1)} = - \gamma_n'$.

The expression \eqref{eq:wk4d-st2d-ANN} is slightly unphysical.
When $z$ moves into the interior of the vectormultiplet
annulus it crosses an infinite number of walls, and hence there will be an
infinite number of wall-crossings applied to \eqref{eq:wk4d-st2d-ANN}.
At any interior point there will be some definite ordering of the
central charges appearing in \eqref{eq:wk4d-st2d-ANN} determining the
true expression. Indeed, we   claim that the possible soliton spectra
and chambers in the vectormultiplet annulus are
precisely those given by the
bi-infinite words in the $\CS$ and $\CK$ factors appearing in  \eqref{eq:wk4d-st2d-ANN}
subject to the following rules:

Define
\begin{equation}
\begin{split}
\CA_n & = \CS_{\gamma_{21} + \gamma_n'},\\
\CC_n & = \CS_{\gamma_{12} + (n+1) \gamma_W}, \\
\CK_n & = \CK_{\gamma_n'},\\
\end{split}
\end{equation}
where $n\in \IZ$. Then the  bi-infinite words are of the form
\begin{equation}\label{eq:binf-word}
\cdots \CS \CK \CS \CK \CS \cdots
\end{equation}
such that:

\begin{enumerate}

\item The $\CS$-factor between $\CK_n$ and $\CK_{n-1}$ can be either    $\CA_{m} $ or $\CC_{m}$ for
some $m$.

\item The word therefore determines three sequences of letters of type $\CK_n$, $\CA_n$ and
$\CC_n$, respectively. The index on all three sequences decreases to the right in steps of one.
Thus we have  three sequences $\cdots \CK_{n+1} \CK_n  \cdots $,   $\cdots \CA_{n+1} \CA_n \cdots $ and
$\cdots \CC_{n+1} \CC_n \cdots $ interwoven in the pattern \eqref{eq:binf-word}.

\end{enumerate}

We will give a strong argument for this claim, although we do not insist it is a completely
rigorous proof.

Before establishing our claim, let us note that it is nicely consistent with wall-crossing. First,
when $z$ crosses the closed WKB curves there is no wall-crossing, which is good since they
foliate the annulus. Indeed every point $z$ lies on one such closed curve and,
as noted in \S \ref{subsec:cp1-sigma}, these give the
two weakly coupled vacua at the north and south poles of the $\IC \IP^1$ sigma model. Next, note that
all the other walls in the interior of the annulus are finite hypermultiplet walls for $\gamma'_n$ for some $n \in \IZ$.
Note that the sum of charges for $\CA_n$ and $\CC_m$ is precisely $\gamma'_{n+m+1}$.
Therefore, when $z$ crosses the wall for $\gamma'_{n+m+1}$ there can be a wall-crossing event of
type \eqref{eq:sks-sks}, involving
\begin{equation}\label{eq:AKC-CKA}
\CA_n \CK_{n+m+1} \CC_m   = \CC_m \CK_{n+m+1} \CA_n.
\end{equation}
It might not at first appear obvious that $\omega(\gamma'_{n+m+1}, \gamma_{21}(z)+\gamma_n')=-1$
or $\omega(\gamma'_{n+m+1}, \gamma_{12}(z)+(m+1)\gamma_W)=-1$ as is required for
\eqref{eq:sks-sks} (which case is realized depends on
which side of the wall $z$ is on). This can be shown by noting that as $z$ approaches the
WKB curve for $\gamma'_{n+m+1}$ the phase of the line for $\gamma_{21}(z)+\gamma_n'$ approaches
that for $\gamma'_{n+m+1}$. On the other hand, WKB walls with the same value of $\vartheta$
cannot intersect except at turning points and
singularities. Therefore, the topology of the WKB curves must be that shown in
the universal situation illustrated in Figure \ref{fig:cross-finite-wkb}.
Equation \eqref{eq:AKC-CKA} now shows how wall-crossings (possibly infinitely many) allow us
to pass between any two words of the type \eqref{eq:binf-word} described above.

\insfig{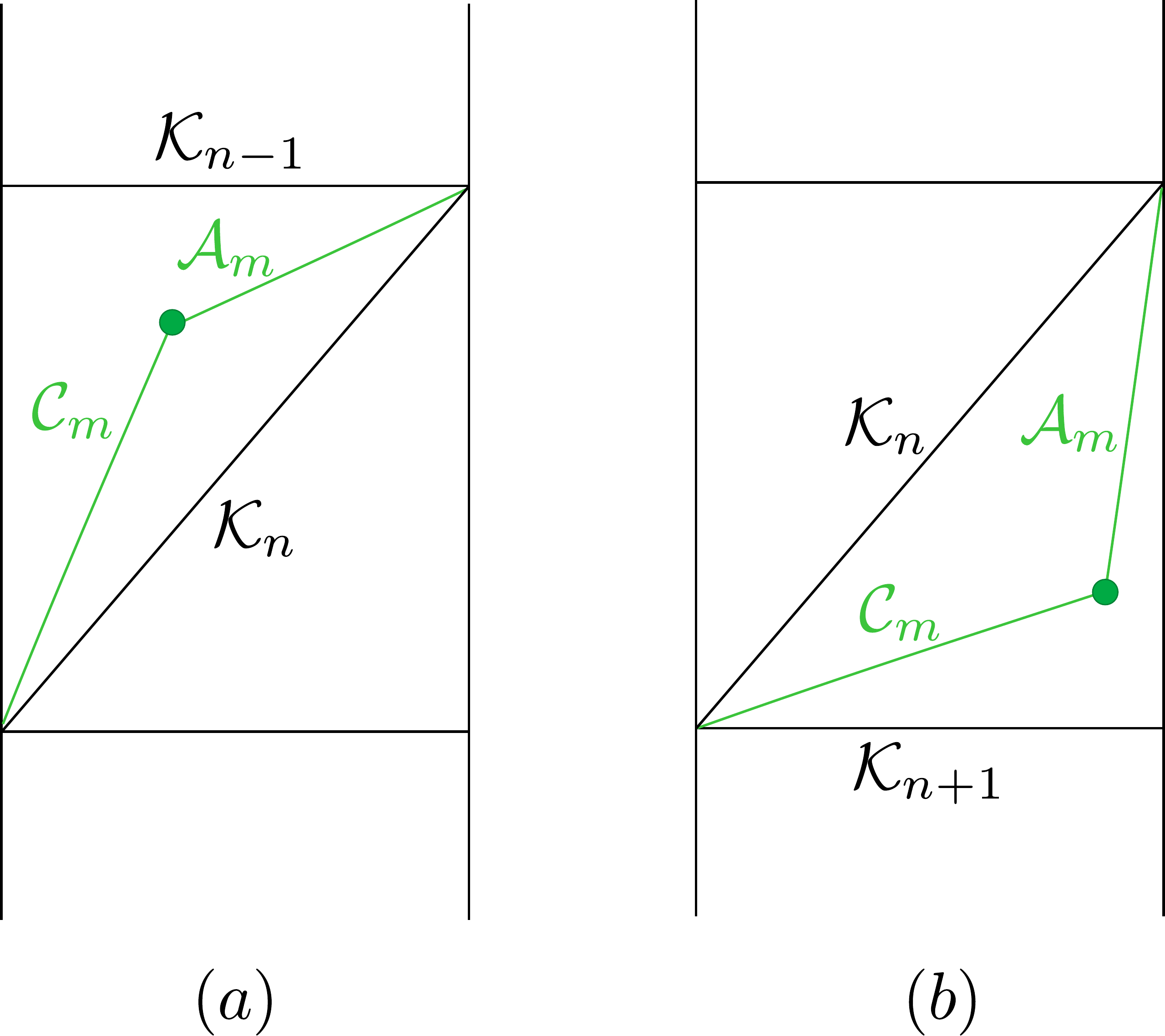}{In case (a) the corresponding sequence of morphisms is
$\cdots \CK_n \CA_{m} \CK_{n-1} \cdots$ because the slope of the $\CA$ ray lies between those of the $\CK$ rays.
In case (b) the corresponding sequence is
$\cdots \CK_{n+1} \CC_{m} \CK_{n} \cdots$.}

Now, to investigate the chamber structure and establish our claim it is first useful to map the annulus to the $w$-plane,
\begin{equation}
w := \int^z_{z_+} \lambda,
\end{equation}
where we integrate from the turning point $z_+$ nearest to $z=0$.  It will be useful to work with an approximate
formula for $w$, valid deep in the weak coupling domain.  Let us introduce the scaled coordinate
\begin{equation}\label{eq:scale-zee}
\tilde z =\left(\frac{\Lambda^2}{u} \right)^{2\alpha} z
\end{equation}
with $0 < 2\alpha < 1$. We consider the weak coupling limit where $u \to \infty$
(or equivalently $\Lambda \to 0$) with $\tilde z$ held fixed.  In this regime  the middle term in the expression
\eqref{eq:spectral-su2} for
 $\lambda^2$ dominates the other two and hence
 we can approximate $\lambda \approx \sqrt{2u} \frac{\de z}{z}$ and hence $w \approx \sqrt{2u} \log z/z_+$.
After further rescaling the real and imaginary parts of $w$ we can therefore take the image of the annulus to be the infinite strip
between $0$ and $1$, with inner turning points at $\I N$ and outer turning points at
$1 + \I N$, $N\in \IZ$.
In $w$-coordinates the WKB curves are straight lines of slope proportional to $\vartheta$. The lines of marginal
stability are the straight lines joining these turning points, and are hence labeled by
pairs of integers, $(m,n)$ and given by $y=(m-n)x + n$. After solving a simple Diophantine
equation one can show that every point in the strip with rational coordinates  is at the intersection
of two such lines. In particular, the walls of marginal stability are dense and
the ``vertices'' of the chambers are the points in the strip with rational coordinates.
Now, any two points not on a WKB wall will be separated by such a wall. (To prove this
draw a straight line segment between the two points.  Choose a point with rational coordinates
sufficiently close to that line segment. There will be a WKB wall through that point.)
Therefore, the ``chambers'' --- defined to be the connected components of the complement of the walls ---
will consist of single points.
The set of walls is a countable union of sets of measure zero and hence has measure zero,
and therefore the chambers have positive measure, and therefore there are uncountably many chambers.

Now suppose that a surface defect is at a point $z_0$ in the vectormultiplet annulus and $z_0$
maps to a point $w_0$. The WKB curves for charges associated with $\CA_n$ are the straight lines
going from $w_0$ to the turning points $\I N$ on the inner boundary while the WKB curves for charges
associated with $\CC_n$ are the straight lines going from $w_0$ to the turning points $1+\I N$
on the outer boundary of the annulus. (Note that this way two such lines can glue together to
give a hypermultiplet WKB curve.) It is now a matter of simple geometry to see that $w_0$
can lie in one of two different kinds of triangles. In each triangle there is
exactly one $\CS$ factor with slope between those of $\CK_{n}$ and $\CK_{n-1}$.
In one type of triangle the slope between
$\CK_{n}$ and $\CK_{n-1}$ will be of $\CA$-type and in the other it will be of $\CC$-type.
See Figure \ref{fig:twotriangles}.

\bigskip

\textbf{Remarks}

\begin{enumerate}

\item   If in Figure \ref{fig:wkbsu2weak} $z$ continues down the imaginary axis then there must be an infinite number of wall-crossings, ending up at a product analogous to \eqref{eq:wk4d-st2d-ANN} for the inner ring of the annulus, with
$\CA$ and $\CC$ exchanged.  This is analogous to a phenomenon we found at strong coupling,
where $z\to 1/z$ had the effect of transforming the soliton spectrum by
the involution $\gamma_{12} \to \gamma_{21}+\gamma$.

\item Our analysis makes some interesting predictions which would be worth checking with
weakly coupled semiclassical field theory methods. The region defined by
the scaling parameter \eqref{eq:scale-zee} defines a regime where the surface defect
may be treated as a Gukov-Witten defect. This may be justified by using the
approximate expression $\lambda \approx \sqrt{2u} \frac{\de z}{z}$ to obtain
approximate expressions for $Z_{\gamma_{12}(z)}$, when $z$ is in the domain
with $u \to \infty$ and $\tilde z$ held fixed. One then finds the IR GW parameter
\begin{equation}\label{eq:scale-IRGW}
t_{\gamma_1}-t_{\gamma_2}    = \frac{\p Z_{\gamma_{12}(z)}}{\p a} =
\begin{cases} \frac{(2\alpha-1)}{2\pi \I } \log \frac{a}{\Lambda} + \cdots & \qquad {\rm outer\ turning\ point}, \\
  \frac{(2\alpha+1)}{2\pi \I } \log \frac{a}{\Lambda} + \cdots & \qquad {\rm inner\ turning\ point}. \\
\end{cases}
\end{equation}
In a semiclassical analysis one would define
 a moduli space of ``ramified monopoles,'' solutions of the bulk BPS equations in the presence of GW boundary conditions
 dictated by \eqref{eq:scale-IRGW}.
Semiclassical quantization of this moduli space for various monopole charges should reproduce the weak-coupling spectrum described in this section: in particular, the spectrum of solitons $\mu(\gamma_{ij})$ is
independent of $z$ but the spectrum of $\omega(\gamma, \gamma_{ij})$ varies strongly with
$z$, and should grow at most linearly in the charges.

\item It should be interesting to study the framed BPS states in this example, but we leave this
for another occasion.

\end{enumerate}

\section{An application: solving Hitchin systems by integral equations}\label{sec:Solve-Hitchin}

Finally let us consider an interesting application of our discussion:  we can use it to give concrete formulas for
solutions of Hitchin equations on punctured surfaces $C$.

Indeed, as we explained in \S \ref{subsec:ClassS-3d}, for theories of class $\CS$,
the $\CY_{\gamma_{i}}(\zeta)$ can be thought of as sections of the universal Higgs bundle over $\CM \times C$.
Allowing $z$
to vary we obtain $\CY_{\gamma_{i}}(\zeta, z)$, a flat section with respect to the
connection $\de + \CA(\zeta)$. This follows from comparing \eqref{eq:lz-exp} and
\eqref{eq:Lvev-Ptrpt}. Alternatively, we can use $\CY$ to define the connection
(see below) and then apply an argument analogous to that used in
Appendix \ref{app:Twistor-HH} to determine that the connection so defined is
the desired one.
 This flat section is concretely computable using the integral
equations \eqref{eq:int-1}, \eqref{eq:int-2}.  The only data one needs to write these equations
are the central charges $Z_\gamma$ and the BPS degeneracies $\omega$, $\mu$.

Suppose we trivialize the ramified cover $\Sigma \to C$
  in a   neighborhood of $z$. Then we can label
the sheets $i=1,2$, and similarly, we can locally trivialize the cover
above $z'$ and label the sheets $i'=1,2$. We can compute   the four
sections  $\CY_{\gamma_{ii'}} $ in a neighborhood of the points $z$ and $z'$,
using the integral equations \eqref{eq:int-1} and \eqref{eq:int-2} together with \eqref{eq:CX-ijprime}.
Any invertible combination of these four sections defines a
fundamental solution $\Psi$ of the flatness equations at $z$.
Then $\Psi^{-1} \de_{z} \Psi = \varphi/\zeta + A + \zeta \bar \varphi$
gives the solution to the Hitchin system, in the form of explicit
$\varphi, A, \bar \varphi$ valid in a neighborhood of $z$ (with the
particular solution and gauge depending on a basepoint $z'$). Different choices
of $\Psi$ are possible, in the form of different linear combinations of the
various $\CY_{ii'}$. They will give different gauge choices for the solution.
For examples, if $z$ and $z'$ are in a common neighborhood with a common
trivialization of the double cover, then at least for sufficiently large $R$, a sum like
\begin{equation}
\Psi = \CY_{\gamma_{12'}} + \CY_{\gamma_{21'}}
\end{equation}
or alternatively a sum like
\begin{equation}
\Psi = \CY_{\gamma_{11'}} + \CY_{\gamma_{22'}}
\end{equation}
will serve as fundamental solutions.

Again, taking $z,z'$ in some common neighborhood
we can also apply the discussion around \eqref{eq:diag-A-01}
of \S \ref{subsec:Integral-Equations} where we take the
diagonal matrix to be
\begin{equation}
 x(z,\zeta;z') =   x_{\gamma_{i2'}} (z,\zeta;z')  e_{ii }
\end{equation}
Here $z'$ and its lift to sheet $2'$ are serving as basepoints.
$e_{ii}$ are matrix units and the sum on $i=1,2$ is understood.
We  can then consider $g(\zeta)x(z,\zeta;z')$
to be a fundamental solution to $(\de_z + \CA) \Psi=0$. It follows that
gauge transformation by $g(\zeta)$
diagonalizes the flat spectral connection for the Hitchin system,
to a diagonal connection
\begin{equation}
 \de_{z} \log x_{\gamma_{i2' }} e_{ii }.
 \end{equation}
We can make this more explicit using \eqref{eq:int-1}:  the connection we obtain
is the same as the semiflat connection
\begin{equation}
\de_{z} \log \CY^{\sf}_{\gamma_{i2'}}  e_{ii}
\end{equation}
plus the correction
\begin{equation}
\de_{z} \left[
 \sum_\gamma \omega(\gamma, \gamma_i;z) \frac{1}{4 \pi \I}
 \int_{\ell_\gamma} \frac{\de \zeta'}{\zeta'} \frac{\zeta' + \zeta}{\zeta' - \zeta}
 \log(1 - \CY_\gamma(\zeta')) \right] e_{ii}.
\end{equation}
The only $z$-dependence in the expression in square brackets
comes from the discontinuities in $\omega(\gamma,\gamma_i;z)$
due to wall-crossing when $z$ crosses finite WKB curves  (see Figure \ref{fig:cross-finite-wkb} in \S
\ref{sec:2d-4ddata}). These give delta-function singularities localized on these curves.
We have argued on general grounds that our solution to the Hitchin equations
should be smooth, so these singularities are expected to cancel against
similar discontinuities in $g(\zeta)$. However, we have not checked this point
in detail.

Because of the simple $\zeta$ dependence of $\CA$,
the information in the full  $\CY_{a b'}(\zeta)$ is rather redundant.
It is sufficient to expand asymptotically for small $\zeta$,
so that
\begin{equation} \label{eq:int-11}
 x_{\gamma_i}(\zeta) \sim \CY_{\gamma_i}^\sf(\zeta) \exp \left[ - \sum_\gamma \omega(\gamma, \gamma_i) \frac{1}{4 \pi \I} \int_{\ell_\gamma} \frac{\de \zeta'}{\zeta'}(1 + 2\frac{\zeta}{\zeta'} + \cdots) \log(1 - \CY_\gamma(\zeta')) \right].
\end{equation}
and
\begin{equation} \label{eq:int-21}
g_i(\zeta) \sim g_i^\sf + \sum_{j \neq i, \gamma_{ji}} \mu_{\gamma_{ji}} \frac{1}{4 \pi \I} \int_{\ell_{\gamma_{ji}}} \frac{\de \zeta'}{\zeta'}(1 + \cdots) g_j(\zeta') x_{\gamma_{ji}}(\zeta'),
\end{equation}

The leading order for $g_i$ and the leading order for $ x_{\gamma_i}$ is sufficient to derive the expression for $\varphi$.  The next to leading order in $ x_{\gamma_i}$ is required to derive $A$. Then $\bar \varphi$ can be
obtained simply by complex conjugation, or from a large $\zeta$ expansion.

\section{Future directions}\label{sec:Future}

In conclusion, we list here some directions for future research
which be believe might be fruitful.

\begin{enumerate}

\item
 The exercise of \S \ref{sec:Hitchin-Expl} appears to require
 genuinely new techniques for higher rank groups.  We are currently exploring
this question.

\item Our construction produces examples of hyperholomorphic bundles over moduli spaces $\CM$
of Higgs bundles.  Such hyperholomorphic bundles fit naturally into the approach of \cite{geom-lang-n4} to the
geometric Langlands correspondence:  they are examples of ``($B$, $B$, $B$) branes'' in that context.
The sections $\tilde\nu_i$ which entered our construction as part of the 2d-4d data define a complex Lagrangian
multi-section of the dual moduli space $\widetilde\CM$,
i.e. a ``($B$, $A$, $A$) brane.''  These two branes
are related by mirror symmetry, i.e. Langlands duality.  In this paper we have obtained a more explicit
picture of the geometry of these ($B$, $B$, $B$) branes than previously known.  It is natural to wonder
whether this picture can be of any use in the geometric Langlands correspondence.

\item In \S \ref{subsubsec:MirrorBranes} we encountered, but did not really address, the question of
systematically describing the structure of our hyperholomorphic bundles near orbifold singularities of $\CM$.
Via mirror symmetry this seems to be related to the same question for branes supported on
complex Lagrangian sections of $\widetilde\CM$ which run into singularities.
For branes supported on a single singular fiber of the Hitchin fibration, the relevant extra structure
has been discussed in \cite{Frenkel:2007tx}; it would be desirable to have a discussion analogous to
that of  \cite{Frenkel:2007tx}
which applies to the branes we met here.

\item In \S \ref{sec:lineops} we sketched a  2d-4d  spin wall-crossing formula.
However, some work remains to be done in connecting this formula to the physical
halo picture in the case when $y\not=-1$. For example, we have left out
the details of how to relate the wall-crossing data $d(i,s)$ etc. to the physical
halo particle degeneracies.

\item It would be interesting to generalize one of the main results of \cite{Gaiotto:2010be}
from line defects to surface defects. In \cite{Gaiotto:2010be} it was shown that the
 formal generating functions for line defects
 $F(L) = \sum_{\gamma} \fro (L_\zeta,y,\gamma) X_{\gamma}$ satisfy an
algebra which is a quantum deformation of the algebra of holomorphic functions on $\CM^\zeta$.
This algebra can be interpreted as the algebra $\CO_{BB}$ of open $BB$ strings of the
canonical coisotropic brane on $\CM$, a certain type of $A$-brane.
Now, in the present paper we have
defined $F^{}(L)= \sum_{\gamma_{ij}} \fro (L,\gamma_{ij}) X_{\gamma_{ij}}$. The algebra of these
functions can be identified with the algebra of holomorphic sections of $\Hom(V_{\bS},V_{\bS})$,
which is the algebra of open string states on a $B$-brane whose Chan-Paton bundle is
$V_{\bS}$. As in \S \ref{sec:lineops}
we can introduce a $y$-deformation $\widehat{F}(L)= \sum_{\gamma_{ij}} \fro (L,\gamma_{ij},y) \widehat X_{\gamma_{ij}}$ to produce a $y$-dependent
deformation of the noncommutative --- but ``classical'' --- algebra $\Gamma\left( \Hom(V_{\bS},V_{\bS})\right)$.
It is natural to conjecture that this $y$-deformed algebra is related to the open string algebra
$\CO_{B_{\IS}, B_{\IS}}$ where $B_{\IS}$ is a higher rank generalization of the canonical coisotropic
brane, (an $A$-brane), which should, moreover  have a natural hyperholomorphic connection. (Higher rank coisotropic branes have been investigated in \cite{Herbst:2010js}.)

\item Recently, Witten  has introduced an approach to
knot homology based on the Hilbert space associated to surface defects in
six-dimensional $(2,0)$ theory \cite{Witten:2011zz}.  It is natural to ask whether the considerations
of this paper can play a role in this approach to knot homology.
Let us consider a theory in the class $\CS$ on $\IR^{4} \times C$. Let us label
coordinates $x^{0,1,2,3}\in \IR^4$ and $z\in C$. Consider
a knot $K\subset \IR \times C$ described by an equation $z = z(x^3)$.
With a suitable topological twisting we can consider the surface
defect $\bS_z$ at $x^1=x^2=0$ with worldsheet coordinates $x^0$ and $x^3$,
with parameter $z=z(x^3)$ varying along $x^3$. Imagine that $z(x^3)$ is
generically constant, but has rapid transitions --- like a soliton.
Then we can consider the surface defect to have several interfaces.   More generally, one could
consider a surface defect with various kinds of interfaces and Janus
defects. It would appear that one would need to generalize surface defect
amplitudes to amplitudes corresponding to several surface defects
located at $(z_1,\dots, z_n)$. Naively, these correspond to having several
M2-branes end on the M5-brane. It would be interesting to generalize the
notion of framed BPS states to such situations. If this can be done it would
then be natural to ask if the spaces of framed BPS states associated
to this setup are related to knot homologies. As a first step one would
want to relate the $q$-grading of knot homology to the $y$-grading of spaces
of framed BPS states.

\item It might be interesting to apply some of our techniques to the study
of branes in Landau-Ginzburg models, a subject which has been extensively
discussed in \cite{Hori:2000ck}. Indeed it was noted at the very end of
\cite{Gaiotto:2009fs} that the period amplitudes $\Pi_i^a$ of  \cite{Hori:2000ck}
should be related to supersymmetric interface amplitudes between a surface
defect and the null surface defect. This in turn might shed further shed
light on the quantum McKay correspondence \cite{Martinec:2002wg,Moore:2004yt,Moore:2005wp}.

\item The integral equations of \cite{Gaiotto:2008cd} for the $\CY_\gamma$ are
formally similar  to the thermodynamic Bethe ansatz. The new integral equations
introduced in \S \ref{subsec:Integral-Equations} are very similar to those
used in the inverse scattering method of integrable systems theory
\cite{MR2348643,MR1995460}. The $x_{\gamma_{ij}}$ are analogous to scattering
data, and the $g_k$ are used to solve the associated linear problem.
In some special cases the link can be made much more explicit. For example,
in a recent paper S. Lukyanov and A. Zamolodchikov
\cite{Lukyanov:2010rn} studied the modified sinh-Gordon equation.
This intersects with the Hitchin systems for AD theories for special values
of the parameter $\alpha$ of  \cite{Lukyanov:2010rn}. Equation (5.15)
of \cite{Lukyanov:2010rn} is related to our proposal for solving the auxiliary
linear problem. It would be useful to sharpen this relation and see whether
there is room for any interesting technology transfer between the theory of
integrable systems and the theory of defects in supersymmetric Yang-Mills theory.

\item Recently the methods of \cite{Gaiotto:2008cd,Gaiotto:2009hg}
have found some surprising applications to scattering in $\N=4$ super
Yang-Mills theory \cite{Alday:2009yn,Alday:2009dv,Alday:2010vh,Alday:2010ku}.
We believe that applying the techniques of this paper to that situation should allow one to study the
shape of the minimal surfaces in $AdS_5$ whose area computes the amplitudes.
Information about the shape of the surface can
be useful in computing mixed correlation functions of a polygonal Wilson loop and local
operators in the strong coupling limit: they should correspond to the integral over
the surface of an appropriate bulk-to-boundary propagator.

\item Another physical context in which this structure appears is in $\CN=2$ supergravity.
Surface defects are replaced by cosmic strings and dyons are replaced by
dyonic black holes \cite{Ooguri:1999bv,Neitzke:2007yw}.  Perhaps the 2d-4d wall-crossing
formula can be useful in this context.  Related to this, we can make surface defects
by wrapping D-branes on Calabi-Yau manifolds.  For example, in Type IIA string theory
we can consider D4-branes wrapping supersymmetric 3-cycles in a Calabi-Yau manifold.
It should be very interesting to apply the present formalism to these surface defects and
we intend to return to this subject in the future.

\item Finally, the 2d-4d BPS degeneracies we have described in this paper should presumably
be identified mathematically as the ``open'' analogue of Donaldson-Thomas invariants.
It would be very interesting to develop this theory. (Some investigation of
such open invariants has been carried out in \cite{Aganagic:2009cg,Sulkowski:2010eg}.)

\end{enumerate}

\section*{Acknowledgements}

We thank David Ben-Zvi, Emanuel Diaconescu, Dan Freed, Martin Rocek, Nathan Seiberg,
Karen Uhlenbeck, and Edward Witten for discussions.

We would like
to thank the Simons Center for Geometry and Physics for hosting several
excellent workshops related to this project and for hospitality at those
workshops. We also thank the Aspen Center for Physics for hospitality during the
initial writing of this paper. GM would like to thank the SCGP for hospitality
during the final stages of writing this paper.

The work of GM is supported by the DOE under grant
DE-FG02-96ER40959.   The work of AN is supported by the NSF under grant number
DMS-1006046.  DG is supported in part by the NSF grant PHY-0503584.
DG is supported in part by the Roger Dashen membership in the Institute for Advanced
Study.

\appendix

\section{Relation of 2d and 4d superalgebras and their multiplets}\label{app:d2d4-multiplets}

The preserved supersymmetries of our surface defect are part of
a subalgebra of the $\CN=2$, $d=4$ superalgebra which can be seen to be
isomorphic to   a $d=2$, $(2,2)$ superalgebra by mapping
\begin{equation}
Q_+ = Q_2^{~2} = - Q_{21} \qquad \qquad \bar Q_+ = \bar Q_{\dot 2 2}
\end{equation}
\begin{equation}
Q_- = \bar Q_{\dot 1 1} \qquad  \bar Q_- = Q_1^{~1} = Q_{12}
\end{equation}
One can then check using the $d=4$ $\CN=2$ algebra relations that
\begin{equation}\label{eq:2d-algebra}
\begin{split}
\{ Q_+, \bar Q_+ \} & =  -2 (P_0 + P_3) \\
Q_+^2 = \bar Q_+^2 & = 0 \\
\{ Q_- , \bar Q_- \} & = - 2 (P_0 - P_3) \\
Q_-^2 = \bar Q_-^2 & = 0 \\
\{ Q_+, Q_- \} & = \{ \bar Q_+ , \bar Q_-\}   = 0 \\
\{ Q_+, \bar Q_- \} & = - 2 \bar Z \\
\{ \bar Q_+,   Q_- \} & = - 2   Z \\
\end{split}
\end{equation}
The $u(1)_V$ symmetry of $(2,2)$ is identified with the action of the
generator of $so(2)_{12}$ (which equals that of $u(1)_r$ acting on the
supercharges) and the $u(1)_A$
symmetry is identified with $u(1)_R$.

An important point is that one does \emph{not} need to set $P_1 = P_2 = 0 $
when deriving \eqref{eq:2d-algebra}. Therefore the field multiplets
of $d=4$, $\CN=2$ decompose as multiplets of $d=2$ $(2,2)$ supersymmetry
without putting any conditions such are normally employed in dimensional
reduction.  In particular, with the above identification, the four-dimensional
fields in a vectormultiplet can be assembled as
\begin{equation}\label{eq:vm-to-tw-cm}
\Upsilon = \varphi - \I \sqrt{2} \vartheta^+ \psi_{21} - \I \sqrt{2}~ \overline{\vartheta^-} \psi_{12}
+  \vartheta^+ \overline{\vartheta^-} ( F_{03}- \I (F_{12}-D_{12})  )
\end{equation}
to produce a twisted chiral multiplet with values in the adjoint representation
of some gauged Lie algebra $\lieg$.  The remaining
degrees of freedom in the 4d vectormultiplet decompose in an infinite tower of
``semi-chiral (2,2) multiplets'' in the language of \cite{Buscher:1987uw,Lindstrom:2005zr,Lindstrom:2007sq}.

With a surface defect we may restrict the four-dimensional field \eqref{eq:vm-to-tw-cm}
to the surface to produce a twisted chiral multiplet on the worldsheet of the
surface defect. The associated vectormultiplet can be used to gauge
a flavor $\lieg$-symmetry of a $(2,2)$ quantum field theory. In particular the
adjoint scalar $\varphi$ plays the role of a twisted mass parameter.

The full supersymmetric expression for the twisted superpotential
contribution to the action is
\begin{equation}
\exp\left[ \frac{\I}{2} \int \de x^3 \de x^0 \, \left( \frac{\p \CW_i}{\p a^I} (\I (D_{12} - F_{12}) + F_{03}^I)
-\frac{\p^2 \CW_i}{\p a^I \p a^J} \psi_+^I \psi^J_- \right) + c.c. \right]
\end{equation}
The variation of the holomorphic term is exactly zero under two supersymmetries $Q_1^{\ 1}$ and $Q_2^{\ 2}$,
and under the other two varies into total derivatives
\begin{equation}
\sim \int \de x^3 \de x^0 (\p_0 \pm \p_3) \left( \frac{\p \CW_i }{\p a^I} \psi_{\mp}^I \right).
\end{equation}
Therefore, when placed on  a half-space $x^0\leq 0$ the supersymmetric variation can
be cancelled by a line defect of the form \eqref{eq:interfaceop}.

Finally, let us comment on the BPS representations and the definition
of the index $\mu(\gamma_{ij})$. In our conventions, in
a massive representation the
little superalgebra is diagonalized by the combinations
\begin{equation}
\begin{split}
\{ Q_+ + e^{\I \rho} Q_- , \bar Q_+ + e^{-\I \rho} \bar Q_-\} & = 4(E-\vert Z \vert), \\
\{ Q_+ - e^{\I \rho} Q_- , \bar Q_+ - e^{-\I \rho} \bar Q_-\} & = 4(E+\vert Z \vert),
\end{split}
\end{equation}
with all other anticommutators vanishing for $e^{\I \rho} Z = \vert Z \vert$.
Therefore, short multiplets when $E = \vert Z \vert$ are spanned by the doublet
$ \vert s \rangle$ and $(\bar Q_+ - e^{-i \rho} \bar Q_-) \vert s\rangle$, where
the Clifford vacuum is defined by
\begin{equation}
(Q_+ + e^{\I \rho} Q_-) \vert s \rangle = (\bar Q_+ + e^{-\I \rho} \bar Q_-) \vert s \rangle
= (Q_+ - e^{\I \rho} Q_-) \vert s \rangle = 0.
\end{equation}
Note that to define the index of \cite{Cecotti:1992rm}\footnote{A
related ``new supersymmetric index'' was defined earlier in
\cite{Cecotti:1992qh}.  The relation of this index to the index $\mu$ is
roughly analogous to the relation between the \hk\ metric on $\CM$ and
the indices $\Omega$ in \cite{Gaiotto:2008cd}.} we must take $F$ to be the $u(1)_V$
symmetry. That is, it must satisfy $[F, Q_+] = Q_+ $  and $[F, Q_-] = Q_-$.  Note that
this leaves $F$ ambiguous by a shift by any operator commuting with the entire superalgebra.
The contribution of the short representation  to the index  $\Tr F e^{\I \pi F}$ is
\begin{equation}
(f - (f-1)) e^{ \I \pi f} = e^{ \I \pi f}
\end{equation}
We may take this short multiplet, denoted $h$, to represent the center of mass degree of
freedom in the soliton multiplet. Factoring this out in a BPS representation
space so that
\begin{equation}
\CH^{BPS}_{\gamma_{ij}} = h \otimes \tilde \CH^{BPS}_{\gamma_{ij}}
\end{equation}
we may write
\begin{equation}\label{eq:muij-def}
\mu(\gamma_{ij}):= \Tr_{ \tilde \CH^{BPS}_{\gamma_{ij}}} e^{\I \pi F}
\end{equation}

\section{A simple Hitchin system}\label{app:N1hitchin}

We consider a rank 2 Hitchin system on $\IC \IP^1$ with a single
irregular singularity at $z = \infty$.  This singularity is invariantly
characterized by the statement that $\Tr \varphi^2$ has a fifth-order pole there.  Since $\Tr \varphi^2$ is regular
everywhere else, this actually fixes it (up to shifts of $z$ and overall rescaling):
\begin{equation}
 \Tr \varphi^2 = 2z \, (\de z)^{2}.
\end{equation}
The precise boundary condition is given by specifying the form of $(A,\varphi)$ near the singularity;
it is given in \S 3 of \cite{Gaiotto:2010be}, but we will not need it.  What we do need to know is that there
is a unique solution obeying this boundary condition.
In this appendix we briefly summarize some facts about this solution.  See \S 9.4.1 of \cite{Gaiotto:2010be} for
its explicit form.

We consider the corresponding flat connection $\nabla(\zeta)$ given by \eqref{eq:hit-conn}.
For any fixed $\zeta$, the asymptotics of the flat sections of $\nabla(\zeta)$
exhibit Stokes phenomena as $z \to \infty$, with three anti-Stokes rays $r_a$, $a = 1, 2, 3$ cyclically.
The $r_a$ are located at $z$ for which $\frac{1}{\zeta} z^{3/2} \in \IR$ (this condition makes sense even though $z^{3/2}$ is
defined only up to sign).
There are three ``small flat sections'' $s_a(\zeta, z)$
defined up to scalar multiple by the condition that $s_a$ decays (exponentially fast) as $z \to \infty$
along $r_a$, indeed $s_a \sim \exp \left[ - \frac{\pi R}{\zeta}\frac{2}{3} z_a^{3/2} \right]$, where ``$z_a^{3/2}$''
means the choice of $z^{3/2}$ making the exponent negative along $r_a$.
This requirement implies that $s_a$ is not a scalar multiple
of $s_{a+1}$, so we can normalize the sections by the requirement that $(s_a, s_{a+1})=1$ where $(,)$ denotes an
$SU(2)$-invariant antisymmetric inner product.  This normalization fixes the $s_a$ up to an overall sign, and
implies that $\sum_a s_a = 0$.  The $s_a$ are not quite single-valued as functions of $\zeta \in \IC^\times$:  rather,
varying $\zeta \to e^{2 \pi \I} \zeta$ permutes the three anti-Stokes rays,
and gives $s_a(e^{2 \pi \I} \zeta) = - s_{a+1}(\zeta)$.\footnote{From what we have explained so
far, one can see easily that $s_a(e^{2 \pi \I} \zeta) = \pm s_{a+1}(\zeta)$.
The sign is subtler to determine.}  On the other hand they are perfectly
analytic in $z$ for fixed $\zeta$.

Now fix a point $z$ and a single root $z^{3/2}$.
From the $s_a$ just discussed, we can assemble
two single-valued but piecewise-continuous sections $\CY_{\pm}(\zeta)$
which have uniform asymptotics as $\zeta \to 0$:
\begin{equation}
\CY_\pm(\zeta) \sim \exp \left[ \pm \frac{\pi R}{\zeta}\frac{2}{3} z^{3/2} \right] e_\pm,
\end{equation}
where $e_\pm$ are $\zeta$-independent.

We have to take a little care since $s_a$ are multivalued as functions of $\zeta$.
Fixing a branch choice is equivalent to fixing a labeling of the three anti-Stokes rays for
every $\zeta$.  First, for $\zeta \in z^{3/2} \R_+$, one of the three anti-Stokes rays actually
contains the point $z$; label this one $r_1$, and then label the other two $r_2$, $r_3$ going around
counterclockwise.  We then vary this picture continuously with $\zeta$, until we reach $\zeta \in z^{3/2} \R_-$,
where we put a branch cut in the $\zeta$-plane.
Having chosen this labeling we now have $s_a(\zeta)$ well defined and single-valued.

We now take $(\CY_+, \CY_-) = (s_3, s_1)$ in the half-plane $\mathrm{Im} \frac{1}{\zeta} z^{3/2}>0$
and $(\CY_+, \CY_-) = (-s_2, s_1)$ in the half-plane $\mathrm{Im} \frac{1}{\zeta} z^{3/2}<0$.
This gives two piecewise-continuous sections $\CY_\pm$ which have the desired asymptotics as $\zeta \to 0$.
What are their discontinuities?
As $\zeta$ crosses the ray $ \frac{1}{\zeta} z^{3/2} \in \IR_+$, the relation $- s_2 = s_1 + s_3$ gives
\begin{equation} \label{eq:stokesfactor-1}
\CY_- \to \CY_-, \qquad \qquad \CY_+ \to \CY_+ + \CY_-,
\end{equation}
while as $\zeta$ crosses $\frac{1}{\zeta} z^{3/2} \in \IR_-$, using $- s_3 = s_1 - (-s_2)$ \ti{and}
taking account of the discontinuities of the $s_i$ themselves across this cut gives
\begin{equation} \label{eq:stokesfactor-2}
\CY_+ \to \CY_+, \qquad \qquad \CY_- \to \CY_- - \CY_+.
\end{equation}

According to the Riemann-Hilbert correspondence, $\CY_\pm(z, \zeta)$ (with $z$ fixed and $\zeta$ varying)
are discontinuous flat sections
for some connection in a rank-2 bundle over $\IC\IP^1$, with irregular singularities
at $\zeta = 0$ and $\zeta = \infty$ (and regular everywhere else.)  The equations
\eqref{eq:stokesfactor-1}, \eqref{eq:stokesfactor-2} give the Stokes factors for this connection, at either irregular
singularity, as $\begin{pmatrix} 1 & 1 \\ 0 & 1 \end{pmatrix}$ and $\begin{pmatrix} 1 & 0 \\ -1 & 1 \end{pmatrix}$.

\section{Two-dimensional bound state radius in Landau-Ginzburg theories}\label{app:LG-Radius}

In this appendix we give a quantitative lower bound for how the
``bound state radius'' of a soliton in a Landau-Ginzburg theory diverges when three critical
points become collinear.

If we take, for simplicity, a trivial Kahler metric $K_{\alpha\beta} = \delta_{\alpha,\beta}$
then the soliton equation for an $(ij)$-soliton is
\begin{equation}
\frac{\de \phi^\alpha}{\de x} = \zeta_{ji} \frac{\partial \bar W}{\partial \bar \phi^\alpha}
\end{equation}
where $W(\phi)$ is the superpotential, the boundary condition is determined by critical
points $\phi_i, \phi_j$:
\begin{equation}
\begin{split}
\lim_{x \to - \infty} \phi(x) & = \phi_{i}\\
\lim_{x \to + \infty} \phi(x) & = \phi_{j}
\end{split}
\end{equation}
and $\zeta_{ji} = \frac{W(j)-W(i)}{\vert W(j)-W(i)\vert} $ is determined by the
corresponding critical values. Here and below $W(i):= W(\phi_i)$, etc.

We can derive immediately
\begin{equation}\label{eq:W-flow}
\frac{\de W}{\de x} = \zeta_{ji} \left| \frac{\partial \bar W}{\partial \bar \phi^\alpha} \right|^2
\end{equation}
so the image of the soliton in the $W$-plane is a straight line of slope $\zeta_{ji}$.
Near each critical point --- assumed to be Morse ---
a quadratic approximation to $W$ applies and the solution
approaches the critical point with a decaying exponential. Therefore, we expect a rapid
transition region from $\phi \approx \phi_i$ to $\phi\approx \phi_j$. This can be estimated
from \eqref{eq:W-flow} to be
\begin{equation}\label{eq:LG-center}
\Delta x_{ji} \sim \frac{\vert W(j) - W(i) \vert}{  \left| \frac{\partial \bar W}{\partial \bar \phi^\alpha} \right|^2_*}
\end{equation}
where in the denominator we use some typical intermediate value from the transition region.
The soliton does not have a well-defined center, but it should be located within a region
of size \eqref{eq:LG-center}.

Now consider a situation where there are at least three critical points $\phi_i, \phi_j, \phi_k$
and the parameters of $W$ are changed so that the critical values $W(i), W(j), W(k)$ are
becoming collinear. We assume moreover that there are $ij$ and $jk$ solitons.
As the critical values become collinear (and we approach the wall of marginal
stability from a ``stable'' region) some of the $ik$ solitons can be viewed as
bound states of $ij$ and $jk$ solitons. We would like to estimate how the
``bound state radius'' of these bound states diverges as the critical values become
collinear.

Along the $x$-axis an $ik$ soliton of the bound state type  will first have
a rapid transition from $\phi\approx \phi_i$ to $\phi \approx \phi_j$ and then will stay
close to $\phi \approx \phi_j$ for a long interval $L$ and then transition rapidly
from $\phi \approx \phi_j$ to $\phi \approx \phi_k$. The width of the two transition regions
can be estimated using \eqref{eq:LG-center} above, and if they are much smaller than $L$,
we can speak of an approximate bound state radius.

We can put a lower bound on the bound state radius as follows.
Only the real part of $\frac{W}{\zeta}$ flows --- the imaginary part is fixed.
In order to get a lower bound on the distance in space corresponding to a
part of the soliton trajectory in the $W$ plane which passes close to the turning point $j$
we can use a quadratic approximation, and integrate the equation
in the finite $W$-plane region of the flow where the quadratic approximation is valid.

As a simple example capturing the spirit of the problem,
consider a one-dimensional situation, with critical point $\phi_j$ at the origin and critical
value $W(j)=0$:
\begin{equation}
W = \frac{1}{2} C \phi^2 + \cdots
\end{equation}
Then we can measure how close the flow is to the critical point
by the small value $\epsilon$ of the imaginary part of $\frac{W}{\zeta}$.
Let us denote the real part of $\frac{W}{\zeta}$ as $t$.
Notice that
\begin{equation}
\left| \frac{\partial \bar W}{\partial \bar \phi^\alpha} \right|^2 \approx |C|^2 |\phi|^2 \approx 2 |C| |W| \approx 2 |C| \sqrt{t^2 + \epsilon^2}+ \cdots
\end{equation}

So the distance in real space $L$  is bounded below by
\begin{equation}\label{eq:Div-Integral}
\int_{t_0}^{t_1} \frac{\de t}{2 |C| \sqrt{t^2 + \epsilon^2}}
\end{equation}
where $t_0, t_1$ are some fixed values far from the transition regions of the $ij$ and $jk$ solitons.
The integral \eqref{eq:Div-Integral} diverges
logarithmically as $\epsilon \to 0$:
\begin{equation}\label{eq:Div-Integral-2}
\int_{t_0}^{t_1} \frac{\de t}{2 |C| \sqrt{t^2 + \epsilon^2}} \sim \frac{1}{\vert C \vert} \log \frac{1}{\epsilon}
\end{equation}
If we consider the triangle formed by $Z_i,Z_j,Z_k$ and let $\epsilon$ be the height of the perpendicular
from $Z_j$ to the segment $[Z_i,Z_k]$, then $\epsilon = {\rm Im}(Z_{ij} \frac{\bar Z_{ik}}{ \vert Z_{ik}\vert} )$,
so we can write the result more invariantly as
\begin{equation}
L \geq \frac{1}{\Lambda} \log \left( \frac{ \vert Z_{ik} \vert }{{\rm Im}(Z_{ij} \bar Z_{ik} )}\right)
\end{equation}
where $\Lambda$ is a nonuniversal number. The argument of the logarithm is essentially the
same quantity that arises in Denef's bound state radius formula for 4dimensional bound state radii.

If the field space is of higher dimension, the above estimate can be given as it is as long as
$|\partial W|^2$ is bounded from above by $|C| |W|$ for some constant $C$ near the critical point. This is trivially true as long as the critical point is non-degenerate.

Thus, one can give a halo-type description to the decay of $1+1$ dimensional solitons, but the
bound state radius formula is not as precise as in the four-dimensional case. The reason is that
the physics responsible for the bound state is rather different, and does not involve the
exchange of massless particles.

\section{Physical models for the affine linearity of $\omega(\gamma, \gamma_i)$}\label{app:Landau-Levels}

The change of the number of chiral multiplets on the surface defect discussed in
\S \ref{subsec:BPS-Deg-WC} is a rather novel
phenomenon.  In this appendix we offer some physical models for understanding
why it happens.

\subsection{A supersymmetric quantum mechanics model}

As a warmup, we consider the following model problem.  Let us thicken
the solenoid so that it carries uniform magnetic flux $B$ in a disk of radius $R$ around the
origin. We search for the ground states of a supersymmetric quantum mechanical
charged particle in the complex plane coupled to this  magnetic field.  We can choose a gauge
\begin{equation}
A = \begin{cases} \half B (z \de \bar z - \bar z \de z ) & \vert z \vert \leq R \\
\half B R^2 \frac{(z \de \bar z - \bar z \de z )}{\vert z \vert^2} &  \vert z \vert \geq  R \\
\end{cases}
\end{equation}
The wavefunctions lie in $\Omega^{0,*}(\IC)$ and the supersymmetry operators are
identified with  $\pbar + \bar A : \Omega^{0,0}(\IC) \to \Omega^{0,1}(\IC)$ and
$(\pbar + \bar A )^\dagger: \Omega^{0,1}(\IC) \to \Omega^{0,0}(\IC)$. For $B>0$ we find that the normalizable
zero-energy wavefunctions in $\Omega^{0,0}$ have the form
\begin{equation}\label{eq:Model-one}
\psi =  \begin{cases} f(z) e^{-\half B \vert z \vert^2} & \vert z \vert \leq R \\
(\frac{z \bar z}{R^2})^{-BR^2/2} e^{-\half BR^2} f(z) &  \vert z \vert \geq  R \\
\end{cases}
\end{equation}
where $f(z)$ is an entire function.  The total
wavefunction is then only normalizable if $f(z)$ is a polynomial of degree $n$ with
$n < BR^2 -1$.  There are no normalizable zero-energy states in $\Omega^{0,1}$.
The situation is reversed for $B<0$ with normalizable wavefunctions in $\Omega^{0,1}$
governed by a polynomial $g(\bar z)$ of degree $n < \vert B \vert R^2 -1$ and
no normalizable wavefunctions of type $\Omega^{0,0}$. One important  point
brought out by this example is that the number of states depends linearly on the flux through the
solenoid,  not just the holonomy around the solenoid.

\subsection{Probe particle in the presence of a solenoid and dyon}

In this section we analyze the quantum mechanics of a halo particle near
a dyon interface in a solenoid in order to determine the structure of the
one-particle Hilbert space which generates the halo Fock space,
as used in \S \ref{subsubsec:4d-halo}.

We model the surface defect with interface as in \S \ref{subsubsec:4d-halo}
as a solenoid with a dyon of charge   $\gamma_{i,j'} \in  \Gamma_{i,j'}$ inside.
One can write the fixed point equations for the preserved supersymmetries. A solution
is provided by the attractor-like equations
\begin{equation}\label{eq:vm-scal}
2 {\rm Im} (\zeta^{-1} Z(\gamma, u(r)) =  - \frac{\langle \gamma, \gamma_{i,j'} \rangle}{r} +
2 {\rm Im} (\zeta^{-1} Z(\gamma, u)).
\end{equation}
In the self-dual formalism the gauge potential is
\begin{equation}\label{eq:sd-potent}
\CA = \half \left( - \cos \theta \de \phi \otimes \gamma_{ij'} - \frac{\de t}{r} \otimes I(\gamma_{ij'}) \right),
\end{equation}
where a projection of $ \gamma_{ij'}$ to $\Gamma_g$ is understood
and $I$ is the complex structure on $\Gamma_g \otimes \IR$ at $u\in \CB$.
As in the Denef halo solutions (for details in this situation see \cite{Gaiotto:2010be}, Appendix C
or \cite{PiTP}),
a test particle of charge $\gamma^h$ is bound at a radius given by the same formula in
the absence of the surface defect. (After all, outside the solenoid, the electromagnetic
fields are zero, and hence the probing dyon feels no extra force from the solenoid. )

The effective Minkowski-signature Lagrangian of a halo particle in the
probe approximation, confined to the bound state radius, is of the form
\begin{equation}
\int \half \mu r^2 (\sin^2 \theta \dot \phi^2 + \dot \theta^2) - \int (\kappa_1 \cos\theta + \kappa_2^{\rm left}) \dot \phi
\end{equation}
or
\begin{equation}
\int \half \mu r^2 (\sin^2 \theta \dot \phi^2 + \dot \theta^2) - \int (\kappa_1 \cos\theta + \kappa_2^{\rm right}) \dot \phi.
\end{equation}
Here $\theta, \phi$ are angular coordinates centered on the dyon. The ``left'' half of the
surface defect is $\theta=0$ and the ``right'' half is $\theta=\pi$. The first
Lagrangian applies for  $0 \leq \theta < \pi$ and  the second for
$0 < \theta \leq \pi$. The two Lagrangians define a smooth quantum measure for
$\kappa_2^{\rm left} - \kappa_2^{\rm right} \in \IZ$. Moreover, $\kappa_2^{\rm left} $ and
$\kappa_2^{\rm right}$ can be shifted by integers.

The Aharonov-Bohm phase of the probe particle of charge $\gamma^h$ around a small circle far
to the left is $\exp[ 2\pi \I \langle \gamma^h, \gamma_i^0 \rangle]$  while that far to the
right is $\exp[ 2\pi \I \langle \gamma^h, \gamma_{j'}^0 \rangle]$. Shifting $\kappa_2$
by appropriate integers we may take the solution
\begin{equation}
\begin{split}
\kappa_1 & = \half \langle \gamma^h, \gamma_{ij'} \rangle, \\
\kappa_2^{\rm left} & = \half \left( \langle \gamma^h, \gamma_i^0 + \gamma_{j'}^0 - \gamma^c \rangle \right), \\
\kappa_2^{\rm right} & = \half \left( \langle \gamma^h, \gamma_i^0 + \gamma_{j'}^0 + \gamma^c \rangle \right).
\end{split}
\end{equation}
Note that
\begin{equation}
\kappa_2^{\rm left}- \kappa_2^{\rm right}= - \langle \gamma^h, \gamma^c \rangle.
\end{equation}

Now let us analyze the ground state wavefunctions of the quantum particle with Lagrangian
\begin{equation}
\int \half \mu r^2 (\sin^2 \theta \dot \phi^2 + \dot \theta^2) - \int (\kappa_1 \cos\theta + \kappa_2) \dot \phi.
\end{equation}
Such wavefunctions are of the form
\begin{equation}
\Psi = e^{\I m \phi} (1 + \cos \theta)^{a } (1- \cos \theta)^{b},
\end{equation}
where $2m\in \IZ$, with $2m$ even or odd according to whether the particle is a boson or a fermion.
The values of $a$, $b$ are
\begin{equation}
a = \frac{ {\rm sign}(\kappa_1) }{2} (\kappa_1 - m'), \qquad
b = \frac{ {\rm sign}(\kappa_1) }{2} (\kappa_1 +m').
\end{equation}
with $m'= m + \kappa_2$. If we require that the wavefunction be nonsingular
everywhere on the sphere, so that $a \geq 0$ and $b \geq 0$, then $m'$ must
lie in the interval
\begin{equation}\label{eq:Range-hatm}
\vert \kappa_1 \vert \geq m'\geq - \vert \kappa_1 \vert
\end{equation}
The energy is $E= \frac{\vert \kappa_1 \vert}{2\mu r^2}$.

Let $N_{\kappa_1, \kappa_2}$ denote the number of solutions of the inequality \eqref{eq:Range-hatm}.
In the case where $\kappa_1$ and $\kappa_2$ are not half-integers and $\kappa_1>0$ its value is given by
\begin{equation}\label{eq:Nkappa-def}
\renewcommand{\arraystretch}{1.3}
\begin{tabular}{|c||c|c|c|c|c|c|}
\hline  $N_{\kappa_1, \kappa_2}$  & $\{ \kappa_1 \} < 1- \{ \kappa_2 \} $ & $\{ \kappa_1 \} \geq  1- \{ \kappa_2 \} $
\\
\hline $\{ \kappa_1 \} \geq    \{ \kappa_2 \} $  &  $2[\kappa_1] + 1$  & $2 [\kappa_1] + 2$
\\
\hline $\{ \kappa_1 \} <    \{ \kappa_2 \} $ & $2[\kappa_1]  $ &  $2[\kappa_1] + 1$
\\
\hline
\end{tabular}
\end{equation}
where $x = [x] + \{ x \} $ is the decomposition of a real number into its greatest integer
and fractional parts.

In a chamber where there are no 2d bound states of charge $\gamma^h_{ii}$ or $\gamma^h_{j'j'}$
we can identify
\begin{equation}\label{eq:omega-Landau}
\omega(\gamma^h, \gamma_{ij'}) := N_{\kappa_1, \kappa_2} \Omega(\gamma^h).
\end{equation}
Note that shifting $\gamma_{ij'} \to \gamma_{ij'} + \Delta \gamma^c$ leads to a
shift in $N_{\kappa_1, \kappa_2}$ reproducing the affine-linearity of $\omega(\gamma^h, \gamma_{ij'})$.

As an aside, we note the curious point that
if we had only demanded that the wavefunctions be $L^2$-normalizable with
respect to the measure $\de(\cos\theta) \de \phi$, then for generic values of $\{ \kappa_1 \}$
and $\{ \kappa_2 \} $ the ground state would in fact be only two-fold degenerate with
ground state energy smaller than $\vert \kappa_1 \vert$ and with the two eigenstates
having diverging wavefunctions at the north and south poles,
respectively.

\section{Twistor construction of hyperholomorphic connections}\label{app:Twistor-HH}

In this appendix we explain how to construct the hyperholomorphic connection $A$ on the bundle
$V$ over $\CM$, from the sections $\CY_{\gamma_i}$ of $V$ over
$\CM \times \IC^\times$.  The construction is local along $\CM$, but global along $\IC^\times$.
It seems to be a close cousin to the construction of self-dual Yang-Mills fields \cite{Ward:1977ta}
and more generally of hyperholomorphic connections from holomorphic vector bundles over
twistor space, although we will not use those constructions explicitly.

Our construction depends crucially on the analytic properties of $\CY_{\gamma_i}$ as a function
of $\zeta \in \IC^\times$ at a fixed point of $\CM$:
\begin{itemize}
 \item $\CY_{\gamma_i}$ is \ti{piecewise} holomorphic in $\zeta$, with discontinuities only
at the BPS rays;
 \item The jumps of $\CY_{\gamma_i}$ across BPS rays are of the form $\CY_{\gamma_i} \to \sum_{k, \gamma_k \in \Gamma_k} c_{\gamma_i,\gamma_k} \CY_{\gamma_k}$,
with $c_{\gamma_i,\gamma_k}$ constants;
 \item The limit of $\CY_{\gamma_i} (\CY^\sf_{\gamma_i})^{-1}$ as $\zeta \to 0$ is finite and smooth;
 \item Finally, to normalize the sections we
  put a reality condition on $\CY_{\gamma_i}$ by demanding that in the
 Hermitian metric  $h$ on $V$, defined by the unitary framing of \S \ref{subsubsec:VB},
 we have the unitarity constraints described in equations \eqref{eq:unitarity-1}, et. seq. in
 \S \ref{subsec:Integral-Equations}.

\end{itemize}

The first three  properties are consequences of the integral equations \eqref{eq:int-1}, \eqref{eq:int-2}.
 Together with the unitarity constraints   all four properties  determine the sections uniquely.

The first step in constructing the hyperholomorphic   connection on $V$
 is to pass from these properties of the $\CY_{\gamma_i}$ to a system of ``Cauchy-Riemann''
equations for sections of $V$.  We first recall that the Cauchy-Riemann equations for ordinary
functions on $\CM$ can be written in the form \cite{Gaiotto:2008cd}
\begin{align}
\frac{\partial}{\partial u^{}} \CY &= \CA_{u^{}} (\zeta) \CY , \label{eq:cr-1} \\
\frac{\partial}{\partial \bar{u}^{}} \CY &= \CA_{\bar{u}^{}} (\zeta)  \CY, \label{eq:cr-2}
\end{align}
where $\CA_{u^{}}(\zeta)$ and $\CA_{\bar{u}^{}} (\zeta)$ are two complex vertical vector fields
on the torus fibers, depending holomorphically on $\zeta \in \IC^\times$.\footnote{ Of course $\CB$ can
have dimension larger than one, so we should understand $u$ to stand for coordinates $u^m$,
$m=1,\dots, \dim \CB$, with the index suppressed. } In particular, these equations hold for
the functions $\CY = \CY_\gamma$.
One can describe the $\zeta$
dependence of these vector fields more precisely:  they are of the form
\begin{align}
\CA_{u^{}}(\zeta) &= \frac{1}{\zeta} \CA^{(-1)}_{u^{}} + \CA^{(0)}_{u^{}}, \\
\CA_{\bar{u}^{}}(\zeta) &=\CA^{(0)}_{\bar{u}^{}} + \zeta \CA^{(1)}_{\bar{u}^{}},
\end{align}
where $\CA^{(-1)}_{u^{}}$ are linearly independent at every point,
and similarly $\CA^{(1)}_{\bar{u}^{}}$.
This form of the Cauchy-Riemann equations is one of the main ingredients in the twistorial construction
of the \hk metric $g$.

Now choose a local trivialization of $V$, so that the
$\CY_{\gamma_j}$ are represented as vector-valued functions.  Also choose a single
$\gamma_j^0 \in \Gamma_j$ for each $j$.  The $\CY_{\gamma^0_j}$ are linearly independent vectors,
at least for $R$ large enough.
We can thus define matrix-valued functions $\CB_{u^{}}(\zeta)$, $\CB_{\bar{u}^{}}(\zeta)$ on $\CM$
by requiring
\begin{align}
\frac{\partial}{\partial u^{}} \CY_{\gamma^0_j} &= \CA_{u^{}}(\zeta) \CY_{\gamma^0_j} + \CB_{u^{}}(\zeta) \CY_{\gamma^0_j}, \label{eq:crv-1} \\
\frac{\partial}{\partial \bar{u}^{}} \CY_{\gamma^0_j} &= \CA_{\bar{u}^{}}(\zeta) \CY_{\gamma^0_j} + \CB_{\bar{u}^{}}(\zeta) \CY_{\gamma^0_j}, \label{eq:crv-2}
\end{align}
for all $j$.  Here each $\CA$ term is a vector field on $\CM$ which acts by differentiation, but is
proportional to the identity matrix as an endomorphism of $V$, while each $\CB$ term is a
matrix-valued function on $\CM$.
It is straightforward to see that the $\CB$ terms are continuous functions of $\zeta$ despite the fact
that $\CY_{\gamma_j^0}$ are not, and moreover that the $\CB$ terms are independent of our choice of
$\gamma_j^0$.  So \eqref{eq:crv-1}, \eqref{eq:crv-2} can be thought of as Cauchy-Riemann equations defining
a holomorphic structure $\bar\partial^{(\zeta)}$ on $V$.

We want to show that the $\zeta$ dependence of $\CB$ has the same form
as that of $\CA$, i.e. that we can decompose
\begin{align}\label{eq:zeta-bee}
\CB_{u^{}}(\zeta) &= \frac{1}{\zeta} \CB^{(-1)}_{u^{}} + \CB^{(0)}_{u^{}},  \\
\CB_{\bar{u}^{}}(\zeta) &=\CB^{(0)}_{\bar{u}^{}} + \zeta \CB^{(1)}_{\bar{u}^{}}.
\end{align}
For this purpose we first consider the semiflat version of \eqref{eq:crv-1} and \eqref{eq:crv-2}.
In that setting we have \cite{Gaiotto:2008cd}
\begin{gather}
\CA^{(-1),\sf}_{u^{}} = -\I \pi R \frac{\partial Z}{\partial u^{}} \cdot \frac{\partial}{\partial \theta},
\quad \CA^{(1),\sf}_{\bar{u}^{}} = -\I \pi R \frac{\partial \bar{Z}}{\partial \bar{u}^{}} \cdot
\frac{\partial}{\partial \theta}, \label{eq:aisf} \\
\CA^{(0),\sf}_{u^{}} = 0, \quad \CA^{(0),\sf}_{\bar{u}^{}} = 0,
\end{gather}
and a direct computation using \eqref{eq:xi-sf} then shows that \eqref{eq:crv-1}, \eqref{eq:crv-2} are satisfied
if we take $\CY = \CY^\sf$ and    $\CB$  to have the form \eqref{eq:zeta-bee} where $\CB^{(-1),\sf}_{u^{}}$
is diagonal  with $jj^{th}$ matrix element given by:
\begin{gather}
\left( \CB^{(-1),\sf}_{u^{}}\right)_{jj} = \pi R \frac{\partial Z_{\gamma^0_j}}{\partial u^{}},
\quad \left( \CB^{(1),\sf}_{\bar{u}^{}}\right)_{jj} = \pi R \frac{\partial \bar Z_{\gamma^0_j}}{\partial \bar{u}^{}},  \label{eq:bisf-1} \\
\CB^{(0),\sf}_{u^{}} = 0, \quad \CB^{(0),\sf}_{\bar{u}^{}} = 0. \label{eq:bisf-2}
\end{gather}
Here to write a definite formula we needed a trivialization of $V$, and we picked the one
determined by the local sections $\gamma^0_j$ of $\Gamma_j$; a $\zeta$-independent change of trivialization
would affect the detailed form of $\CB$ but not the general structure of its $\zeta$ dependence.

Away from the semiflat approximation, to see that $\CB$ still has the desired form, we use our asymptotic
condition on $\CY_{\gamma^0_j}$.  An efficient way to get the result is as follows.  For any fixed $u$, we
can consider the torus $\CM_u$ as the locus of real values of $\theta_\gamma$, sitting inside its
complexification $\CM_u^\IC$ where $\theta_\gamma$ are allowed to be arbitrary.  The functions
$\CY^\sf_\gamma$ have natural continuations to this complexification.
Then we can define a map $\Upsilon(\zeta):  \CM_u \to \CM_u^\IC$ by
requiring $\CY_\gamma = \CY^\sf_\gamma \circ \Upsilon$.  The asymptotics of $\CY_\gamma$ \cite{Gaiotto:2008cd}
say that $\Upsilon(\zeta)$ is finite in the limit $\zeta \to 0$ or $\zeta \to \infty$.  Our asymptotic condition
on $\CY_{\gamma^0_j}$ above just says that the natural extension of $\Upsilon(\zeta)$ to a map of \ti{bundles}
is also finite as $\zeta \to 0, \infty$.  In particular we have $\CB(\zeta) = \Upsilon(\zeta)^* \CB^\sf(\zeta)$,
so we get the desired form for $\CB$, with $\CB^{(-1)}$ and $\CB^{(1)}$ given by the pullbacks of their
semiflat values \eqref{eq:bisf-1}, \eqref{eq:bisf-2}.

Now we are ready to define the unitary connection $A$.
We will engineer $A$ so that \eqref{eq:crv-1} and \eqref{eq:crv-2}
say
\begin{equation} \label{eq:dazeta}
 [(\de+A) \CY_{\gamma_j^0}]^{(0,1)_\zeta} = 0.
\end{equation}
In other words, \eqref{eq:crv-1} and \eqref{eq:crv-2} are
the $\zeta$-dependent Cauchy-Riemann equations for sections of a hyperholomorphic bundle.
The equivalence between \eqref{eq:crv-1}, \eqref{eq:crv-2} and \eqref{eq:dazeta} means
\begin{align}
 A \left( \CA_{u^{}}^{(-1)} \right) &= \CB_{u^{}}^{(-1)}, \\
 A \left( - \partial_{u^{}} + \CA_{u^{}}^{(0)} \right) &= \CB_{u^{}}^{(0)}, \\
 A \left( - \partial_{\bar{u}^{}} + \CA_{\bar{u}^{}}^{(0)} \right) &= \CB_{\bar{u}^{}}^{(0)}, \\
 A \left( \CA_{\bar{u}^{}}^{(1)} \right) &= \CB_{\bar{u}^{}}^{(1)}.
\end{align}
These requirements uniquely define $A$.

Note that

\begin{enumerate}

\item Applying this procedure in the semiflat case, i.e. using \eqref{eq:aisf}-\eqref{eq:bisf-2},
we quickly recover $A^\sf$ given in \eqref{eq:a-sf}.

\item It would have been impossible to construct an $A$
with the desired property if $\CB$ had contained, say, a $1 / \zeta^2$ term.

\item The transformations of $\CY_{\gamma_i}$ across the BPS walls $W_{a}$ involve multiplication
by holomorphic quantities and  do not
lead to discontinuities in $A$.

\item
Our reality conditions show that
\begin{equation} \label{eq:unitarity}
A^\dagger = -A
\end{equation}
i.e. $\de + A$ is a \ti{unitary} connection with respect to the metric $h$ in $V$.

\item Following the discussion in \cite{Gaiotto:2008cd} we expect that the sections
$\CY_{\gamma_i}$ also satisfy interesting equations in $R$ and $\zeta$ expressing
anomalous Ward identities. We leave a full discussion of this for another occasion.

\end{enumerate}

\section{$\CY_{\gamma_{ij'}}$ in $A_1$ theories}\label{app:a1-signs}

In this appendix we give a precise definition of the sections $\cY_{\gamma_{ij'}}$ in $A_1$ theories.
The main point of this discussion is to check carefully the sign $\sigma$ appearing in the multiplicative
law, and to see that the discontinuities of $\cY_{\gamma_{ij'}}$ are as expected.

\subsection{Defining $\cY$}

We choose $(u,\zeta)$ generic (not on any BPS wall), and also choose the edges of the WKB triangulation so that $z$ does not lie on any edge.  Let $z_i$ and $z_j$ denote the two preimages of $z$ on $\Sigma$.

Recall from \cite{Gaiotto:2009hg} that given an edge $E$ of a quadrilateral $Q_E$ of the WKB triangulation,
we defined $\gamma_E$ to be the odd sum of lifts of a
path in $Q_E$ connecting the two turning points.  Its overall orientation is fixed by requiring
$\inprod{\gamma_E, \hat{E}} = 1$, where $\hat{E}$ denotes either of the lifts of $E$ to $\Sigma$, with its standard
WKB orientation (we will use this orientation several times below.)
Then with vertices $1234$ going around $Q_E$ counterclockwise, and using the shorthand $ab$
for $s_a \wedge s_b$, we defined
\begin{equation}
 \cY_{\gamma_E} = -\cX_{\gamma_E} = \frac{12\ 34}{23\ 41}.
\end{equation}
More general $\cY_\gamma$ were defined using the rule
\begin{equation} \label{eq:sigma-1}
\cY_{\gamma + \gamma'} = (-1)^{\inprod{\gamma,\gamma'}} \cY_{\gamma} \cY_{\gamma'}.
\end{equation}

Now let $S$ denote the ``sector'' where $z$ sits (a triangle
bounded by an edge of the WKB triangulation and two separating WKB curves; two vertices of this triangle
are vertices of the WKB triangulation, while the third is a turning point.)
Let $\gamma_{ij,S}$ denote a path from $z_i$ to $z_j$,
obtained as the odd sum of lifts of a path from $z_i$ to the turning point on the boundary
of $S$.  Also let
\begin{equation}
 s_{i,S} = s_a (bc)
\end{equation}
where $a$ is the vertex reached by flowing along a lifted WKB curve from $z_i$,
and vertices $abc$ go around the triangle counterclockwise.  Similarly define $s_{j,S}$.

Define $\cY_{\gamma_{ij,S}}$ to be an endomorphism of the fiber over $z$ of our rank-2 bundle, which maps $s_{i,S} \mapsto 0$
and $s_{j,S} \mapsto \nu_{i,S} s_{i,S}$, where $\nu_{i,S} = +1$ if the lifted WKB curve through $z_i$
goes around the triangle counterclockwise, and $\nu_{i,S} = -1$ if it goes around clockwise.
For more general $\gamma_{ij}$ we define $\cY_{\gamma_{ij}}$ using the rule
\begin{equation} \label{eq:sigma-2}
\cY_{\gamma_{ij} + \gamma} = \sigma(\gamma) \cY_\gamma \cY_{\gamma_{ij}},
\end{equation}
where $\sigma$ is the ``canonical quadratic refinement'' of \cite{Gaiotto:2009hg}.

Let $\gamma_{ii,0}$ denote the element in $\Gamma_{ii}$ corresponding to $0 \in \Gamma$.
Define $\cY_{\gamma_{ii,0}}$ to be the endomorphism which maps $s_i \mapsto s_i$ and $s_j \mapsto 0$.
For more general $\gamma_{ii}$ we use the rule
\begin{equation} \label{yii}
\cY_{\gamma_{ii,0}+\gamma} = \cY_\gamma \cY_{\gamma_{ii,0}}.
\end{equation}

The fact that $\sigma$ is indeed a quadratic refinement implies that the product we have defined is associative.

\subsection{Multiplication laws}

Directly from the definitions we have the relation
\begin{equation}
 \cY_{\gamma_{ij,S}} \cY_{\gamma_{ji,S}} = - \cY_{\gamma_{ii,0}}.
\end{equation}
(The minus sign comes from the fact that $\nu_{i,S} \nu_{j,S} = -1$.)

Multiplying both sides by general $\cY_\gamma$ this becomes
\begin{equation}
\sigma(\gamma) \cY_{\gamma_{ij,S}+\gamma} \cY_{\gamma_{ji,S}} = - \cY_\gamma \cY_{\gamma_{ii,0}},
\end{equation}
i.e.
\begin{equation}
\cY_{\gamma_{ij,S}+\gamma} \cY_{\gamma_{ji,S}} = - \sigma(\gamma) \cY_{\gamma_{ii,0} + \gamma}.
\end{equation}
Multiplying in another $\cY_{\gamma'}$ gives
\begin{equation}
\sigma(\gamma') \cY_{\gamma_{ij,S}+\gamma} \cY_{\gamma_{ji,S}+\gamma'} = - \sigma(\gamma) \cY_{\gamma'} \cY_{\gamma_{ii,0} + \gamma},
\end{equation}
i.e.
\begin{equation}
\cY_{\gamma_{ij,S}+\gamma} \cY_{\gamma_{ji,S}+\gamma'} = - \sigma(\gamma) \sigma(\gamma') (-1)^{\inprod{\gamma,\gamma'}} \cY_{\gamma_{ii,0} + \gamma + \gamma'},
\end{equation}
which is
\begin{equation} \label{eq:sigma-3}
\cY_{\gamma_{ij}} \cY_{\gamma_{ji}} = - \sigma(\gamma_{ij}+\gamma_{ji}) \cY_{\gamma_{ij}+\gamma_{ji}}.
\end{equation}

Finally, multiplying \eqref{yii} by $\cY_{\gamma'}$ gives
\begin{equation}
\cY_{\gamma'} \cY_{\gamma_{ii,0}+\gamma} = \cY_{\gamma'} \cY_\gamma \cY_{\gamma_{ii,0}},
\end{equation}
i.e.
\begin{equation}
\cY_{\gamma'} \cY_{\gamma_{ii,0}+\gamma} = (-1)^{\inprod{\gamma,\gamma'}} \cY_{\gamma'+\gamma+\gamma_{ii,0}},
\end{equation}
i.e. (writing $\gamma_{ii} = \gamma_{ii,0} + \gamma$ and then relabeling $\gamma' \to \gamma$)
\begin{equation} \label{eq:sigma-4}
\cY_{\gamma} \cY_{\gamma_{ii}} = (-1)^{\inprod{\gamma_{ii},\gamma}} \cY_{\gamma+\gamma_{ii}}
\end{equation}
where in writing $\inprod{\gamma_{ii},\gamma}$ we use the isomorphism $\Gamma_{ii} \simeq \Gamma$.

The equtaions \eqref{eq:sigma-1}, \eqref{eq:sigma-2}, \eqref{eq:sigma-3}, \eqref{eq:sigma-4} give the definition of
the $\sigma$ appearing in the multiplicative law for $\CY$:
\begin{align}
 \sigma(\gamma, \gamma') &= (-1)^{\inprod{\gamma,\gamma'}}, \\
 \sigma(\gamma, \gamma_{ii}) &= (-1)^{\inprod{\gamma, \gamma_{ii}}}, \\
 \sigma(\gamma, \gamma_{ij}) &= \sigma(\gamma), \\
 \sigma(\gamma_{ij}, \gamma_{ji}) &= - \sigma(\gamma_{ij} + \gamma_{ji}).
\end{align}

\subsection{Morphism:  crossing a separating WKB curve}

Now suppose we displace from $z$ in sector $S$ clockwise to $z'$ in sector $S'$, while remaining in the same triangle.
In what follows we drop the subscript $S$, and for quantities which should have subscript $S'$ we just put a prime on top.
We are free to choose $i$ to be the sheet with $\nu_{i} = 1$, i.e. the WKB curves passing through $z_i$
go around counterclockwise.  Then $\nu'_{i} = -1$.

Number the vertices of the triangle so that the WKB curve through $z_i$ runs from vertex $1$ to $2$.
We have
\begin{equation}
\cY_{\gamma_{ij}} (s_i) = 0, \quad \cY_{\gamma_{ij}} (s_j) = s_i
\end{equation}
i.e.
\begin{equation}
\cY_{\gamma_{ij}} (s_2) = 0, \quad \cY_{\gamma_{ij}} (s_1) = s_2 \frac{31}{23},
\end{equation}
and also
\begin{equation}
\cY_{\gamma_{ji}} (s_j) = 0, \quad \cY_{\gamma_{ji}} (s_i) = -s_j
\end{equation}
i.e.
\begin{equation}
\cY_{\gamma_{ji}} (s_1) = 0, \quad \cY_{\gamma_{ji}} (s_2) = -s_1 \frac{23}{31}.
\end{equation}

Then the WKB curve through $z'_i$ runs from $1$ to $3$, and we have
\begin{equation}
\cY'_{\gamma_{ij}} (s'_i) = 0, \quad \cY'_{\gamma_{ij}} (s'_j) = -s'_i,
\end{equation}
i.e.
\begin{equation}
\cY'_{\gamma_{ij}} (s_3) = 0, \quad \cY'_{\gamma_{ij}} (s_1) = -s_3 \frac{12}{23}.
\end{equation}
Using the identity
\begin{equation} \label{id}
(12) s_3 + (23) s_1 + (31) s_2 = 0
\end{equation}
this becomes
\begin{equation}
\cY'_{\gamma_{ij}} ((23)s_1 + (31)s_2) = 0, \quad \cY'_{\gamma_{ij}} (s_1) = \left( \frac{23}{12} s_1 + \frac{31}{12} s_2 \right) \frac{12}{23},
\end{equation}
i.e.
\begin{equation}
\cY'_{\gamma_{ij}} ((23)s_1 + (31)s_2) = 0, \quad \cY'_{\gamma_{ij}} (s_1) = s_1 + \frac{31}{23} s_2,
\end{equation}
i.e.
\begin{equation}
\cY'_{\gamma_{ij}} (s_2) = -\frac{23}{31} s_1 - s_2, \quad \cY'_{\gamma_{ij}} (s_1) = s_1 + \frac{31}{23} s_2,
\end{equation}
which means
\begin{equation}
\cY'_{\gamma_{ij}} = \cY_{\gamma_{jj}} - \cY_{\gamma_{ii}} + \cY_{\gamma_{ij}} + \cY_{\gamma_{ji}}
\end{equation}
or finally,
\begin{equation}
\cY'_{\gamma_{ij}} = (1 - \cY_{\gamma_{ji}}) \cY_{\gamma_{ij}} (1 + \cY_{\gamma_{ji}}).
\end{equation}
We also have
\begin{equation}
\cY'_{\gamma_{ji}} (s'_j) = 0, \quad \cY'_{\gamma_{ji}} (s'_i) = s'_j,
\end{equation}
i.e.
\begin{equation}
\cY'_{\gamma_{ji}} (s_1) = 0, \quad \cY'_{\gamma_{ji}} (s_3) = s_1 \frac{23}{12}.
\end{equation}
Using \eqref{id} again this is
\begin{equation}
\cY'_{\gamma_{ji}} (s_1) = 0, \quad \cY'_{\gamma_{ji}} (s_2) = -s_1 \frac{23}{31},
\end{equation}
i.e.
\begin{equation}
\cY'_{\gamma_{ji}} = \cY_{\gamma_{ji}}
\end{equation}
which is also
\begin{equation}
\cY'_{\gamma_{ji}} = (1 - \cY_{\gamma_{ji}}) \cY_{\gamma_{ji}} (1 + \cY_{\gamma_{ji}}).
\end{equation}
So in summary we have the expected transformation law
\begin{equation}
 \cY' = \cS^{\mu=1}_{\gamma_{ji}} \cY.
\end{equation}
Note $\gamma_{ji}$ is the path from $z_j$ to $z_i$ through the turning point --- in particular, when $\zeta$ sits exactly on the BPS ray
$\ell_{\gamma_{ji}}$
$\gamma_{ji}$ can be represented by a lifted WKB path, with the \ti{opposite} of the WKB orientation
as expected (since $Z_{\gamma_{ji}} / \zeta \in \R_-$.)

\subsection{Morphism:  crossing an edge}

We can also displace $z$ across an edge $E$ to $z'$ in a neighboring triangle.  We keep the labeling as above,
so $E$ is the $12$ edge, and let the other vertex of the neighboring triangle be $4$ (so $142$ go
around counterclockwise).  Then going around the quadrilateral $Q_E$ counterclockwise we have $1423$, i.e.
\begin{equation}
 \cY_{\gamma_E} = \frac{14\ 23}{42\ 31}.
\end{equation}
We also have
\begin{equation}
s'_i = s_i \frac{14}{31}, \quad s'_j = s_j \frac{42}{23}, \quad \nu' = -\nu
\end{equation}
which says
\begin{equation}
\cY'_{\gamma'_{ij}} = - \cY_{\gamma_E} \cY_{\gamma_{ij}} = \cY_{\gamma_E + \gamma_{ij}}, \qquad \cY'_{\gamma'_{ji}} = - \cY_{\gamma_E}^{-1} \cY_{\gamma_{ji}} = \cY_{-\gamma_E+\gamma_{ji}}.
\end{equation}
But indeed $\gamma'_{ij} = \gamma_{ij} + \gamma_E$ and $\gamma'_{ji} = \gamma_{ji} - \gamma_E$ so this just says
\begin{equation}
 \cY' = \cY
\end{equation}
as it should since crossing an edge does not correspond to any BPS ray.

\subsection{Morphism:  flips and juggles}

If the triangulation undergoes a flip or a juggle somewhere far away from $z$, then $\cY_{\gamma_{ij,S}}$ is unaffected.
Together with the multiplication laws and the known action of flips and juggles on $\cY_\gamma$ from \cite{Gaiotto:2009hg},
we believe this is enough to show that the action of flips and juggles on $\cY$
is by $\cK_{\gamma}^\omega$ as it should be.

\section{A review: hidden flavor symmetries in 3d Coulomb branches}\label{app:Flavor-Twisted}

Consider some three-dimensional field theory which admits an effective low energy description
as a non-linear sigma model on a smooth moduli space of vacua $\CM$.
(In this Appendix, $\CM$ is a general target space, not necessarily
the space \eqref{eq:M-intro}.)
Flavor symmetries of the UV theory which are spontaneously broken by the vacuum are realized as isometries of $\CM$.
Flavor symmetries of the UV theory which are unbroken at all points of
$\CM$ might still manifest themselves in the IR description, but in a
more subtle way.

Let $\varphi: \IR^{1,2} \to \CM$ and consider
a closed 2-form $b$ on $\CM$.
The pullback of $b$ is a closed 2-form in the 3d spacetime, hence
at least classically a conserved current:
\begin{equation}
J = \varphi^* b.
\end{equation}
Notice that two closed 2-forms which differ by an exact form
give rise to currents which differ by an ``improvement'' term, and measure the same conserved charges.
Thus, the sigma model has an abelian group of global symmetries whose Lie algebra may be
identified with the de Rham cohomology $H^2_{DR}(\CM)$. The corresponding
conserved charges cannot be changed by quantum effects in the IR sigma model, though they
very well might not exist in the full UV theory.

Now note that the Hilbert space can be graded by
$\pi_2(\CM)$ because a map  $\varphi: \IR^2 \to \CM$ at a fixed time
maps the boundary at infinity in $\IR^2$ to the vacuum, i.e. a point in $\CM$.
Elementary excitations are in the trivial homotopy class and
solitonic excitations are in nontrivial homotopy classes.
Suppose $\exp: H^2_{DR}(\CM) \to \CF$ is the exponential map of the
flavor group $\CF$.
If $f = \exp(b)$ is in the connected component of the identity
then the action of $f$ on wavefunctions in a sector $[\varphi] \in \pi_2(\CM)$  is
\begin{equation}\label{eq:charge-act}
f\cdot \Psi = e^{2 \pi \I \int \varphi^* b }\Psi
\end{equation}
The charge thus defines a character in ${\rm Hom}(\pi_2(\CM), U(1))$, which
actually only depends on the image under the
Hurewicz map, i.e. it descends to ${\rm Hom}(H_2(\CM,\IZ), U(1))$.

Now, $ {\rm Hom}(H_2(\CM,\IZ), U(1)) \simeq H^2(\CM, U(1))$ so it is  natural
to suspect that $\CF \simeq H^2(\CM,U(1))$ is the correct expression for the full flavor
group.  $H^2(\CM,U(1))$  is a compact abelian group whose Lie algebra
may be identified with $H^2_{DR}(\CM)$.
The connected component of the
identity of this group can be identified with a torus $\CT = H^2_{DR}(\CM)/H_2(\CM,\IZ)$,
in harmony with \eqref{eq:charge-act}.  However,
$\CF$ might well have a nontrivial discrete group $\CD$ of components,
leading to discrete flavor symmetries not continuously connected to the identity and
equation \eqref{eq:charge-act} generalizes naturally to this case.
In general, $\CF$ is a semidirect product of the connected torus $\CT$ with a discrete group $\CD$.
(In fact $\CD$ is isomorphic to the torsion subgroup of $H^3(\CM, \IZ)$.)
If $\CM$ also has an isometry group $G$, it will act
on the space of closed 2-forms and more generally on $\CF$;
hence the full flavor symmetry group of the IR theory will
be a semidirect product of $G$ and $\CF$.

It is an interesting problem to relate the flavor symmetry group of
the UV theory and of the IR sigma-model description.
A sufficient amount of supersymmetry, say $\CN=2$ in three dimensions ($4$ supercharges), makes the task easier.
To every $U(1)$ subgroup of the UV flavor group, one has a real mass parameter
$m$ in the UV theory, which would be the vev of
the scalar superpartner of a background gauge field gauging the $U(1)$ flavor subgroup.
In the IR, the parameter $m$ for a flavor symmetry which is unbroken everywhere on the Coulomb branch
only affects the metric of the sigma model, and hence enters in a very specific way in the IR Lagrangian.
Supersymmetrization of this coupling gives a coupling to the background gauge field,
i.e. to the IR conserved current for the $U(1)$ flavor symmetry.
In superspace, the kinetic term arises from integrating the K\"ahler potential
\begin{equation}
\int \de^4 \theta\,\CK(\Phi, \bar \Phi, M).
\end{equation}
Here $M$ is a linear multiplet whose lowest component is the real mass $m$.
(Thus $M$ is a sum $\Psi + \bar \Psi$ of chiral multiplets, $m$ is the real part of the
complex scalar, and the imaginary part is dualized to a flavor gauge field with fieldstrength $F_{fl}$.)
Expanding, we find a coupling
\begin{equation}
\int (\partial - \bar \partial) \partial_m \CK \wedge F_{fl}.
\end{equation}
As the K\"ahler potential and its first derivatives are not globally defined, we should integrate by parts to an expression involving the K\"ahler form:
\begin{equation}
\int  \partial_m \omega \wedge A_{fl}.
\end{equation}
Hence the relevant 2-form $b$ on $\CM$ is simply the variation of the K\"ahler form under a change
in the real mass parameter.

We can extend the above statements, valid for $U(1)$ subgroups of the flavor group,
to discrete flavor symmetry groups which
manifest themselves in the IR in non-trivial components of $\CF$.   If $f \in \CF$ then we can
modify the path  integral on $\IR^{1,1}\times S^1$
by using $f$-twisted boundary conditions  around $S^1$ -- that is, we can
insert a ``flavor Wilson line'' corresponding to $f$.
If the circle is sufficiently large that we can use the effective
three-dimensional IR theory, but sufficiently small that we can express
the partition function as that of an effective two-dimensional theory,
then this twisting is equivalent to the insertion of a $B$-field
amplitude in the two-dimensional sigma model. For example, if
$f=\exp(b)$ is an element of the connected component of
the identity in $\CF$ then the twisted partition function
has an insertion of an operator
\begin{equation} \exp 2 \pi \I \int_{2d} \varphi^* b \end{equation}
which survives the limit of taking a small circle.
Hence the flavor Wilson line induces a $B$-field $b$ in the
effective 2d sigma model. This makes perfect sense for a
$U(1)$ flavor symmetry:  the 3d K\"ahler parameter $m$ becomes a
complexified 2d K\"ahler parameter built out of $m$ and $b$.
For a general $f\in \CF$, we get a coupling for the corresponding flat $B$-field in $H^2(\CM, U(1))=\CF$.

Finally, it is useful to consider line defects in the 3d theory which introduce a monodromy given by a flavor group element $f$ in $\CF$.
Such defects would appear naturally if we were to gauge a (possibly discrete) subgroup of $\CF$ containing $f$.
Now, a line defect stretching along $\ell \subset \IR^{1,2}$ can be viewed as the
boundary of a domain wall $S$, i.e. $\p S = \ell$. The fields across the domain wall are
related by a flavor group transformation $f$. If $f = \exp b  $ is in the connected
component of the identity then the domain wall carries a coupling
\begin{equation}\label{eq:B-amp}
 \exp 2 \pi \I \int_{S} \varphi^* b.  \end{equation}
More generally, it has a coupling pairing  the class $\varphi^*(f)$    with $S$.
Because $S$ has a boundary, this pairing does not define a complex number,
but rather a section of a line bundle on the loop space of spacetime,
as in the theory of D-branes in the presence of a $B$-field.
In order to define a good operator in the sigma model we need the line operator at the boundary of $S$ to
be a section of the dual line bundle on loop space.

As in the theory of D-branes in the presence of a topologically
nontrivial $B$-field there  is a natural class of such line operators
related to ``connections on twisted bundles.'' See \cite{Kapustin:1999di}
for a concrete discussion. The flat $B$-field $f \in H^2(\CM, U(1))$
determines a (torsion) twisting class in ${\mathrm{Tors}}(H^3(\CM, \IZ))$.
The twisted class can be viewed as an 't Hooft flux, thus determining --- for example ---
an  $SU(N) / \IZ_N$ bundle which does not lift to an $SU(N)$ bundle.
A connection on the $SU(N) / \IZ_N$ bundle has well-defined holonomy
in the adjoint representation, but the holonomy in representations of
$SU(N)$ transforming nontrivially under the center must be regarded as
sections of line bundles over the loop space of $\CM$. For an appropriately
twisted vector bundle and connection, then, ${\rm Hol}(A, \ell)$ -- the trace of the
holonomy in the fundamental representation --  will
live in the dual line bundle (over loop space) to that where \eqref{eq:B-amp} is valued,
and hence the product
\begin{equation}
 {\rm Hol}(A, \ell) \cdot \exp 2 \pi \I \int_{S} \varphi^* b
\end{equation}
will be a well-defined function. In the present context this gives a well-defined
line operator.

\bibliography{D2D4_Paper}

\end{document}